\documentclass{aa}  

\usepackage{graphicx}
\usepackage{txfonts}
\usepackage{amssymb}
\usepackage{amsmath}
\usepackage[export]{adjustbox}
\usepackage{float}
\usepackage{lscape}
\begin{document} 

%
\newcommand{\herschel}{\emph{Herschel}}
\newcommand{\spitzer}{\emph{Spitzer}}
\newcommand{\planck}{\emph{Planck}}
\newcommand{\akari}{\emph{Akari}}
\newcommand{\swift}{\emph{Swift}}
\newcommand{\iras}{IRAS}
\newcommand{\iso}{ISO}
\newcommand{\alma}{ALMA}
\newcommand{\galex}{GALEX}
\newcommand{\wise}{WISE}
\newcommand{\twomass}{2MASS}
\newcommand{\lsun}{\mbox{L$_\odot$}}     
\newcommand{\msun}{\mbox{M$_\odot$}}     
\newcommand{\msunyr}{\mbox{M$_\odot$~yr$^{-1}$}}
\newcommand{\rsun}{\mbox{R$_\odot$}}     
\newcommand{\lbat}{$L_{\rm 14-195keV}$}  
\newcommand{\lbol}{$L_{\rm bol}$}        
\newcommand{\lagn}{$L_{\rm AGN}$}        
\newcommand{\lbolagn}{$L_{\rm bol,AGN}$} 
\newcommand{\ltorus}{$L_{\rm Torus}$}    
\newcommand{\lfir}{$L_{\rm FIR}$}        
\newcommand{\lir}{$L_{\rm IR}$}          
\newcommand{\reblue}{$R_{\rm e,70}$}     
\newcommand{\regreen}{$R_{\rm e,100}$}   
\newcommand{\rered}{$R_{\rm e,160}$}     
\newcommand{\sigfir}{$\Sigma_{\rm FIR}$} 
\newcommand{\sigsfr}{$\Sigma_{\rm SFR}$} 
\newcommand{\deltare}{$\Delta_{\rm Re}$} 
\newcommand{\tdust}{$T_{\rm dust}$}      
\newcommand{\sco}{$S_{\rm CO}$}          
\newcommand{\alpco}{$\alpha_{\rm CO(1-0)}$} 
\newcommand{\mstellar}{$M_{\ast}$}       
\newcommand{\vout}{$v_{\rm out}$}        
\newcommand{\vohabs}{$v_{84}{\rm (abs)}$}
\newcommand{\rout}{$R_{\rm out}$}        
\newcommand{\mout}{$M_{\rm out}$}        
\newcommand{\mdotout}{$\dot{M}_{\rm out}$}     
\newcommand{\tflow}{$t_{\rm flow}$}      
\newcommand{\kms}{km~s$^{-1}$}           
\newcommand{\jykms}{Jy~km~s$^{-1}$}      
\newcommand{\ergs}{erg~s$^{-1}$}         
\newcommand{\ergscm}{erg~s$^{-1}$~cm$^{-2}$} 
\newcommand{\mum}{\mbox{$\mu$m}}         
\newcommand{\cii}{[C{\sc ii}]}
\newcommand{\oi}{[O{\sc i}]}

\title{Molecular outflows in local galaxies: Method comparison and a role 
  of intermittent AGN driving}

   \subtitle{}

   \author{D.~Lutz\inst{1}
          \and E.~Sturm\inst{1}
          \and A.~Janssen\inst{1,2}
          \and S.~Veilleux\inst{3,4,5}
          \and S.~Aalto\inst{6}
          \and C.~Cicone\inst{7,8}
          \and A.~Contursi\inst{1,9}
          \and R.I.~Davies\inst{1}
          \and C.~Feruglio\inst{10}
          \and J.~Fischer\inst{11}
          \and A.~Fluetsch\inst{12,13}
          \and S.~Garcia-Burillo\inst{14}
          \and R.~Genzel\inst{1}
          \and E.~Gonz\'alez-Alfonso\inst{15}
          \and J.~Graci\'a-Carpio\inst{1}
          \and R.~Herrera-Camus\inst{1,16}
          \and R.~Maiolino\inst{12,13}
          \and A.~Schruba\inst{1}
          \and T.~Shimizu\inst{1}
          \and A.~Sternberg\inst{17}
          \and L.J.~Tacconi\inst{1}
          \and A.~Wei\ss\/\inst{18}
          }

   \institute{Max-Planck-Institut f\"ur extraterrestrische Physik,
              Giessenbachstrasse 1, 85748 Garching, Germany\\
              \email{lutz@mpe.mpg.de}
         \and NOVA-ASTRON, Oude Hoogeveensedijk, 7991 PD Dwingeloo,
              The Netherlands
         \and Department of Astronomy and Joint Space-Science Institute, 
              University of Maryland, College Park, MD 20742 USA
         \and Institute of Astronomy and Kavli Institute for Cosmology 
              Cambridge, University of Cambridge, Cambridge CB3 0HA, UK
         \and Space Telescope Science Institute, Baltimore, MD 21218, USA
         \and Department of Earth and Space Sciences, Chalmers University
              of Technology, Onsala Observatory, 439 94 Onsala, Sweden
         \and INAF -- Osservatorio Astronomico di Brera, via Brera 28, 
              I-20121 Milano, Italy
         \and Institute of Theoretical Astrophysics, University of Oslo,
              P.O. Box 1029, Blindern, 0315 Oslo, Norway
         \and IRAM, 300 Rue de la Piscine, F-38406 Saint Martin D'H\`eres, 
              Grenoble, France
         \and INAF -- Osservatorio Astronomico di Trieste, via G.B. Tiepolo 11,
              I-34143 Trieste, Italy
         \and George Mason University, Department of Physics \& Astronomy,
              MS 3F3, 4400 University Drive, Fairfax, VA 22030, USA
         \and Cavendish Laboratory, University of Cambridge, 
              19 J.J.~Thomson Ave., Cambridge CB3 0HE, UK
         \and Kavli Institute for Cosmology, University of Cambridge, 
              Madingley Road, Cambridge CB3 0HA, UK
         \and Observatorio Astron\'omico Nacional (OAN-IGN) -- Observatorio de
              Madrid, Alfonso XII, 28014 Madrid, Spain
         \and Universidad de Alcal\'a, Departamento de F\'isica y 
              Matem\'aticas, Campus Universitario, E-28871 Alcal\'a de Henares,
              Madrid, Spain
         \and Departamento de Astronom\'ia, Universidad de Concepci\'on,
              Avenida Esteban Iturra s/n, Concepc\'ion, Chile
         \and School of Physics \& Astronomy,
              Tel Aviv University, Ramat Aviv 69978, Israel
         \and Max-Planck-Institut f\"ur Radioastronomie, Auf dem H\"ugel 69,
              53121 Bonn, Germany
             }

   \date{Received 28 September 2019; accepted 7 November 2019}

 
  \abstract{
  We report new detections and limits from a NOEMA and ALMA CO(1-0) search 
  for molecular outflows in 13 local galaxies with high 
  far-infrared surface brightness, and combine these with local universe
  CO outflow results from the literature. CO line ratios and spatial outflow 
  structure of our targets provide some constraints on the conversion steps 
  from observables to physical quantities such as molecular mass outflow 
  rates. Where available, ratios between 
  outflow emission in higher J CO transitions and in CO(1-0) typically are 
  consistent with excitation $R_{\rm i1}\lesssim 1$. For IRAS~13120-5453, 
  however, $R_{31}=2.10\pm 0.29$ 
  indicates optically thin CO in the outflow. Like much of the outflow
  literature, we use \alpco = 0.8, and we present arguments for using C=1 in 
  deriving molecular mass outflow rates 
  $\dot{M}_{\rm out} = C \frac{M_{\rm out}  v_{\rm out}}{R_{\rm out}}$. 
  We compare the two
  main methods for molecular outflow detection: CO mm interferometry
  and \herschel\ OH-based spectroscopic outflow searches. For 26 sources
  studied with both methods,
  we find an 80\% agreement in detecting \vout\/~$\gtrsim$~150~\kms\ 
  outflows, and 
  non-matches can be plausibly ascribed to outflow geometry and 
  signal-to-noise ratio.
  For the \citet{gonzalez-alfonso17} sample of 12 bright ultraluminous 
  infrared galaxies (ULIRGs) with
  detailed OH-based outflow modeling, CO outflows are detected in all but 
  one. Outflow masses, velocities, and sizes for 
  these 11 sources agree well between the two methods, and modest 
  remaining differences may relate to the different but overlapping regions 
  sampled by CO emission and OH absorption. 
  Outflow properties correlate better with active galactic nucleus (AGN) 
  luminosity and
  with bolometric luminosity than with far-infrared surface brightness.
  The most massive outflows are found for systems with
  current AGN activity, but significant outflows in non-AGN systems must 
  relate to star formation or to AGN activity in the recent past.
  We report scaling relations for the increase of outflow mass, rate, 
  momentum rate, and kinetic power with bolometric luminosity.  
  Short $\sim 10^6$~yr flow times and some sources with resolved 
  multiple outflow episodes support a role of intermittent driving, 
  likely by AGN.

  }

     \keywords{galaxies: active,
             galaxies: ISM,
             galaxies: kinematics and dynamics
               }

   \maketitle
%

\section{Introduction}

\begin{table*}
\caption{NOEMA and ALMA CO(1-0) observations}
\begin{tabular}{lrrrcccc}\hline
Source&cz$_{\rm h}$& RA  &  DEC&Date(s)&Min, Max    &RMS (20km/s)&Beam\\
      &km/s        &J2000&J2000&       &baseline [m]&mJy/beam    &arcsec\\ 
 (1)  & (2) & (3) &  (4)  &   (5)   &   (6)      &  (7)       &(8) \\\hline
III Zw 035       & 8221&01:44:30.71& 17:06:07.6&07/2016 - 02/2018&18.4, 760.0&0.30&1.42$\times$0.94\\
NGC 2623         & 5524&08:38:24.08& 25:45:16.7&05/2017 - 09/2017&15.0, 174.1&0.75&4.03$\times$3.33\\
IC 860           & 3889&13:15:03.53& 24:37:07.9&08/2016 - 11/2016&15.0, 303.4&0.92&2.81$\times$1.79\\
NGC 7479         & 2387&23:04:56.65& 12:19:22.3&05/2016 - 06/2016&15.0, 262.0&1.15&2.98$\times$2.27\\ \hline
IRAS F05189-2524 &12843&05:21:01.47&-25:21:45.4&09/2015          &15.1, 1600 &0.37&0.52$\times$0.47\\
IRAS F09111-1007W&16276&09:13:36.38&-10:19:28.2&12/2016          &15.1, 460.0&0.33&1.84$\times$1.47\\
IRAS F10173+0828 &14687&10:20:00.26& 08:13:33.6&12/2016          &15.1, 460.0&0.34&1.84$\times$1.68\\
IRAS F12224-0624 & 7938&12:25:03.91&-06:40:52.0&12/2016          &15.1, 491.0&0.37&1.76$\times$1.53\\
NGC 4418         & 2099&12:26:54.65&-00:52:38.4&01/2017          &15.1, 460.0&0.58&1.62$\times$1.19\\ 
IRAS 13120-5453  & 9308&13:15:06.20&-55:09:24.0&07/2015 - 08/2016&15.1, 1800 &0.65&0.54$\times$0.52\\
IRAS F14378-3651 &20417&14:40:59.01&-37:04:32.0&07/2015          &43.3, 1600 &0.67&0.57$\times$0.51\\ 
IRAS F17207-0014 &12825&17:23:21.98&-00:17:01.7&01/2017          &15.1, 452.8&0.46&2.52$\times$1.87\\ 
IRAS F20551-4250 &12891&20:58:26.79&-42:39:00.3&06/2015          &43.3, 1600 &0.64&0.57$\times$0.45\\\hline
\end{tabular}
\tablefoot{Observations obtained with NOEMA (top) and ALMA (bottom).\\
(1) Source name\\
(2) Heliocentric radial velocity based on our CO(1-0) data\\
(3), (4) Coordinates of adopted phase center. See Table~\ref{tab:continuum} for the measured 
position of mm continuum.\\
(5) Observation period\\
(6) Interferometer baselines used\\
(7) RMS sensitivity for a 20~km/s channel width\\
(8) Beam size}
\label{tab:observations}
\end{table*}

\begin{table*}
\caption{Continuum properties}
\begin{tabular}{lrrccccr}\hline
Source&  RA  &  DEC  &E(RA)  &E(DEC)&FWHM  &$\nu_{\rm Obs}$&S$_{\rm \nu ,Obs}$\\ 
                 &J2000 &J2000  &arcsec &arcsec&arcsec&GHz            &mJy\\
(1)&(2)&(3)&(4)&(5)&(6)&(7)&(8)\\\hline
III Zw 035       &01:44:30.538& 17:06:08.68&0.01&0.01&0.43$\pm$0.01&112.2& 3.84$\pm$0.03\\
IRAS F05189-2524 &05:21:01.401&-25:21:45.28&0.01&0.01&0.17$\pm$0.03&104.1& 2.40$\pm$0.03\\
NGC 2623         &08:38:24.087& 25:45:16.67&0.01&0.02&1.22$\pm$0.06&113.2& 6.40$\pm$0.08\\
IRAS F09111-1007W&09:13:36.452&-10:19:30.00&0.01&0.01&0.76$\pm$0.11&102.3& 1.33$\pm$0.03\\
IRAS F09111-1007E&09:13:38.829&-10:19:19.98&0.01&0.01&$<$3.12      &102.3& 0.62$\pm$0.12\\
IRAS F10173+0828 &10:20:00.203& 08:13:33.93&0.01&0.01&$<$0.75      &102.9& 1.72$\pm$0.03\\
IRAS F12224-0624 &12:25:03.909&-06:40:52.63&0.01&0.01&$<$0.48      &105.8& 2.33$\pm$0.33\\
NGC 4418         &12:26:54.610&-00:52:39.45&0.01&0.01&0.17$\pm$0.05&107.7&11.13$\pm$0.04\\
IC 860           &13:15:03.505& 24:37:07.70&0.01&0.01&$<$0.54      &113.8& 6.64$\pm$0.08\\ 
IRAS 13120-5453  &13:15:06.328&-55:09:22.80&0.01&0.01&0.89$\pm$0.03&105.3&10.07$\pm$0.16\\
IRAS F14378-3651 &14:40:59.013&-37:04:31.91&0.01&0.01&0.37$\pm$0.05&101.4& 1.47$\pm$0.06\\
IRAS F17207-0014 &17:23:21.959&-00:17:00.91&0.01&0.01&0.66$\pm$0.06&104.1& 8.70$\pm$0.06\\
IRAS F20551-4250 &20:58:26.792&-42:39:00.29&0.01&0.01&0.25$\pm$0.03&104.1& 2.78$\pm$0.05\\
NGC 7469         &23:04:56.634& 12:19:22.67&0.04&0.04&0.89$\pm$0.15&114.4& 1.77$\pm$0.09\\
\hline
\end{tabular}
\tablefoot{Results are from UV fitting a Gaussian model to the NOEMA 
continuum visibilities, and from fitting a Gaussian model to the ALMA 
continuum image. IRAS F09111-1007E is not part of our 
sample but detected along with IRAS F09111-1007W.\\
(1) Source name\\
(2), (3) Continuum position\\
(4), (5) Uncertainties of position. These include only the formal fit errors, 
  to which 2--5\% of the beam size should be added to account for phase 
  calibration stability \citep{almatechhandbook18}\\
(6) Intrinsic source full width at half maximum FWHM (beam-deconvolved)\\
(7) Observed Frequency\\
(9) Total continuum flux density from Gaussian fit.
} 
\label{tab:continuum}
\end{table*}

Gas outflows triggered by intense star formation and/or active galactic nuclei
are important agents in the evolution of galaxies, in regulating the fraction
of baryons inside a dark matter halo that are converted to stars, and
for  the metal enrichment of the circumgalactic and intergalactic medium 
\citep[see, e.g., reviews by][]{veilleux05,fabian12}. A significant step
forward in characterizing the {\em  molecular} phase of outflows from
local galaxies has been made recently via detection of P-Cygni profiles of 
far-infrared OH lines 
\citep[e.g.,][]{fischer10,sturm11,veilleux13,spoon13,gonzalez-alfonso17} 
and detection of broad line wings to the mm emission lines of CO  
\citep[e.g.,][and references therein]{feruglio10,alatalo11,cicone14,pereira-santaella18,fluetsch19}. 
Molecular outflows have been detected preferentially 
but not exclusively in dusty (ultra)luminous infrared galaxies ((U)LIRGs). 
Outflow rates \mdotout\ up to several 100~\msunyr\ and mass loading factors 
\mdotout\//SFR around and significantly above 1 have been observed. In
extreme cases and if velocities are high enough to escape the galaxy, 
extrapolation of the current outflow rate would clear out 
most of a galaxy's  gas content within $10^7$~yr. While both star 
formation and active galactic nuclei (AGN) may contribute to driving 
outflows, the OH outflow velocity in ULIRGs as well as the 
CO-based outflow rates seem to be best correlated with
AGN luminosity \citep{veilleux13,cicone14,fiore17,fluetsch19}. 
Ionized, atomic, and 
molecular gas phases may all significantly contribute to the outflow rates 
\citep{contursi13,rupke13,janssen16,rupke17}, in this paper we focus on 
the usually dominant molecular phase. There is a continuum of 
properties from the fast and energetic outflows reported in the studies 
mentioned above, towards smaller scale and typically slower and less energetic
molecular outflows. These are increasingly reported from high spatial and 
spectral resolution ALMA and NOEMA interferometric studies of nearby galaxies
\citep[e.g.,][]{combes14,querejeta16,salak16,slater19,ramakrishnan19,alonsoherrero19,combes19}, 
along with indications for inflow, warps, and for non-circular motions in 
complex, for example barred, potentials. 
Here, we focus on fast and energetic outflows with potential non-local impact.

\begin{table*}
\caption{Measured CO(1-0) outflow parameters}
\begin{tabular}{lrrrrrr}\hline
Source&\sco\ (gauss)&\sco (wings)&$\Delta v$&$FW_{10\% }$&\vout&\rout  \\ 
                 &\jykms &\jykms &\kms &\kms &\kms &\arcsec \\ \hline
III Zw 035       &  41.05&  14.25&   23&  579&  313& 4.00\\
IRAS F05189-2524 &   2.85&   1.11&   60&  771&  446& 1.24\\
NGC 2623         &   3.21&   0.92&  -33& 1108&  587& 1.62\\
NGC 2623 fossil  &       &   0.37&     &     &  400&14.00\\
IRAS F09111-1007W&   3.34&   0.95&   -9&  644&  331& 0.58\\
IRAS F10173+0828 &       &$<$0.22&     &     &     &     \\
IRAS F12224-0624 &   0.84&   0.35&    2& 1128&  566& 0.50\\
NGC 4418         &   2.53&   0.43&    9&  630&  324& 0.23\\
IC 860           &       &$<$0.58&     &     &     &     \\
IRAS 13120-5453  &  34.50&   4.29&    1&  588&  301& 0.63\\
IRAS F14378-3651 &       &$<$0.43&     &     &     &     \\
IRAS F17207-0014 &   0.84&   0.49& -514& 1203& 1116& 0.45\\
IRAS F20551-4250 &   0.98&   0.40&   96&  787&  490& 0.27\\
NGC 7479         &       &$<$0.73&     &     &     &     \\ \hline
\hline
\end{tabular}
\tablefoot{See Sect.~\ref{sect:outprop} for definitions and methods to 
measure these quantities.} 
\label{tab:outobs}
\end{table*}

Thanks to efficient \herschel\ spectroscopy, the best statistics
up to now is available from searches for OH P-Cygni absorptions. For those,
the step from measuring outflow kinematics \citep[e.g.,][]{veilleux13,stone16} 
to deriving outflow masses and outflow rates involves complex modeling
\citep[e.g.,][]{sturm11,gonzalez-alfonso17,stone18}. For the CO-based searches,
the obvious and highly desirable next steps are to extend the local 
galaxy samples under study, and to broaden the range of 
galaxy parameters covered. Additional value can come from truly spatially 
resolved studies of very nearby outflow sources. 

Three major sources of 
uncertainty currently affect the derivation of molecular outflow rates from
interferometric CO data: (i) the emission from outflowing gas has to be
separated from that of the host; (ii) a CO conversion factor has to be adopted
to convert from luminosity to gas mass, but may not necessarily be the same
as for typical Galactic molecular clouds; (iii) if only an outflow 
velocity and overall size are known, the conversion from outflow mass to 
outflow rate is uncertain by at least a factor of three due to different 
possible geometrical assumptions 
\citep[see discussions for example in][and Sect.~\ref{sect:outrates}]{cicone14,veilleux17}.
Poorly known inclination of the outflow can add to these geometrical 
uncertainties.

Even if separated reliably from gas orbiting in the host, the outflowing 
gas will face a different fate depending on its velocity and the depth of the
potential. There is a continuum from `fountains' that will 
re-join the host in much less than a Gyr, to gas staying in the 
circumgalactic medium for longer 
periods until being re-accreted, and to gas able to escape the galaxy's dark 
matter halo. In Sect.~\ref{sect:outprop} below we conservatively
identify outflowing gas with line wings only. This emphasizes the second and 
third path, but at typical escape velocities of order 600~\kms\ the fraction 
of gas truly escaping a galaxy's potential is typically small if 
ballistic motion is assumed \citep[e.g.,][]{fluetsch19}. Focus on fast wings 
outside the galaxy's line core also avoids ambiguities between more
modest velocity outflow and other non-circular motions such as inflow or 
flows in the potential of a stellar bar.  

Using \herschel\ 70~\mum\ data, \citet{lutz16} and \citet{lutz18} derive 
sizes of the far-infrared emission region for large samples 
of local galaxies, in order to derive scaling
relations of far-infrared size and surface brightness with salient galaxy 
properties, and establish that AGN hosts preferentially have more compact 
star formation than non-active galaxies. This sample can be used for an 
objective selection of 
compact far-infrared sources \citep[Table~1 in][]{lutz16}, based on
observations in the \herschel\ archive that are not affected by bright 
neighbors. The 18 sources
in that table include several for which CO outflows have been recently 
detected, as one might expect for their high far-infrared surface 
brightness
\sigfir\/~$\gtrsim 10^{11.75}$~\lsun\/~kpc$^{-2}$. 
If star formation powers the far-infrared, this is well beyond thresholds 
in star formation 
rate surface density above which star formation-driven outflows from
galaxy disks tend to be observed in the local universe \citep{heckman02} 
or at high redshift \citep{newman12,davies19}.
Alternatively, if an AGN significantly contributes to the observed 
far-infrared emission, it must be luminous and could drive an 
outflow. The likelihood of hosting outflows, and the sample's proximity 
(z=0.0072 to 0.0676 with median 0.027) make it an obvious 
basis for a search for local outflow laboratories, well suited for detailed
study.

We here report on a NOEMA and ALMA search for \mbox{CO(1-0)} outflows in 
galaxies 
from the \citet{lutz16} sample of compact far-infrared galaxies that lack
previous deep CO data, and discuss the results in conjunction with 
literature results for the remainder of the sample and outside the sample's
definition criteria. 
We adopt an $\Omega_m =0.3$, $\Omega_\Lambda =0.7$ and 
$H_0=70$ km\,s$^{-1}$\,Mpc$^{-1}$ cosmology, redshift-independent distances
from the NASA Extragalactic Database (NED), if available, 
for z$<$0.01 galaxies, a \citet{chabrier03} initial mass function (IMF), 
a conversion for far-infrared luminosity to star formation rate (SFR)
${\rm SFR} = 1.9\times 10^{-10}~L_{\rm FIR=40-120\mum}$ as appropriate for the
\citet{kennicutt98} conversion scaled to Chabrier IMF, and 
\lir\/(8--1000~\mum\/)/\lfir\/(40--120~\mum\/)=1.9.

\begin{figure}
\includegraphics[width=\hsize]{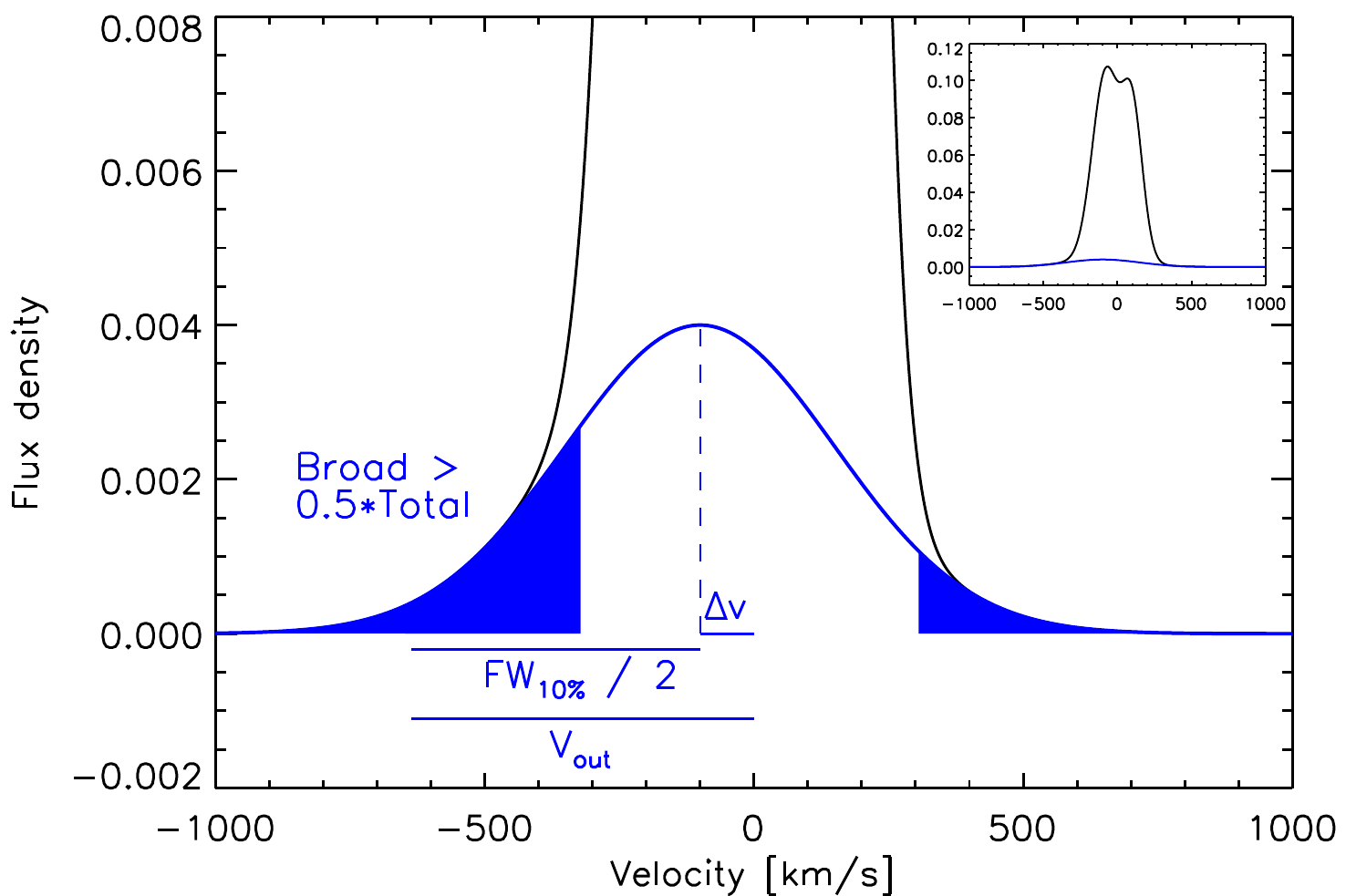}
\caption{Visualization of adopted conventions for outflow velocity
and outflow CO flux. $v_{\rm out}= |\Delta v| + {FW_{10\%}}/{2}$
(Eqn.~\ref{eq:voutdef}) is derived from the shift of the broad Gaussian 
outflow component with respect to systemic velocity and its full width 
at a tenth 
of its peak. \sco (gauss) is the integral of this broad component. 
The flux in the wings of the broad component \sco (wings), indicated by the 
blue shaded area, integrates the broad component only over the velocity ranges 
for which it contributes at least 50\% of the total flux density of 
the line. }
\label{fig:outflowgauss_visu}
\end{figure}

\section{Sample and Observations}

\subsection{NOEMA observations}
Observations for the DEC$>$10\degr\ targets were taken with the IRAM
NOrthern Extended Millimeter Array (NOEMA) on Plateau de Bure between 
May 2016 and February 2018 (Projects s16bh, s17ar, w17cp). 
CO(1-0) was observed in the 3~mm band,
mostly using the WideX correlator (3.6~GHz bandwidth) and 
typically 6 or 7 antennae in configurations between C and D. 
These low 2--4\arcsec\ spatial
resolution configurations were chosen to initially provide maximum 
detection sensitivity, and not over-resolve faint and possibly 
extended high velocity outflow components.
 
Table~\ref{tab:observations} lists the baselines used, achieved RMS 
sensitivity per beam for a 20~\kms\ channel, and beam size. For III~Zw~035, 
a strong 
outflow was detected and already spatially resolved in the low resolution
data. This source was followed up in the higher resolution A configuration
using the new PolyFix correlator and new receivers. The combined data yield 
the smaller beam listed for this source in Table~\ref{tab:observations}.

Calibration and mapping followed standard procedures in the 
GILDAS\footnote{http://www.iram.fr/IRAMFR/GILDAS} environment 
\citep{guilloteau00}.
For CO-bright and compact sources (III~Zw~035, IC~860, NGC~2623), 
self calibration was able
to improve the dynamic range in the channels with bright line emission, this 
is of little impact for the detection of outflow wings and continuum, though. 
Velocities 
quoted below are relative to our CO-based heliocentric redshifts as listed in
Table~\ref{tab:observations}. Millimeter continuum emission was clearly 
detected for all sources. Positional offsets quoted below are relative to the 
continuum positions that are listed along with other continuum properties in 
Table~\ref{tab:continuum}. The continuum was subtracted from the visibilities,
and all line information is based on continuum free visibilities or
spectral cubes. The observations were targeted at CO(1-0), for which we derived
spectral cubes at 20, 60, and 200~\kms\ resolution and occasionally used
in addition data for velocity ranges specifically tailored to the outflow 
properties of a given source. Total fluxes for CO and some other transitions 
detected in the wide spectral bandwidth are listed in 
Table~\ref{tab:applines}. 

\subsection{ALMA observations}

The DEC$<$10\degr\ targets were observed with the Atacama Large 
Millimeter/submillimeter Array (ALMA) Cycle 4 program 2016.1.00177.S in 
late 2016 to early 2017, using between 38 and 46 antennae. Again, the CO(1-0) 
line was observed in a relatively compact configuration yielding an 
$\approx$2\arcsec\ beam in ALMA Band 3, and observational 
parameters are listed in Table~\ref{tab:observations}. 
The pipeline-calibrated data were mapped using standard CASA 
\citep{mcmullin07} procedures.  
Three more sources from Table~1 of 
\citet{lutz16} were already observed in CO(1-0) in ALMA program 
2012.1.00306.S (PI E.~Sturm) using $\sim$0.5\arcsec\ beams. For 
IRAS~13120-5453, we combined our 2012.1.00306.S CO(1-0) data with a 
similar dataset from 2015.1.00287.S (PI K.~Sliwa), in order to improve 
signal-to-noise. Again, we 
mapped the calibrated data in CASA and include the fourth target of 
2012.1.00306.S, IRAS~F20511-4250, which falls just below the criteria of 
Table~1 of \citet{lutz16}. 

In reducing the ALMA data, 
we adopted the same channel widths as for the NOEMA data. 
Self calibration was 
applied for IRAS F09111-1007W, NGC 4418, IRAS F17207-0014, IRAS F20551-4250
but not the other sources for which we found no noticeable improvement.

\section{Outflow properties of the targets}
\label{sect:outprop}

First, we determine whether an outflow is present, and then measure its 
CO flux, velocity, 
and size. Outflow identification can be relatively straightforward
in cases where an extremely broad outflow line component is superposed on a 
core of velocity width that is typical for undisturbed gas moving in a 
galaxy's 
gravitational potential. This distinction is more difficult if the 
velocity range with observed `broad' CO 
emission does not strongly exceed plausible non-outflow rotational velocities
within the galaxy's potential. Different 
radial distributions of molecular gas disks rotating in the potential, 
misaligned disk components, bar flows, or gas moving in the potential of a 
merger all can 
lead to quite complex total line profiles over a velocity range that is
similar to local circular velocities. For outflows from spatially resolved 
disks in regular rotation, the 
kinematic separation of outflowing and non-outflowing gas can be improved
by removing the regular velocity field \citep[e.g.,][]{newman12}, but this
approach is less suited for a sample of compact nuclei that also includes 
mergers. 
Needing two Gaussian components for a satisfactory fit of the line profile
is alone not sufficient to ascribe the broader component to an outflow. 

We hence continue with a brief discussion of individual targets on the 
basis of total CO maps and selected channel maps, spectra, positions in 
different velocity channels, and position-velocity diagrams, and 
visualize in Figures~\ref{fig:iiizw035} to 
\ref{fig:ngc7479} such basic CO(1-0) properties. As appropriate for the
observed spatial extent of an outflow, we extract nuclear or aperture
spectra and fit multiple Gaussians to the spectrum which, is the sum of host 
emission and (if present) broad 
outflow. Visual inspection is used to choose the number of Gaussians 
needed to fit the host emission.  Fig.~\ref{fig:outflowgauss_visu} 
illustrates the adopted conventions for outflow velocity and flux.
For intrinsically somewhat asymmetric outflows and little 
obscuration, the shift $\Delta v$ of the broad component can take positive as
well as negative values, this is indeed observed (Table~\ref{tab:outobs}).
Unless mentioned otherwise, we adopt as outflow velocity  

\begin{equation}
v_{\rm out}= |\Delta v| + \frac{FW_{10\%}} {2}
\label{eq:voutdef}
\end{equation} 

computed from the shift of the centroid of the broad Gaussian line component
for the outflow with respect to the systemic velocity, and the
full width at a tenth of the maximum of this broad outflow component. 
With this definition, \vout\ will be in the 
line profile wings that are dominated by the outflow. Because outflow
components at velocities that also show host CO emission are hard to
quantify, we quote below in Table~\ref{tab:outobs} two measured CO fluxes
for the outflow: \sco\ (gauss) is the integral of the full Gaussian broad
line component, while \sco (wings) integrates this Gaussian only over the 
range(s) where the outflow dominates emission (i.e., contributes more 
than 50\% of the total 
fitted profile). In further analysis, we adopt the more conservative
\sco (wings). This is similar to the integration of wings only  
\citep[e.g.,][]{cicone14} or to removal of host flux on the basis of a 
disk model \citep[e.g.][]{veilleux17,herrera-camus19}, but different 
from adopting the full broad Gaussian flux as in several other references
in the literature. 
For objects without outflow detection, we assign limits for the 
outflow CO flux:
$S_{\rm CO,wings} < 10\times RMS_{200} \times 200$,  
based on the RMS noise (Jy/beam) 
over a 200~\kms\ channel. Using here a factor 10 rather
than 3 considers that undetected outflows may be spatially extended or 
spectrally broad.
The outflow radius is directly taken from the maps
for a few well resolved outflows, otherwise we derive a radius 

\begin{equation}
R_{\rm out}= |\Delta R| + FWHM/2
\label{eq:routdef}
\end{equation}

Here, $|\Delta R|$ is the spatial shift between line position and continuum 
position and FWHM refers to the line emission. Both line position and FWHM
are derived from a Gaussian spatial fit in the UV plane using a 
velocity range that is dominated by outflow, 
i.e. by the broad component of the spectral multi-Gaussian fit. 
\rout\ is effectively a beam-deconvolved intrinsic radius. For 
symmetric bipolar outflow we use average properties of the two lobes. 
We do not attempt to correct the measured \rout\ or \vout\/, or the 
literature results adopted in Sect~\ref{sect:literature}, for the effects of 
outflow opening angle and inclination. This emphasizes homogeneity of 
results with respect to more complete information that may be available for
a few of the best spatially resolved and strongly collimated outflows.    
Observed outflow quantities or limits are listed in Table~\ref{tab:outobs}.

\subsection{Properties of individual targets}
\label{sect:individualtargets}

{\bf III~Zw~035} (Fig.~\ref{fig:iiizw035}): Extended CO emission is 
centered on the brighter northern lenticular-shaped component of this 
galaxy pair \citep[for an optical image see, e.g.,][]{kim13}. The second 
fainter optical component, offset 
roughly along the major axis of the brighter galaxy by $\sim$7\arcsec\ toward 
PA $\sim$200\degr\/, does not stand out in CO emission
or kinematics. On top of this extended disk-like CO emission with a weak SW-NE 
velocity gradient that does not strictly follow the galaxy's major axis, we 
detect a 
strong and asymmetric spatially resolved bipolar CO outflow up to 
velocities of $\pm$500 to 600~km/s. The outflow has PA$\sim$-45\degr\ and
we adopt as \rout\ the 4\arcsec\ extent of its stronger NW (redshifted) side. 

The line profile in an r=3\arcsec\ aperture, offset by 1\arcsec\ each to N 
and W, includes both sides of the outflow and is well fit by the sum of 
three Gaussians. One is narrow and the other two are broad and (on average) 
redshifted, with FWHM 270~\kms\ and 633~\kms\/. We assign both broad 
components to the outflow. The sum of the two broad Gaussians
dominates the flux within this aperture at all velocities. Hence, it is not 
possible to follow our standard procedure and separate the outflow line wings 
from the line core at a velocity where the outflow starts to dominate the 
line profile. Instead, guided by the PV diagrams, we adopt the velocity range 
from -120~\kms\ to 120~\kms\ as line core and include into our `wings' 
CO flux only emission from both broad components that is outside that range. 
The adopted $\Delta v$, $FW_{10\%}$ and \vout\ are the 
flux-weighted means of the respective values for the two 
broad components.

{\bf IRAS~F05189-2524} (Fig.~\ref{fig:f05189}): Extended CO emission 
centered on the nucleus shows only little velocity change on 1--2\arcsec\ 
scales, consistent with a mostly face-on view of this Type~1 AGN host. 
A gradient from approaching (W) to receding (E) velocities is seen in the 
inner 0.3\arcsec\/. Beyond the $\sim\pm$200~\kms\ covered by these 
host structures, a 
clear red wing reaches out to $\sim$500~\kms\ and traces an outflow. This 
emission is extended and we use an r=1\arcsec\ aperture to extract its
flux (green component in the multi-Gaussian fit of Fig.~\ref{fig:f05189} 
top right). There may 
be a hint of near-nuclear blueshifted emission down to $\sim$-1000~\kms\/, but 
this nowhere reaches $>$3$\sigma$, and hence is not discussed further.      
We adopt $v_{\rm out} = 446$~\kms\ from the Gaussian fit to the 
r$<$1\arcsec\ outflow spectrum, and  
$R_{\rm out} = 0.41\arcsec +0.5\times 1.66$\arcsec\ 
from a Gaussian spatial fit to the 210--510~\kms\ visibilities. 

{\bf NGC~2623} (Fig.~\ref{fig:ngc2623}): The inner 
2\arcsec\ shows a roughly E/W velocity gradient from gas rotating around 
the nucleus of this merger. At $|v|>$300~\kms\/, line wings trace a bipolar 
outflow in approximately N/S (polar) direction. We adopt 
$v_{\rm out} = 587$~\kms\ from the fit to the r$<$4\arcsec\ 
outflow spectrum and $R_{\rm out} = 1.62$\arcsec\ from averaging the 
properties of the blue and red outflows at $|v|=[300,500]$~\kms\/.

In addition to the bipolar outflow with its blue component shifted to the N, 
the [-500,-300]~\kms\ channel shows emission at $\sim$14\arcsec\ from the 
nucleus, at PA$\sim$130\degr\/. This emission peaks at 4.2$\sigma$ and has
total flux 0.37~\jykms\/. Given 
the quite distinct morphology, we do not assign this emission to the main 
bipolar outflow but interpret it as the result of an earlier outflow
episode with different orientation (see also Sect.~\ref{sect:intermittent}). 
We list this outflow as `NGC~2623 fossil'.

{\bf IRAS~F09111-1007W} (Fig.~\ref{fig:f09111}): The inner 
2\arcsec\ shows a roughly E/W velocity gradient from gas rotating around 
the nucleus of this galaxy, which is part of a wide pair. 
At $|v|\gtrsim 200$~\kms\/, line wings trace a bipolar 
outflow in approximately N/S (polar) direction. We adopt 
$v_{\rm out} = 331$~\kms\ from the fit to the nuclear
outflow spectrum and $R_{\rm out} = 0.58$\arcsec\ from averaging the 
properties of the blue and red outflows at $|v|=[210,510]$~\kms\/.

{\bf IRAS~F10173+0828} (Fig.~\ref{fig:f10173}): A rotating disk 
with kinematic major axis at PA$\sim$45\degr\ is seen out to radius
$\sim$8\arcsec\/. The bright and compact nuclear CO emission does not 
obviously connect to this rotating disk and has wings out to $\pm$300\kms\/,
beyond the $\pm$100~\kms\ LOS velocities of the disk. While this could
indicate a modest velocity outflow, we cannot firmly
exclude bound motion in the near-nuclear potential and hence
assign an upper limit to the emission of a faster molecular outflow.

{\bf IRAS~F12224-0624} (Fig.~\ref{fig:f12224}): Weak blue- and redshifted 
outflow components are 4$\sigma$ detected in addition to bright spatially 
compact CO that has a velocity gradient along PA$\sim$-30\degr\/.  We adopt 
$v_{\rm out} = 566$~\kms\ from the fit to the nuclear
outflow spectrum. The spatial width of the weak outflow is very poorly 
constrained, we adopt a fiducial $R_{\rm out} = 0.5$\arcsec\/.

{\bf NGC~4418} (Fig.~\ref{fig:ngc4418}): CO emission shows a compact
core that is on average redshifted by $\approx$15~\kms\ relative to the 
extended 
emission. The redshift in Table~\ref{tab:observations} refers to extended 
emission at radii 3--8\arcsec\/. At first glance, the nuclear spectrum and 
PV diagram suggest a 
strong red and a weak blue wing to CO(1-0). But the compact, warm and dense
nuclear region of NGC~4418 is known to have a rich chemistry and a 
complex mm/submm spectrum \citep[e.g.][]{sakamoto13,costagliola15}, 
so that sensitive ALMA data 
detect rare molecules in addition to CO and CN, which are detected
for all sources of the sample within the frequency range of our data. 
Specifically, there is emission
near $\sim$-740~\kms\ relative to CO, which we identify with emission from
the nitrogen sulfide radical NS,
 $J=5/2-3/2$ at $\sim$115.55~GHz rest frequency
\citep[see also line surveys such as][]{aladro15,meier15}. 
This includes six hyperfine components, and $\Lambda$-doubling means that
similar strength NS emission is 
superposed on the red wing of CO at $\sim$290~\kms\/, with another
 six hyperfine components at $\sim$115.15~GHz rest. We use the `blue'
complex and local thermal equilibrium line ratios for the 6+6 lines in the 
blue and red complex 
(M.~Drosdovskaya, priv. comm.) to infer and subtract contamination by NS for 
the red wing of CO. The adopted intensities are for T=150~K but the relative
strength of blue and red complex varies by less than a percent between 
100 and 200~K, the range of central dust temperatures suggested 
for NGC~4418 by \citet{sakamoto13}. We fix relative fluxes of the 
NS line components and their velocities relative to CO, but vary 
the overall flux normalization and 
line width to fit the blue nitrogen sulfide complex, and then subtract both 
complexes from the spectrum (see Fig.~\ref{fig:ngc4418}).

Indeed much of the wing red-wards of the CO(1-0) line is contributed by 
nitrogen sulfide, but CO line wings out to $\gtrsim\pm$300\kms\ remain on 
both the blue and red side after subtraction of that contamination. 
We proceed with the NS-subtracted spectrum and adopt 
$v_{\rm out} = 324$~\kms\ from a 3 host + 1 outflow component 
multi-Gaussian fit. As geometrical size we use
$R_{\rm out} = 0.23$\arcsec\ from data that are averaging 
$v=[-330,-190]$~\kms\ in the blue wing only, because the red wing is 
contaminated by nitrogen sulfide.
\citet{fluetsch19} present CO(2-1) data for NGC~4418 with a line
profile that is very similar to our NS-subtracted CO(1-0) profile
(see also Fig.~\ref{fig:excitation} below), but adopt a simple
two Gaussian decomposition that assigns a larger fraction of the total
line flux
to an outflow that dominates at $|v|\gtrsim 100$~\kms\/. We maintain our
more conservative decomposition but note that none of the available data
exclude that such a large fraction of low velocity emission is due to slow
outflow-like motion, rather than gravitationally bound motion.

{\bf IC~860} (Fig.~\ref{fig:ic860}): CO emission shows a prominent compact
core with intrinsic diameter about 1.5\arcsec\/. We find no significant 
emission 
at $|v|>250$~\kms\/. The $|v|\lesssim 200$~\kms\ line wings near the nucleus are 
broader than the velocity of larger scale rotation and the positions of these 
channels fold back to near the nucleus rather than extending the 
PA$\sim$20--30\degr\ rotation pattern. We cannot discriminate some modest
outflow motion (unable to escape the galaxy) from bound motion in the 
near-nuclear potential. We assign a limit of 0.58~\jykms\ to a fast outflow.

{\bf IRAS~13120-5453} (Fig.~\ref{fig:i13120}): This merger shows prominent
CO emission \citep[see also][]{sliwa17}. Our data indicate a clear
rotating disk extending to at least 4\arcsec\/, with projected rotation 
velocity $\sim$150~\kms\/. Channel maps near systemic velocity  show a 
depression at the center of this disk.  The central line profile 
is double-horned with peak velocities similar to the larger scale rotation, 
and shows additional outflow wings in the blue and red. Wings suggesting
outflow are also seen in HCN(4-3) \citep{privon17} and CO(3-2) 
\citep[][see also Sect.~\ref{sect:convfact}]{fluetsch19}. Spectral 
decomposition using 4 host plus 2 outflow Gaussian components results in 
the outflow Gaussians contributing roughly half of the r$<$1\arcsec\ 
CO flux (15\%\ of the total flux), if integrating over the full profile. 
This type of profile decomposition is backed by the need for a strong 
optically thin central outflow component, in order to explain
the relation of CO(3-2) and (1-0) emission in this source 
(see discussion in Sect.~\ref{sect:convfact} below, including 
Figs.~\ref{fig:excitation} and \ref{fig:r31_i13120}). We adopt 
$v_{\rm out} = 301$~\kms\ from this fit to the r=1\arcsec\
outflow spectrum and $R_{\rm out} = 0.63$\arcsec\ from averaging the 
properties of the blue and red outflows at $v=[-690,-270]$~\kms\ and 
$v=[270,690]$~\kms\/.
The adopted $\Delta v$, $FW_{10\%}$ and \vout\ are the 
flux-weighted means of the respective values for the two 
broad components.

{\bf IRAS~F14378-3651} (Fig.~\ref{fig:f14378}): The data show CO emission
extended over a radius $\sim$2\arcsec\/, with a velocity gradient along 
PA$\sim$20\degr\/. The nuclear line profile is well fitted with two Gaussians.
While there may be a slight flux excess over this fit at $|v|\sim 250$~\kms\/,
we do not consider this clear evidence for an outflow and assign a flux limit
of 0.43~\jykms\ to a faster outflow.

{\bf IRAS~F17207-0014} (Fig.~\ref{fig:f17207}): The inner arc second of
this merger shows a disk-like velocity gradient along PA$\sim$-50\degr\ but 
larger
scales do not follow this trend. The spatial and velocity structure of CO
in this merger is complex, including a tidal arm that is detected in CO 
out to at 
least 15\arcsec\/. The line profile of the galaxy is extremely broad and 
reaches out to $\sim\pm$500~\kms\/. Beyond that, an outflow is detected on 
the blue side only, out to $v\gtrsim -1000$~\kms\/. An outflow covering
a similar velocity range has been reported for CO(2-1) 
\citep[][see also Sect.~\ref{sect:convfact}]{garcia-burillo15}. 
Our lower spatial resolution data are consistent with their PA$\sim$160\degr\
for the blue outflow. We adopt 
$v_{\rm out} = 1116$~\kms\ from the fit to the nuclear
outflow spectrum and an uncertain $R_{\rm out} = 0.45$\arcsec\ from the 
properties of the blue outflow at $v=[-1000,-500]$~\kms\/.

{\bf IRAS~F20551-4250} (Fig.~\ref{fig:f20551}): The velocity field
of this merger is complex with a general NE/SW velocity gradient in the inner
arc second. A clear red outflow is detected out to $v\sim 500$~\kms\/, 
spatially offset 
towards the opposite direction as the red `disk' emission. A weaker blue
counter-outflow is indicated, at $>$4$\sigma$ in the contours for the 
[-510,-210]~\kms\ range as shown in Fig.~\ref{fig:f20551}, but only  
$\sim$3$\sigma$ for ranges ending at -270~\kms\/. We base quantitative
analysis on the red outflow only, but stay aware of the tentative blue outflow
when comparing to other outflow indicators.  We adopt 
$v_{\rm out} = 490$~\kms\ from the spectral fit at the
position of the red outflow, and $R_{\rm out} = 0.27$\arcsec\/.

{\bf NGC~7479} (Fig.~\ref{fig:ngc7479}): There is strong CO(1-0) emission
near the center and along the bar of this galaxy. The spectrum inside 
r=3\arcsec\ shows no evidence for fast outflow at $|v|>300$~\kms\/.
The centroid positions of different channel maps as well as the PV-diagram
indicate a pronounced bipolar CO structure at $|v|\sim 200$\kms\/, at 
PA$\sim$30\degr\/. We do not ascribe this structure to outflow, but follow
previous works \citep[e.g.,][]{sempere95,laine99,baker00}, which on the 
basis of 
shallower data ascribed CO kinematics in the inner region of NGC~7479 to
gravitationally bound gas flows within the barred potential of this 
galaxy. We assign a limit of 0.73~\jykms\ to a fast outflow.

\subsection{Outflow data from the literature}
\label{sect:literature}

\begin{table*}
\caption{Adopted CO(1-0) outflow parameters for literature sources}
\begin{tabular}{lrrrrl}\hline
Source&J$_{\rm up}$&\sco (wings)&\vout&\rout&Reference  \\ 
                 & &\jykms &\kms &\arcsec& \\ \hline
NGC~253          &1& 707.00&   50& 13.35&\citet{bolatto13}\\
I~Zw~1           &1&$<$0.30&     &      &\citet{cicone14}\\ 
PG~0157+001      &3&   0.86&  402&  0.23&\citet{fluetsch19}\\
NGC~1068         &3&  98.00&  100&  1.50&\citet{garcia-burillo14}\\
NGC~1266         &1&  26.70&  322&  2.75&\citet{alatalo11}\\
IRAS~F03158+4227 &1&   1.11&  876&  2.60&\citet{gowardhan18}\\
NGC~1377         &2&  29.77&  138&  1.96&\citet{aalto12a}\\
NGC~1433         &3&  16.10&  100&  2.00&\citet{combes13}\\
NGC~1614         &1&   3.55&  360&  1.73&\citet{garcia-burillo15}\\
IRAS~F05081+7936 &1&   0.90&  800&  0.96&\citet{leroy15}\\
NGC~2146         &1& 1450.0&  200& 25.00&\citet{tsai09}\\  
IRAS~F08572+3915 &1&   2.91&  800&  0.73&\citet{cicone14}\\
M~82             &1&2413.33&  100& 53.00&\citet{walter02}\\
IRAS~F10035+4852 &1&$<$1.80&     &      &\citet{leroy15}\\
NGC~3256 N       &1&   7.65&  700&  2.40&\citet{sakamoto14}\\
NGC~3256 S       &1&   5.65&  500&  4.00&\citet{sakamoto14}\\
IRAS~F10565+2448 &1&   3.22&  450&  1.30&\citet{cicone14}\\  
IRAS~F11119+3257 &1&   0.40& 1000&  2.22&\citet{veilleux17}\\
NGC~3628         &1&  82.00&   90& 11.00&\citet{tsai12}\\
ESO~320-G030     &2&   5.21&  340&  3.71&\citet{pereira-santaella16}\\
IRAS~F12112+0305 SW&2& 0.57&  324&  0.55&\citet{pereira-santaella18}\\
IRAS~F12112+0305 NE&2&10.70&  465&  1.15&\citet{pereira-santaella18}\\
Mrk~231          &1&   4.16&  700&  0.72&\citet{cicone12}\\
Mrk~273          &1&   3.11&  650&  0.73&\citet{cicone14}\\
4C~12.50         &1&$<$0.18&  640&  0.07&\citet{dasyra12,dasyra14}\\
SDSSJ135656.10+102609.0&3&0.99&500& 0.14&\citet{sun14}\\
IRAS~F14348-1447 SW&2& 7.60&  419&  1.05&\citet{pereira-santaella18}\\
IRAS~F14348-1447 NE&2& 1.48&  373&  0.60&\citet{pereira-santaella18}\\
Zw~049.057       &2&   5.00&  300&  0.30&\citet{falstad18}\\
Arp~220          &1&   0.88&  380&  0.20&\citet{barcos-munoz18}\\
Mrk~876          &1&   3.75& 1050&  1.54&\citet{cicone14}\\
NGC~6240         &1&  17.80&  450&  1.33&\citet{feruglio13}\\
IRAS~F17020+4544 &1&   1.37& 1670&  0.86&\citet{longinotti18}\\
IRAS~F17132+5313 &1&$<$1.50&     &      &\citet{leroy15}\\
PDS~456          &3&   1.42&  900&  1.00&\citet{bischetti19}\\
NGC~6764         &1&   1.40&   53&  7.00&\citet{leon07}\\
IRAS~F20100-4156 &1&   1.12&  929&  1.00&\citet{gowardhan18}\\
IC~5063          &2&  12.30&  500&  2.00&\citet{dasyra16}\\
IRAS~F22491-1808 &2&   2.00&  325&  0.35&\citet{pereira-santaella18}\\
IRAS~F23060+0505 &1&   2.26&  700&  1.38&\citet{cicone14}\\
IRAS 23365+3604  &1&   0.86&  450&  1.00&\citet{cicone14}\\
\hline\hline
\end{tabular}
\tablefoot{These parameters may differ from those published in the
original references. They are reconstructed from the published
information in order to be consistent
with our adopted separation between line core and line wings, and our 
definitions for \vout\ and \rout\/. For consistency, they are not corrected 
for outflow inclination, even if this is constrained in some original 
references. See also Sect.~\ref{sect:literature} for details.} 
\label{tab:litobs}
\end{table*}

\begin{figure*}[ht]
\includegraphics[width=\hsize]{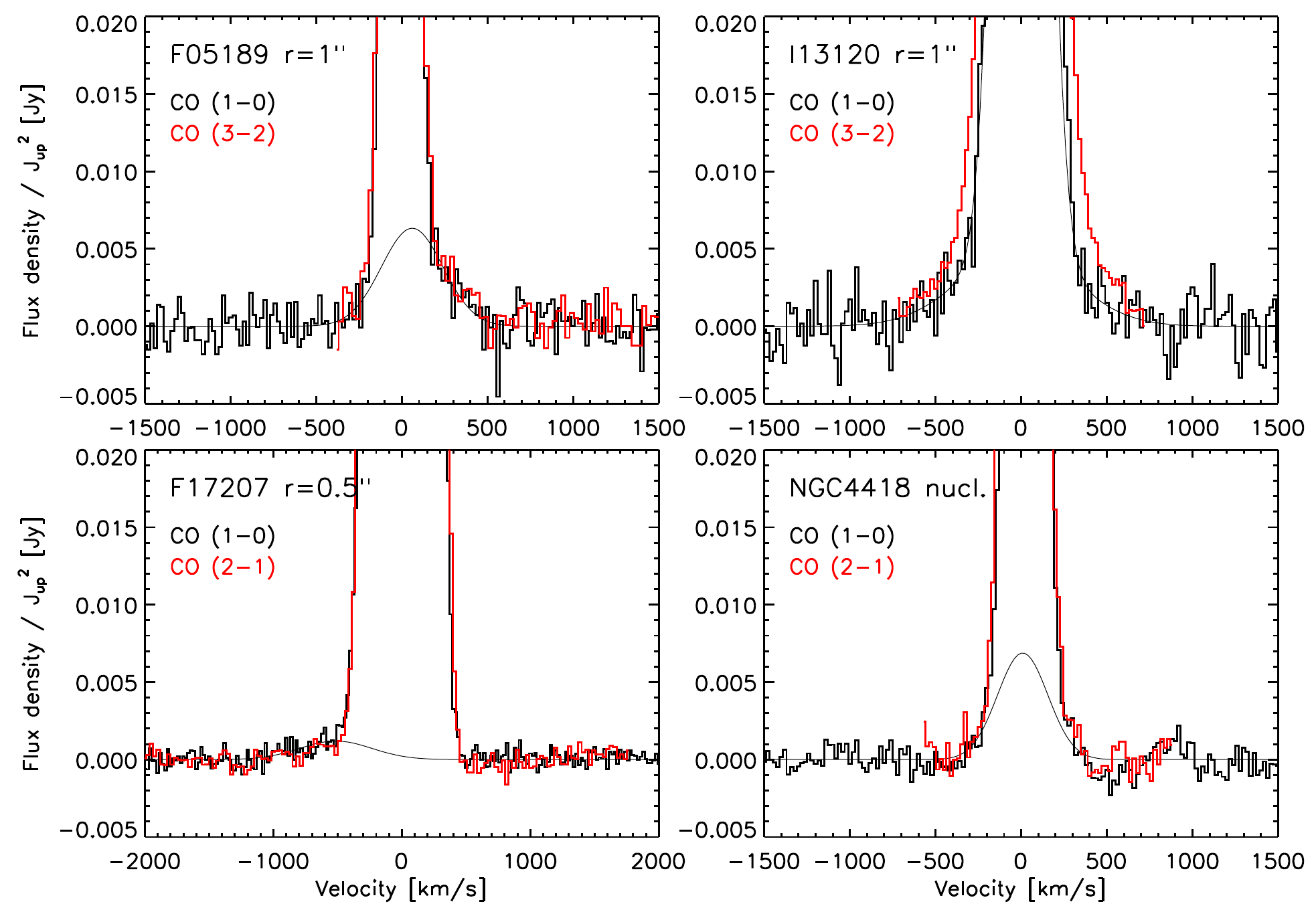}
\caption{Excitation of outflowing molecular gas, comparing our CO(1-0) results
(black, the fitted outflow component is also indicated) with higher J
results from \citet{garcia-burillo15} and \citet{fluetsch19}. Spectra are 
divided by J$_{\rm up}^2$ so that they will overlap for excitation 
$R_{\rm ij}$=1, i.e., same brightness temperature.}
\label{fig:excitation}
\end{figure*}

We have searched the literature for local galaxies with interferometric 
low-J CO outflow detections or limits, starting with the sources 
from Table~1 of \citet{lutz16} that we did not observe, but also including 
galaxies outside that
parameter range. 
As noted in the introduction, our selection is not complete towards 
some of the smallest scale and least energetic outflows that were 
recently reported from very high resolution interferometric studies.
These  would escape detection with 
`outflow wing' methods similar to ours, and be of lower impact beyond the 
local (nuclear) region of the host.
Since different references follow a range of approaches
for measuring outflow properties,
we use the published results to reconstruct `observed' 
\sco\ (wings), \vout\/, \rout\  that are homogenized to our adopted 
conventions and consider differences in adopted distance and CO
conversion factor where applicable. Where full parameters of multiple
Gaussian fits are reported in the literature, our approach as described in 
Sect.~\ref{sect:outprop}
can be directly followed for deriving \sco\ (wings) and \vout\/.
For references that only quote outflow fluxes for a velocity range
outside the host-dominated line core \citep[e.g.,][]{cicone14}, we adopt 
those fluxes and the average velocity of the outflow bin. If only
the total \sco\ (gauss) for a broad Gaussian component is available,
we assume \sco\ (wings) to be 1/3 of  \sco\ (gauss), where the factor 1/3
is adopted from the typical value measured for our own data.
Again, we do not attempt to correct for outflow opening angle and inclination.
This provides a homogeneous sample but does not make full use of all 
information that is in hand for some of the best studies cases 
\citep[e.g.,][]{barcos-munoz18,aalto16}.
In a few cases we use published plotted spectra to estimate \vout\/.
For 4C~12.50 we adopt the \citet{dasyra14} limit for CO outflow in emission
but we add their outflow velocity from CO(3-2) absorption and their 
size estimate.
For the single dish outflow detection of IRAS~F17020+4544 and the low 
resolution interferometry of IRAS~05083+7936 we adopt a fiducial 
1~kpc for \rout\/. Our adopted homogenized literature data are summarized in 
Table~\ref{tab:litobs}.

From this point on, we treat our observational results and the adopted 
literature values jointly and consistently, as described in 
Sect.~\ref{sect:derivprop}, and call this the `combined' sample.  
For literature sources observed in CO transitions other than
CO(1-0), we adopt the brightness temperature ratios in Table~4 of 
\citet{bothwell13} for conversion to CO(1-0). 

\begin{figure}
\includegraphics[width=\hsize]{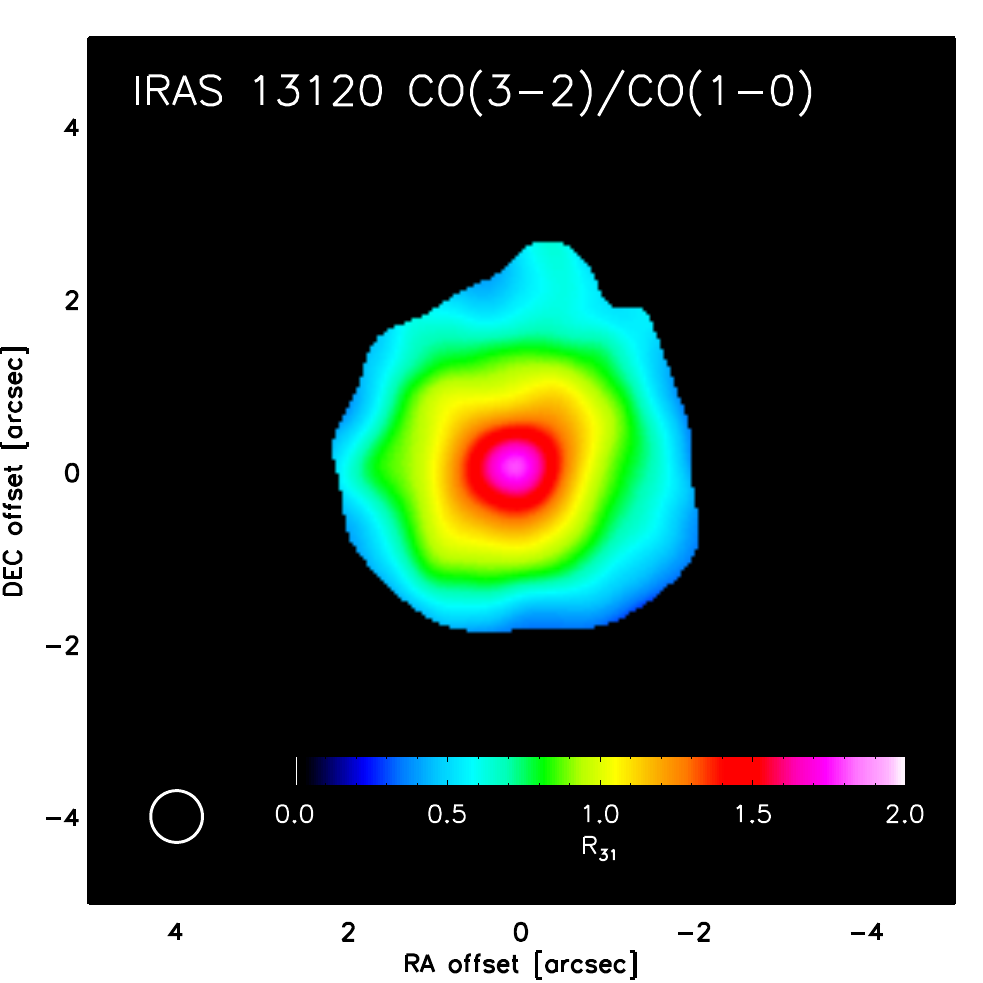}
\caption{Brightness temperature ratio $R_{31}$ of CO(3-2) and CO(1-0) 
emission for the  $\pm$200~\kms\ line core of IRAS~13120-5453. 
Data have been matched
to the same beam. The nuclear region exceeds the maximum of 1 for optically 
thick thermal emission.}
\label{fig:r31_i13120}
\end{figure}

\section{Outflow structure and derived properties}
\label{sect:derivprop}

Deriving outflow masses \mout\/, mass outflow rates \mdotout\/, and
energy and momentum flows from the direct observed quantities 
(Sect.~\ref{sect:outprop}, Table~\ref{tab:outobs}, Table~\ref{tab:litobs}) 
is a multi-step procedure. 
We describe here our assumptions, and to what extent our data constrain 
the uncertainties in the conversions involved.

\subsection{Outflow masses and CO conversion factor}
\label{sect:convfact}

Outflowing molecular gas may not necessarily be in self-gravitating
molecular clouds for which the standard `Galactic' conversion factor
from CO luminosity to molecular gas mass is originally derived 
\citep[see, e.g.,][]{bolatto13a}. 
A range of possible conversion factors have been discussed for molecular 
outflows, ranging from a Galactic 
\alpco\/~=~4.36~\msun\/~/~(K~km~s$^{-1}$~pc$^2$)
to \alpco\/~=~0.80~\msun\/~/~(K~km~s$^{-1}$~pc$^2$) as often used for
ULIRGs (adopted partly due to the first outflow 
detections being in ULIRGs), and finally yet lower values for \alpco\ that 
may arise in gas with optically thin CO emission. In that case,  
\alpco\/~$\approx$~0.34~\msun\/~/~(K~km~s$^{-1}$~pc$^2$) for an excitation 
temperature $T_{\rm ex}=30$~K \citep{bolatto13a}, and the
ratio of a rotational transition i and a lower transition j (in brightness 
temperature units) may
exceed the value $R_{\rm ij}\lesssim$1 for thermalized optically thick gas.
Ratios may reach
higher values up to $R_{21}\approx 2.8$ and $R_{32}\approx 1.3$, for
$T_{\rm ex}=30$~K \citep[e.g.,][]{leroy15a}. 

Indeed, high CO line ratios and optically thin CO have been suggested for 
the outflow of
IC~5063 \citep{dasyra16}. For four sources in our sample we can compare
our CO(1-0) outflow results to published spectra for higher J CO transitions, 
which we have re-extracted for consistent apertures 
(Fig.~\ref{fig:excitation}). Values consistent
with $R_{\rm i1}\lesssim 1$ are found for three targets: 
$R_{31}=1.35\pm 0.23$ for IRAS~F05189-2524 \citep[where we combine with 
CO(3-2) from][]{fluetsch19}, $R_{21}=0.66\pm 0.15$ for 
IRAS~F17207-0014, \citep[CO(2-1) from][]{garcia-burillo15}), 
and $R_{31}=1.34\pm 0.23$ for NGC~4418 \citep[CO(2-1) from][]{fluetsch19}).
We have derived these brightness temperature ratios by fitting multiple 
Gaussians to the higher J 
transition where mean velocity and width of the outflow component(s) are fixed
to the CO(1-0) derived values, and only their normalization is varied. The 
error
estimates combine statistical errors and an assumed 10\%\ calibration 
uncertainty for each line. These results do not point toward the low 
conversion factor of the optically thin 
case. This is consistent with similar ratios reported for other outflows
\citep{weiss05,cicone12,feruglio15,zschaechner18,cicone18}, with
\alpco\/~=~$2.1\pm 1.2$~\msun\/~/~(K~km~s$^{-1}$~pc$^2$) suggested by 
\citet{cicone18} on the basis of CO and [CI] in the N6240 outflow,
 and with an attempt to 
constrain the outflow conversion factor for M~82, leading
to \alpco\ 2 to 4 times below the Galactic value  \citep{leroy15a}.

The clear exception in our sample is IRAS~13120-5453. The CO(3-2) spectrum 
\citep[see also][]{fluetsch19}, divided by 9 to get consistent brightness 
temperature scale with CO(1-0),  clearly has wings above
the CO(1-0) spectrum (Fig.~\ref{fig:excitation}). We derive 
$R_{31}=2.10\pm 0.29$, above the maximum of 1 that is possible for 
thermalized optically thick CO emission, and consistent
with a large contribution of optically thin gas that may have a moderate 
$T_{\rm ex}\sim 30$~K. This high excitation
is not only observed in the line wings. Fig.~\ref{fig:r31_i13120} shows the
beam-matched ratio map of CO(3-2) and CO(1-0) flux (in brightness temperature
units $R_{\rm ij}$), integrated over the line core 
within $\pm$200~\kms\/. Values corresponding to $R_{31}\sim 0.6$ are observed
in the disk, but $R_{31}>1$ is found in the nuclear region. Combined with 
the general CO morphology of the source (see moment 0 map and PV diagrams
in Fig.~\ref{fig:i13120}), the source is best described as combination
of a rotating disk with normal `molecular cloud' CO properties and radius 
at least 4\arcsec\/,
and a largely optically thin CO outflow that is dominating the nuclear region 
over the full line width.  

Three of the four targets studied here and the vast majority of
literature results are consistent with optically thick CO emission in the
outflows. But the clear
detection of a high ratio (optically thin CO) for IRAS~13120-5453 is a
strong warning that this is not true for all outflows, and that associated
uncertainties on conversion factor and on outflow masses cannot be fully
removed with the present data.
Below, we follow much of the local universe molecular outflow literature and 
adopt \alpco\/~=~0.8~\msun\/~/~(K~km~s$^{-1}$~pc$^2$) 
but note that our CO data alone do not determine a 
preference for this `ULIRG' value compared to a Galactic one. For a Galactic 
\alpco\/~=~4.36~\msun\/~/~(K~km~s$^{-1}$~pc$^2$), recently found to be 
consistent also with ULIRG gas/dust comparisons \citep[e.g.,][]{tacconi18}, 
derived outflow masses and rates would increase more than five-fold. 
It is also important to recall that outflows may be complex and multi-phase.
For detection and current constraints from high density tracers in the 
Mrk~231 and NGC~253 outflows, see \citet{aalto12,aalto15,walter17}.

In the absence of a reliable source-specific conversion factor, we also apply 
\alpco\/~=~0.8~\msun\/~/~(K~km~s$^{-1}$~pc$^2$)
to IRAS~13120-5453, noting that the associated overestimate
of outflow mass for the gas that is optically thin in CO and the 
underestimate due to including only line wings may partly compensate. 
Concerning the full sample, 
the good agreement that we report below with OH-based outflow masses 
(Sect.~\ref{sect:ohcomp}) is sensitive to our adopted CO conversion factor.

\subsection{Outflow rates}
\label{sect:outrates}

\begin{figure*}
\includegraphics[width=0.27\hsize]{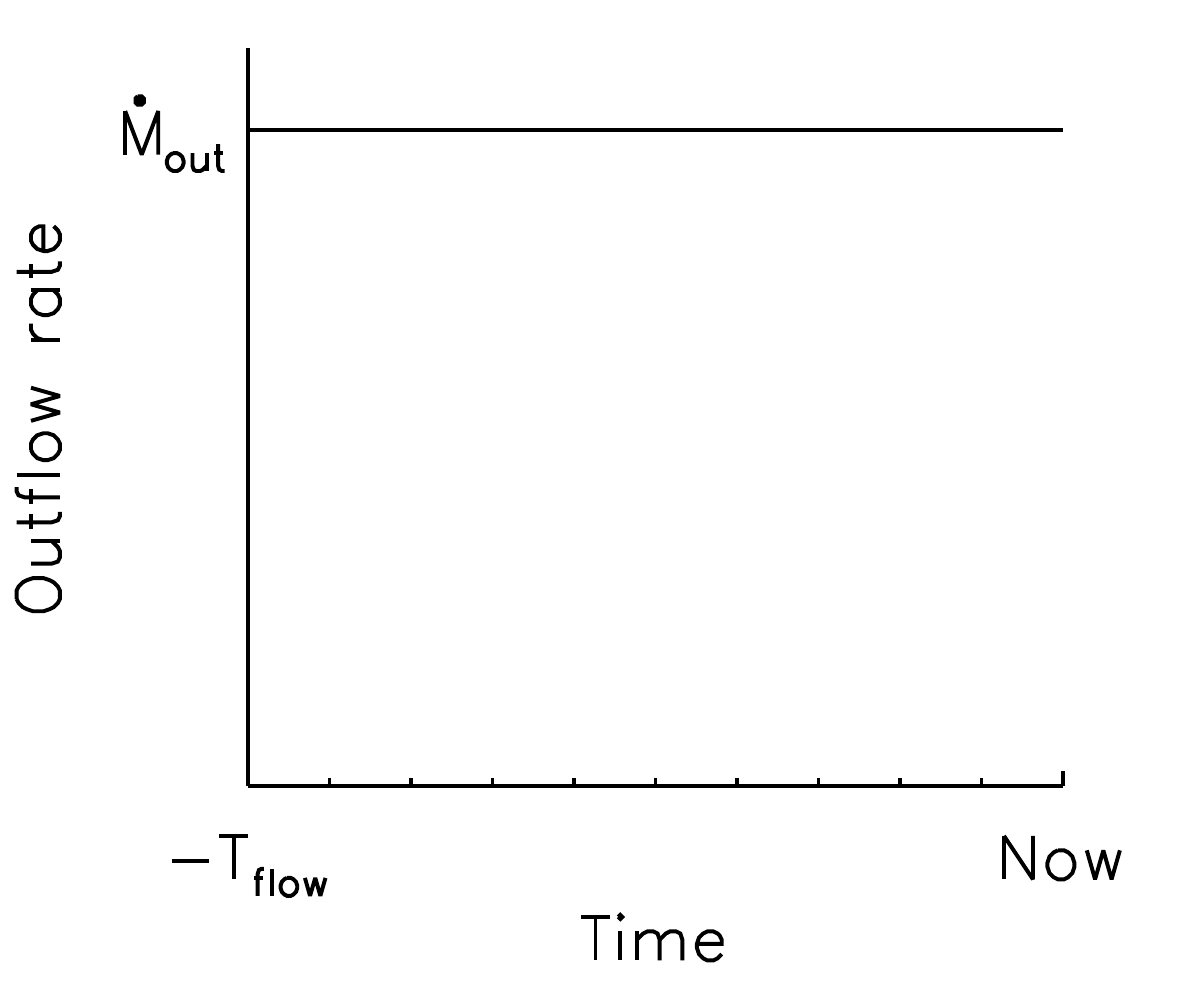}
\includegraphics[width=0.27\hsize]{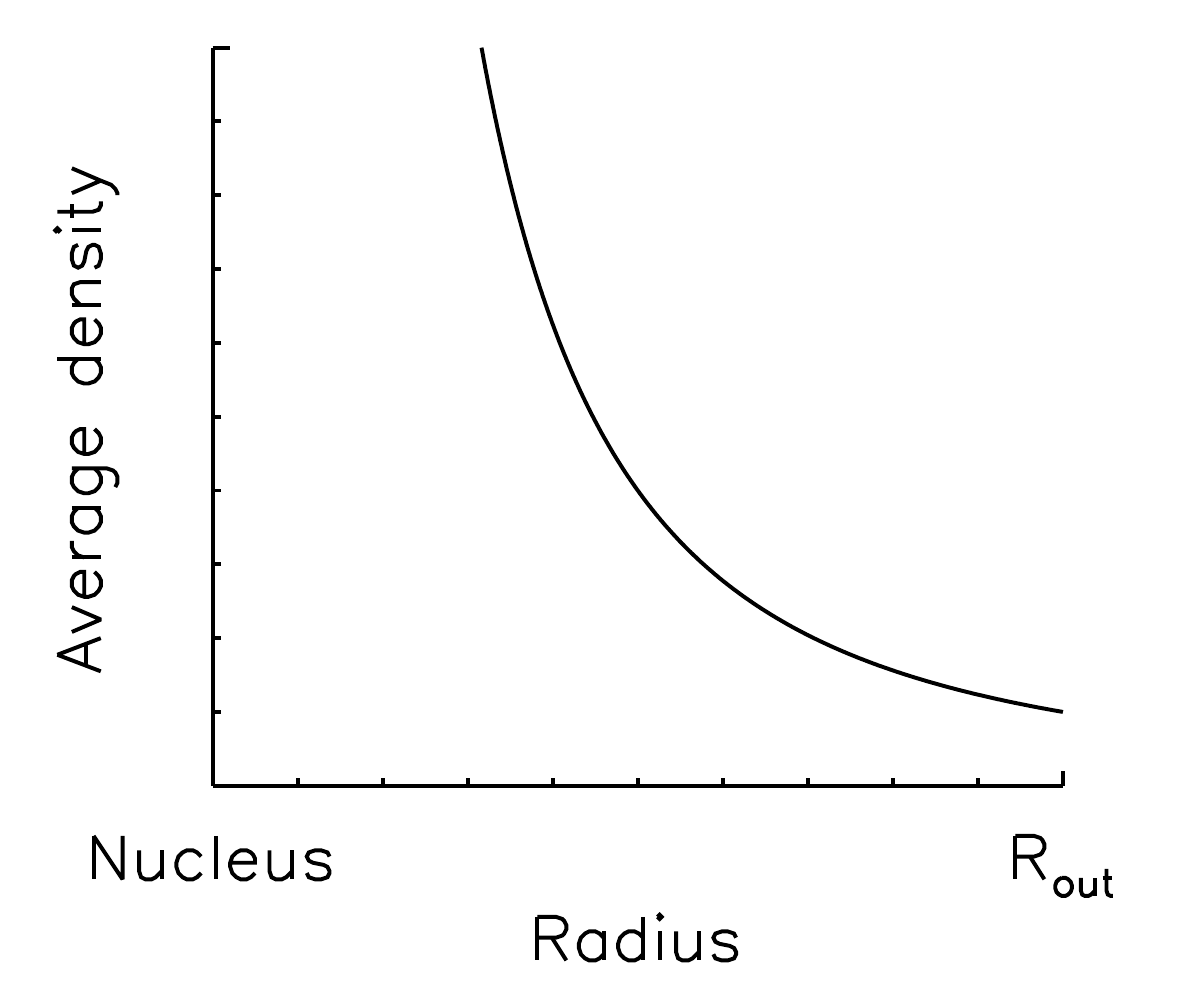}
\includegraphics[width=0.22\hsize]{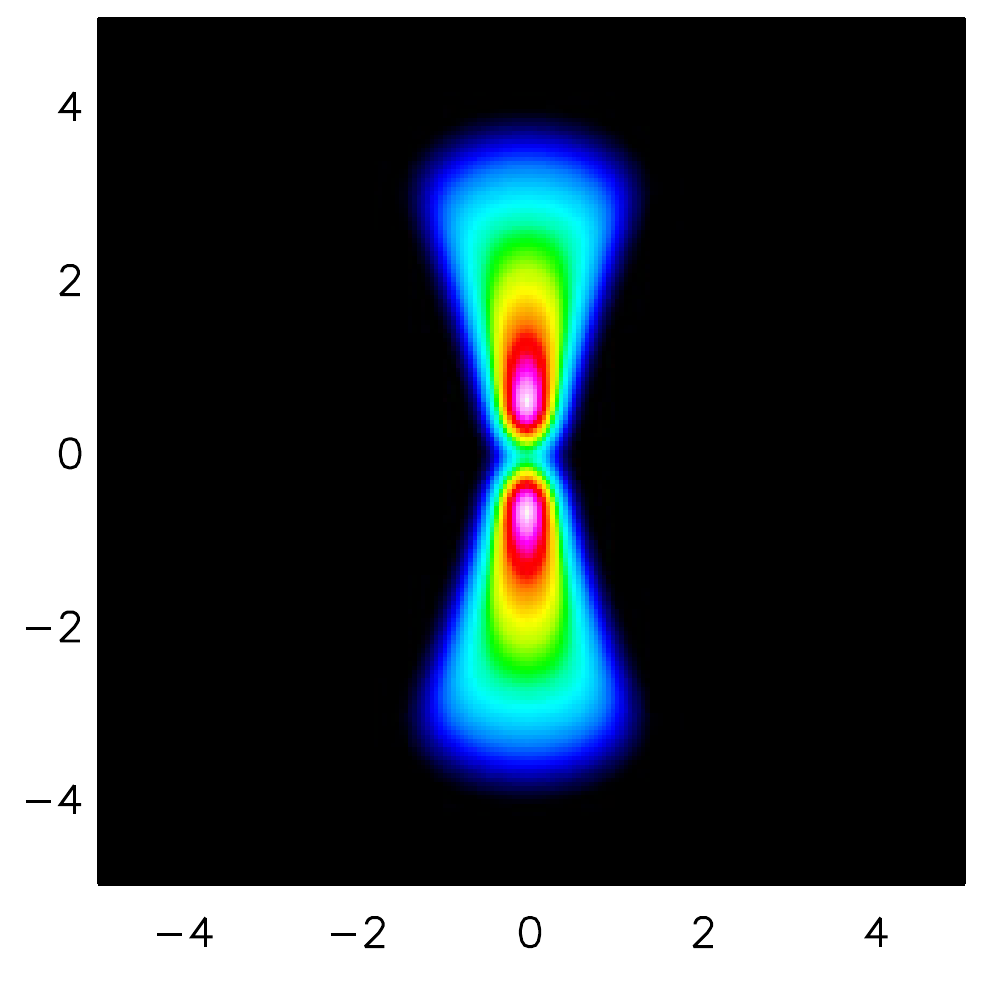}
\includegraphics[width=0.22\hsize]{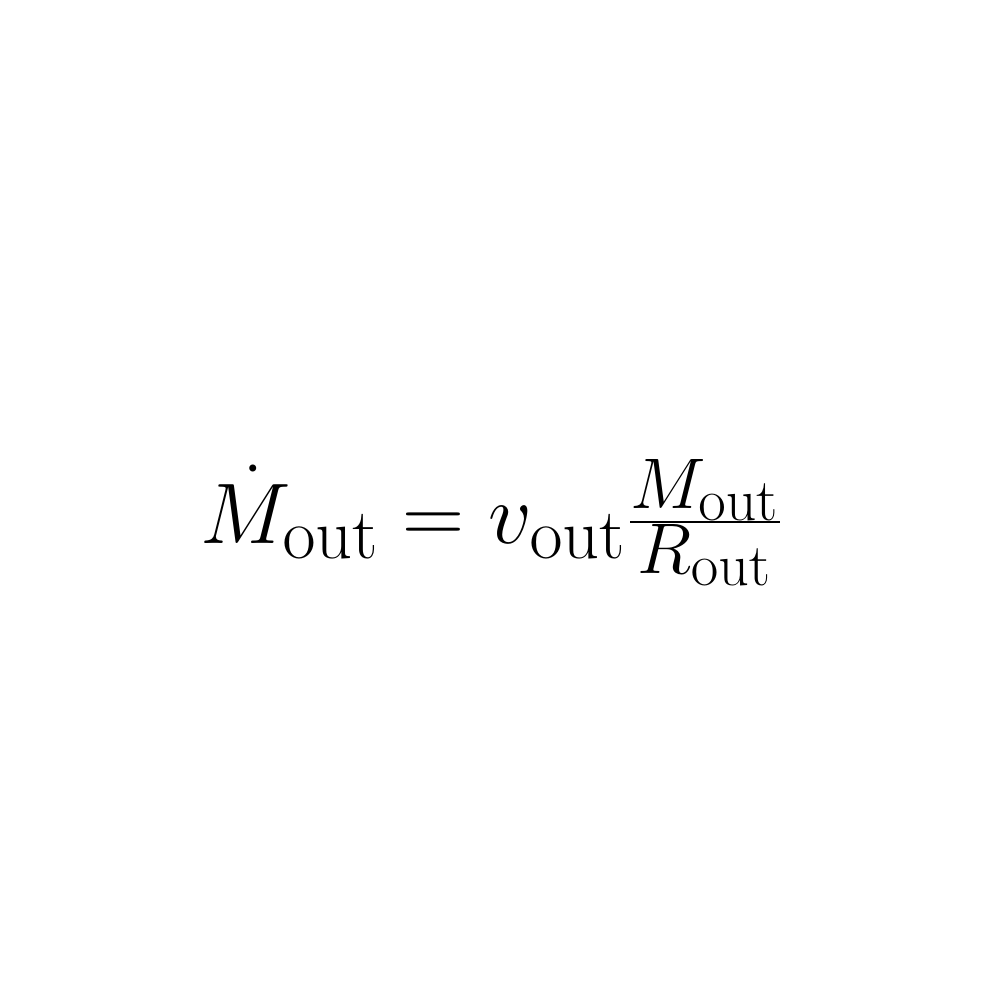}
\includegraphics[width=0.27\hsize]{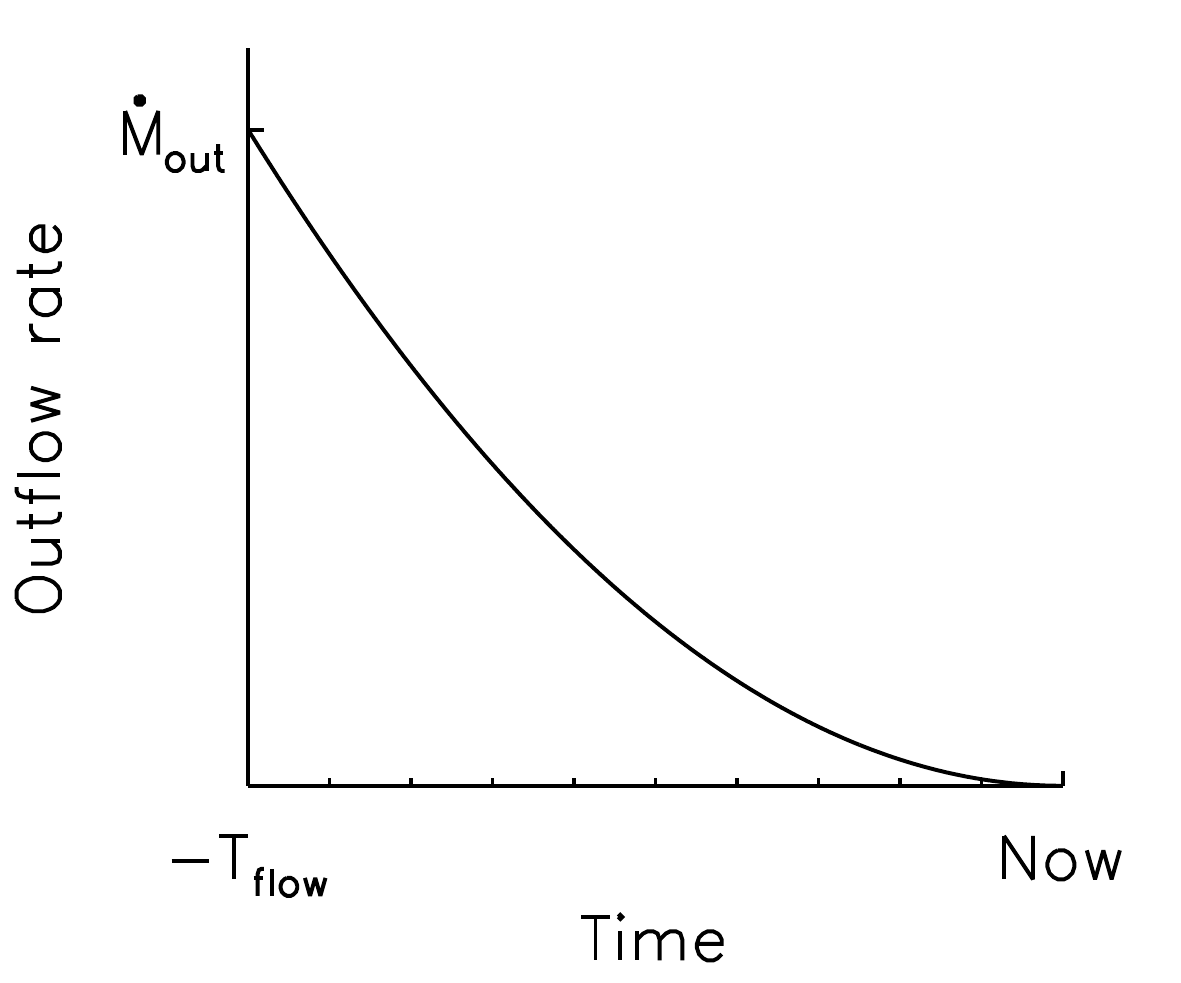}
\includegraphics[width=0.27\hsize]{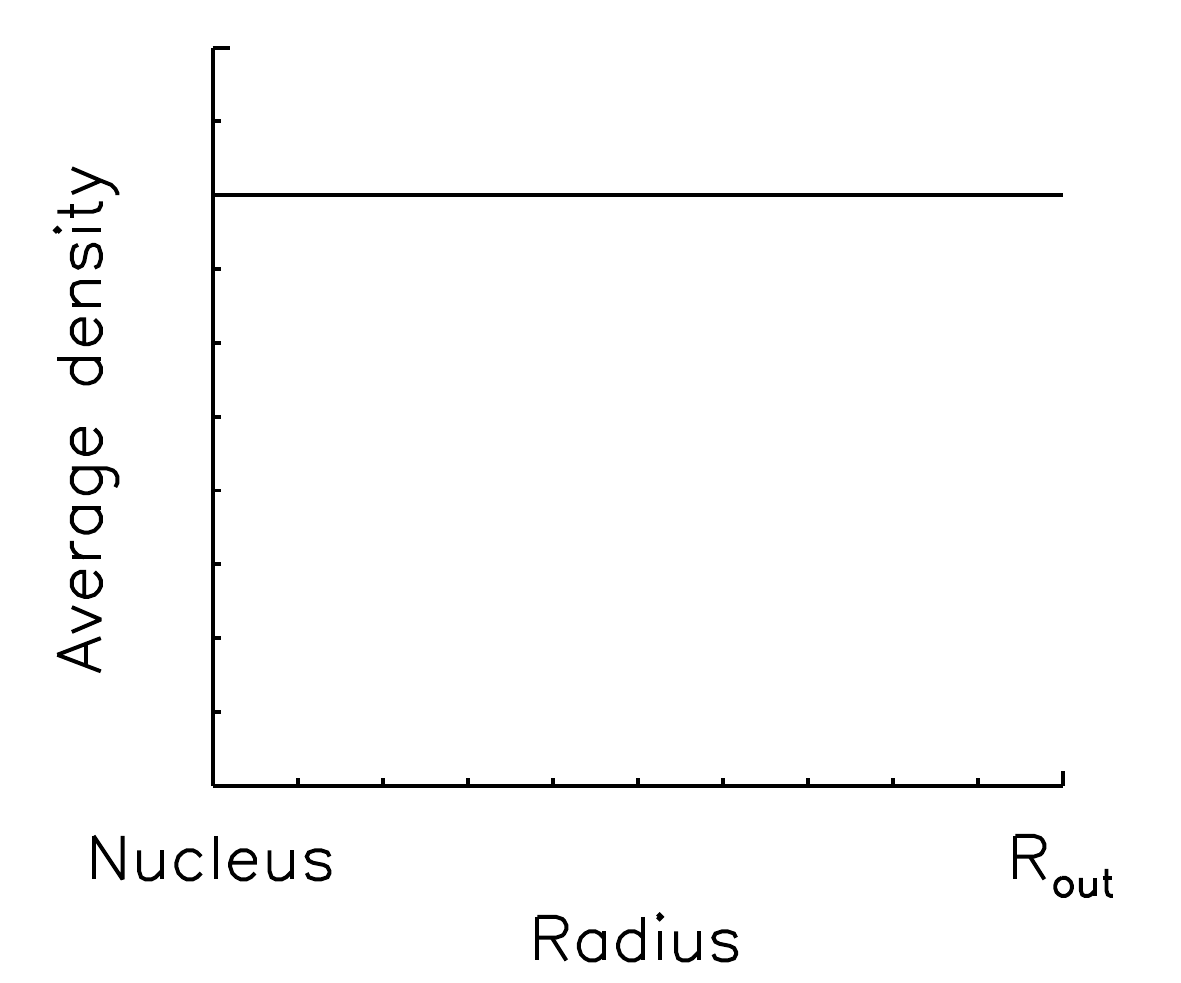}
\includegraphics[width=0.22\hsize]{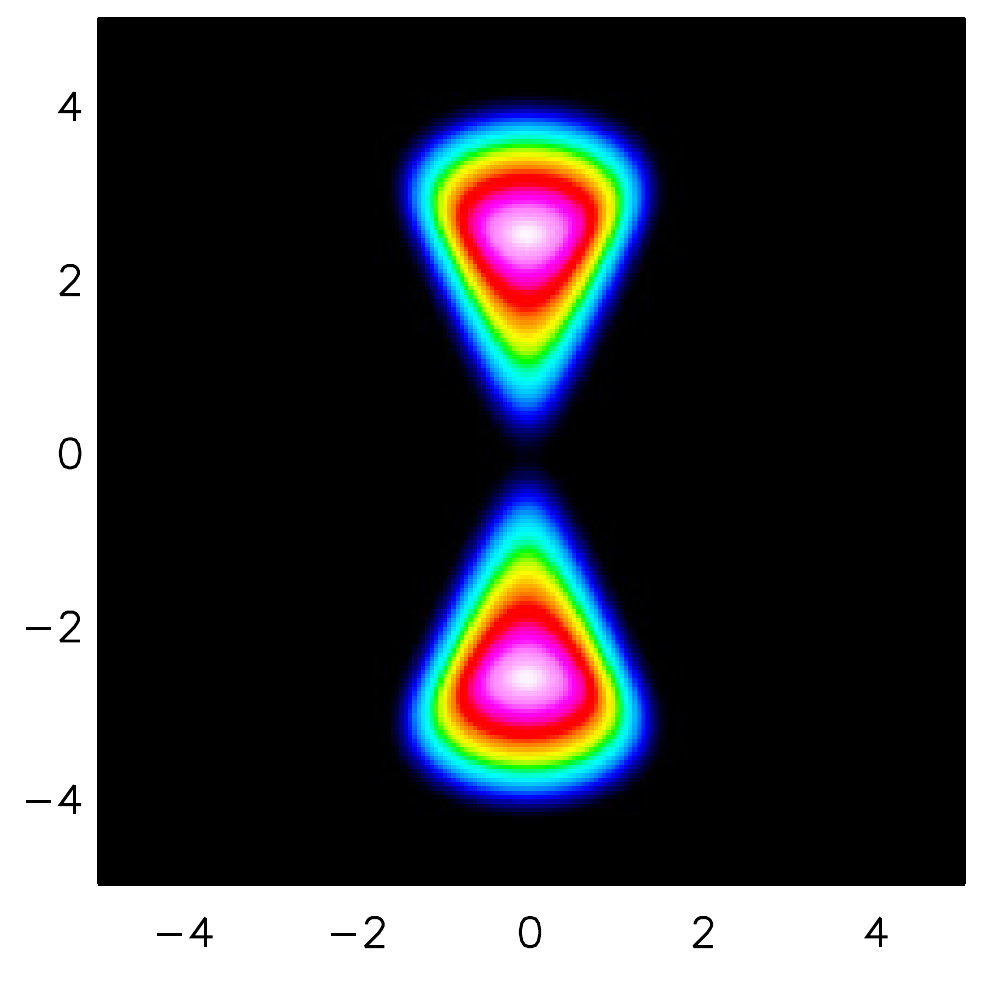}
\includegraphics[width=0.22\hsize]{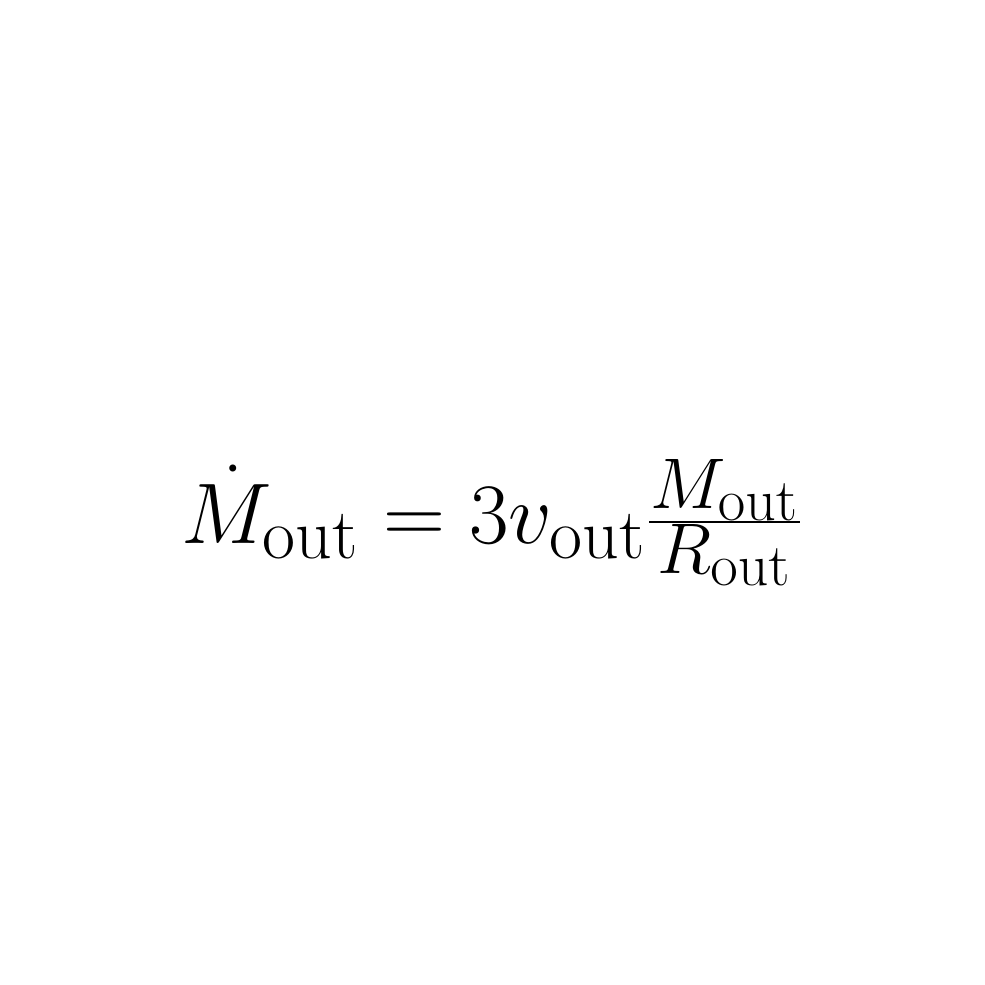}
\includegraphics[width=0.27\hsize]{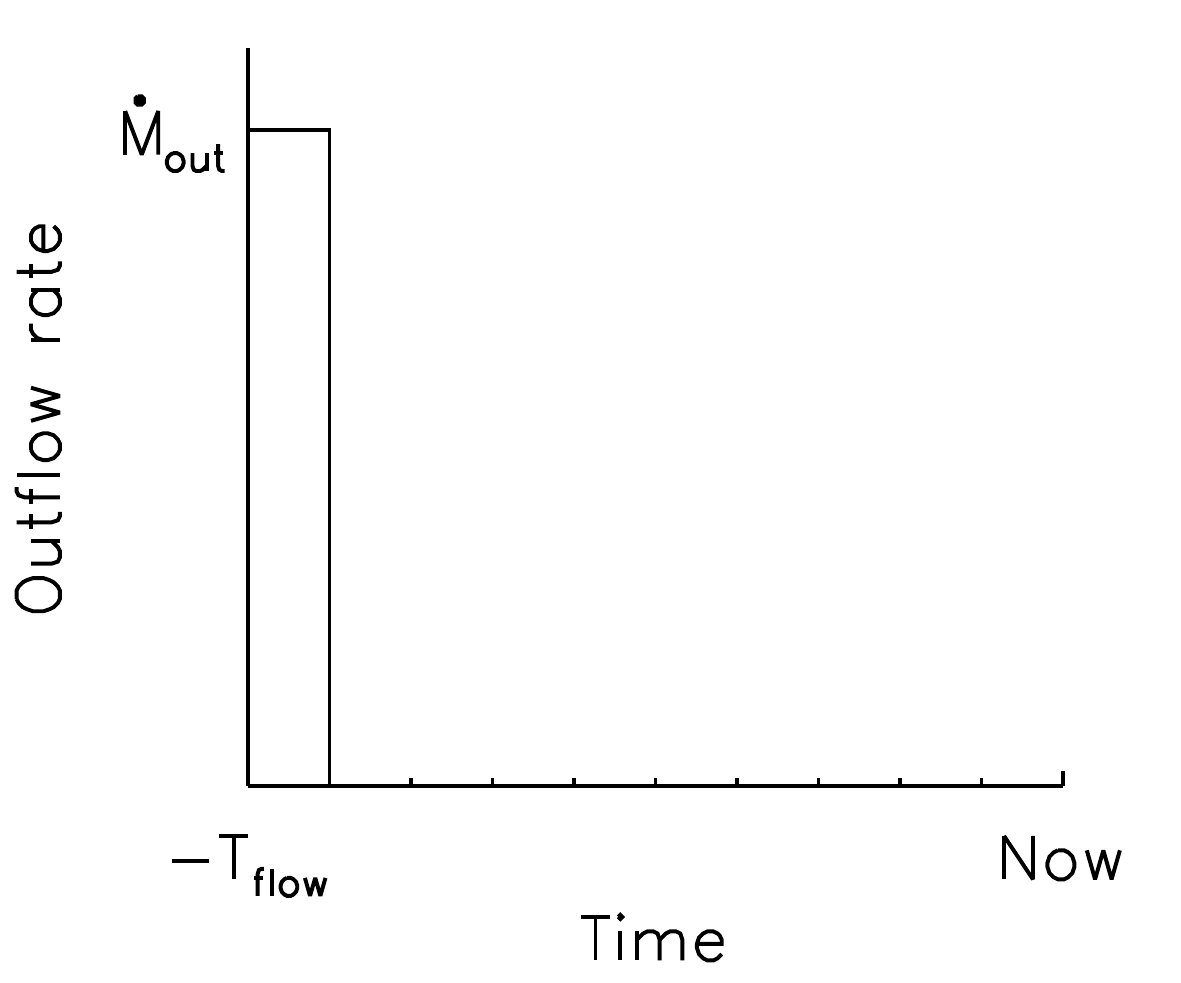}
\includegraphics[width=0.27\hsize]{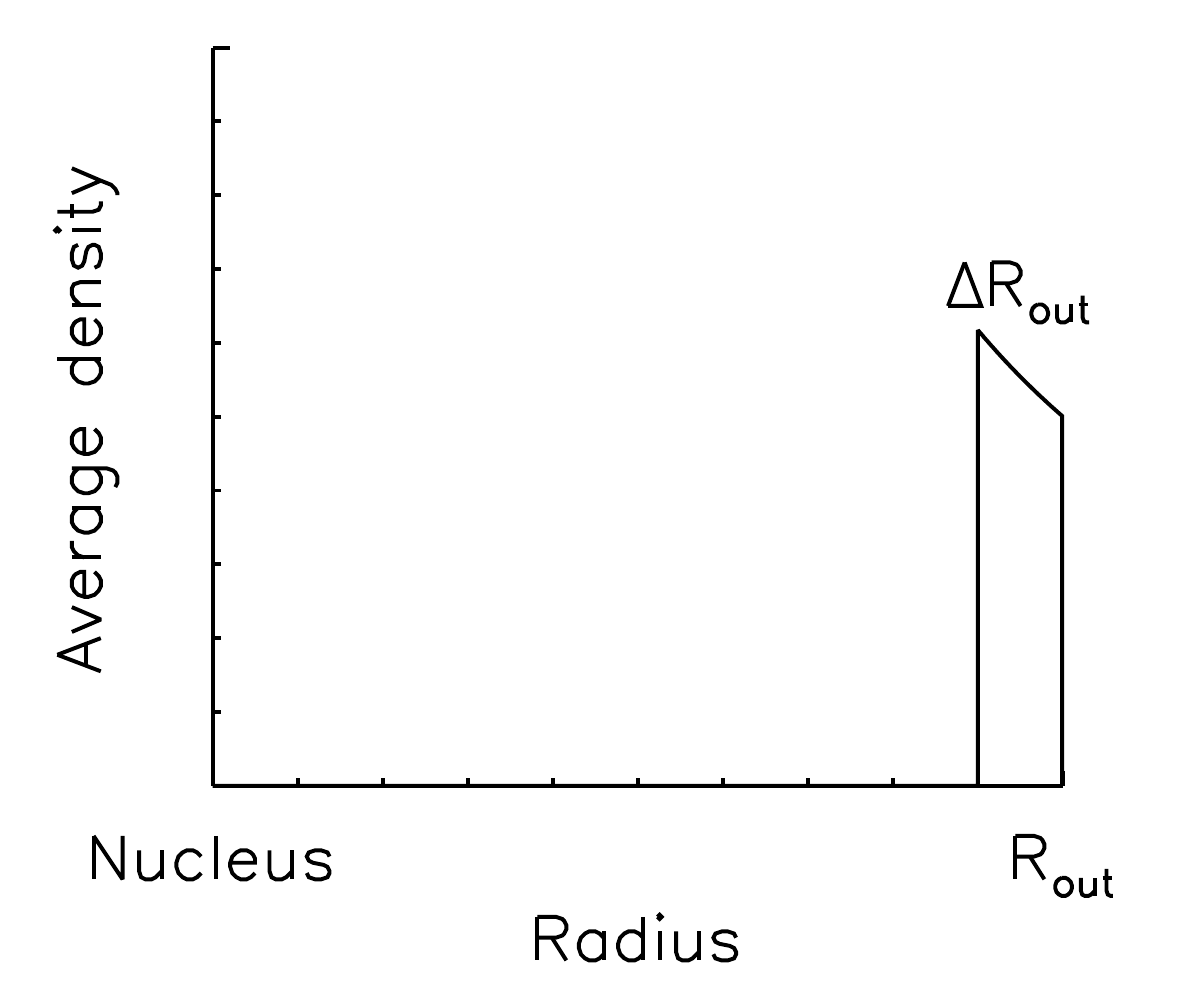}
\includegraphics[width=0.22\hsize]{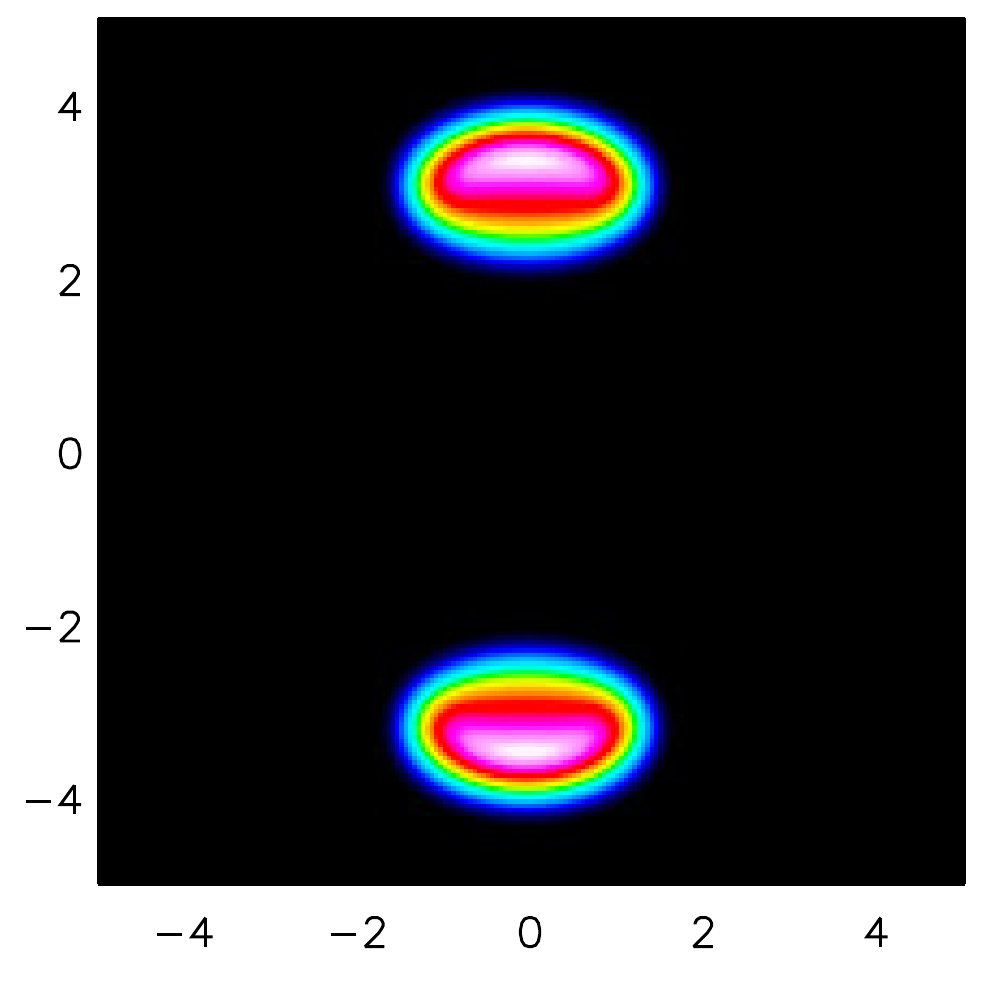}
\includegraphics[width=0.22\hsize]{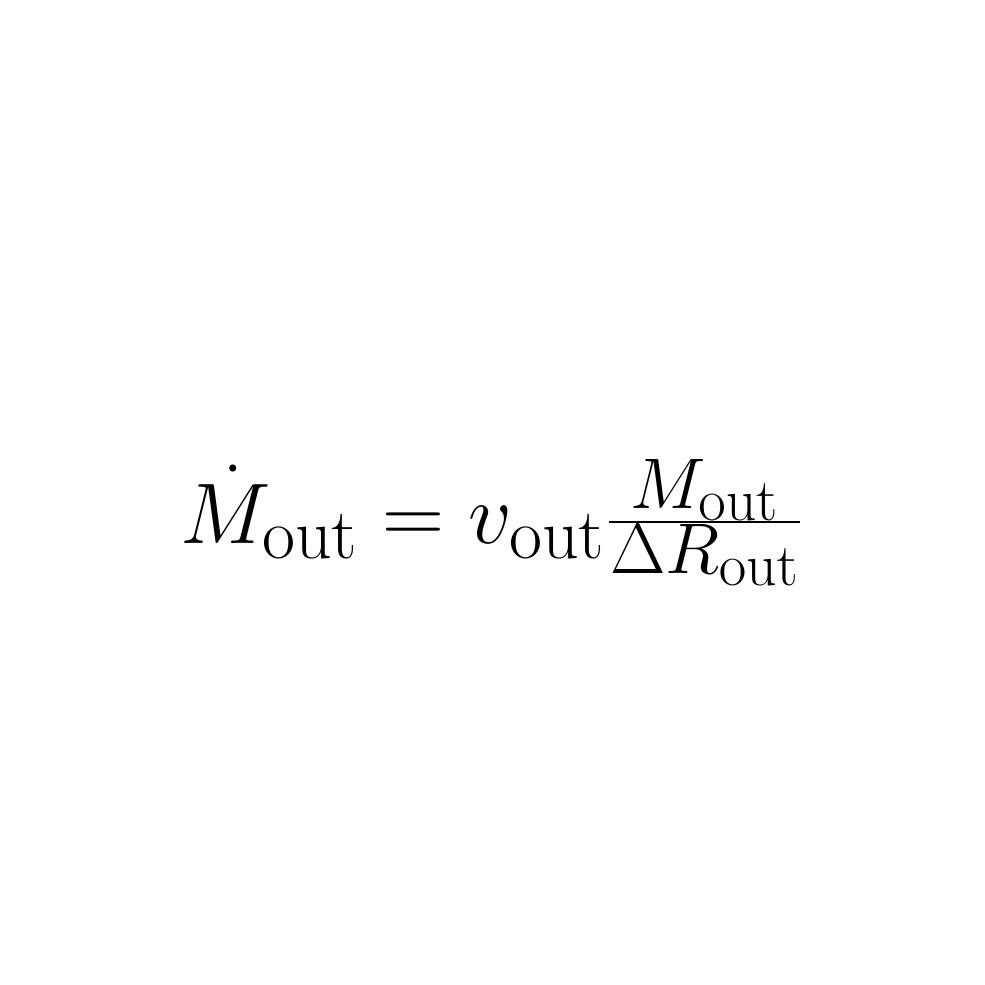}
\caption{Schematic visualization of three commonly adopted outflow histories
and their effects on radial density profile and conversion from outflow mass 
to outflow rate. From left to right, panels show 
(i) outflow history, 
(ii) average radial density profile (local cloud conditions may differ 
from this average),
(iii) simulated moment 0 map, for a modest 20\degr\ half opening angle flow 
seen with flow axis inclined by 60\degr\ with respect to the line of sight, 
and
(iv) equation for conversion from outflow mass to outflow rate.
From top to bottom, the three cases are (a) constant outflow history,
(b) constant average volume density in cone, requiring a decaying outflow
history, and (c) thin outflowing shell.}
\label{fig:schemastruc}
\end{figure*}

Given an observed outflow mass \mout\/, radius \rout\ and velocity \vout\/,
mass outflow rates are commonly derived by assuming a spherical or (multi-)cone
geometry with constant velocity within that volume, that is with the
simplifying assumption of no temporal or spatial acceleration or deceleration 
of the flow. This is a nontrivial assumption given that a multitude of 
mechanisms such as ballistic deceleration in the galaxy's potential, 
acceleration by radiation pressure, or interaction with an entraining 
non-molecular wind can all affect velocities, but it is appropriate for
interpretation of the current body of observations which is typically not 
highly resolved. The outflow rate is then derived via

\begin{equation}
\dot{M}_{\rm out} = C  \frac{M_{\rm out} v_{\rm out}}{R_{\rm out}}
\label{eq:mdot}
\end{equation}

The value of C depends on the adopted outflow history 
(see also illustration in Fig.~\ref{fig:schemastruc}). Assuming that the 
outflow has started at a point in the past 
at $-t_{\rm flow} = -R_{\rm out}/v_{\rm out}$ and has continued with a constant
mass outflow rate until now directly leads to C=1. This is what we adopt, and
agrees with the `time averaged thin shell' approach often used to derive
outflow rates from absorption line data \citep[e.g.,][]{rupke05b,
gonzalez-alfonso17}.
In this case, the average volume density of the outflowing gas decreases 
outwards $\rho\propto R^{-2}$ but local cloud or filament densities may not 
follow this drop. 

Several recent works have derived molecular outflow rates assuming C=3
\citep[e.g.,][]{cicone14,fiore17}. This is obtained when one assumes that the
outflowing gas fills the spherical or (multi-)cone volume with constant
average volume density. This corresponds to a mass outflow rate that 
starts with 
its maximum in the past at -\tflow\/, when the material was emitted that is
now reaching \rout\/, and for  which \mdotout\ is quoted. After the
start at -\tflow\ the mass outflow rate drops in time 
$\propto (t/t_{\rm flow})^2$, reaching zero now.

Adopting a fixed CO conversion factor means CO light traces molecular 
gas mass. This is 
plausible if the outflow is seen as an ensemble of flowing clouds or filaments
with conversion factor reflecting local cloud conditions rather than the
larger scale average density of the flow. This is an important assumption for
both cases, but in particular for the first case (C=1) with average density 
decreasing outwards.

Both scenarios are strongly simplified, but for resolved outflows that are
strongly inclined with respect to the line of sight and have moderate 
opening angle, the radial profiles can provide insights into the outflow
history. 
Fig.~\ref{fig:schemastruc} schematically shows outflow history, radial 
profile of the average density, and projected surface brightness, for 
these two scenarios 
and for a thin shell case. Most CO outflows in our sample
and in the current literature are barely spatially resolved, preventing
any related conclusions. An exception is III~Zw~035 (Fig.~\ref{fig:iiizw035}).
Its data suggest an inclined outflow of modest opening angle, lacking
the brightening at large radius that is expected for the constant volume 
density or thin shell cases (Fig.~\ref{fig:schemastruc}). Similar arguments 
can be made
for the strongly inclined outflows of M~82 \citep{walter02,leroy15a},
NGC~253 \citep{bolatto13} and NGC~3256S \citep{sakamoto14}. For
ESO~320-G030 substructure is observed, but again no brightening 
at large radii \citep{pereira-santaella16}.

We hence proceed computing outflow rates with C=1 in Eqn.~\ref{eq:mdot}. 

\section{Widespread molecular outflows}

\subsection{Consistency of \herschel\ OH-based  and interferometric CO outflow detections}
\label{sect:ohcomp}

Samples of dozens of galaxies each have now been studied using the two 
complementary techniques for detecting molecular outflows in local galaxies: 
P-Cygni profiles of far-infrared OH lines, enabled by the \herschel\ mission, 
and CO line wings, enabled by wide-band receivers at mm interferometers.
Comparing results from these two approaches is mandated in the first place 
by their basic difference. OH absorption probes columns along the lines of 
sight towards the far-infrared emitting region, inducing a need for 
model derivations of covering factor and radial location, under the assumption
of overall spherical symmetry. 
CO emission traces all emitting gas, but only to the extent it can be 
kinematically and/or spatially separated from the host CO emission. Both 
methods invoke conversions from observables to total outflowing molecular 
gas mass that include explicit conversion factors -- OH abundance 
and \alpco\/, respectively.
We compare the two approaches in two ways. First, we use our combined CO 
sample and a large set of OH data from the literature 
\citep{veilleux13,spoon13,falstad15,stone16} to check for consistency 
of detections and of the measured outflow velocities. Second, we compare
the outflow properties that \citet{gonzalez-alfonso17} derive
for a dozen well-characterized OH outflow sources with CO results -- 
interferometric data are available for their entire sample from our 
observations and from the literature. 

\begin{figure}
\includegraphics[width=\hsize]{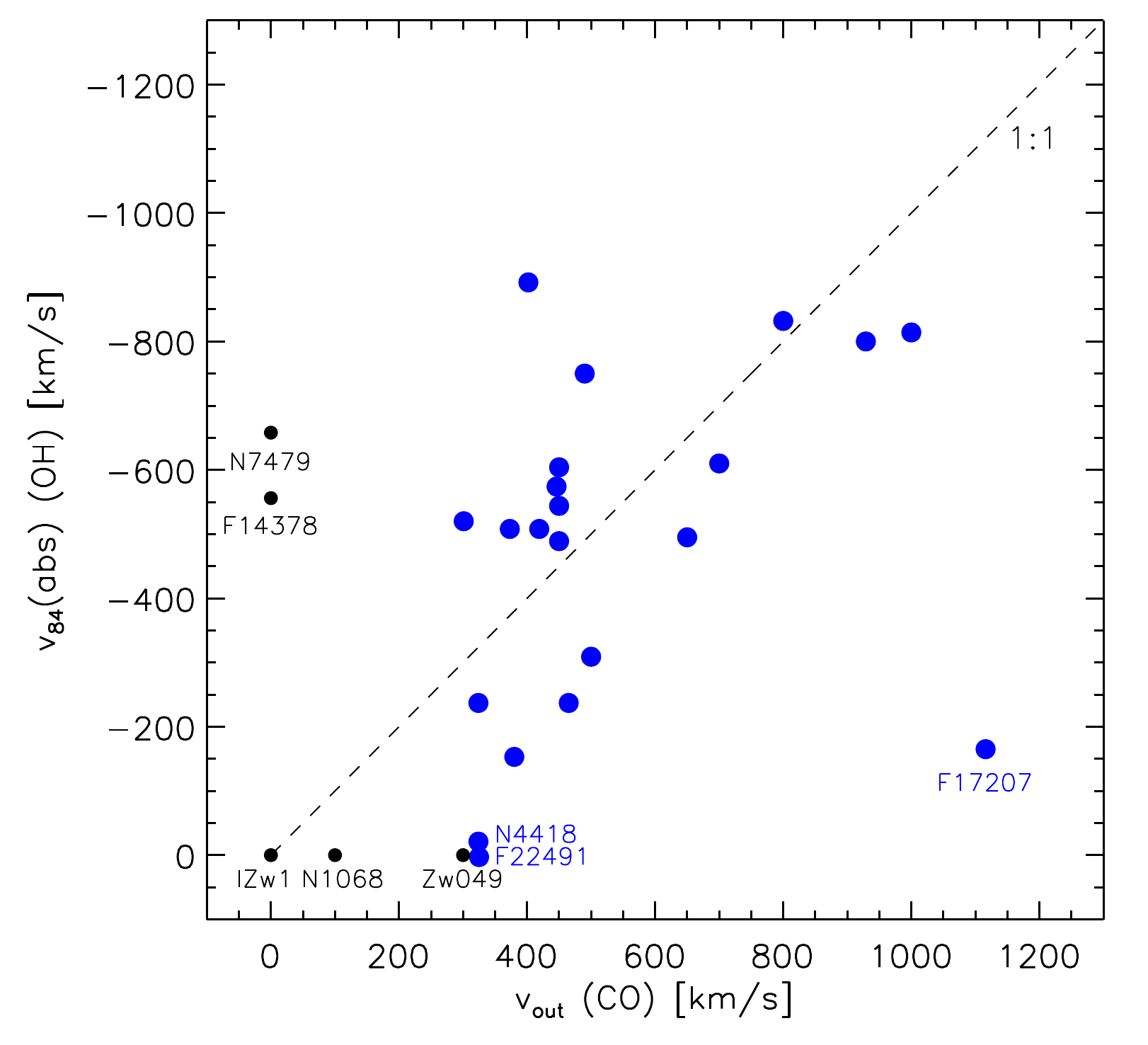}
\caption{Comparison of outflow velocities measured from CO mm interferometry
and from OH far-infrared P-Cygni profiles. For objects without outflow 
detection in one of the tracers, we use black symbols and arbitrarily 
set $v$ to 0 for the tracer without outflow detection.}
\label{fig:co_oh}
\end{figure}

Figure~\ref{fig:co_oh} compares CO- and OH-derived outflow velocities.
\vout\/(CO) is derived as described is Sect.~\ref{sect:derivprop}. 
Our \vout\/(CO) (not corrected for outflow inclination) are
compared with the OH absorption in the line of sight.
For OH, several different 
conventions are in use. We adopt \vohabs\ as defined by \citet{veilleux13},
since it is more robust at limited S/N than the alternative $v_{\rm max}$, and
better matched to our \vout\ convention for CO which samples the outflow wings
but not the maximum detected velocity. In one case where the original OH
reference quotes only $v_{\rm max}$ \citep[IRAS F20100-4156,][]{spoon13}, 
we estimate \vohabs$\sim$800~\kms\ from their Fig.~10. 

The agreement between the two approaches is good: Of 26 sources with data 
from both methods, 21 show both CO outflow and absorption dominated or 
P-Cygni OH spectra with available \vohabs\/.
CO- and OH-based velocities are in broad agreement, but there is considerable
$\sim$0.36~dex scatter around the 1:1 line. Correlations of outflow 
velocities with
galaxy properties will be subject to related scatter, likely for either of the
two methods in use.

Some insights can be gained also from the few galaxies for which CO and OH 
disagree in reported outflows. NGC~7479 shows a clear fast OH outflow
with \vohabs$\sim$-650~\kms\ \citep{stone16} but no detectable CO outflow 
at such velocities (Fig.~\ref{fig:ngc7479}). The line of sight sampling the
OH outflow against the FIR continuum
must have conditions that cannot be extrapolated to a larger scale and 
large covering factor outflow. Conversely, NGC~1068 exhibits no OH outflow
\citep{stone16} but molecular outflows have been suggested in CO studies.
This discrepancy may be due to its moderate velocity CO outflow occurring 
mostly in the disk plane \citep[e.g.,][]{garcia-burillo14}, and not affecting
our line of sight. The faster outflow reported by \citet{gallimore16} might
be closer to the line of sight, but restricted to a much smaller volume and
covering only a small fraction of the far-infrared continuum.
For NGC~4418, \citet{gonzalez-alfonso12} and \citet{veilleux13} report 
OH absorption but the absorption is slightly redshifted, suggesting 
molecular gas in slow $\approx 100$~\kms\ inflow. Inflow is also suggested
for atomic gas by an inverse P-Cygni profile in \oi\ 63~\mum\ 
\citep{gonzalez-alfonso12}.
Any CO emission from this slow inflow component would be hard to separate 
from the line
core in Fig.~\ref{fig:ngc4418}. Finally, the modest OH outflow velocity for
IRAS~F17207-0014 clearly differs from the fast (but low flux density) outflow 
seen both in CO(1-0) (this paper) and CO(2-1) \citep{garcia-burillo15}, 
indicating that the OH column at these fast velocities must be low towards 
our line of sight.

As in the explanation of these mismatches in OH and CO 
outflow detection, we believe that for the comparison of OH and CO velocities 
of the outflows in Fig.~\ref{fig:co_oh} that are detected in both tracers, 
systematic factors and geometry are main contributors to the scatter. 
Similar arguments apply to the comparisons of outflow masses, 
velocities, and rates below in Fig.~\ref{fig:ga17}. Concerning CO outflow
velocities, the nominal errors of typically $\lesssim 10$\% for the fitted 
width of the broad CO outflow component do not consider the effects of 
possibly non-Gaussian true profiles, nor the impact of varying flux and width 
of host CO 
emission on the splitting off of the outflow wings that are adopted in our 
analysis (Fig.~\ref{fig:outflowgauss_visu}). Concerning the OH \vohabs\/, 
\citet{veilleux13} estimate a typical uncertainty of 50~\kms\ that is dominated
by continuum placement uncertainties.
Both estimates do not consider the different response of 
the CO emission and OH absorption methods to variations in alignment of 
complex outflow geometries with respect to the line of sight, among other 
systematics such as the properties of the background FIR source against which
OH is observed.

For the vast $\approx$80\%\ majority of the objects with both OH and CO 
diagnostics in hand, both OH absorption and CO outflow are detected, with 
plausible mainly geometrical explanations for the remaining mismatches. For 
a sample of obscured nuclei characterized by HCN mm emission from 
vibrationally excited levels, \citet{falstad19} discuss in more detail 
geometrical and evolutionary factors that might lead to absence of 
far-infrared OH outflow signatures.
The different sampling of the outflow geometry by emission and by absorption, 
as well as the modest S/N of many detections certainly also contribute to 
the scatter around the 1:1 line in the velocity comparison of 
Fig.~\ref{fig:co_oh}. This good agreement would be too optimistic if dominated 
by CO followup of OH outflows, with preferential reporting
of detections in the literature. But CO detections/limits for our 
high FIR surface brightness sample,
and the analysis of the \citet{gonzalez-alfonso17} sample in the next 
paragraph argue that such effects are not dominant. 

\begin{figure*}
\includegraphics[width=0.33\hsize]{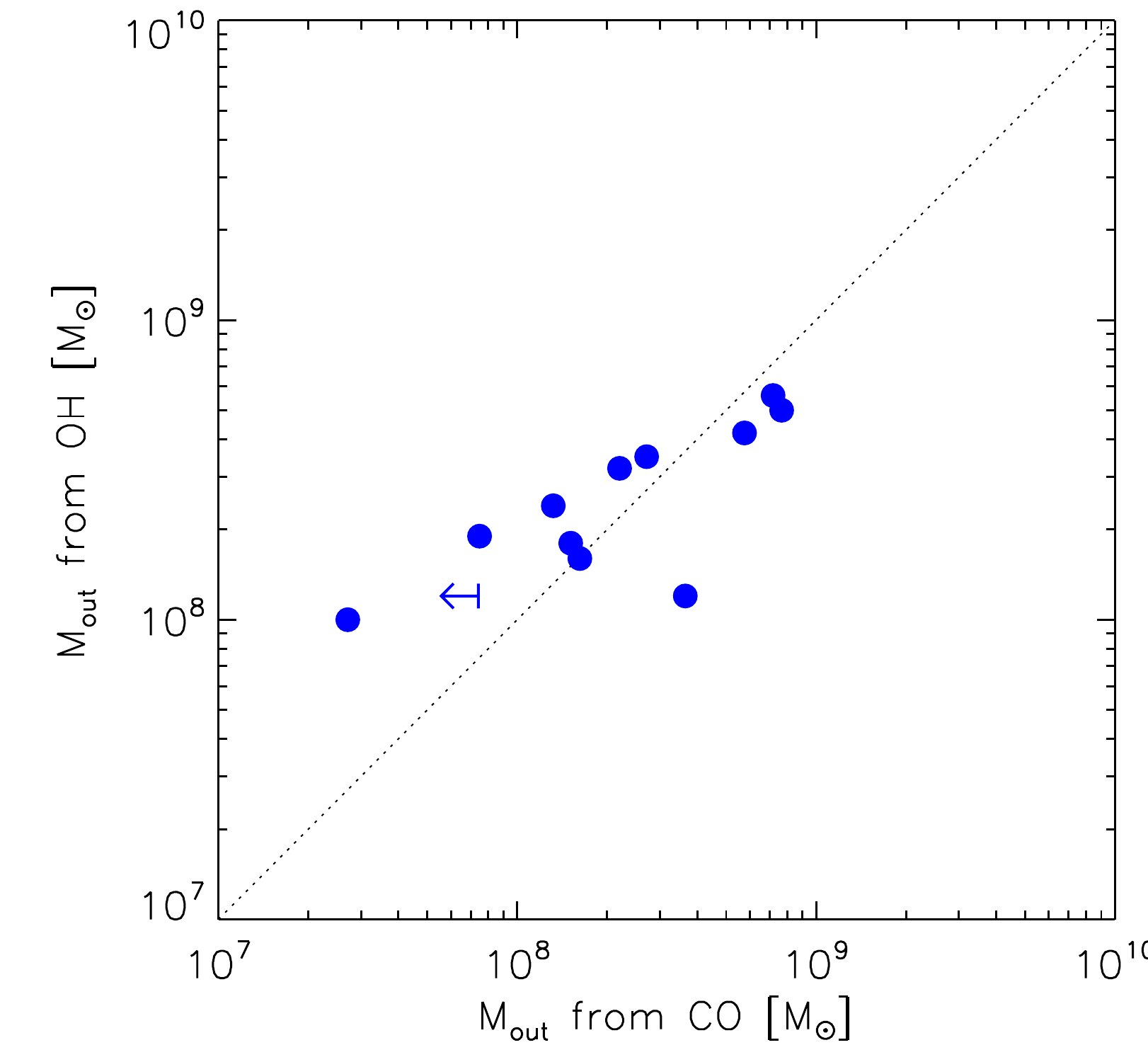}
\includegraphics[width=0.33\hsize]{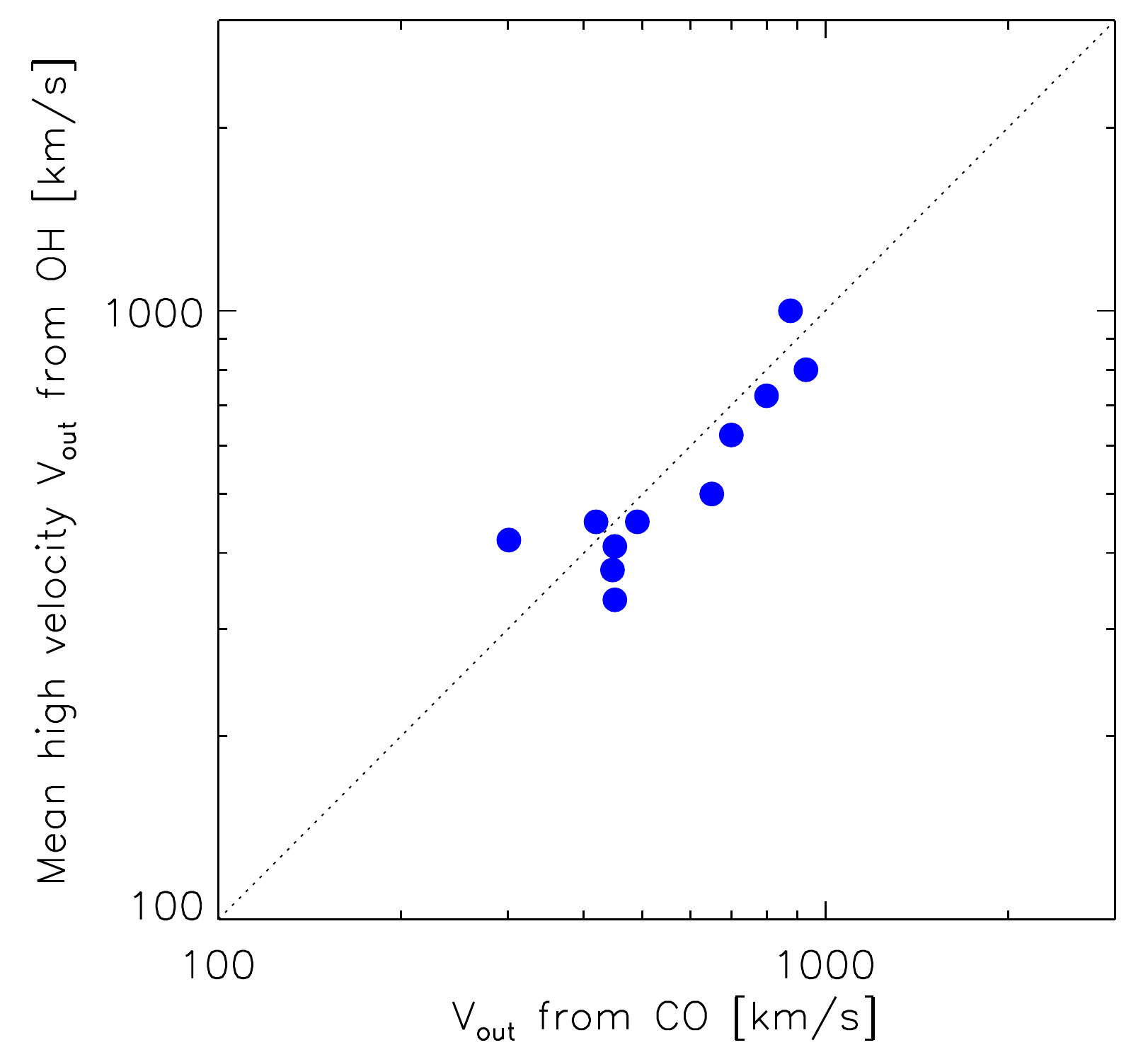}
\includegraphics[width=0.33\hsize]{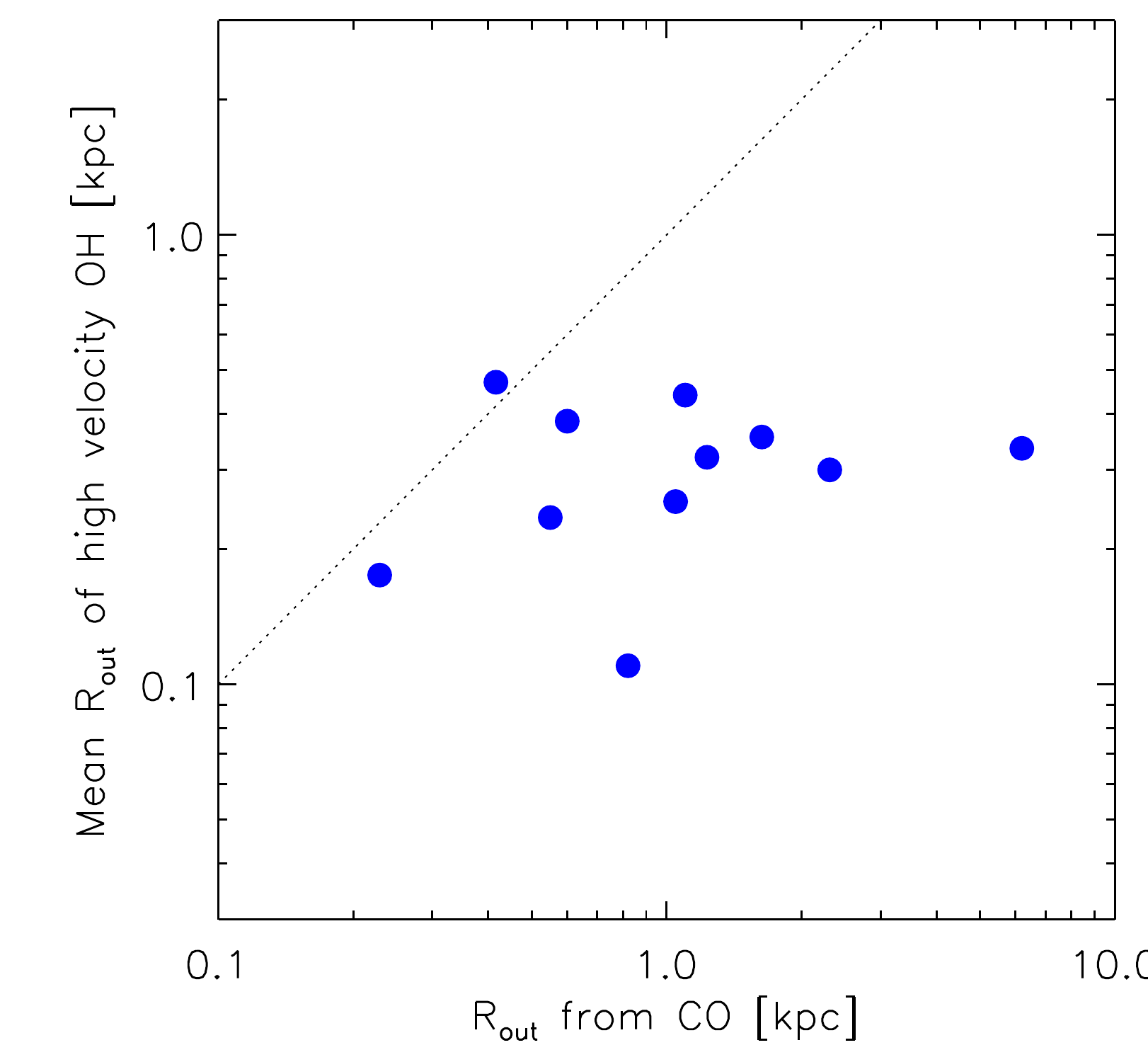}
\includegraphics[width=0.33\hsize]{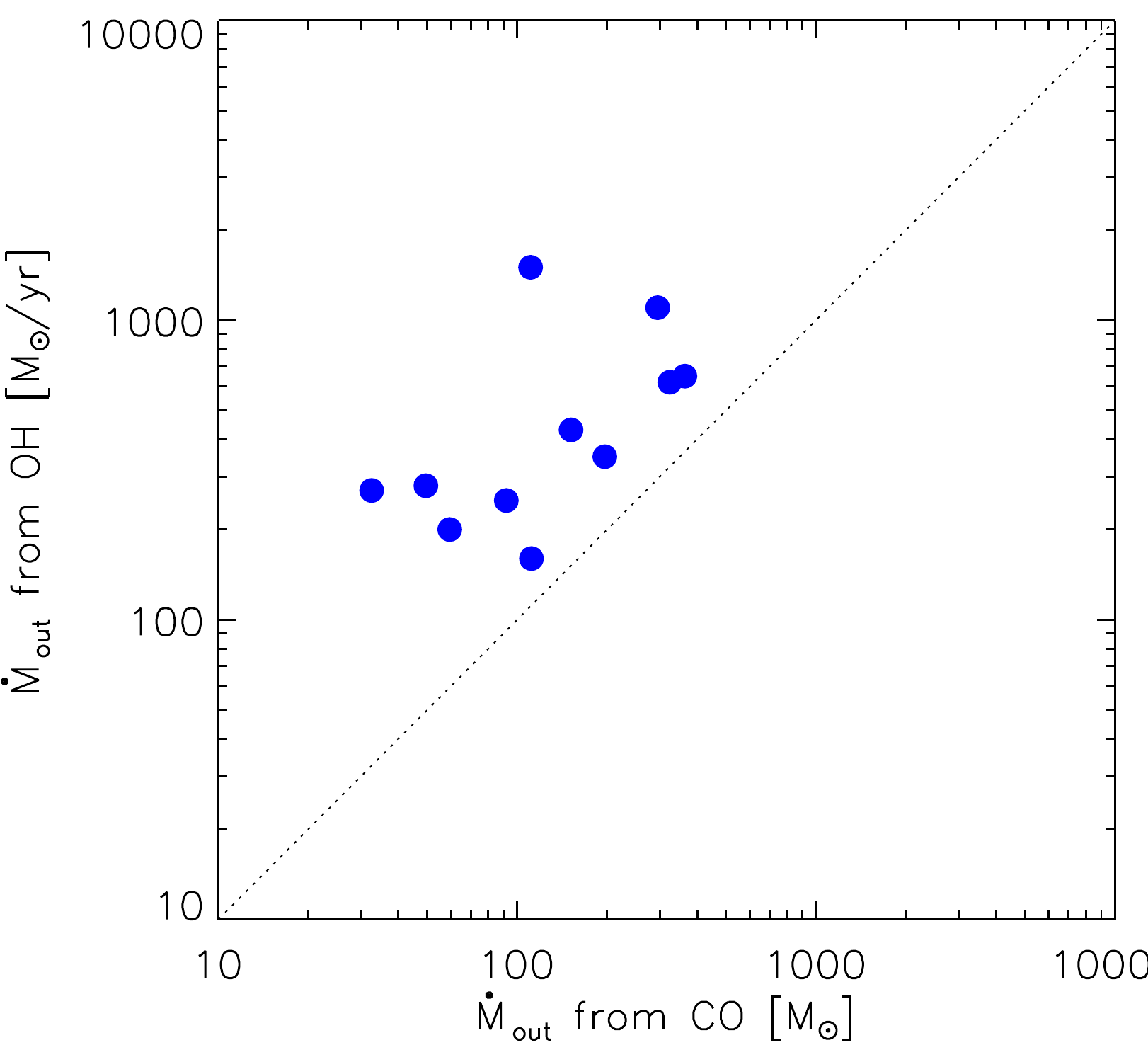}
\includegraphics[width=0.33\hsize]{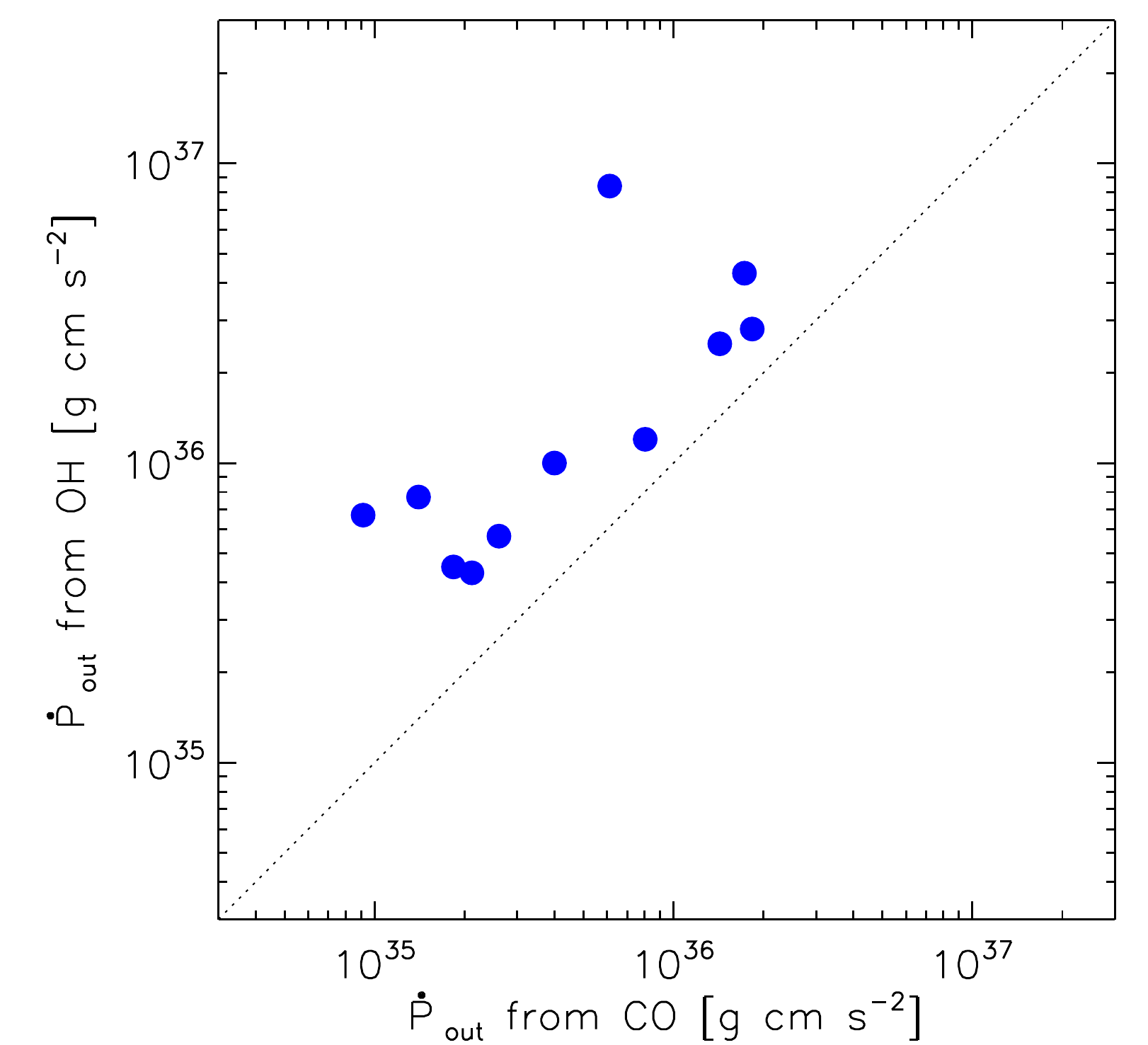}
\includegraphics[width=0.33\hsize]{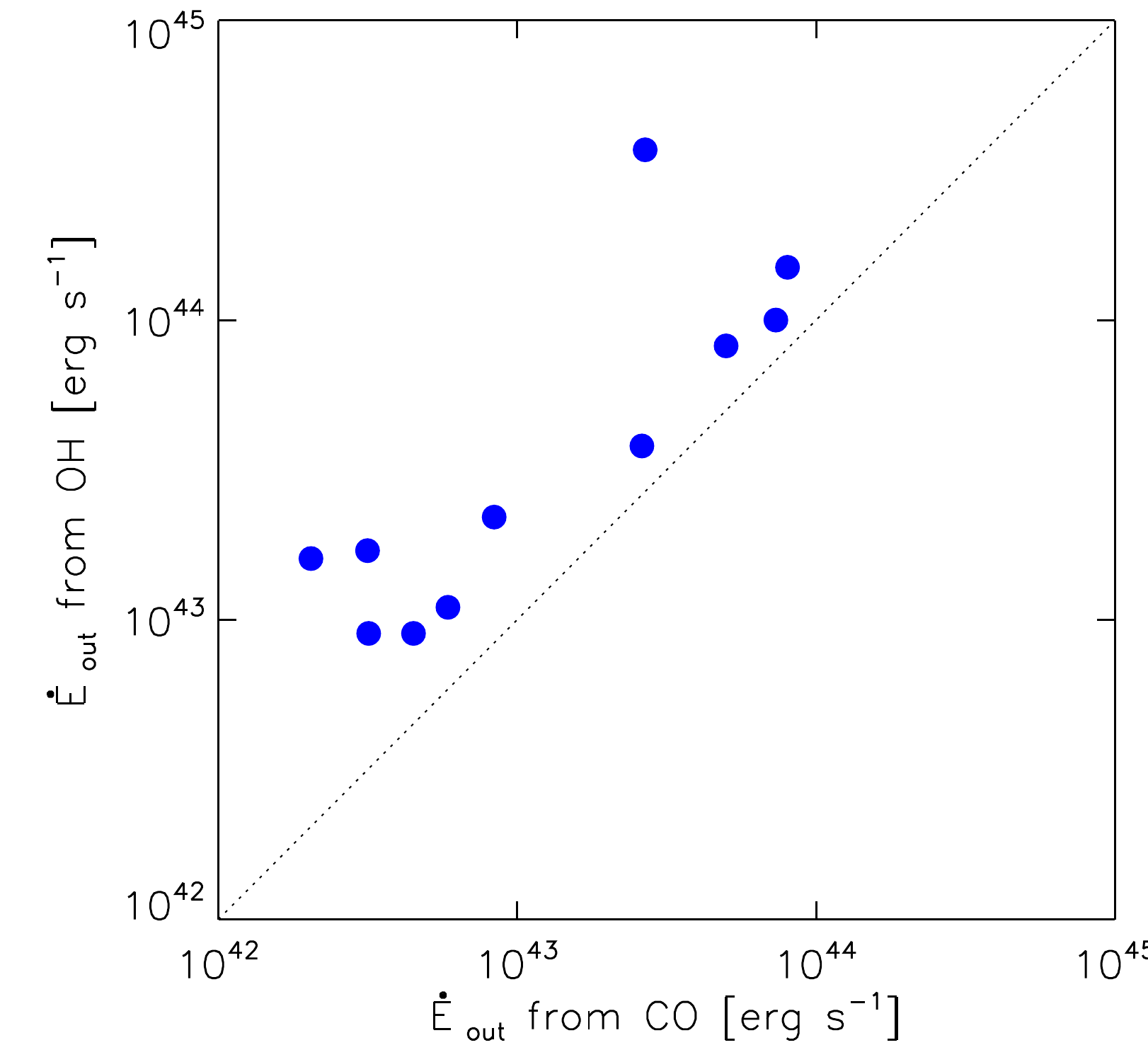}
\caption{Comparison of CO-based outflow quantities with results from
OH-based outflow models for 12 ULIRGs as derived by 
\citet{gonzalez-alfonso17}. Dotted lines indicate a 1:1 relation. See 
also discussion in Sect.~\ref{sect:ohcomp}
on how this is influenced by differences in the regions sampled by OH 
absorption and CO emission.}
\label{fig:ga17}
\end{figure*}

A deeper comparison of CO-based outflow masses, rates, and 
energetics is possible for the 12 ULIRGs for which \citet{gonzalez-alfonso17} 
were able to derive reliable spherically symmetric outflow models 
that are constrained by both ground state and excited level 
transitions of OH, observed with \herschel\/-PACS. All twelve sources are 
part of our combined CO sample. 
Fig.~\ref{fig:ga17} compares basic outflow mass, radius, and velocity
as well as derived mass, momentum, and energy outflow rates. The 
OH models include up to three components
per source, we follow \citet{gonzalez-alfonso17} in comparing CO data only
to fast $v\geq 200$~\kms\ OH components, since the CO equivalent of the slower
OH components will be blended with the host line core. Masses are the sum
of values for individual high velocity OH components ($v\geq 200$~\kms\/), 
while radii and velocities are averages. Concerning outflow history,
the assumptions of \citet{gonzalez-alfonso17} agree with our use of C=1 in 
Eqn.~\ref{eq:mdot}. 
For IRAS~F14348-1447, \citet{pereira-santaella18} report
two distinct CO outflows from the two nuclei while the \herschel\ beam 
includes both. We use the dominant southwest CO outflow for comparison to OH.

Of the 12 ULIRGs, only IRAS~F14378-3651 lacks a CO outflow detection and
our assigned upper limit for CO-based outflow mass is within a factor 2 
of the OH result. In fact, Fig.~\ref{fig:f14378} indicates some excess at 
$v\sim$250~\kms\ above the fit with two Gaussians, but we have conservatively
opted in Sect.~\ref{sect:outprop} to not assign this excess at modest velocity
to an outflow. Also, any weak CO flux excess at the velocities 
-500 to -1000~\kms\ which are reached by the OH outflow is not 
significant.
For the remaining 11 ULIRGs, outflow masses agree very well
(top left of Fig.~\ref{fig:ga17}),
with mean log$(M_{\rm out,OH}/M_{\rm out,CO})= 0.06$~dex and
dispersion 0.29~dex. This agreement is noteworthy given the 
independent derivation of these masses, including in particular the 
conversion factor $\alpha_{\rm CO} = 0.8$~\msun\/~/~(K~km~s$^{-1}$~pc$^2$)
on the CO side, and an interstellar medium OH abundance 
$X_{\rm OH} = 2.5 \times 10^{-6}$ on the OH side. A  consistent lower 
limit on $X_{\rm OH}$ has been obtained by \citet{stone18}. 
The good agreement can be regarded as support of the adopted \alpco\/, 
but see also the discussion below in the context of outflow sizes.

Outflow velocities are in excellent agreement for these bright and 
well-studied galaxies (top center), with mean  
log$(v_{\rm out,OH}/v_{\rm out,CO})= -0.03$~dex and dispersion 0.08~dex.

Larger but still modest differences are found 
for the outflow radii. Outflow radii from the OH models are around 0.3~kpc
for all ULIRGs, while the CO observations vary more widely and detect 
radii up to several kpc.
In the mean,  log$(R_{\rm out,OH}/R_{\rm out,CO})= -0.54$~dex 
with dispersion 0.39~dex. Similar 
$\approx$0.5~dex differences are found for the outflow rates for mass, 
momentum, and mechanical energy (bottom panels of Fig.~\ref{fig:ga17}).
While the OH radii are based on spherically symmetric models and the 
CO radii measured, it is improbable to ascribe all differences to 
modeling uncertainties. This is in particular the case given the observed 
OH absorptions in the 84~\mum\ and/or 65~\mum\ transitions that are 
radiatively excited by intense FIR radiation fields close to an optically 
thick and necessarily small FIR source
\citep{gonzalez-alfonso17}. Breakup into several individually optically thick 
sources could enlarge the sizes but is not supported by the usually compact 
structure of high resolution (sub)mm images \citep[e.g.,][]{barcos-munoz17}.

Assuming that much of the size difference between OH and CO outflow 
(top right of Fig.~\ref{fig:ga17}) is real, one is led to a 
scenario where a more compact and embedded molecular outflow region 
traced by the OH absorptions connects to a larger and older component that can 
still be traced by the collisionally excited CO emission. In contrast, the
covering by clumps absorbing OH will decrease with outward
motion. In that case,
the good agreement of outflow masses (top left of Fig.~\ref{fig:ga17}) will be 
partly fortuitous or erroneous since the two molecules trace different 
though overlapping
regions. A further factor in this comparison arises from the definition 
of outflow masses. OH-based outflow masses from \citet{gonzalez-alfonso17}
reflect all directions of a spherically symmetric outflow. 
Emission line components of such a flow that are moving close to the plane 
of sky blend with 
the bright core of the line and are excluded by our definition of 
\sco\/~(wings), see also Fig.~\ref{fig:outflowgauss_visu}. Toy model 
calculations for a spherically symmetric case suggest \sco\/~(wings) would
measure a mass 2--3 times smaller than the 4$\pi$ outflow, but this correction
is obviously dependent on outflow velocity, width of the line core, and 
may be an upper limit if replacing the spherically symmetric flow with a wide 
angle flow with large line of sight component.   

A non-unique scenario that further reduces the small tensions 
in Fig.~\ref{fig:ga17} between OH- and CO-based results  can be described 
as follows:
(i) An outflow that is constant in rate and velocity over the flow time
derived from CO data is sampled in its entirety by CO emission, and in its 
inner half or somewhat less by the OH absorptions. 
The molecular outflow masses traced by CO and by OH should reflect the 
different fractions of the flow that are sampled -- the CO-based 
molecular outflow mass should be larger than the OH-based one. 
(ii) The similar actually derived outflow masses (Fig.~\ref{fig:ga17}) 
then must be due to an additional correction. Depending on 
actual outflow geometry, this 
could be achieved by considering the fraction of an about spherically 
symmetric CO outflow that is missed by the emission line wings (see above). 
Alternatively, increased \alpco\ or increased OH  abundance 
(decreased molecular hydrogen mass traced 
by OH) could help achieving an appropriate ratio of masses traced by CO 
and by OH. (iii) Either of the corrections described in (ii) would also 
bring the outflow rates in the bottom panels of Fig.~\ref{fig:ga17} closer to 
agreement, consistent with the assumed constant outflow rate.
   
In summary, the comparisons provide a very successful cross-validation
of the OH P-Cygni and CO interferometric methods of characterizing
molecular outflows. There is no indication for a need to modify
the methods and conversions adopted for CO outflows in 
Sect.~\ref{sect:outprop} and \ref{sect:derivprop}, nor for modifying
the OH modeling methods of \citet{gonzalez-alfonso17}. At least for 
ULIRG-like systems which constitute the bulk of current outflow samples 
and definitely of the comparisons above, the comparison also backs the large 
molecular outflow covering factor reported in OH outflow studies. 
This finding is originally based on the high incidence of P-Cygni absorptions 
in ULIRGs \citep[e.g.,][]{sturm11,veilleux13}, and is supported by the model
results of \citet{gonzalez-alfonso17}. Current knowledge of outflow 
geometries and histories for large samples is however insufficient to 
clarify whether remaining modest differences between OH- and CO-based 
results can find a purely geometrical explanation.
Use of the time-efficient
OH spectroscopy for high-z outflow searches is encouraged, be it 
with \alma\ \citep{spilker18,herrera-camus19b} or with planned future far-infrared/submm 
space facilities \citep[e.g., SPICA,][]{roelfsema18}.   

\subsection{Relation of outflow properties to FIR surface 
brightness, AGN luminosity, and bolometric luminosity}

Several recent papers have studied scaling relations of molecular outflow 
masses, 
mass outflow rates, momentum flows and energy flows with star formation 
rate, AGN luminosity and bolometric luminosity of the hosting galaxies 
\citep{cicone14,fiore17,gonzalez-alfonso17,fluetsch19}, and to what extent 
the flows may be momentum or energy driven. We here focus on the additional
comparison with far-infrared surface brightness, and briefly revisit the
relations to AGN and bolometric luminosity. Table~\ref{tab:multi} summarizes 
these and related properties for the combined sample.
As far as other quantities are concerned, our sample concurs with 
observational results from these references with modest modifications due 
to different assumptions (e.g. C = 1 or 3 in
Eqn.~\ref{eq:mdot}, outflow line fluxes corresponding only to wings or to 
entire broad components). One 
should note that all these studies have overlapping samples, since
they are enlarging their original samples from the literature 
(Sect.~\ref{sect:literature} for this paper), or are entirely literature
based.

\longtab{
\begin{landscape}
\begin{longtable}{lrrrrrrrrrl}
\\ \hline
\\
\caption{Other properties of combined sample}\\ \hline
\\
Source           &D$_{\rm L}$&log(\mout\/)&\tflow&
$\dot{M}_{\rm out}$&${\rm log} (P_{\rm kin})$&${\rm log} (\dot{M}_{\rm out}v)$&
log($\Sigma_{\rm FIR}$)&log($L_{\rm Bol}$)&log($L_{\rm AGN}$)&Ref. (AGN)\\ 
                 &Mpc        &\msun       &Myr&
\msun\ yr$^{-1}$   &erg s$^{-1}$          &g cm s$^{-2}$           &
\lsun\ kpc$^{-2}$ &\lsun    &\lsun    &\\ 
\\ \hline
\endfirsthead
\caption{continued.}\\
Source           &D$_{\rm L}$&log(\mout\/)&\tflow&
$\dot{M}_{\rm out}$&${\rm log} (P_{\rm kin})$&${\rm log} (\dot{M}_{\rm out}v)$&
log($\Sigma_{\rm FIR}$)&log($L_{\rm Bol}$)&log($L_{\rm AGN}$)&Ref. (AGN)\\ 
\\
\endhead
\\
NGC~253               &  3.40&   7.20& 4.30&  3.71&39.47&33.07&   12.14&10.58& $<$7.07&\citet{cicone14}\\
I~Zw~1                &273.98&$<$7.62&     &      &     &     &   10.87&12.01&   11.96&\citet{veilleux13}\\ 
III~Zw~035            &119.91&   8.59& 6.88& 56.70&42.24&35.05&   11.81&11.71& $<$7.76&\citet{gonzalez-martin15}\\
PG~0157+001           &781.80&   7.99& 1.54& 63.81&42.51&35.21&$>$11.18&12.68&   12.50&\citet{veilleux13}\\ 
NGC~1068              & 14.40&   6.93& 1.02&  8.27&40.42&33.72&   11.36&11.37&   10.15&\citet{ricci17}\\
NGC~1266              & 29.90&   7.67& 1.21& 38.31&42.10&34.89&$>$11.89&10.54&    9.73&\citet{cicone14}\\
IRAS~F03158+4227      &632.90&   8.88& 6.92&110.83&43.43&35.79&        &12.68&   12.36&\citet{fluetsch19}\\
NGC~1377              & 21.00&   6.88& 1.42&  5.36&40.51&33.67&$>$11.49&10.09&    9.35&\citet{cicone14}\\
NGC~1433              &  9.90&   5.82& 0.94&  0.70&39.34&32.64&        & 9.24&    8.32&\citet{fluetsch19}\\
NGC~1614              & 68.88&   7.51& 1.52& 21.33&41.94&34.68&   11.21&11.73& $<$8.94&\citet{brightman11}\\
IRAS~F05081+7936      &239.20&   7.98& 1.22& 28.23&43.20&35.60&   10.61&12.05& $<$8.79&\citet{fluetsch19}\\
IRAS~F05189-2524      &189.47&   7.87& 2.30& 32.56&42.31&34.96&   11.78&12.22&   12.09&\citet{veilleux13}\\ 
NGC~2146              & 17.20&   7.94&10.20&  8.49&41.03&34.03&   10.46&11.16& $<$7.51&\citet{cicone14}\\
NGC~2623              & 80.03&   7.05& 1.01& 11.21&42.09&34.62&   11.93&11.62&   10.62&\citet{evans08}\\ 
NGC~2623 fossil       &      &   6.66&12.80&  0.36&40.26&32.96&&&&\\
IRAS~F08572+3915      &260.32&   8.56& 1.00&363.80&43.87&36.26&   11.73&12.20&   12.06&\citet{veilleux13}\\ 
IRAS~F09111-1007W     &242.11&   8.01& 1.81& 57.12&42.29&35.08&   11.64&12.01&$<$10.80&\citet{nardini10}\\ 
M~82                  &  3.90&   7.86& 9.78&  7.34&40.36&33.67&   11.08&10.90& $<$7.96&\citet{cicone14}\\
IRAS~F10035+4852      &291.20&$<$8.45&     &      &     &     &        &12.06&$<$11.52&\citet{fluetsch19}\\
IRAS~F10173+0828      &217.64&$<$7.29&     &      &     &     &   11.77&11.90& $<$8.18&\citet{iwasawa11}\\
NGC~3256 N            & 35.00&   7.26& 0.57& 31.93&42.69&35.15&   10.79&11.60& $<$8.39&\citet{cicone14}\\
NGC~3256 S            &      &   7.13& 1.33& 10.11&41.90&34.50&\\          
IRAS~F10565+2448      &190.70&   8.34& 2.39& 91.90&42.77&35.42&   11.54&12.12&   11.79&\citet{veilleux13}\\ 
IRAS~F11119+3257      &920.20&   8.75& 6.85& 81.43&43.41&35.71&        &12.72&   12.63&\citet{veilleux13}\\ 
NGC~3628              &  7.70&   6.98& 4.46&  2.13&39.74&33.08&   11.13&10.08& $<$7.21&\citet{cicone14}\\
ESO~320-G030          & 46.60&   6.81& 2.36&  2.76&41.00&33.77&   11.28&11.33& $<$9.33&\citet{alonso-herrero12}\\
IRAS~F12112+0305 SW   &330.34&   7.53& 2.31& 14.62&41.68&34.47&        &12.38&   11.64&\citet{veilleux13}\\ 
IRAS~F12112+0305 NE   &      &   8.80& 3.36&188.41&43.11&35.74&&&&\\
IRAS~F12224-0624      &115.71&   6.95& 0.46& 19.42&42.29&34.84&   11.83&11.35&$<$10.75&This work\\
NGC~4418              & 25.35&   5.73& 0.09&  6.29&41.32&34.11&   12.50&10.94&   10.68&\citet{veilleux13}\\ 
Mrk~231               &186.42&   8.43& 0.84&323.84&43.70&36.15&$>$12.26&12.61&   12.51&\citet{veilleux13}\\ 
IC~860                & 56.11&$<$6.55&     &      &     &     &   12.07&11.19& $<$9.19&\citet{alonso-herrero12}\\ 
IRAS~13120-5453       &136.14&   8.18& 1.35&111.53&42.50&35.33&   11.99&12.33&   11.85&\citet{veilleux13}\\ 
Mrk~273               &166.43&   8.21& 0.83&196.30&43.42&35.91&$>$12.13&12.24&   11.77&\citet{veilleux13}\\ 
4C~12.50              &568.50&$<$8.11&     &      &     &     &        &12.41&   12.11&\citet{fluetsch19}\\
SDSS~J135646.10+102609.0&574.70& 8.09& 0.59&207.49&43.21&35.82&        &12.43&   12.41&\citet{sun14}\\
IRAS~F14348-1447 SW   &375.26&   8.76& 3.80&151.34&42.92&35.60&        &12.42&   11.66&\citet{veilleux13}\\ 
IRAS~F14348-1447 NE   &      &   8.05& 2.44& 45.91&42.30&35.03&&&&\\
IRAS~F14378-3651      &306.70&$<$7.87&     &      &     &     &   11.81&12.26&   11.59&\citet{veilleux13}\\ 
Zw~049.057            & 56.40&   6.96& 0.26& 35.09&42.00&34.82&   11.90&11.26& $<$9.26&\citet{alonso-herrero12}\\ 
Arp~220               & 78.70&   7.02& 0.19& 54.93&42.40&35.12&   12.53&12.24&   11.01&\citet{veilleux13}\\ 
Mrk~876               &605.21&   9.38& 3.31&720.05&44.40&36.68&   10.70&12.30&   12.17&\citet{ricci17}\\
NGC~6240              &106.81&   8.59& 1.43&272.00&43.24&35.89&   11.28&11.91&   11.66&\citet{veilleux13}\\ 
IRAS~F17020+4544      &270.52&   8.27& 0.59&315.95&44.44&36.52&        &11.61&   11.13&\citet{giroletti17}\\
IRAS~F17132+5313      &226.60&$<$8.16&     &      &     &     &        &12.00&    8.70&\citet{iwasawa11}\\
IRAS~F17207-0014      &189.20&   7.52& 3.33& 98.96&43.59&35.84&   11.97&12.48&$<$11.18&\citet{veilleux13}\\ 
PDS~456               &898.60&   8.61& 3.37&120.03&43.39&35.83&        &13.42&   13.41&\citet{bischetti19}\\ 
NGC~6764              & 32.00&   6.44&20.00&  0.14&38.09&31.67&        &10.46&    8.65&\citet{cicone14}\\
IRAS~F20100-4156      &608.20&   8.86& 2.43&295.08&43.90&36.24&        &12.73&   12.16&\citet{spoon13}\\ 
IC~5063               & 48.71&   7.23& 0.90& 18.61&42.17&34.77&   10.24&10.92&   10.84&\citet{ricci17}\\
IRAS~F20551-4250      &190.20&   7.43& 0.46& 59.44&42.65&35.26&   11.64&12.11&   11.87&\citet{veilleux13}\\ 
IRAS~F22491-1808      &352.74&   8.13& 1.55& 86.73&42.46&35.25&   11.26&12.26&   11.42&\citet{veilleux13}\\ 
NGC~7479              & 34.31&$<$6.22&     &      &     &     &   11.31&10.89&    9.95&\citet{ricci17}\\
IRAS~F23060+0505      &834.25&   9.42& 5.66&463.76&43.85&36.31&        &12.59&   12.48&\citet{cicone14}\\
IRAS~23365+3604       &289.17&   8.12& 2.67& 49.46&42.50&35.15&   11.65&12.27&   11.92&\citet{veilleux13}\\ 
\\ \hline
\\
\footnote{Molecular outflow masses, flow times and outflow rates for the combined
sample, as computed using the 
observed quantities from Tables~\ref{tab:outobs} and \ref{tab:litobs}. We also list adopted distances, 
far-infrared surface brightnesses \citep[from][]{lutz16,lutz18}, bolometric and AGN luminosities. 
Some references quote AGN fractions to the bolometric luminosity which we
have converted to $L_{\rm AGN}$ for consistency. The limit for 
IRAS~F12224-0624 was estimated from Spitzer-IRS data with methods as used
by \citet{veilleux13}. Note that AGN luminosities are a mix of X-ray-based 
(reacting rapidly to accretion variations) and mid-infrared-based (averaging
over typically tens of years).} 
\label{tab:multi}
\end{longtable}
\end{landscape}
}

\begin{figure*}
\includegraphics[width=0.5\hsize]{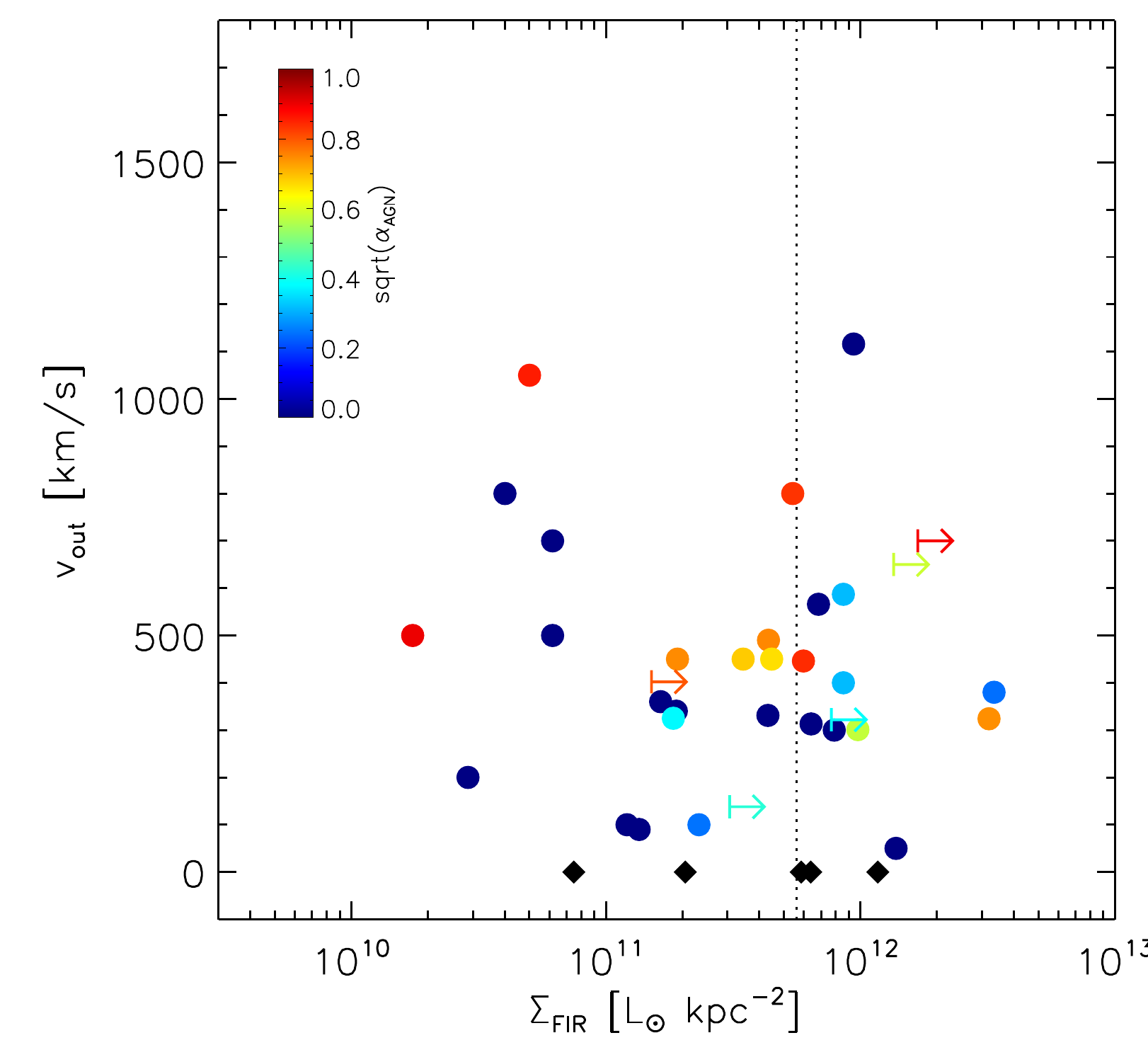}
\includegraphics[width=0.5\hsize]{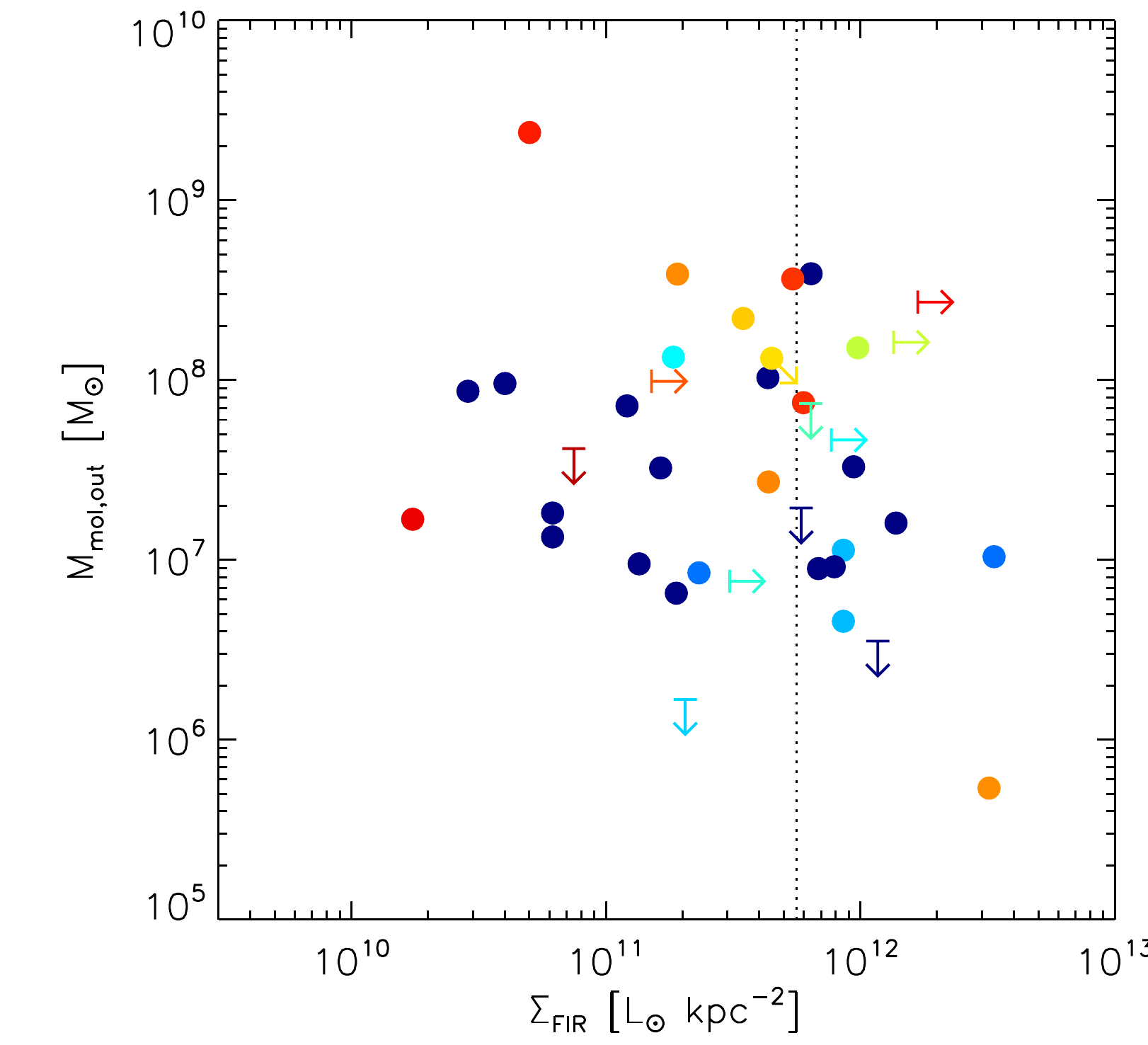}
\includegraphics[width=0.5\hsize]{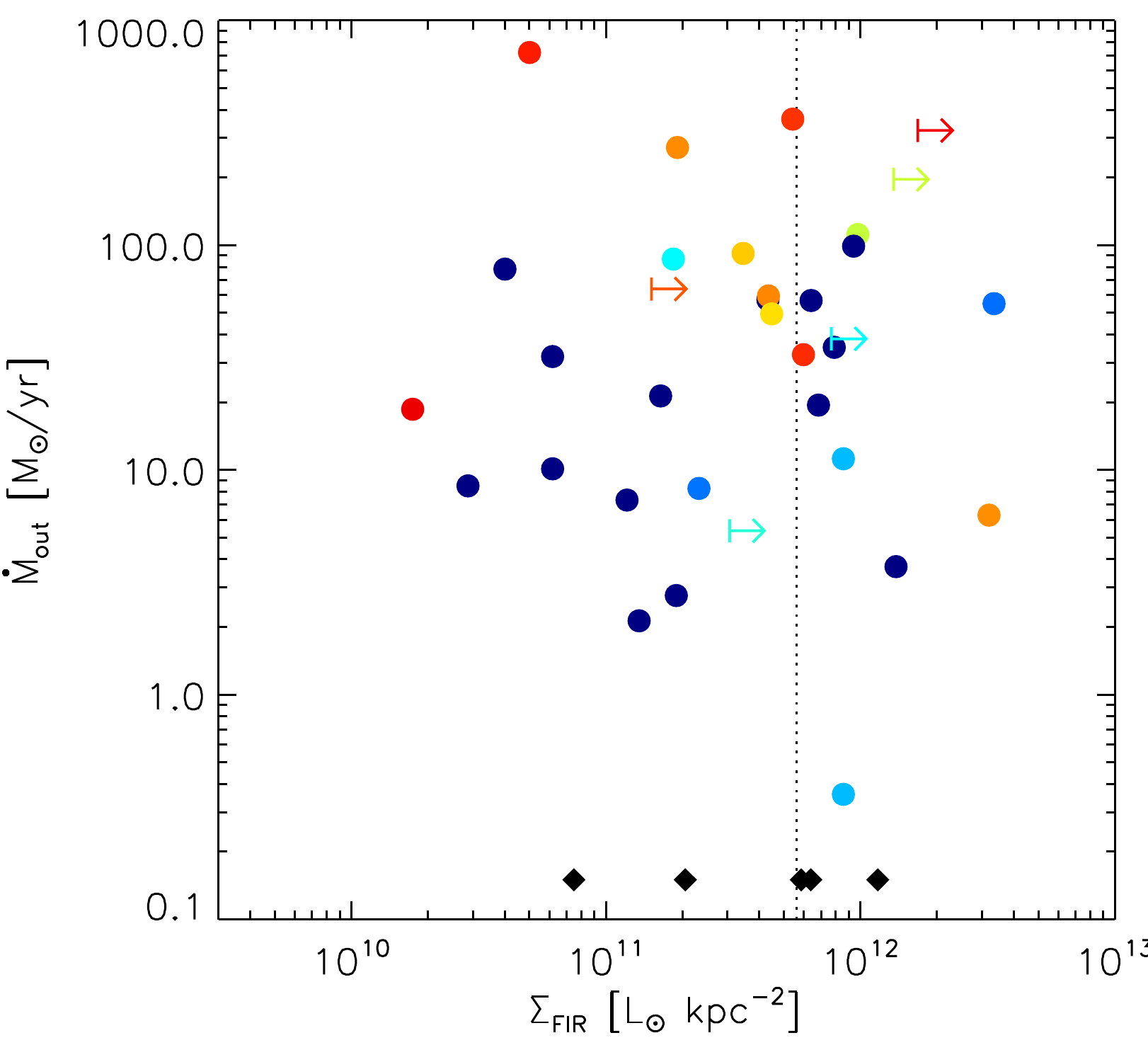}
\includegraphics[width=0.5\hsize]{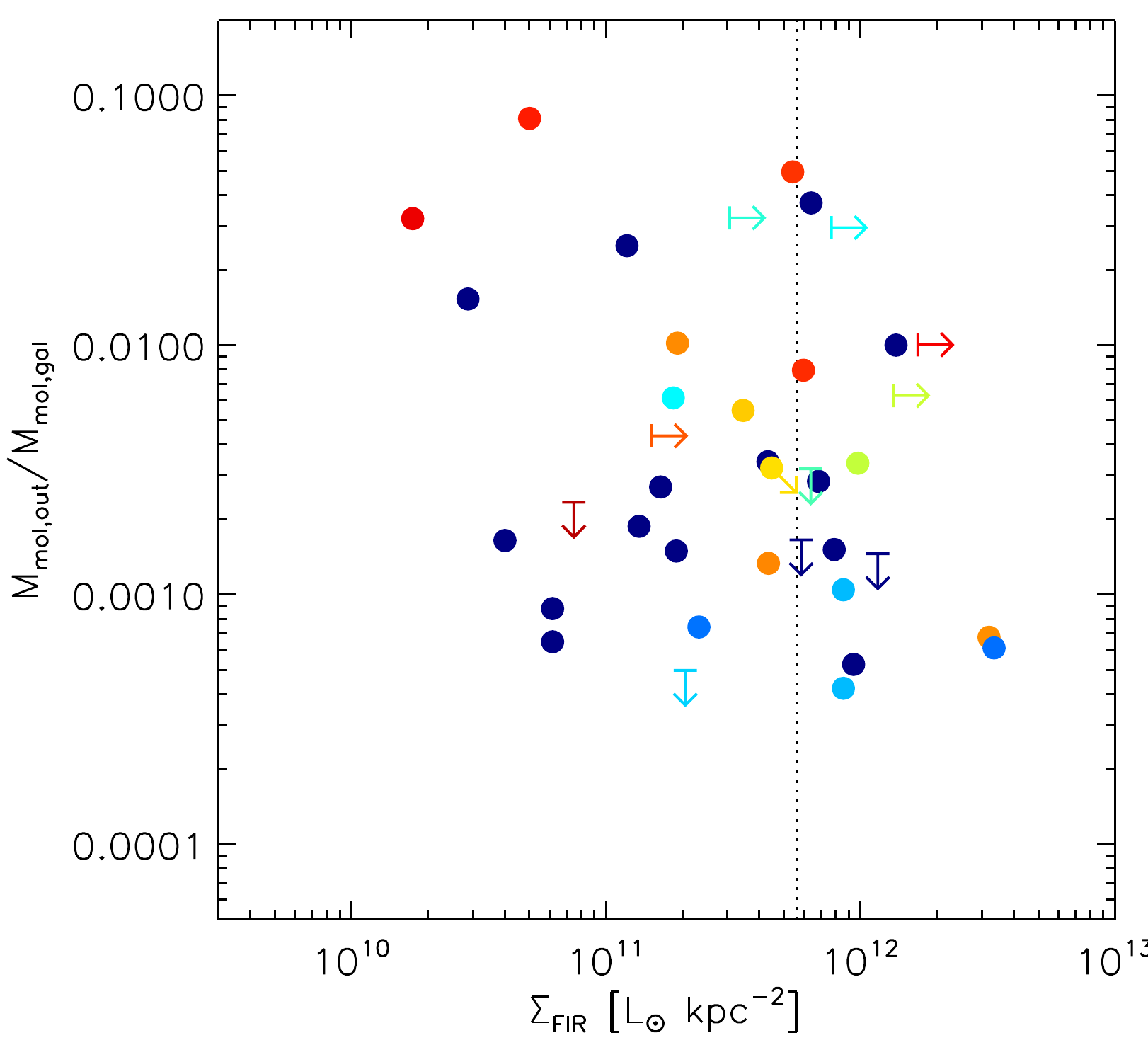}
\caption{Relation of outflow properties to far-infrared surface brightness.
The AGN contribution to the bolometric luminosity is color-coded on a 
square root scale between 0 (dark blue) and 1 (dark red).
Black symbols in the left panels, plotted at an arbitrary low value, refer
to sources lacking an outflow detection (upper limits in the right panels).
High far-infrared surface brightness sources from Table~1 of \citet{lutz16}
are at log(\sigfir\/) $\gtrsim$~11.75~\lsun\/~kpc$^{-2}$ 
(vertical dotted lines). CO observations are close to complete here (17/18),
unlike the incomplete literature results for lower surface brightness.}
\label{fig:sigfir}
\end{figure*}

Fig.~\ref{fig:sigfir} shows outflow velocity, outflow molecular mass, 
mass outflow rate, and the ratio of outflow and host molecular gas mass 
as a function of far-infrared surface brightness for 39 sources in our 
combined sample, which have measurements or limits for far-infrared surface 
brightness \citep{lutz16,lutz18}. None of these quantities shows
a significant trend with far-infrared surface brightness, but see the note 
below on selection effects at low surface brightness. 
If we first focus on the high surface brightness end with 
log(\sigfir )$\gtrsim$11.75~\lsun\/~kpc$^{-2}$, we indeed find
the expected very high incidence of molecular outflows that motivated our
observations. Of the 18 sources 
with log(\sigfir )$\geq$11.75~\lsun\ kpc$^{-2}$ and/or close to optically 
thick FIR emission that are listed in
Table~1 of \citet{lutz16}, all but ESO~173-G015 have deep CO observations
probing for outflows, from our own data or from the literature in the combined
sample. Of these 17 objects, CO outflows have been detected in 13 sources 
while upper limits have been derived for IRAS~F10173, IC~860, IRAS~F14378
and NGC~7479. In several cases, these limits are conservative and 
a contribution of an outflow to the observed moderate velocity line wings 
remains a possibility.
As expected, high far-infrared surface brightness galaxies
are excellent hunting grounds for molecular outflows. This is reflected 
in the high detection rate, at least 13/18 sources. On the other hand,
the cases with sensitive upper limits demonstrate that far-infrared
surface brightness alone is not sufficient to produce strong and fast
outflows in {\em all} cases, despite strongly exceeding SFR thresholds above 
which outflows from galaxy disks are prevalent 
\citep{heckman02,newman12,davies19}. 

Given its selection above an already high surface brightness threshold 
and the modest 
sample size, it is not surprising that no trends of outflow properties 
with surface brightness are detected within this high surface brightness
sample. More noteworthy
is the absence of trends in Fig.~\ref{fig:sigfir} over the full range of 
the combined sample, which extends to more than an order of magnitude
lower surface brightness. A high incidence of molecular
outflows is found also among the low surface brightness
members of this sample. Severe and poorly known selection effects
may, however, be at work here. Much of the molecular outflow literature is
in papers studying single or very few sources, and selection 
criteria are diverse. Upper limits may often remain unpublished. We 
certainly cannot exclude
abundant existence of low surface brightness galaxies that have only weak or 
no molecular outflows. An obvious speculation is that the molecular outflow 
detection rate is smaller in this regime than for the high surface 
brightness sample. But unbiased studies of larger samples are needed
to establish such a difference, or detect any trends in outflow properties 
over a large range of FIR surface brightness.

\begin{figure*}
\includegraphics[width=0.5\hsize]{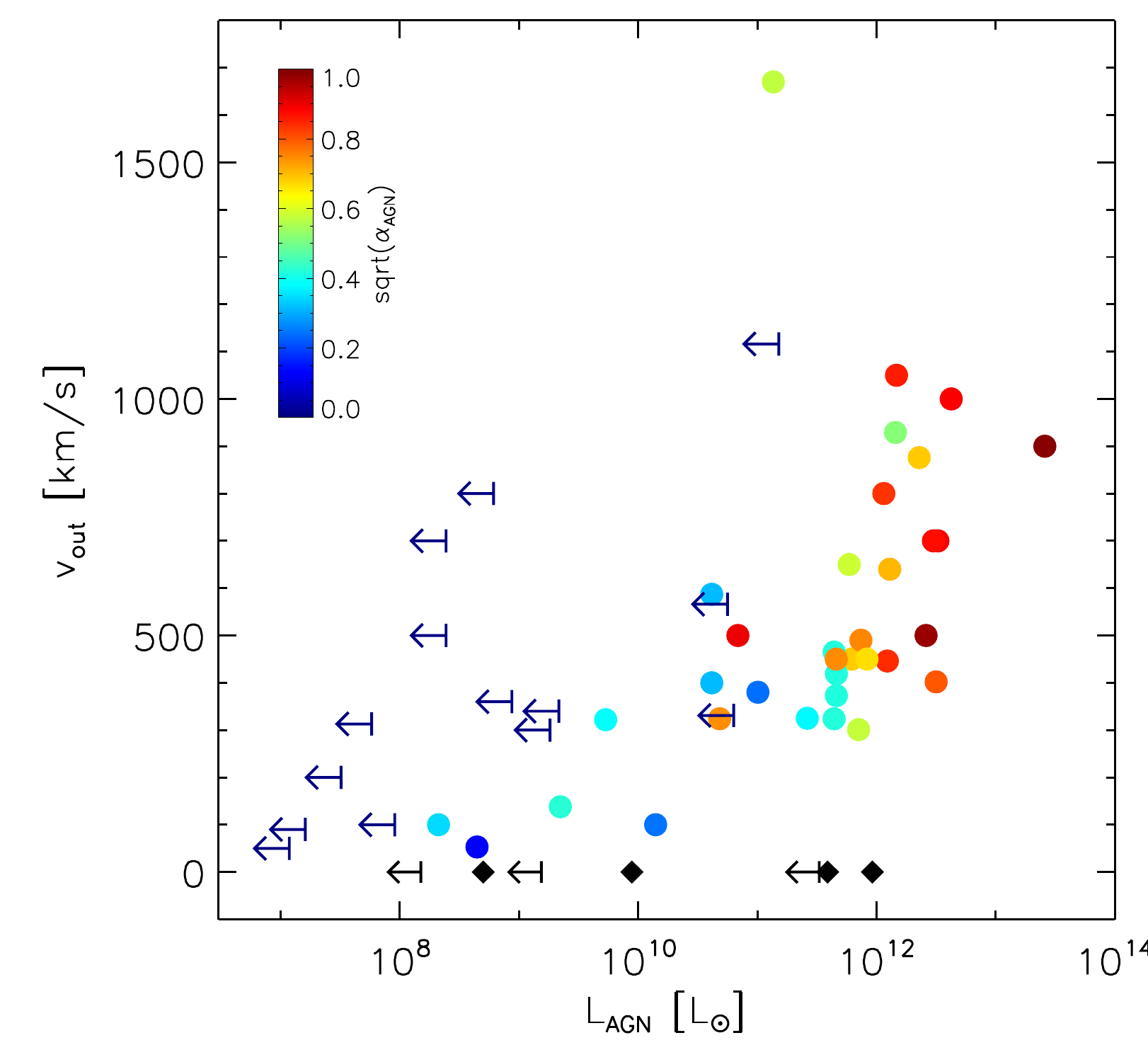}
\includegraphics[width=0.5\hsize]{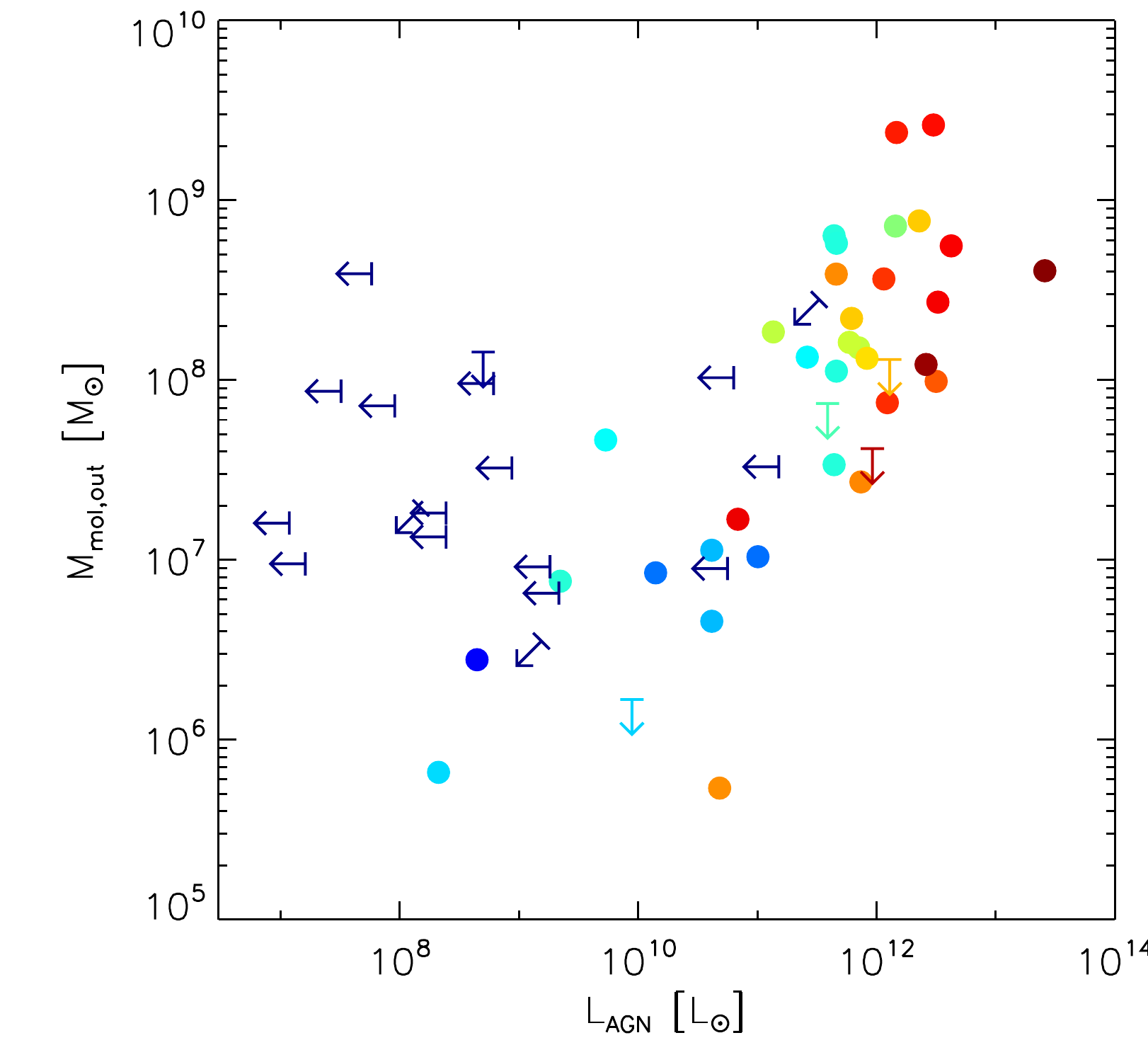}
\includegraphics[width=0.5\hsize]{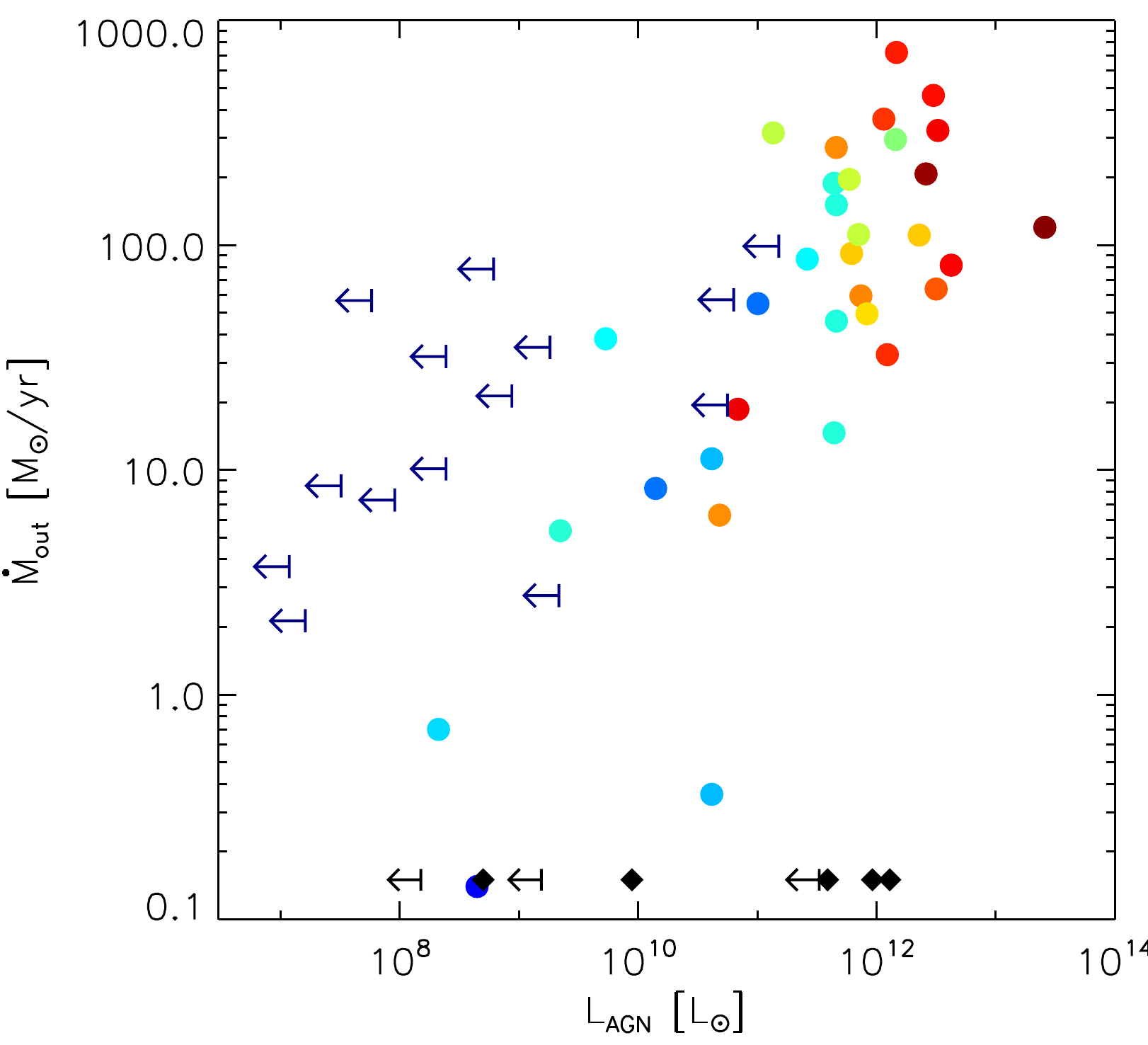}
\includegraphics[width=0.5\hsize]{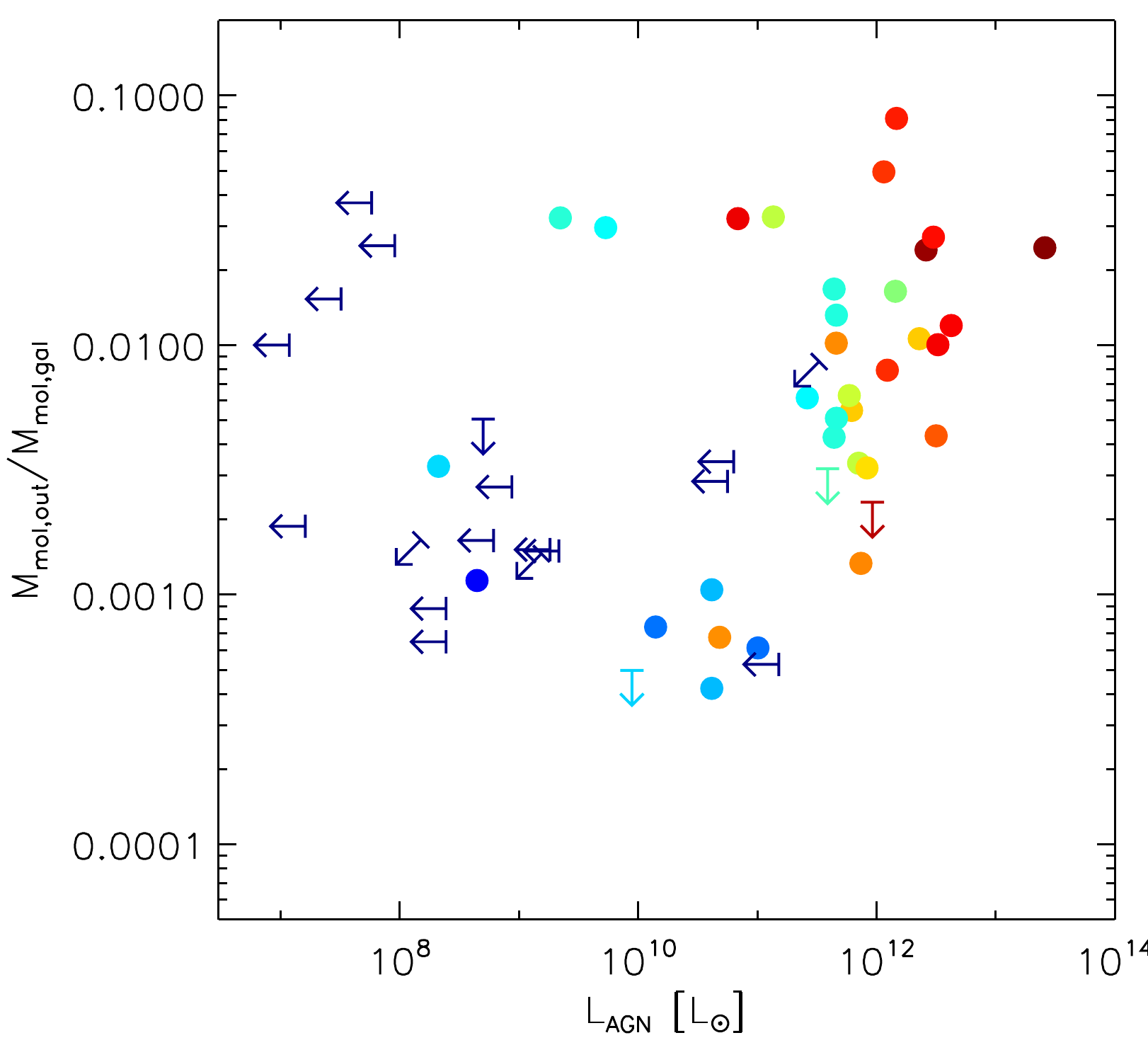}
\caption{Relation of outflow properties to AGN luminosity. The AGN 
contribution to the bolometric luminosity is color-coded on a square root 
scale between 0 (dark blue) and 1 (dark red). Black symbols 
in the left panels, plotted at an arbitrary low value, refer
to sources lacking an outflow detection (upper limits in the right panels).}
\label{fig:lagn}
\end{figure*}

\begin{figure*}
\includegraphics[width=0.5\hsize]{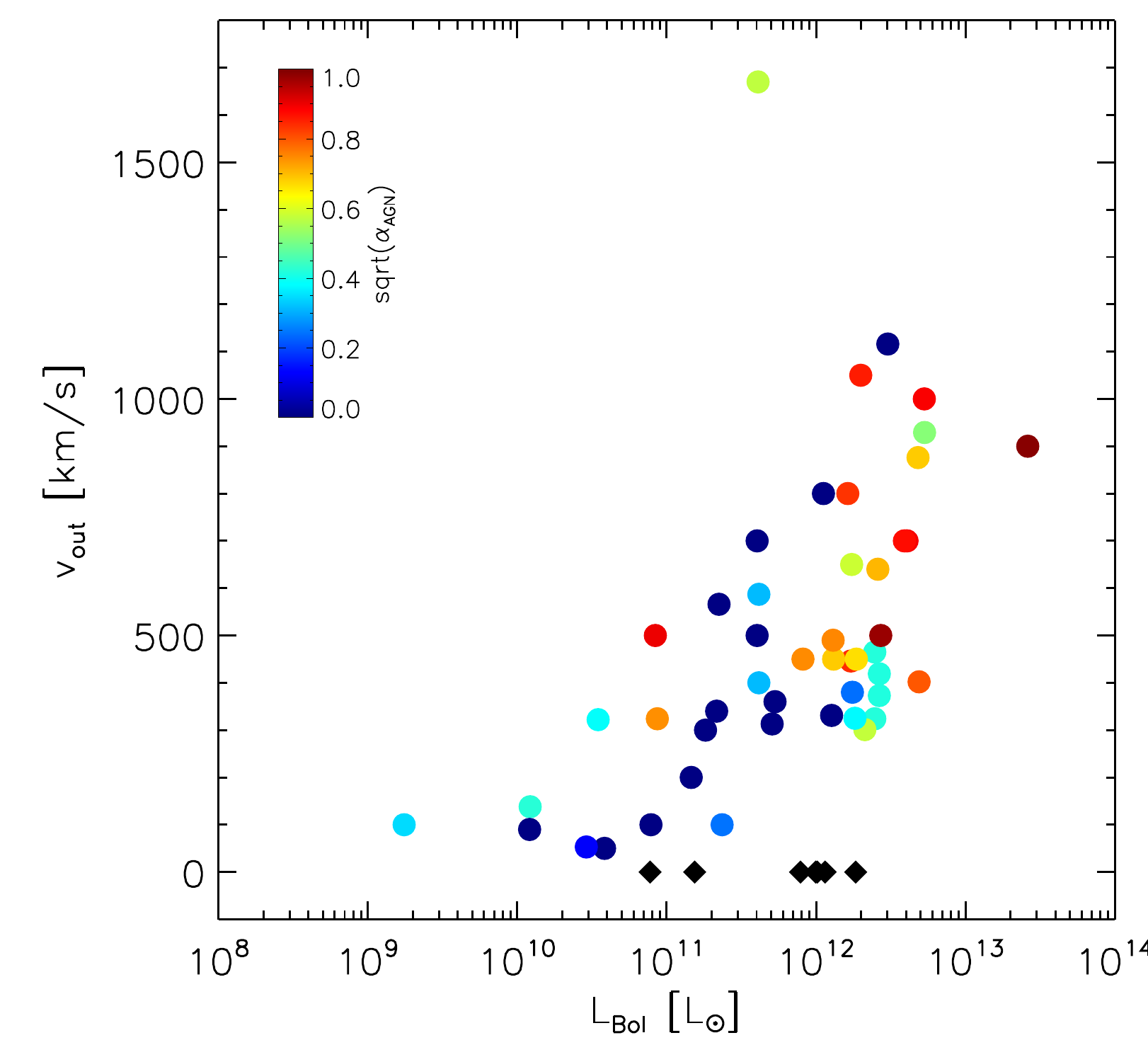}
\includegraphics[width=0.5\hsize]{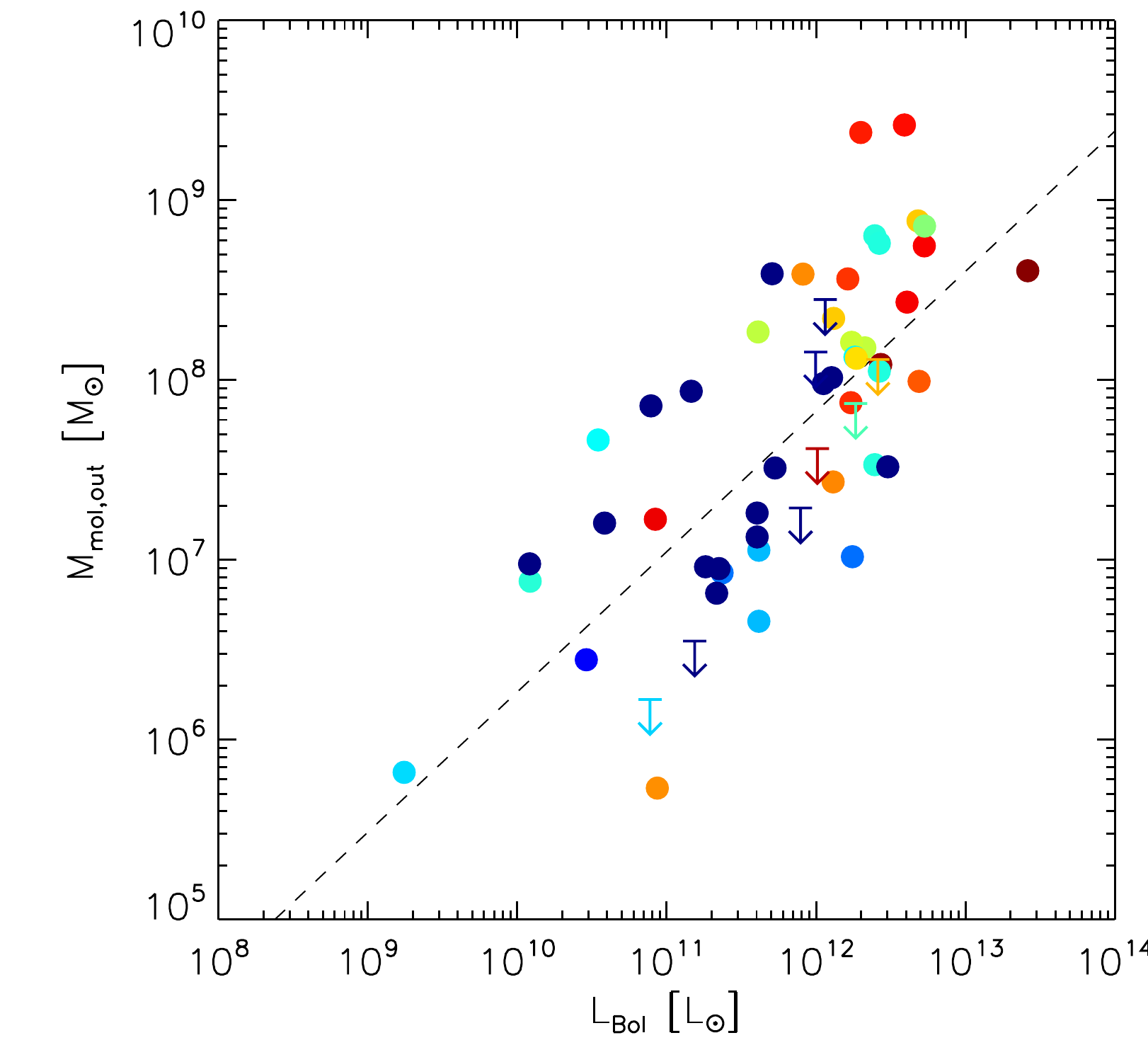}
\includegraphics[width=0.5\hsize]{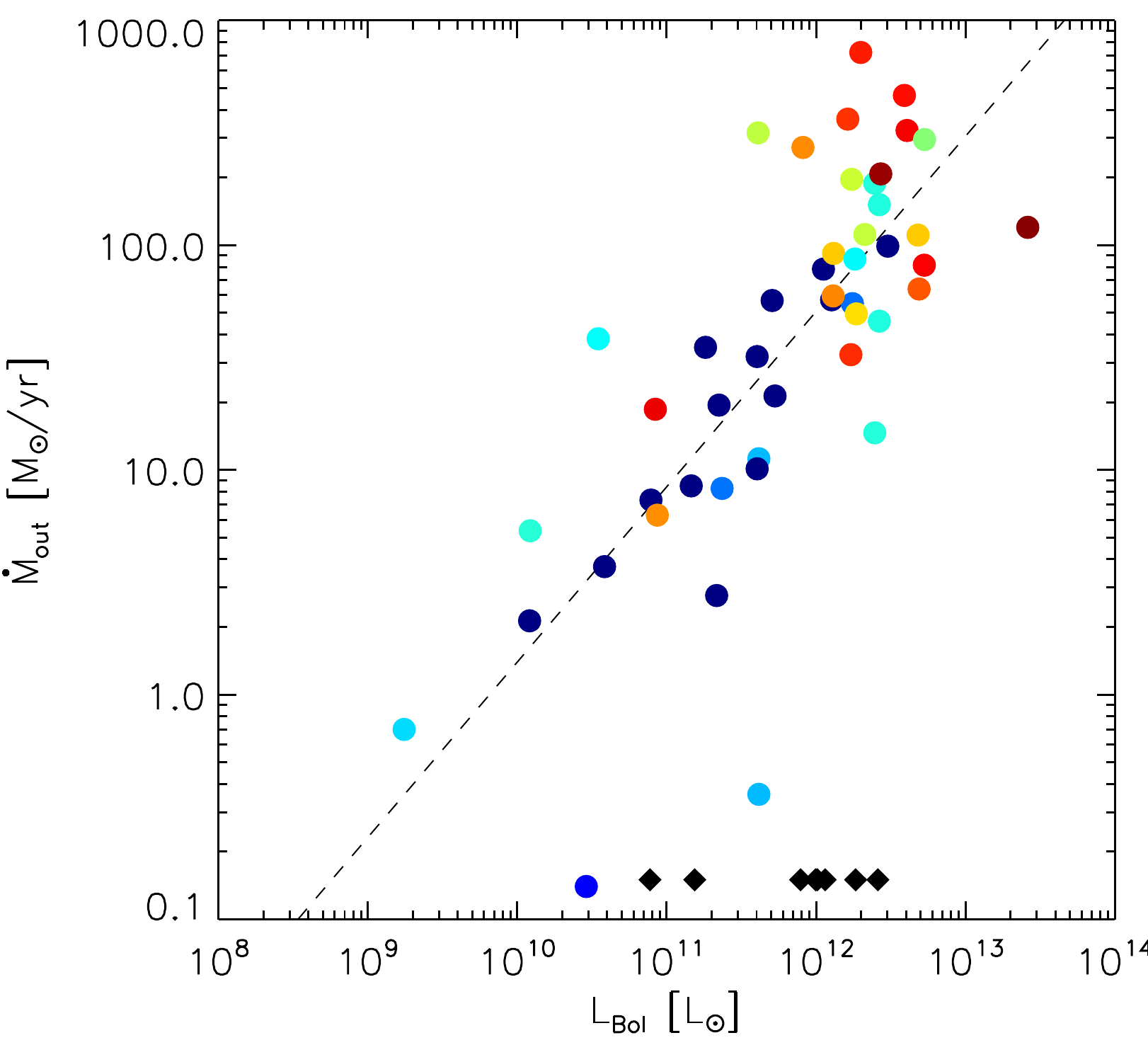}
\includegraphics[width=0.5\hsize]{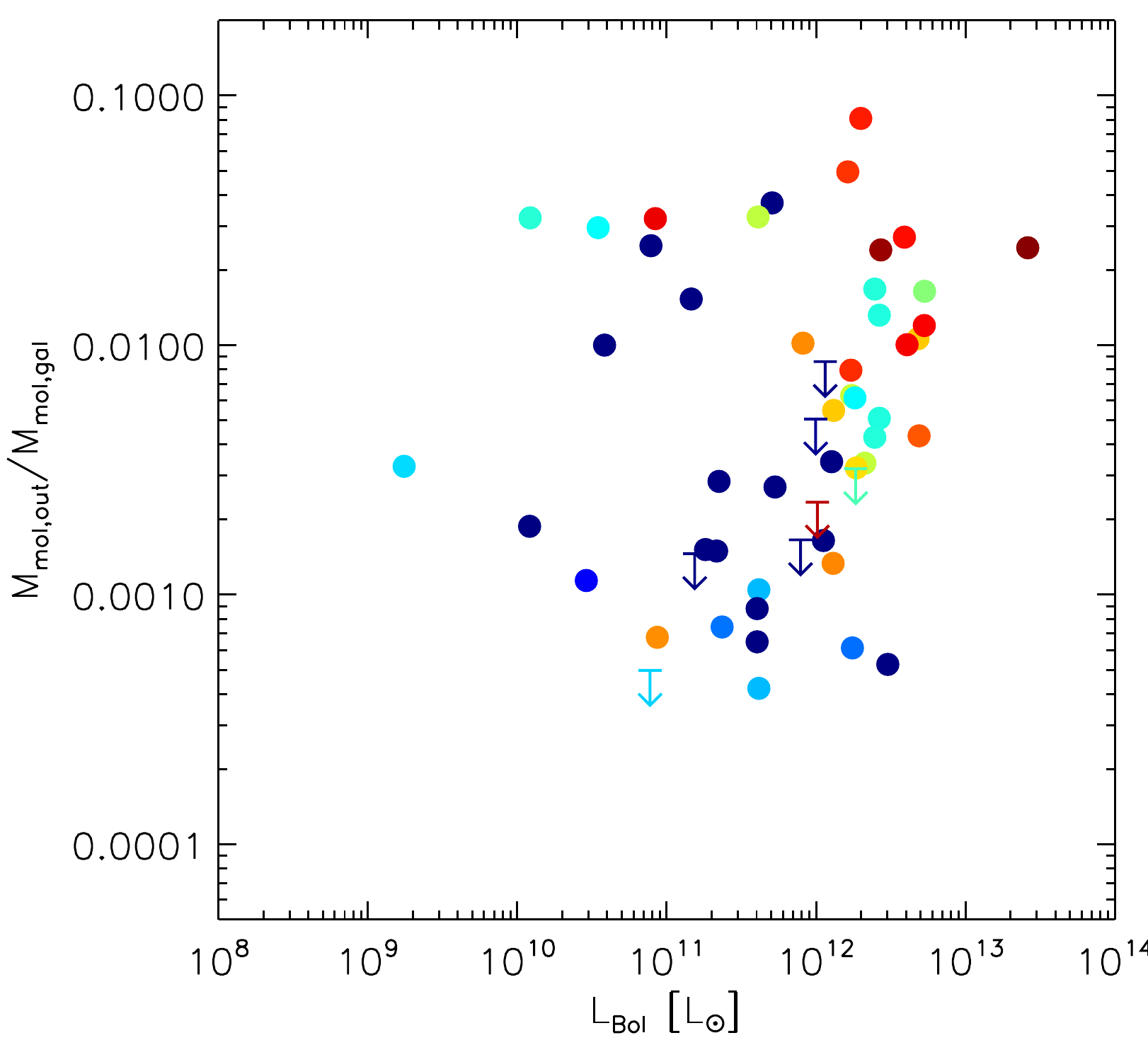}
\caption{Relation of outflow properties to bolometric luminosity. The AGN 
contribution to the bolometric luminosity is color-coded on a square root 
scale between 0 (dark blue) and 1 (dark red). Black diamonds 
in the left panels, plotted at an arbitrary low value, refer
to sources lacking an outflow detection (upper limits in the right panels).
Dashed lines reflect the fitting relations from Eqns.~\ref{eq:moutlbol}
and \ref{eq:mdotlbol}}
\label{fig:lbol}
\end{figure*}

\begin{figure*}
\includegraphics[width=0.5\hsize]{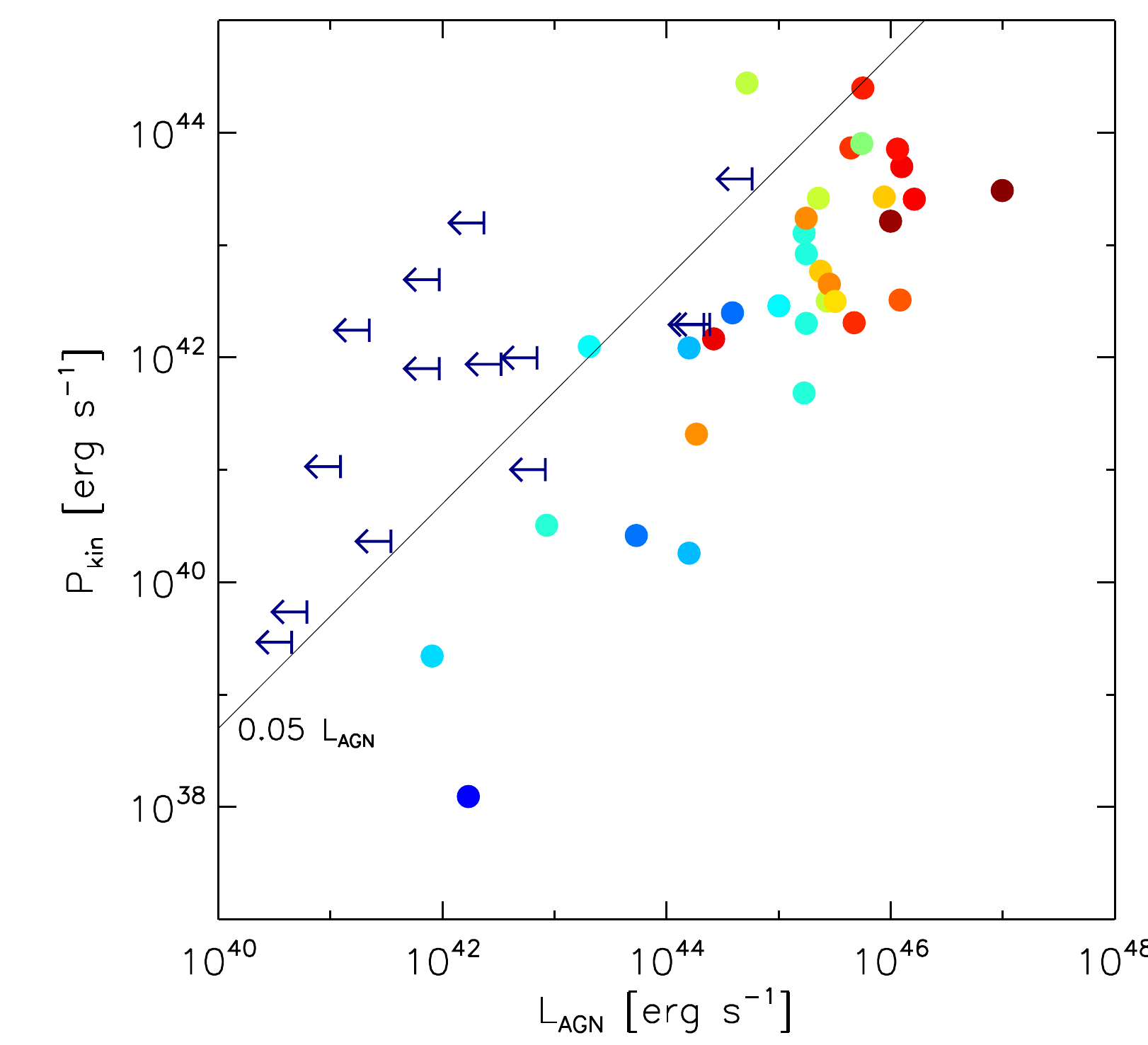}
\includegraphics[width=0.5\hsize]{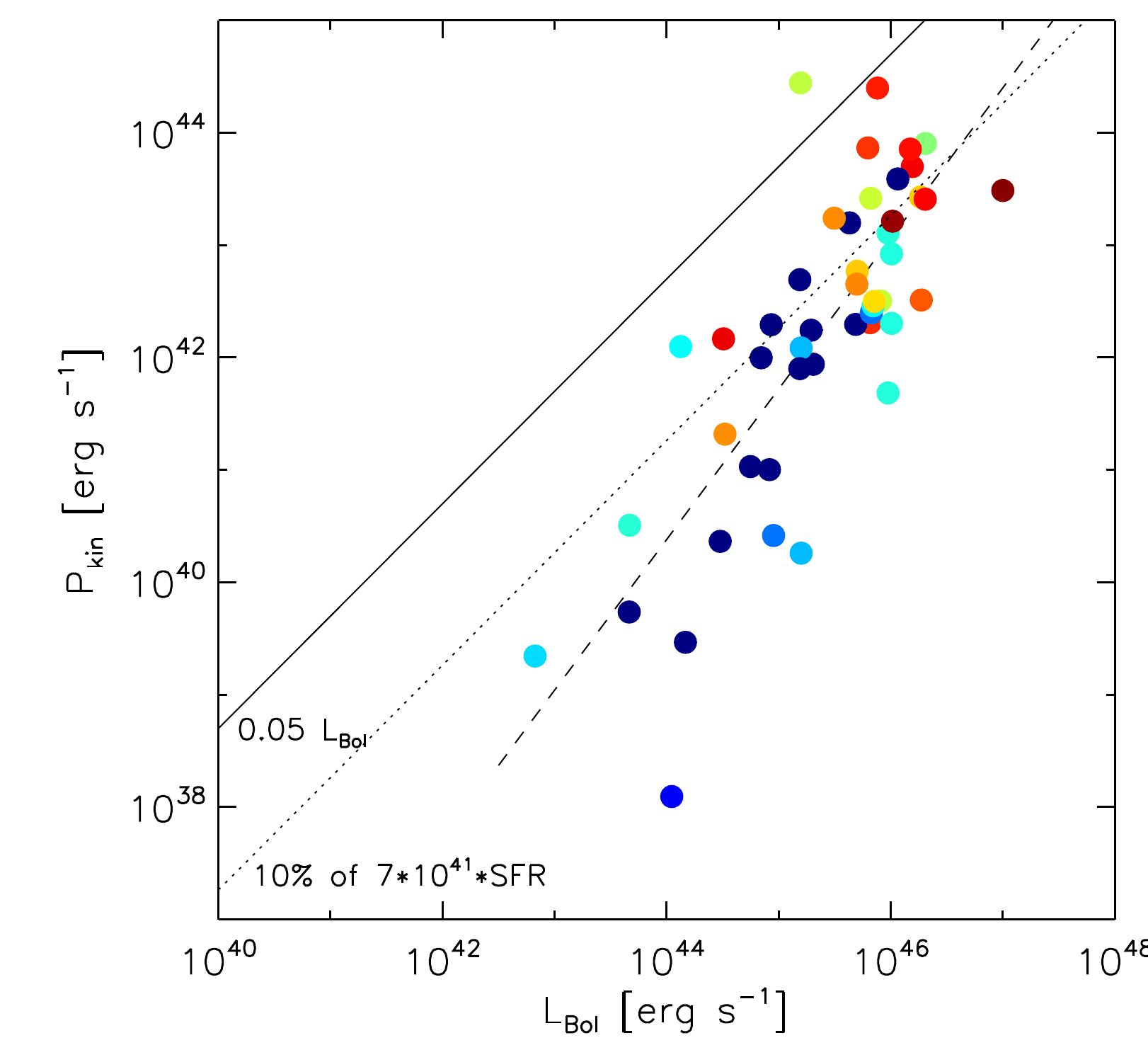}
\caption{Relation of outflow kinetic power to AGN luminosity (left) and
bolometric luminosity (right). The AGN 
contribution to the bolometric luminosity is color-coded on a square root 
scale between 0 (dark blue) and 1 (dark red). The continuous line indicates 
coupling of 5\%\ of the luminosity into outflow power, while the dotted
line indicates coupling of 10\%\ of the supernova kinetic power into outflow
power (assuming the bolometric luminosity is SFR dominated). The dashed 
line in the right panel reflects the fit from Eqn.~\ref{eq:pkinlbol}.}
\label{fig:energy}
\end{figure*}

\begin{figure*}
\includegraphics[width=0.5\hsize]{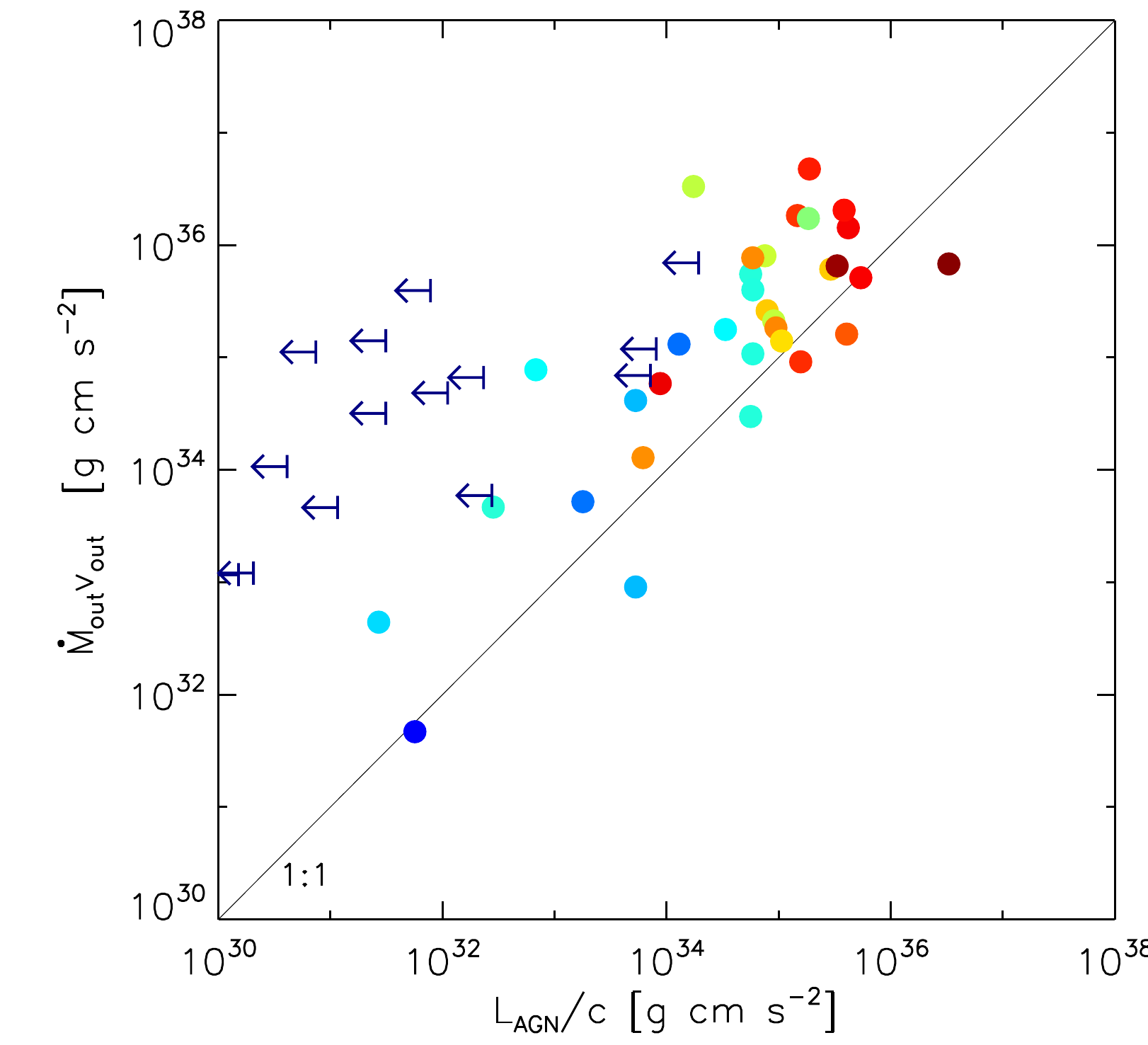}
\includegraphics[width=0.5\hsize]{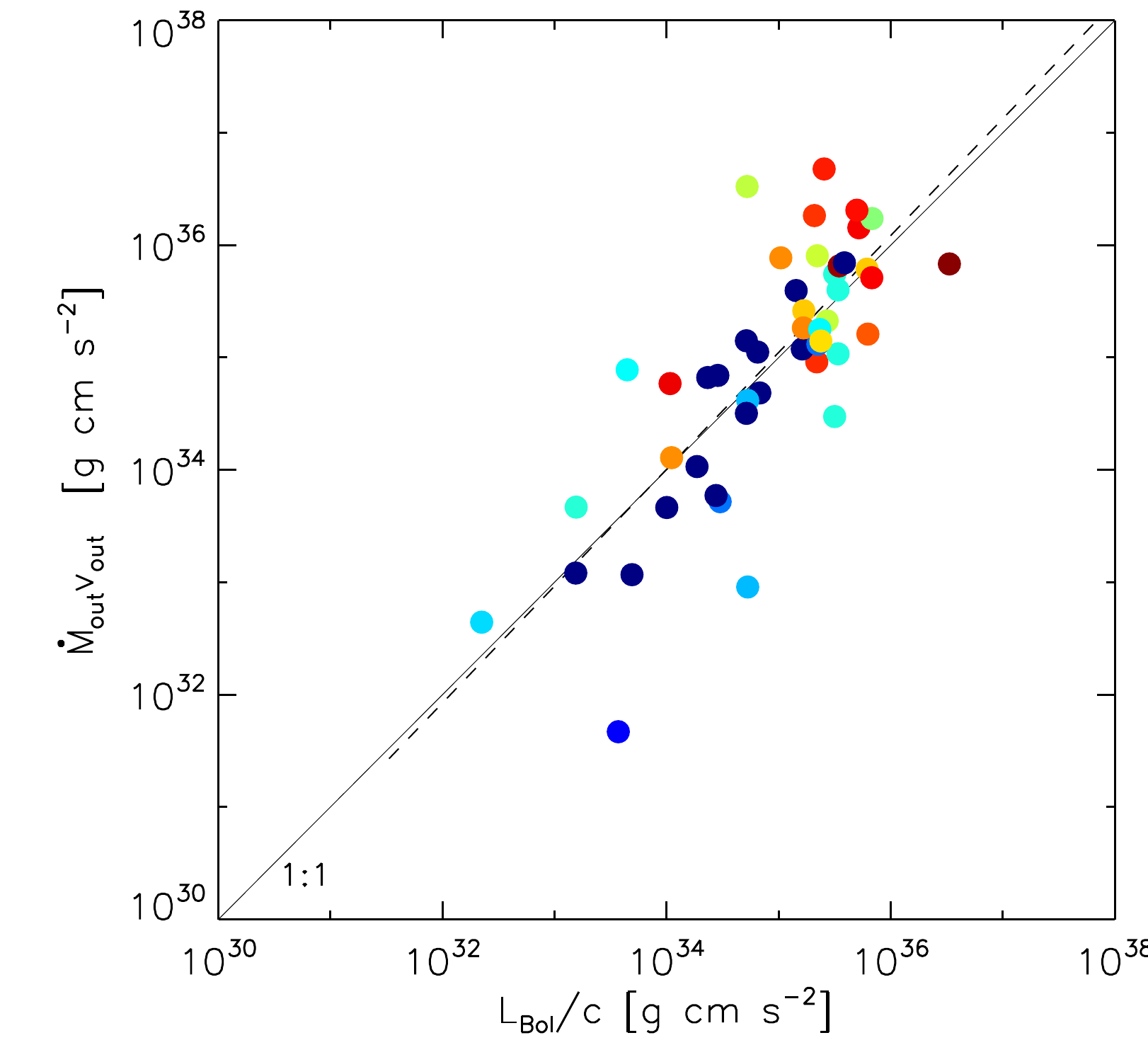}
\caption{Relation of outflow momentum rate to AGN luminosity 
momentum rate (left) and bolometric luminosity momentum rate (right). 
A 1:1 relation is over-plotted. The dashed line in the right panel reflects
the fit from Eqn.~\ref{eq:mrlbol}. The AGN 
contribution to the bolometric luminosity is color-coded on a square root 
scale between 0 (dark blue) and 1 (dark red).}
\label{fig:momentum}
\end{figure*}

Fig.~\ref{fig:lagn} and  Fig.~\ref{fig:lbol} show the same outflow 
quantities for the combined sample, now in relation
to AGN luminosity and to bolometric luminosity, as also listed in 
Table~\ref{tab:multi}. Consistent with previous 
results that are based on samples that are partly overlapping with our 
combined sample \citep{sturm11,veilleux13,spoon13,cicone14,fiore17,
gonzalez-alfonso17,fluetsch19},
we find higher outflow velocities, outflow masses, and outflow rates 
for sources that host luminous AGN. But 
clearly to be noted in our larger sample (Fig.~\ref{fig:lagn})
is the presence of some quite important outflows also in sources with
low values or upper limits for L$_{\rm AGN}$ (e.g. III Zw 035), that 
were not included in early work. There is no obvious link
of outflow properties to \lagn\ at low \lagn$\lesssim$$10^{10}$~\lsun\/,
the observed trends are driven by the high \lagn\/ sources only.  
Such increased scatter in the relation to AGN luminosity was also noted 
by \citet{fluetsch19}. A further contribution to variation in outflow
properties for given \lagn\ will arise from different geometries affecting
the coupling of AGN wind to the larger scale interstellar medium. 
In combination, these findings suggest an 
important though not exclusive role of driving that is linked to
present AGN activity. 

In fact, quite clear relations of molecular outflow mass and mass outflow 
rate with bolometric luminosity are observed (Fig.~\ref{fig:lbol}), 
which we quantify as:

\begin{multline}
{\rm log} (M_{\rm out} / {\rm M_\odot}) = \\ (7.76\pm 0.09) + (0.78\pm 0.12)\times ({\rm log}(L_{\rm Bol} / {\rm erg\ s^{-1}})
-45.5).
\label{eq:moutlbol}
\end{multline}

\begin{multline}
{\rm log} (\dot{M}_{\rm out} / {\rm M_\odot\ yr^{-1}}) = \\ (1.64\pm 0.09) + (0.75\pm 0.10)\times ({\rm log}(L_{\rm Bol} / {\rm erg\ s^{-1}})
-45.5).
\label{eq:mdotlbol}
\end{multline}

Here we have used linmix\_err \citep{kelly07} and adopted errors of 0.15~dex 
for the bolometric luminosities and 0.3~dex for the outflow 
quantities. 0.15~dex is conservatively larger than 
uncertainties in  measured \lir\ or optical AGN continuum, in 
order to consider also uncertainties in the extrapolation
from these to bolometric luminosity, and possible AGN variability.
The fit for outflow mass is derived including both detections and upper 
limits, while the fit for mass outflow rates
includes only the outflow detections since \rout\ and \vout\ are needed
to derive outflow rate or limit. We have verified that adopting fiducial 
\rout\/~=~0.7~kpc and  \vout\/~=~450~km/s to tentatively include the limits on 
outflow mass also for outflow rate leaves the slope of Eqn.~\ref{eq:mdotlbol} 
effectively unchanged, and reduces the normalization by $\sim$0.1~dex. 
While these relations can be helpful in assessing the likely importance of 
a molecular outflow from a given object, or for 
estimating total outflow rates for ensembles, we caution again about possible
selection effects in the combined sample that is significantly literature
based. In addition, bolometric luminosity, AGN luminosity, and SFR are 
all correlated 
for this sample, causing a link of outflow properties to all three, including 
the SFR that we do not discuss in detail here.

There are two obvious ways to reconcile the preference of strong molecular
outflows for AGN hosts with the detection of outflows also in galaxies
that are lacking an AGN, and with the increased scatter at low \lagn\ in 
Fig.~\ref{fig:lagn}. First, intense star formation is a significant contributor
to driving molecular outflows, which becomes dominant in the absence of 
a powerful AGN. Second, AGN driven outflows may not be recognized as such 
because they persist for some time even if rapid accretion rate variations 
lower the AGN luminosity. A third possibility is driving by highly obscured 
AGN activity, but we note that for many cases the AGN luminosities in
Table~\ref{tab:multi} use infrared spectroscopy which is sensitive to AGN
heated dust even for obscured cases. 
We believe that both star formation driving and intermittent AGN activity 
are at work here, and address
the second option in some more detail in Sect.~\ref{sect:intermittent}. 
But first we discuss constraints from kinetic power and momentum rate balance. 
  
Figure~\ref{fig:energy} compares the outflow kinetic power
$P_{\rm kin} = 0.5 \dot{M}_{\rm out} v_{\rm out}^2$ with AGN luminosity and
bolometric luminosity. Most AGN hosts stay below a wind kinetic power of 5\%\ 
of the AGN luminosity, which is often considered the limit for coupling
AGN radiative power into outflows via `energy-conserving' winds 
\citep[e.g.,][]{king15}. Three AGN hosts (NGC~1266, Mrk~876, IRAS~F17020+4544)
hover around or slightly above that line, still consistent with plausible 
uncertainties of wind parameters or with modest AGN variability. 
Notable is that limits on AGN luminosity place many of the sources with 
non-detected AGN well above that line. Also, their outflow momentum rate
would sometimes require `momentum boosts' far above the maximum AGN radiative 
momentum that corresponds to the limit on \lagn\/. This boost reaches up to
more than a factor 100 (Fig.~\ref{fig:momentum}), beyond plausibility even 
for energy conserving flows \citep{faucher-giguere12,zubovas12}. These two 
findings quantify that some of the observed molecular outflows cannot be 
driven by the currently observed level of AGN activity.

The right panels of Figs.~\ref{fig:energy} and \ref{fig:momentum} compare 
outflow power and momentum rate to bolometric 
luminosity and associated momentum rate, which for the non-AGN systems are
both dominated by star formation. Comparing outflow kinetic power to the
total mechanical luminosity delivered by supernovae 
$P_{\rm kin,SN} = 7\times 10^{41}$~($SFR/{\rm M}_\odot$~yr$^{-1})$~erg~s$^{-1}$
\citep{veilleux05}, a quite high conversion efficiency $\gtrsim$10\%\ 
would be needed in some cases. Similarly, the momentum rate of some of the
stronger outflows from non-AGN hosts exceeds the radiative momentum rate.
While this is in principle possible via multiple scattering in heavily obscured
regions \citep[e.g.,][]{murray05}, we consider it more likely that we are 
facing a mix of driving by star formation and by past AGN activity.

Given the quite clear relations of outflow kinetic power and momentum 
rate with bolometric luminosity, we quote fitting relations, again with 
a caveat about possible selection effects from the literature selection: 

\begin{multline}
{\rm log} (P_{\rm kin} / {\rm erg\ s^{-1}}) = \\ (42.39\pm 0.14) + (1.34\pm 0.17)\times ({\rm log}(L_{\rm Bol} / {\rm erg\ s^{-1}})
-45.5).
\label{eq:pkinlbol}
\end{multline}

\begin{multline}
{\rm log} (\dot{M}_{\rm out}v / {\rm g\ cm\ s^{-2}}) = \\ (35.07\pm 0.11) + (1.04\pm 0.13)\times ({\rm log}(L_{\rm Bol} / {\rm erg\ s^{-1}})
-45.5).
\label{eq:mrlbol}
\end{multline}

\subsection{Evidence for a role of intermittent AGN-driving}
\label{sect:intermittent}

We next present arguments that the contribution of AGN to the driving
of outflows 
must have a short timescale and intermittent component, in order to match 
the observed outflow properties.

\begin{figure}
\includegraphics[width=\hsize]{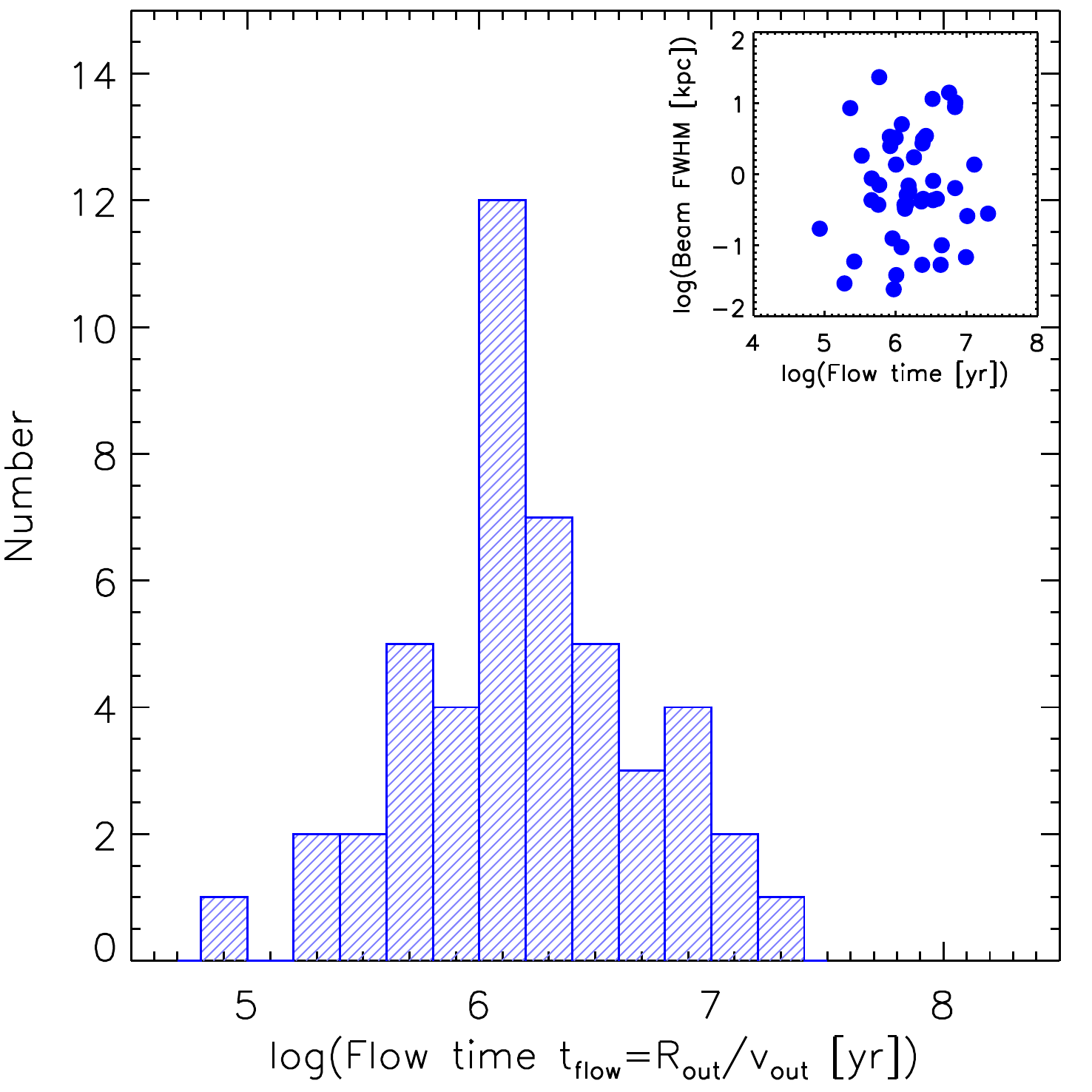}
\caption{Histogram of flow times for the combined sample. The insert compares
flow time with physical beam size.}
\label{fig:flowtime}
\end{figure}

Figure~\ref{fig:flowtime} shows that typical flow times 
$t_{\rm flow}=R_{\rm out}/v_{\rm out}$ for the combined
sample are short, just above a million years (median 1.5~Myr). For an outflow 
which is so massive that it needs to be driven not only by single young 
cluster or star formation region, but rather by a large part of a galaxy's 
total (SFR+AGN) energetics and/or momentum,
such a short timescale is difficult to achieve with star formation events.
Given the star formation history and the smoothing effect of massive star 
lifetimes, 
star formation events would naturally lead to a $\gtrsim 10^7$~yr timescale,
even in merger-induced starbursts with peaked star formation
histories \citep[e.g.,][]{mihos96,hopkins08,hopkins13,hayward14}.
In a suite of 112 hydrodynamic major merger simulations from \citet{cox06} and
\citet{wuyts10}, the star formation rate stays above half of its peak value 
for a median duration of $\sim$70~Myr, with a minimum of 5~Myr and only 
three cases with duration below 10~Myr. Short $\sim$Myr outflow events 
in massive galaxies hence are much easier 
explained by the highly and rapidly variable accretion onto an AGN, even 
if considering that observed outflows may average to some extent over 
rapid repetitive AGN `flickering' \citep{zubovas16}.

This argument assumes that the \rout\/, \vout\ values obtained from
CO interferometry and used to derive the flow time are representative for
the full duration of the outflow event, having constant velocity 
and starting at the center of the galaxy. This assumption would be wrong
if molecular gas from earlier phases of this outflow event would remain 
undetected due to observational limitations, or due to
being converted to atomic or ionized gas phases. What
might happen to such molecular gas? If just hiding at low surface density and
CO surface brightness, observations with too small beams might miss an 
important fraction of the outflow. While such effects surely exist, 
we infer from the lack of a trend in the inset of Fig.~\ref{fig:flowtime}
that beam size effects are not dominating the observed flow time distribution.
Also, accumulated fast molecular gas from long outflow episodes has not been 
reported in large aperture single dish data that are sensitive to low surface
brightness CO.
In a situation where a large fraction of the outflowing gas will not finally 
escape the halo \citep[e.g.,][]{fluetsch19}, flow times as derived here 
are also limited by deceleration. Such effects will however work on 
$\gtrsim$10$^7$~yr timescales related to the dynamical timescale, longer 
than the short typical \tflow\ observed here. 

Several studies have compared molecular outflows to outflows in the atomic 
phase and the ionized $\sim 10^4$~K phase and concluded that, while the 
mass in those phases may sometimes approach the outflowing molecular mass,
it does not exceed it by large factors \citep{contursi13,rupke13,leroy15a,
janssen16,rupke17,fiore17,fluetsch19}. Specifically for M~82, 
\citet{leroy15a} suggest a 
molecular fraction in the outflow that is decreasing with radius, and 
falling below half at $\sim$1~kpc, but with overall similar molecular and 
atomic outflowing mass. Less observationally
constrained is a path to very hot $\gtrsim 10^7$~K dilute gas that is not
easily detected in common X-ray data \citep{strickland09}. However, only
for the fastest observed molecular outflows could the post-shock temperature
$T=\frac{3}{16} \frac{\mu v_s^2}{k}$ reach beyond $10^7$~K, and potentially 
high local densities may favor re-cooling. Finally, some outflowing 
molecular gas could be converted to stars \citep{zubovas14,elbadry16,wang18}, 
but observational confirmation is still scarce 
\citep{maiolino17,gallagher19,rodriguezdelpino19}, 
not yet having reached unambiguous detection of young stars that are both 
moving at the outflow speed and trapping a significant fraction of the 
outflow rate. Very large masses of young stars above the molecular outflow 
mass would build up, and should be little 
obscured, since the outflows typically extend beyond the dusty galaxy centers. 
In summary, there seems
to be no obvious storage of molecular outflow in other phases that would 
help to drastically shorten the observed molecular gas flow times. 

A second argument for time-dependent AGN driving rests on the energy and 
momentum budget. For about 10\% of their sample, \citet{fluetsch19} argue
that outflow kinetic power and momentum cannot be driven by the combination
of the currently available AGN luminosity and SFR -- a past episode of
higher AGN luminosity must have been present.

Finally and most directly, in a few spatially well resolved cases distinct 
outflow episodes can be inferred from the morphology. NGC~2623 shows
both a clear bipolar outflow with the blue lobe north of the nucleus,
and a second `fossil' blue component at larger offset to the southeast. 
(see Sect.~\ref{sect:outprop}, Fig.~\ref{fig:ngc2623}). Flow times for
these two differ by more than an order of magnitude: 1~Myr vs. 13~Myr.
Rapidly evolving near-nuclear conditions of this advanced merger 
or driving by different nuclei \citep[but note the single component 
appearance in K-band,][]{rossa07}
might be responsible for the different orientation of these two flows.
For IRAS 08572+3915, deep IRAM-NOEMA data detect a second spatially separated
outflow cloud that is offset by $\sim$6~kpc 
\citep{janssen16b,herrera-camus19a} in 
addition to the previously known outflow \citep{cicone14}. On a smaller
scale, spatially resolved ALMA data of ESO 302-G030 
\citep[][their Fig.~6]{pereira-santaella16} show several clumps
over a total scale of $\sim$0.8~kpc, with a rather well defined blue/red
symmetry. Driving by individual supernovae may be just energetically possible 
in this case, but intermittent AGN emission appears equally plausible.

Accretion rate variations on many timescales are a characteristic of the AGN
phenomenon. Beyond the timescales of direct flux monitoring programs or 
archival light curve studies, their imprint on local AGN phenomena has been 
demonstrated in various ways. Light echoes of past activity can be seen in 
the form of (extended) narrow line regions that are too powerful for the 
currently observed AGN luminosity (e.g., \citealt{lintott09,keel12,schirmer13};
see also \citealt{sartori18}).
Conversely, AGN with deficient NLR compared to their X-ray luminosity
may trace the onset of activity \citep{schawinski15}. Perhaps most directly,
X-ray light echoes trace accretion variations of the (much less luminous)
activity in our Galactic Center \citep[e.g.,][]{clavel13}. Recent simulations 
start to reproduce the observed `flickering' AGN accretion variations 
\citep[e.g.,][]{novak11,gabor13}.
At high redshift, the smoothing of accretion rate variations provided
by the observed outflow can explain the higher incidence of AGN-driven
nuclear outflows, compared to direct AGN indicators such as X-rays
\citep{genzel14,foersterschreiber19}.

Molecular outflow statistics in a complete mass selected sample could
provide further insights into the timescales of AGN driving.
Limitations arise from the additional role of star formation driven 
outflows that can be hard to discern from fossil AGN-driven outflows, 
and from the diverse target selection of outflow studies in the current
literature. Tentatively, both the identification of multiple episodes
in a few targets and the large 13/17 outflow incidence for the \citet{lutz16}
high surface brightness sample argue that the $\sim$1~Myr
outflow events may be episodes in a longer outflow history. However,
the latter sample is a special population that favors a 
compact dense circumnuclear ISM, and the low incidence of 
OH outflows in the Seyfert sample of \citet{stone16} clearly shows that
its statistics cannot be transferred to AGN in general.  
In any case, `depletion times' needed to expel a galaxy's entire molecular
gas mass have to be viewed with great caution, if ones assumes that 
the current outflow rate persists unchanged over a very long time.

\section{Conclusions}

We have presented new CO(1-0) interferometry from ALMA and NOEMA, probing 
molecular outflows in 13 local galaxies with high far-infrared surface 
brightness, and combined our results with literature data on local universe 
molecular outflows. Our main findings are:

(1) Galaxies with high far-infrared surface brightness 
log(\sigfir )$\gtrsim$11.75~\lsun\/~kpc$^{-2}$ show the expected
high incidence of molecular outflows. 17 of 18 sources in the \citet{lutz16} 
sample now have suitable CO data, and 13 of these exhibit molecular 
outflows. Several of
these  nearby targets, in particular III~Zw~035, are well-suited to 
spatially resolved studies.  

(2) Partly constrained by observed line ratios and by spatial structure of the
best cases, we adopt in the conversion of CO observables to outflow 
physical properties \alpco\/~=~0.8~\msun\/~/~(K~km~s$^{-1}$~pc$^2$) and C=1 in 
  $\dot{M}_{\rm out} = C \frac{M_{\rm out}  v_{\rm out}}{R_{\rm out}}$.
The high $R_{31}=2.1$ in IRAS~13120-5453 clearly indicates that this outflow
is optically thin in CO. This cautions that \alpco\ may vary between 
outflows, but
the good agreement with independent OH-based results globally supports the 
adopted CO conversion factor for outflows. We adopt conservative 
definitions of outflow 
flux and outflow velocity that focus on the line wings. While this may miss 
slow outflow components, which are likely to fall back, it improves robustness
against confusion of outflow with undisturbed host gas. 

(3) We compare our results with outflow properties 
that are based on \herschel\ far-infrared spectra of OH absorption, with
special focus on the 12 ULIRGs with detailed OH models by 
\citet{gonzalez-alfonso17}. 
The good agreement in derived properties is an important 
cross-validation of the two main methods for characterizing molecular 
outflows. Modest remaining differences may relate to different but overlapping 
regions sampled by the two methods: Smaller for the OH absorption vs. 
larger for CO emission.

(4) Outflow properties correlate better with AGN luminosity and with 
bolometric luminosity than with far-infrared surface brightness.
The outflows with largest mass and mass outflow rate prefer systems with
current luminous AGN activity, suggesting an important role of AGN driving,
but significant outflows in some non-AGN systems must relate to star 
formation or to AGN activity in the recent past.
  
(5) We report scaling relations of outflow properties with bolometric 
luminosity (Eqns.~\ref{eq:moutlbol} to \ref{eq:mrlbol}), but 
note that these might still be influenced by selection effects.
 
(6) Short $\sim 10^6$~yr flow times and some sources with spatially 
resolved multiple outflow episodes support a role of intermittent/episodic 
driving, likely by AGN.

\begin{acknowledgements}
We thank the referee for a helpful report. This work is based on 
observations carried out under project numbers
S16BH, S17AR, and W17CP with the IRAM NOEMA interferometer. IRAM is 
supported by INSU/CNRS (France), MPG (Germany) and IGN (Spain). It is also 
based on ALMA projects ADS/JAO.ALMA\#2016.1.00177.S and 
ADS/JAO.ALMA\#2012.1.00306.S. ALMA is a 
partnership of ESO (representing its member states), NSF (USA) and 
NINS (Japan), together with NRC (Canada), MOST and ASIAA (Taiwan), and 
KASI (Republic of Korea), in cooperation with the Republic of Chile. 
The Joint ALMA Observatory is operated by ESO, AUI/NRAO and NAOJ.
We acknowledge reduction support support by IRAM astronomers Miguel 
Montarges, Michael Bremer, and Melanie Krips. We thank
Maria Drosdovskaya and Stefano Facchini for help concerning emission of 
the NS radical. S.V. acknowledges support from a Raymond and Beverley Sackler
Distinguished Visitor Fellowship and thanks the host institute, the
Institute of Astronomy, where this work was concluded. S.V. also
acknowledges support by the Science and Technology Facilities Council
(STFC) and by the Kavli Institute for Cosmology, Cambridge. R.M. acknowledges
ERC Advanced Grant 695671 ``QUENCH''.

\end{acknowledgements}

\bibliographystyle{aa}
\bibliography{outflow18}

\begin{appendix}

\section{Figures for individual sources}

\begin{figure*}[ht]
\centering
\includegraphics[width=0.33\hsize]{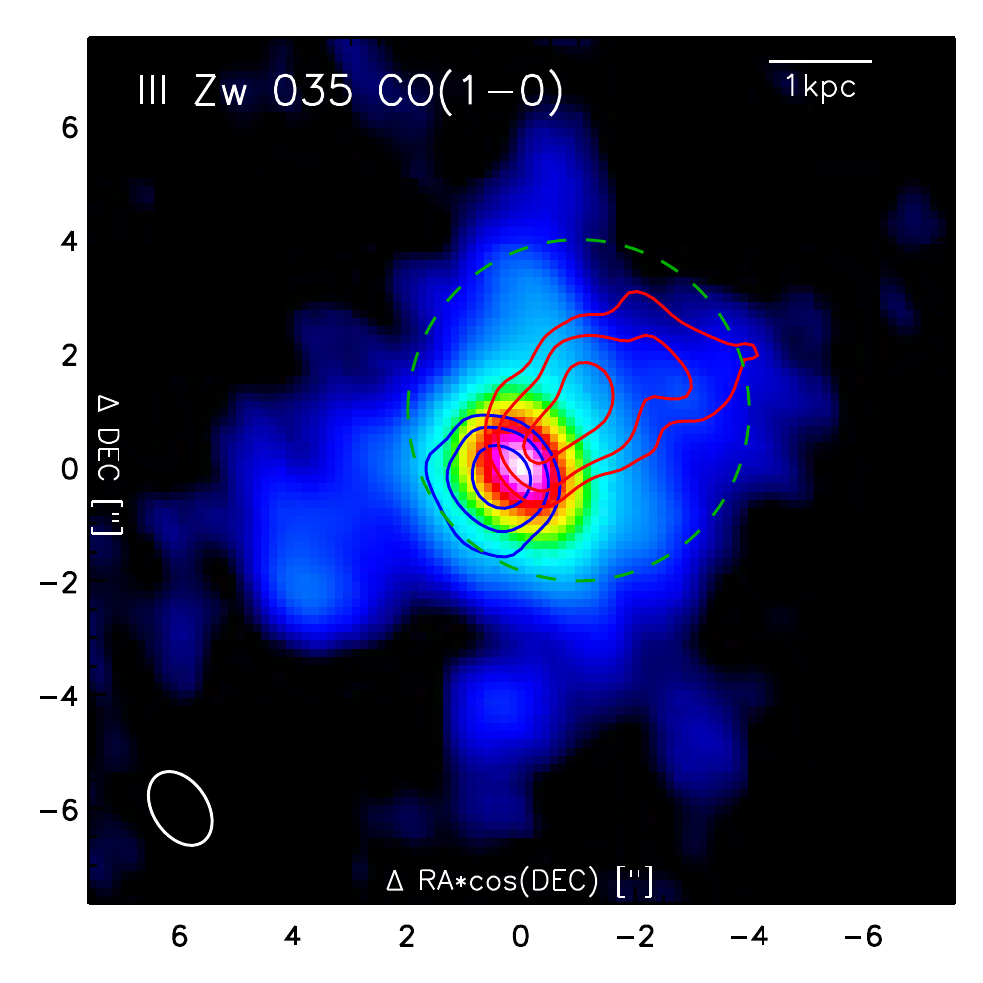}
\includegraphics[width=0.66\hsize]{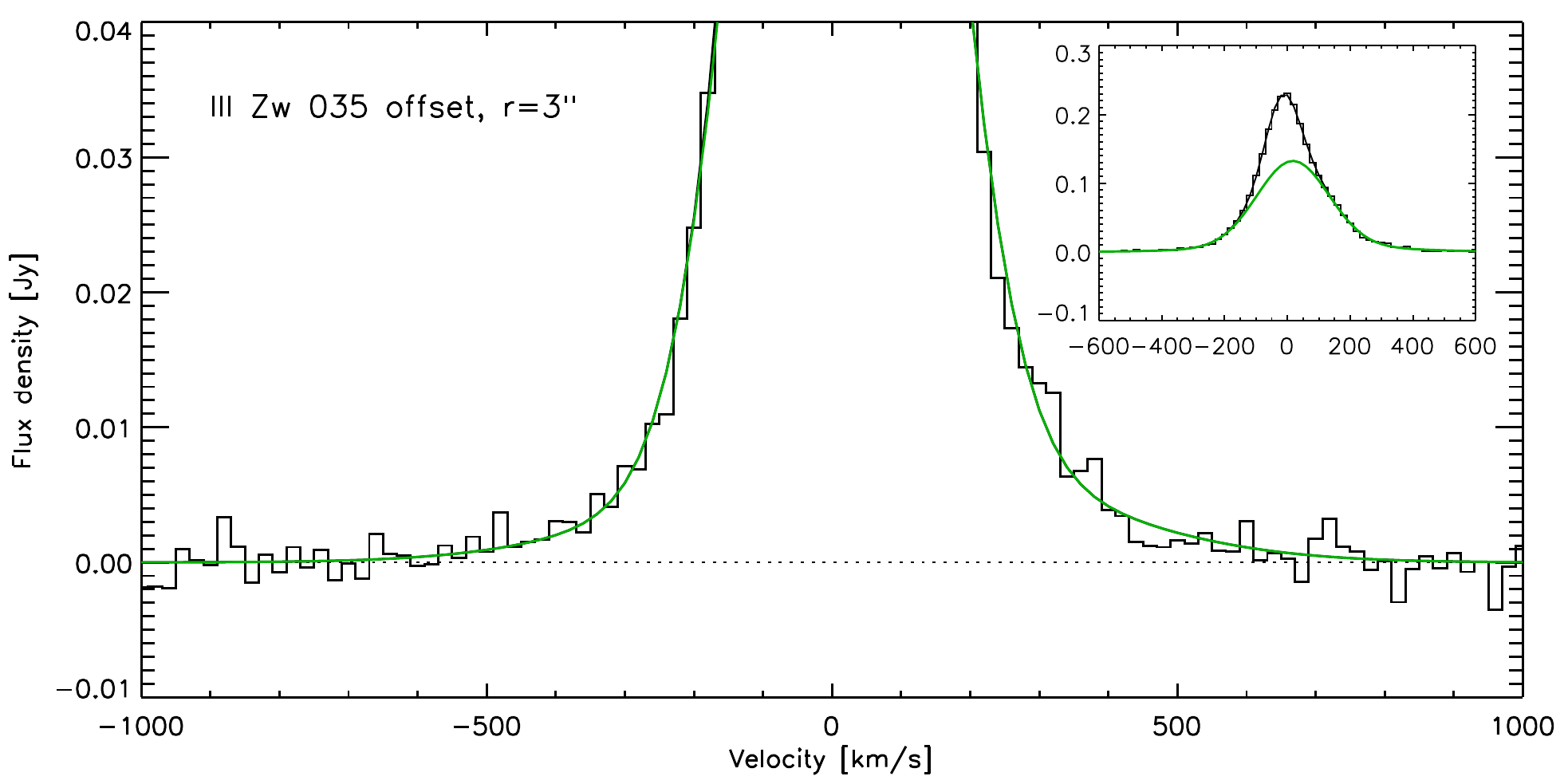}
\includegraphics[width=0.33\hsize]{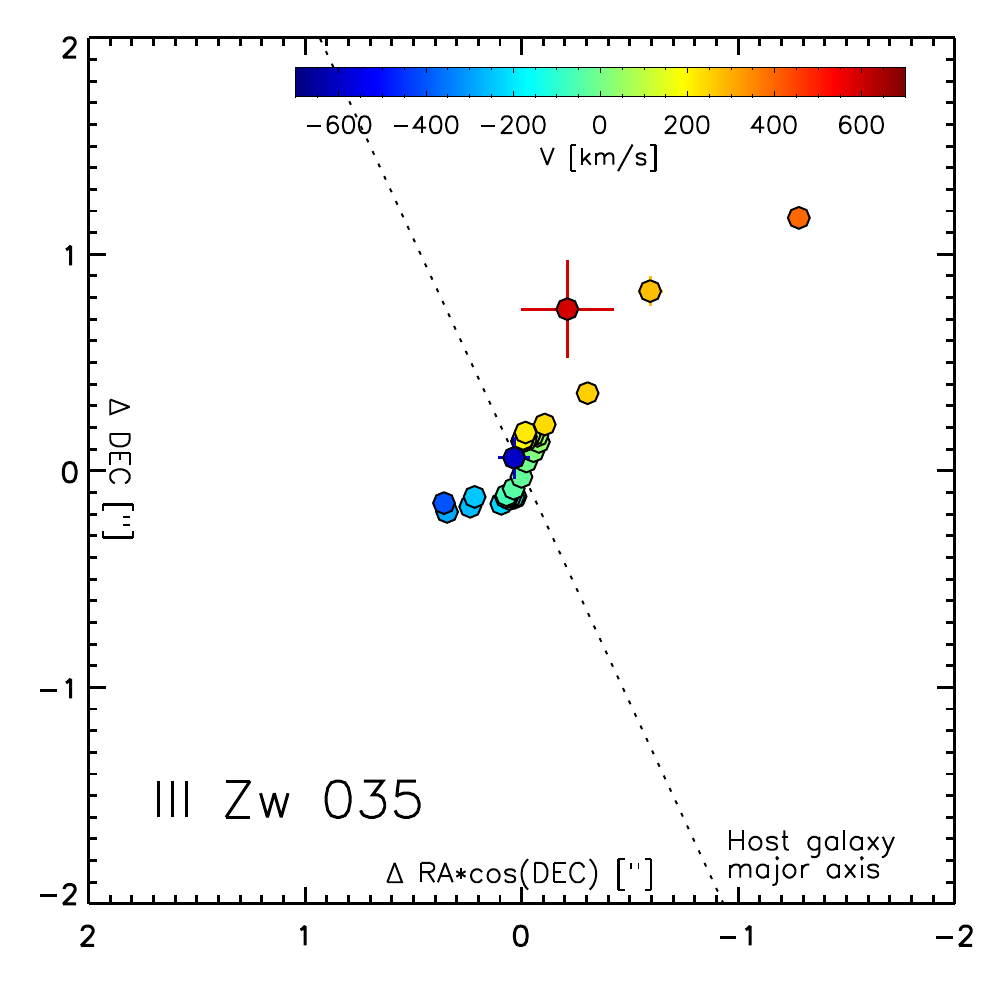}
\includegraphics[width=0.66\hsize]{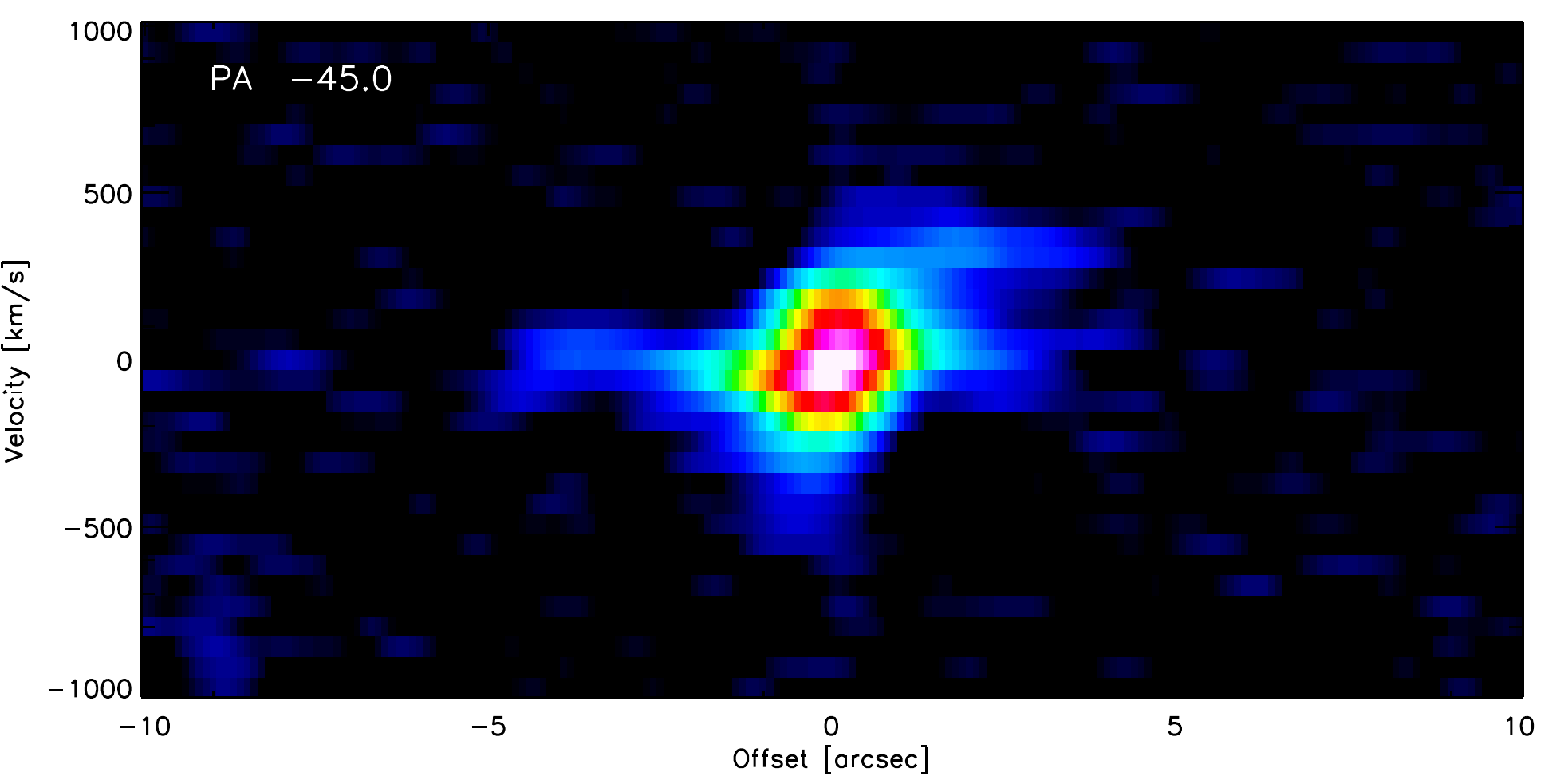}
\includegraphics[width=0.66\hsize,right]{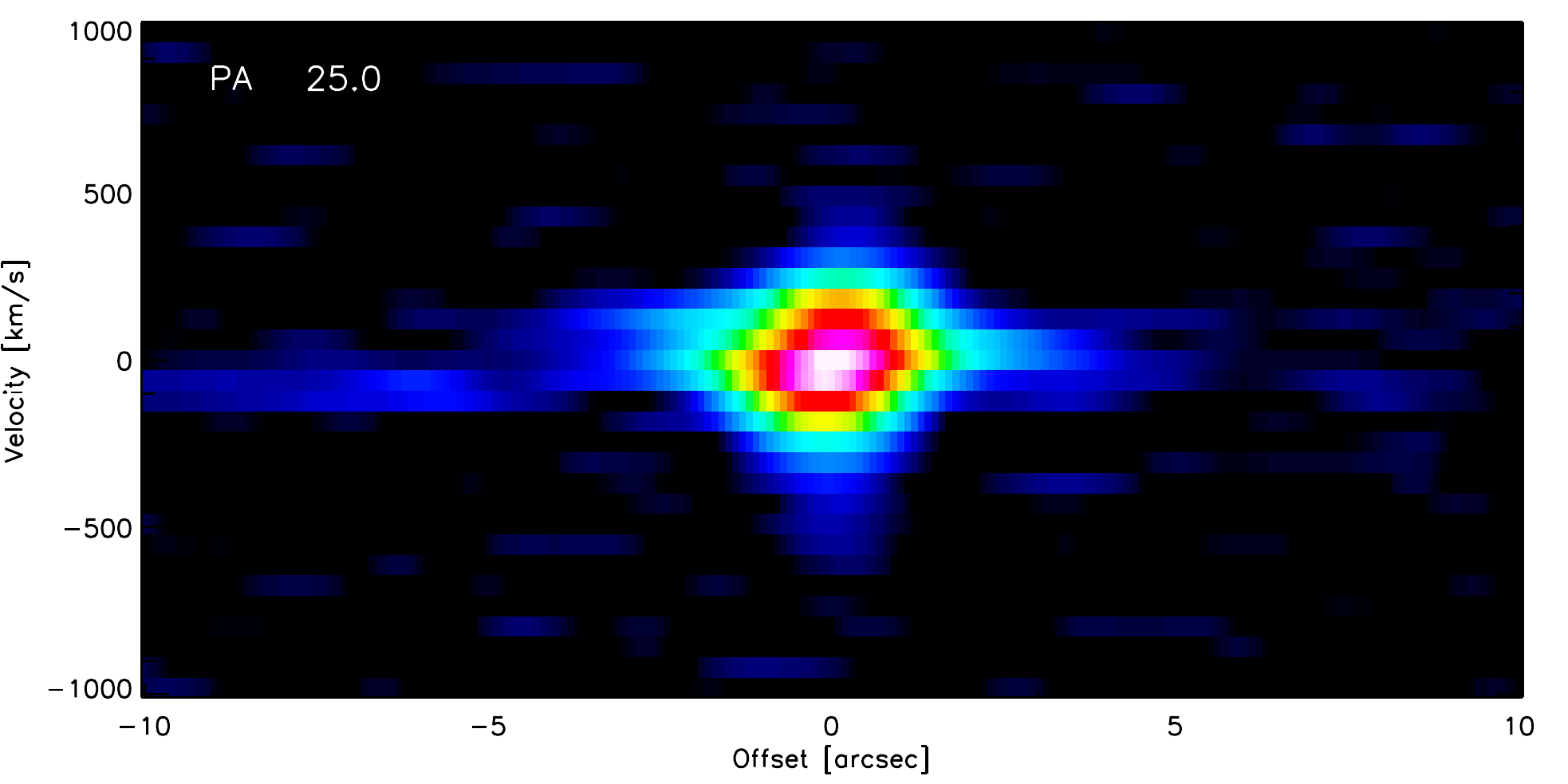}
\caption{NOEMA CO(1-0) data for III~Zw~035.
Top left: Moment 0 map. All spatial offsets are in arc seconds relative to the 
position of the 3~mm continuum (Table~\ref{tab:continuum}). 
N is to the top and E to the left. Overlaid contours
are for outflow in the velocity range [-570,-270]~\kms\ (blue) and 
[270,570]~\kms\ (red), with contours at [0.3, 0.6, 1.2]~mJy/beam. 
Top right: Spectrum in a r=3\arcsec\ aperture offset from the continuum
nucleus by -1\arcsec\ in RA$\times$cos(DEC) and +1\arcsec\ in DEC (green dashed circle
in top left panel). The 
spectrum is decomposed into three Gaussians, one for host and 2 for outflow
(green). The black line includes all three Gaussian components.
Center left: Center positions and their errors for selected 
velocity channels. These are derived by fitting a Gaussian 
model to the UV data for the respective velocity channel.
N is to the top and E to the left. 
The dotted line marks the galaxy major axis 
position angle $\sim$25\degr\ \citep{kim13}. 
Center right: Position-velocity diagram along the direction of the 
strong red outflow at PA -45\degr\/.
Bottom right: Position-velocity diagram along the galaxy major axis 
PA 25\degr\/. The 
fainter galaxy of the pair is located near offset -7\arcsec\ in this PV 
diagram but does not stand out clearly from the major galaxy's emission. 
We detect a strong and asymmetric bipolar outflow.
}
\label{fig:iiizw035}
\end{figure*}

\begin{figure*}
\centering
\includegraphics[width=0.33\hsize]{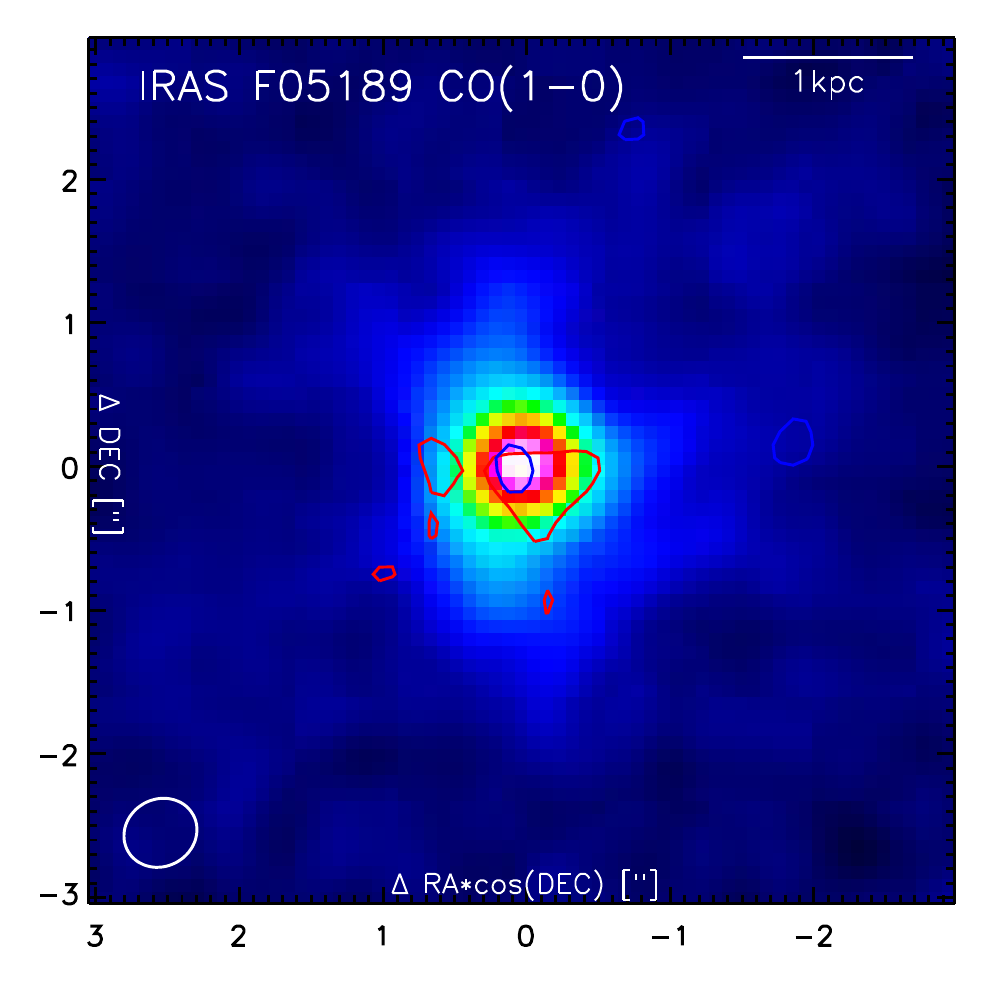}
\includegraphics[width=0.66\hsize]{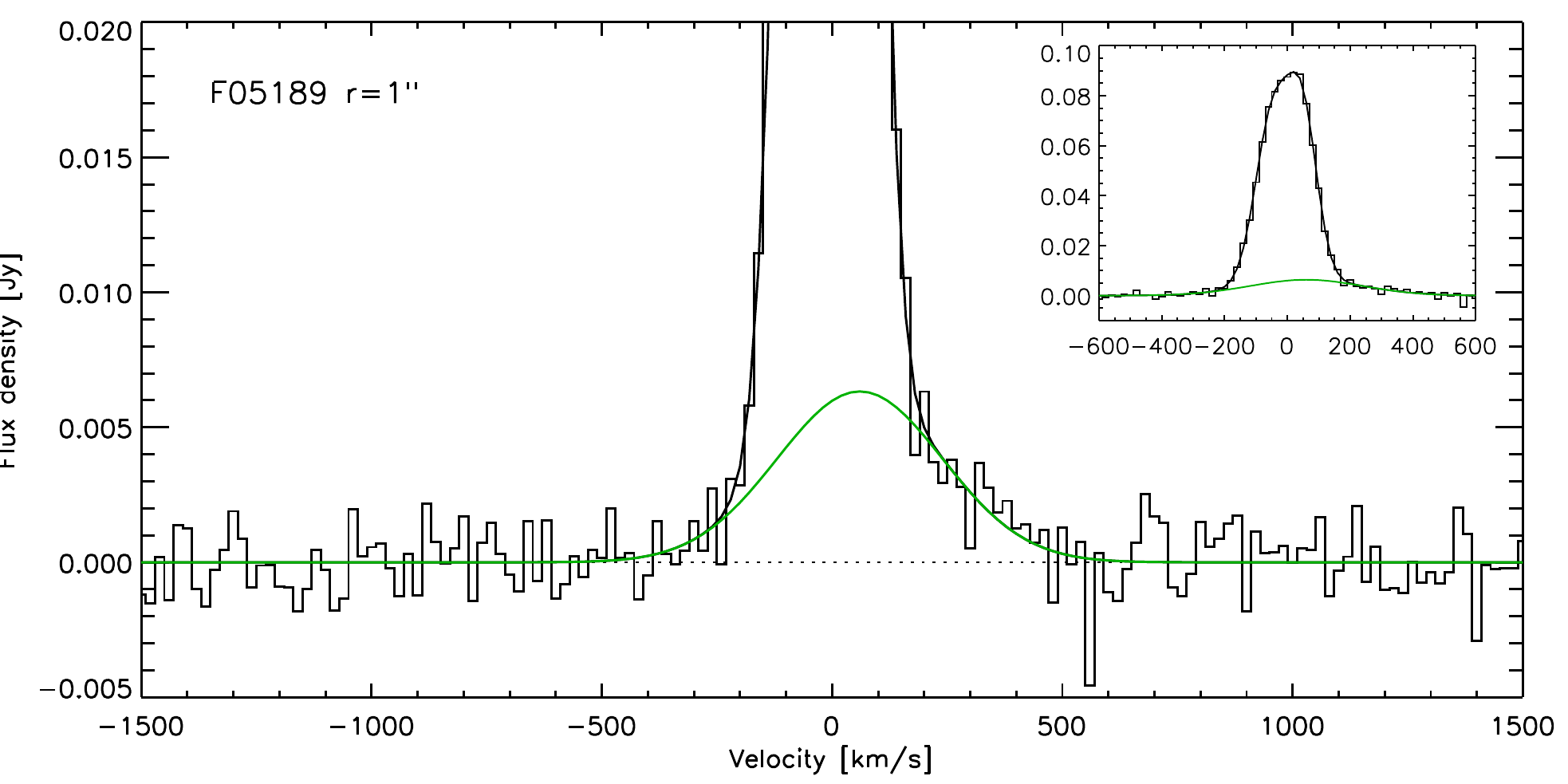}
\includegraphics[width=0.33\hsize]{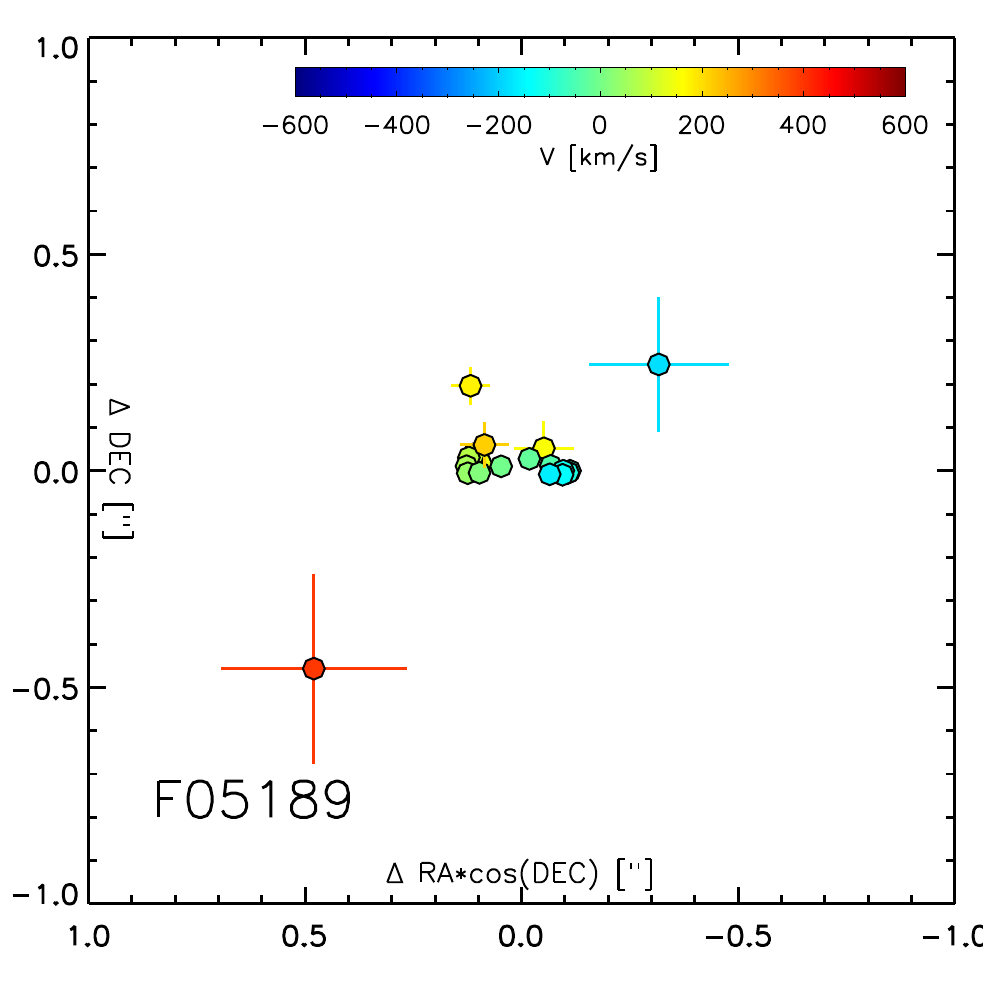}
\includegraphics[width=0.66\hsize]{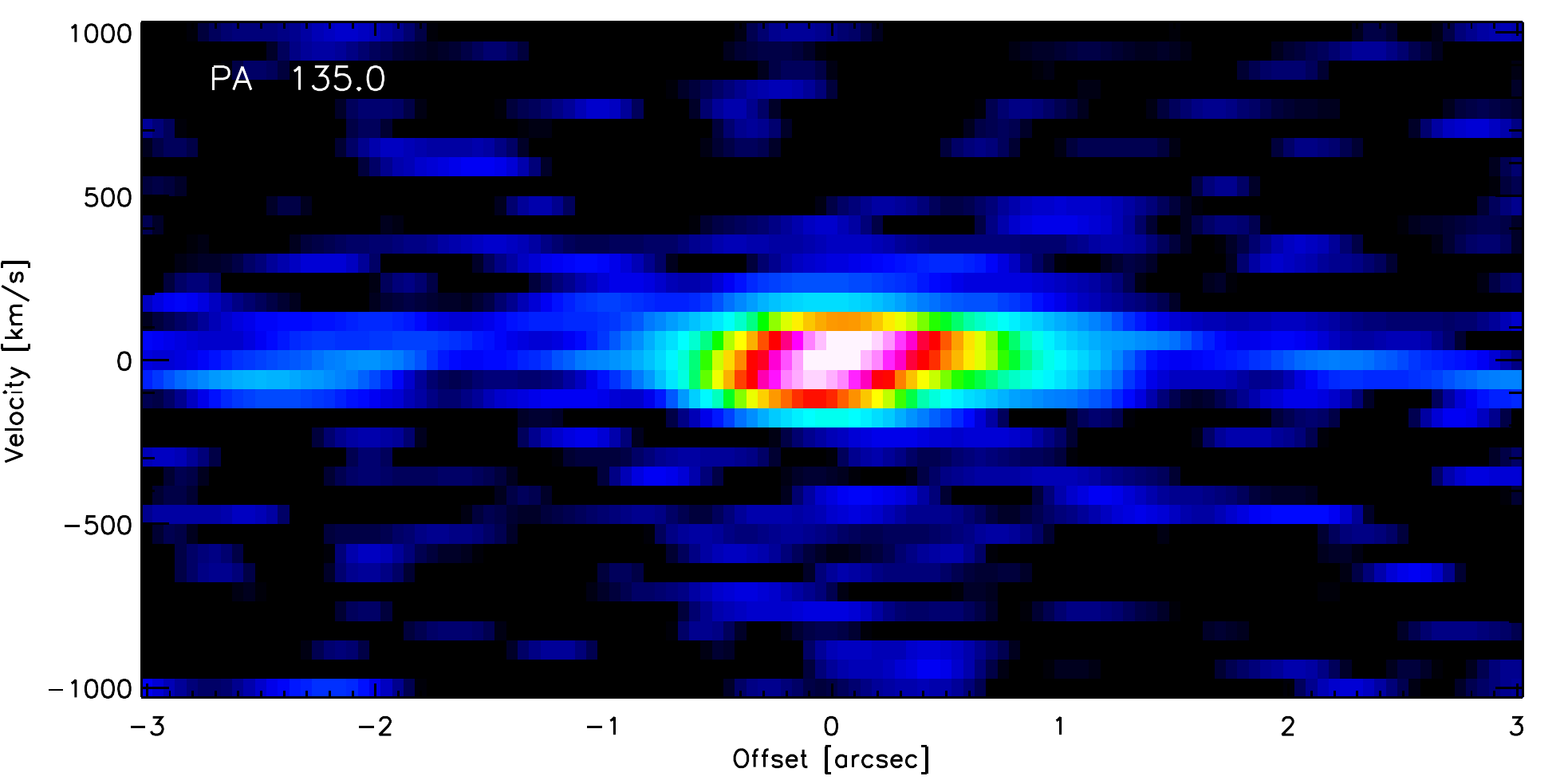}
\includegraphics[width=0.66\hsize,right]{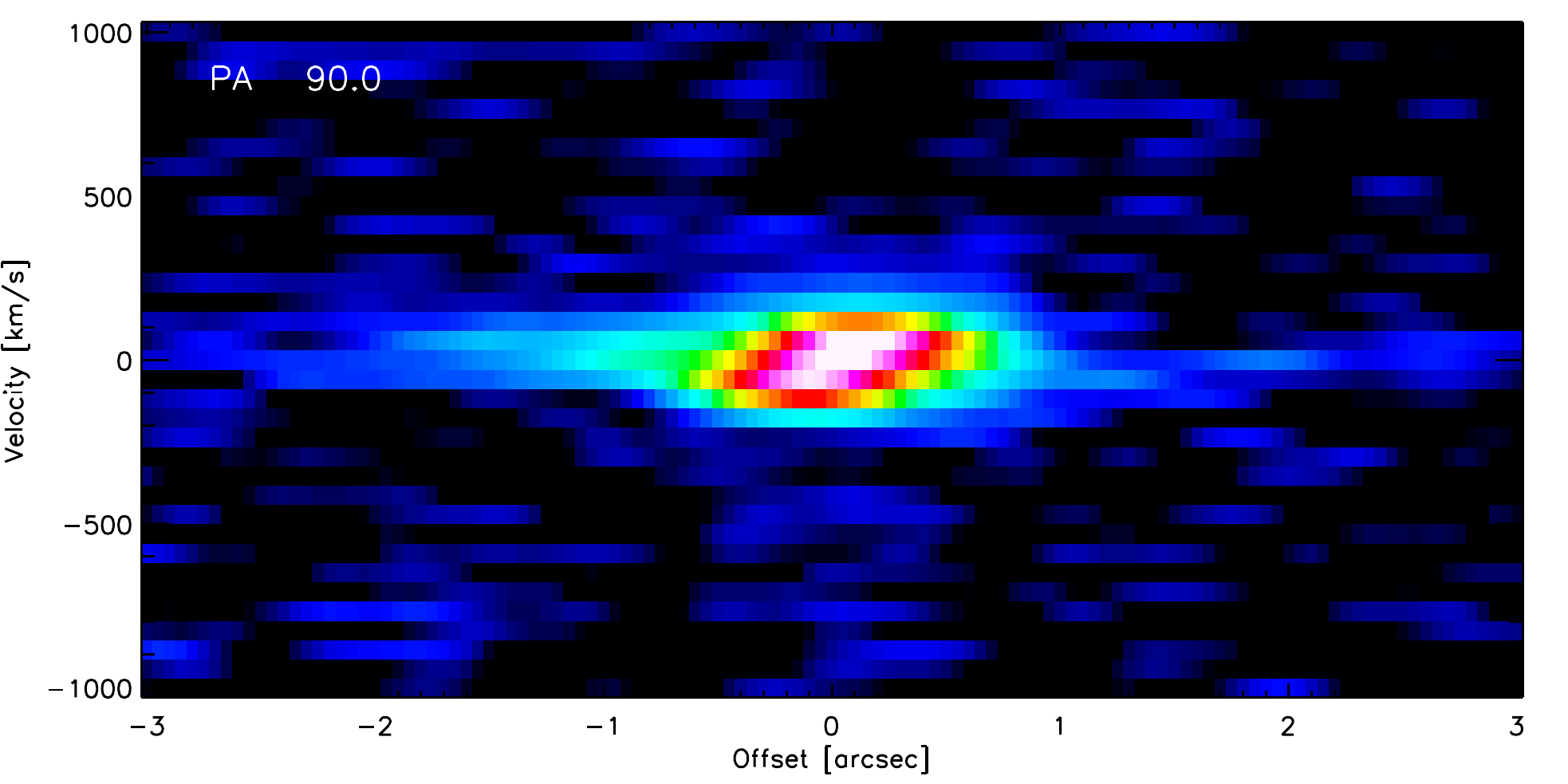}
\caption{ALMA CO(1-0) data for IRAS~F05189-2524.
Top left: Moment 0 map. Overlaid contours
are for the outflow in the velocity range [210,510]~\kms\ (red, 0.3~mJy/beam) 
and the marginal emission at [-990,-270]~\kms\ (blue, 0.2~mJy/beam). 
Top right: Spectrum in a r=1\arcsec\ aperture centered on the continuum
nucleus. The 
spectrum is decomposed into three Gaussians, two for host and one for outflow
(green). The black line includes all three Gaussian components. Outflow is 
clearly detected in the redshifted wing, and tentatively indicated for
 the blue wing.
Center left: Center positions and their errors, from UV fitting a Gaussian model to individual
channels. 
Center and bottom right: Position-velocity diagrams.  PA +135\degr\ approximately follows the diffuse red outflow 
component towards the southeast or south, while PA +90\degr\ traces the 
velocity gradient of the line core. 
}
\label{fig:f05189}
\end{figure*}

\begin{figure*}
\includegraphics[width=0.33\hsize]{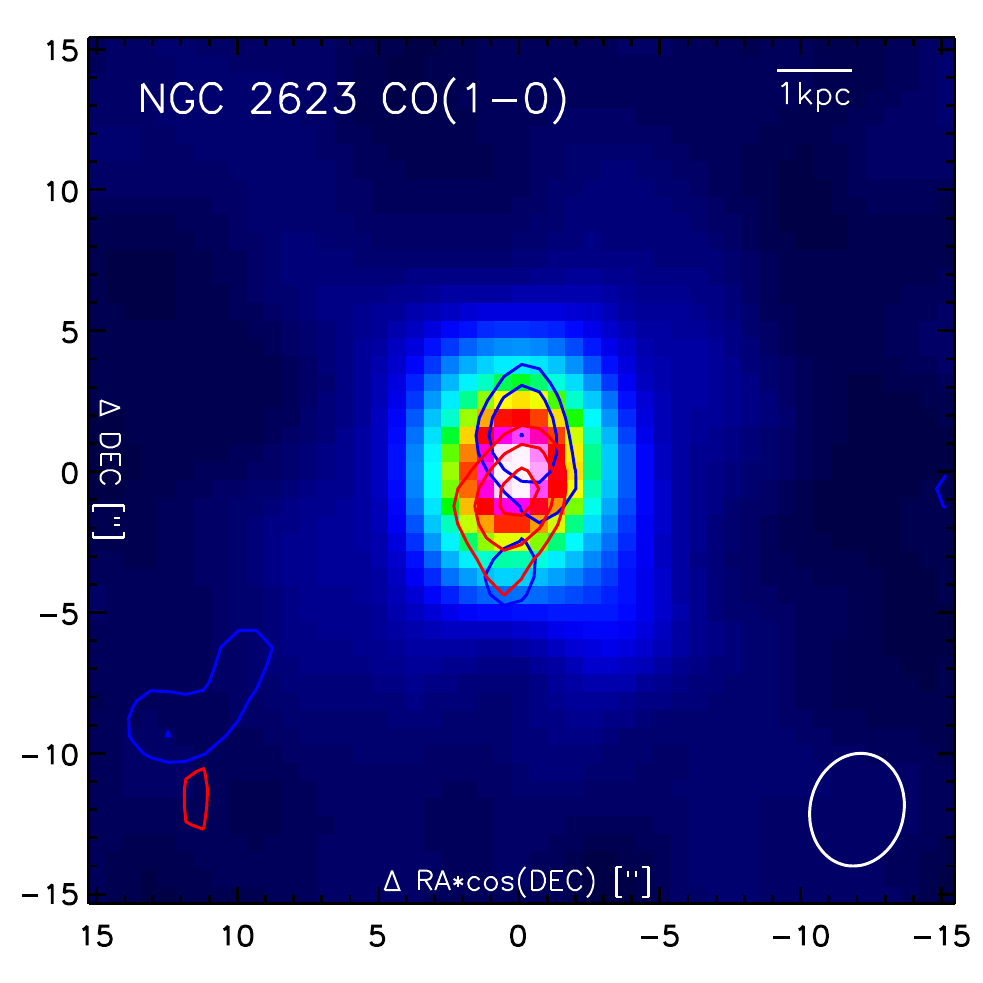}
\includegraphics[width=0.66\hsize]{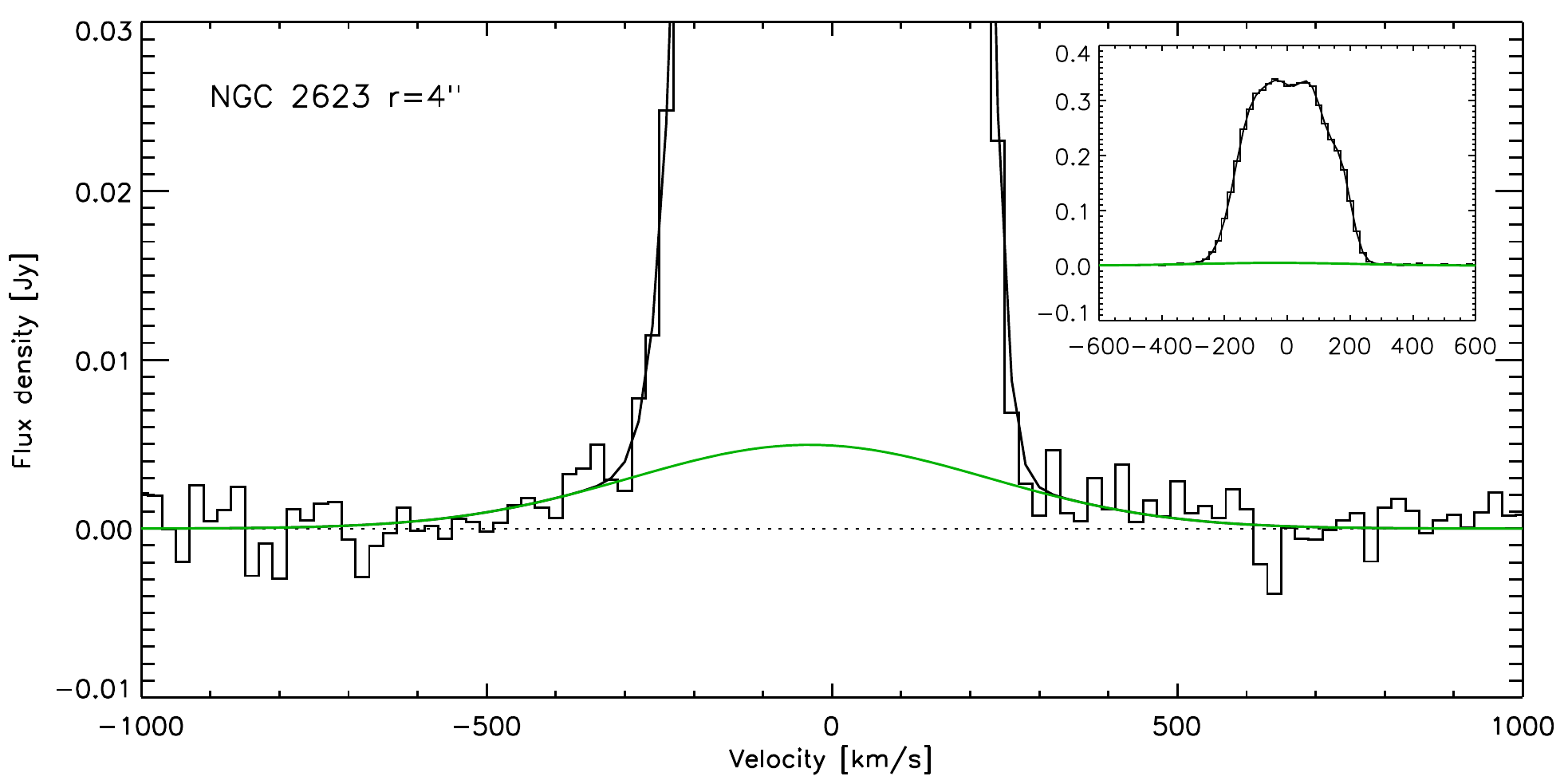}
\includegraphics[width=0.33\hsize]{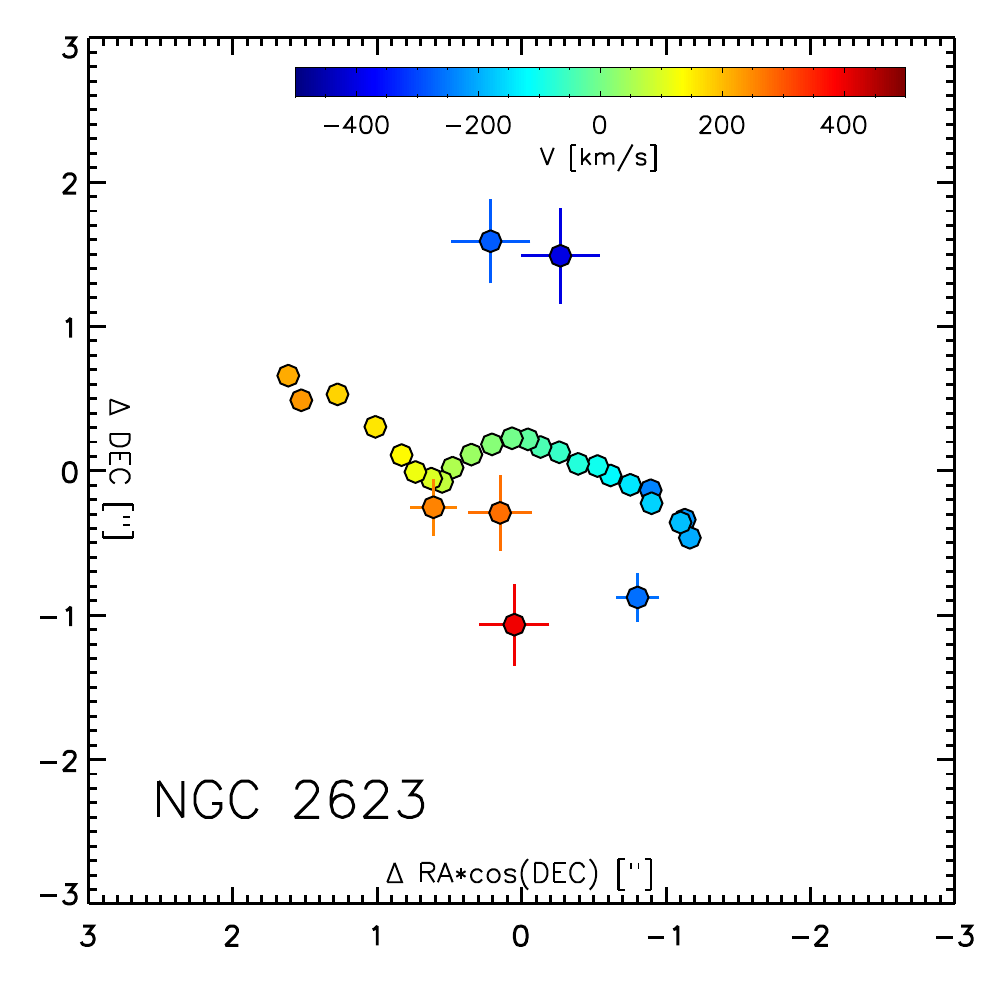}
\includegraphics[width=0.66\hsize]{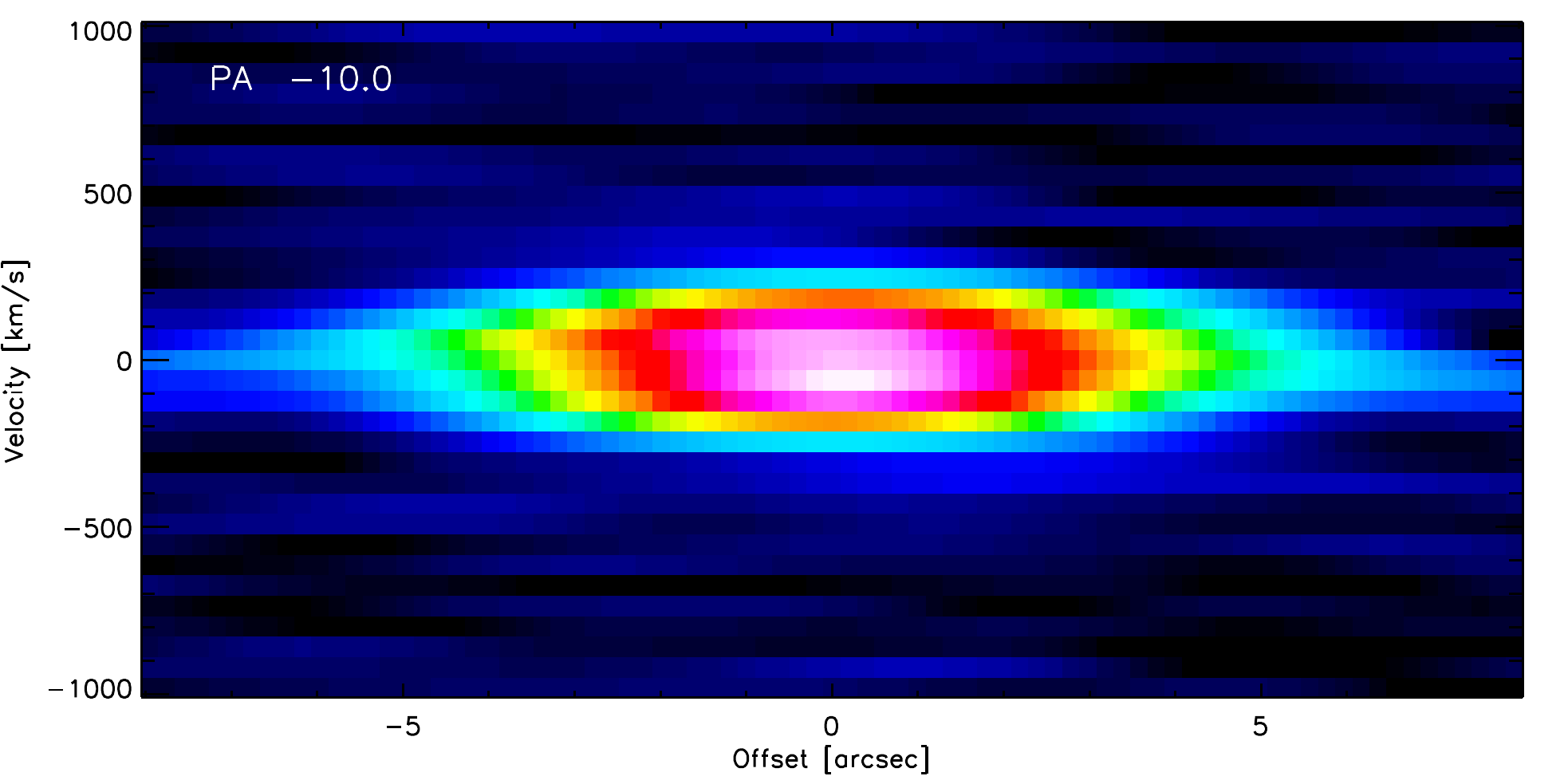}
\includegraphics[width=0.33\hsize]{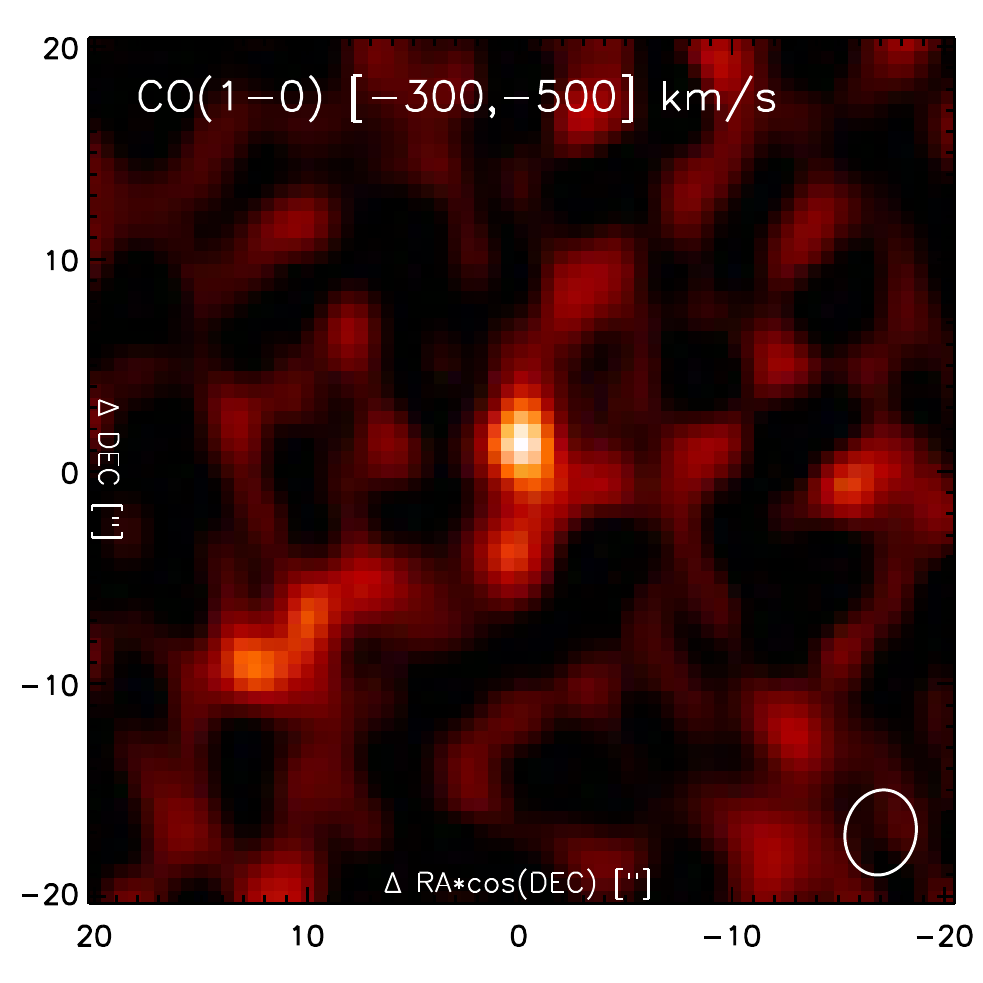}
\includegraphics[width=0.66\hsize]{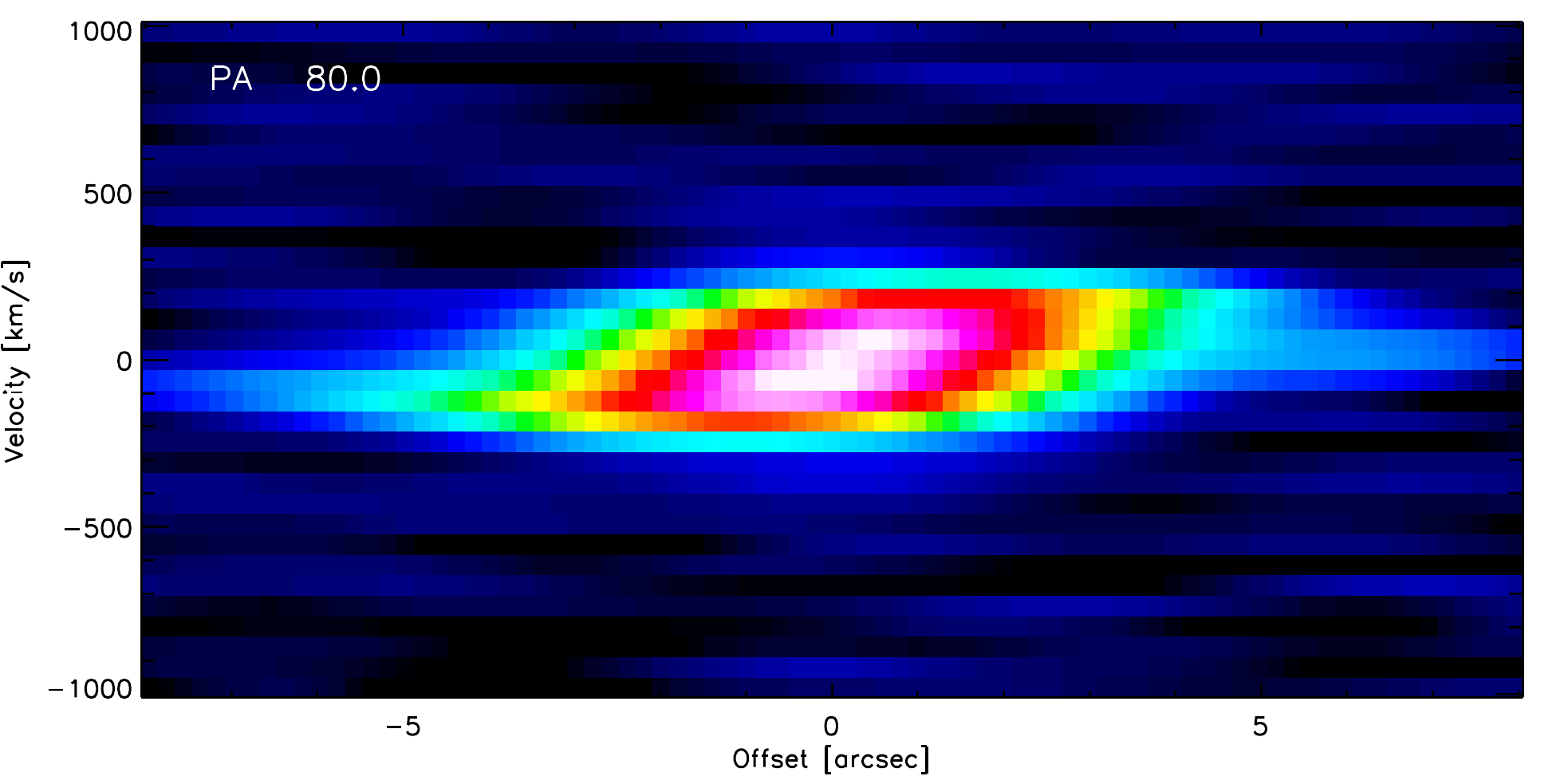}
\caption{NOEMA CO(1-0) data for NGC 2623. Top left: Moment 0 map. 
Overlaid contours are for the outflow in the velocity range [-500,-300]
(blue) and [300,500]~\kms\ (red), with contours at [0.7, 1.0, 1.5]~mJy/beam. 
Top right: Spectrum in a r=4\arcsec\ 
aperture centered on the continuum nucleus. The 
spectrum is decomposed into five Gaussians, four for the complex
host line profile and one for outflow (green). The black line includes 
all five Gaussian components.
Center left: Center positions and their errors, from UV fitting a Gaussian model to individual
channels. Center and bottom right: Position-velocity diagrams. 
PA~+80\degr\ traces the velocity gradient of the line core. The orthogonal
PA~-10\degr\ is also close to the orientation of the roughly N-S bipolar 
outflow seen in the top left and center left panels.
Bottom left:
[-500,-300]~\kms\ channel map showing the SE `fossil' outflow that is also 
indicated in the blue contours of the top left panel, in addition to the 
blue component of the main bipolar outflow.
}
\label{fig:ngc2623}
\end{figure*}

\begin{figure*}
\centering
\includegraphics[width=0.33\hsize]{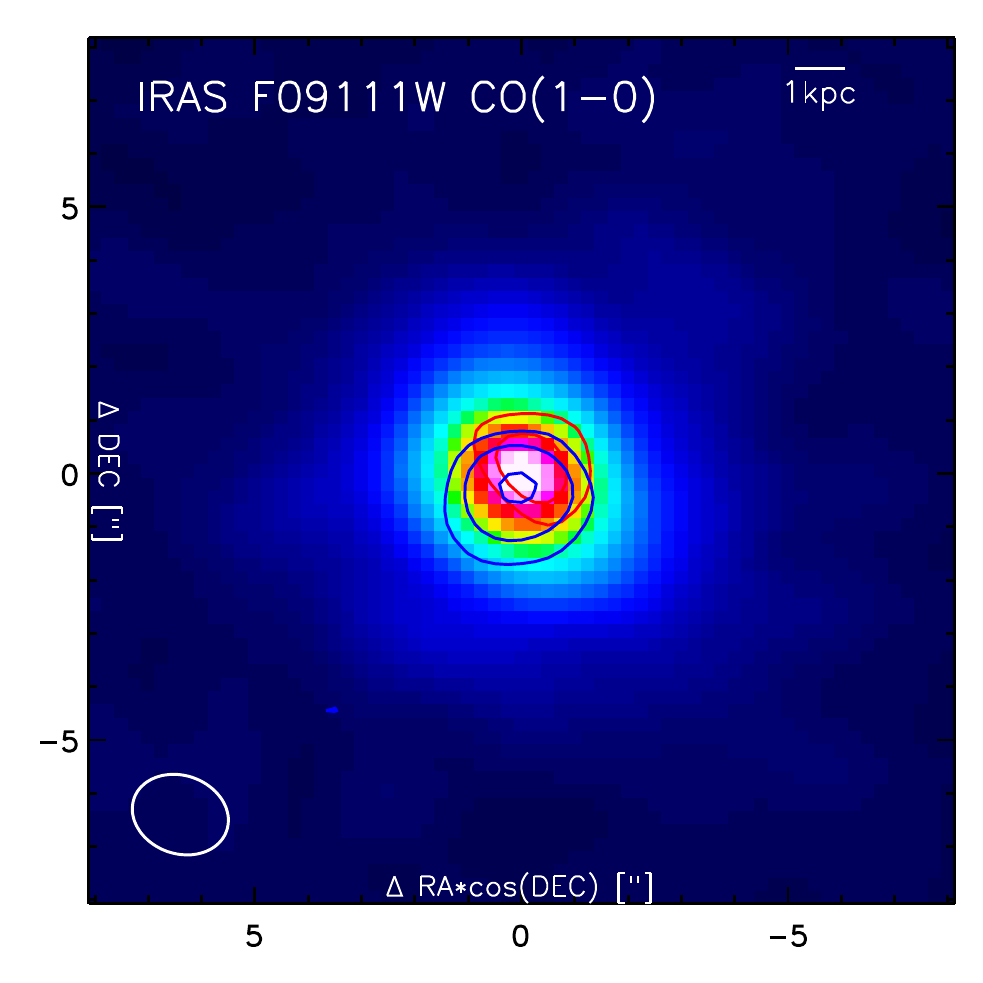}
\includegraphics[width=0.66\hsize]{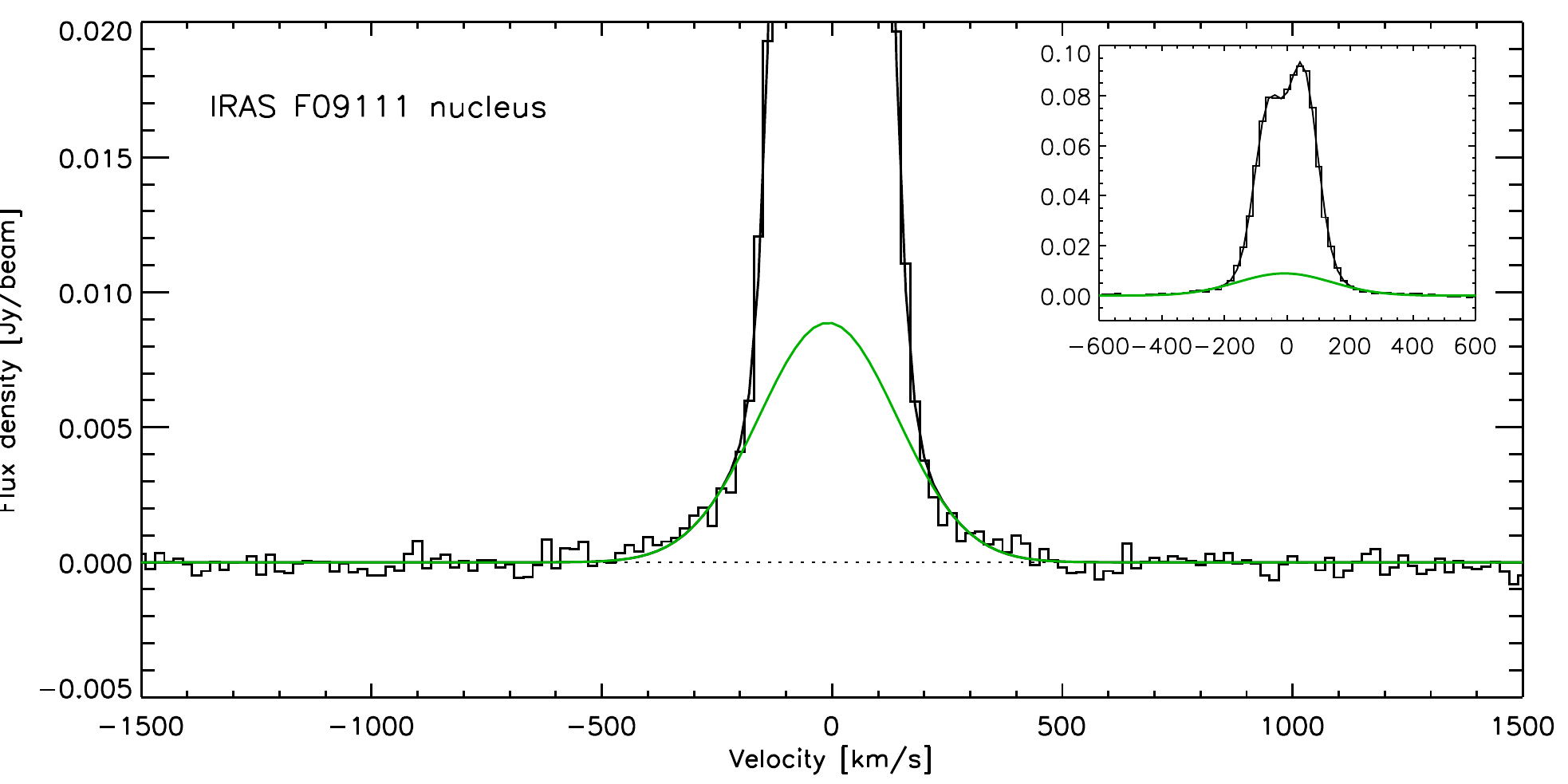}
\includegraphics[width=0.33\hsize]{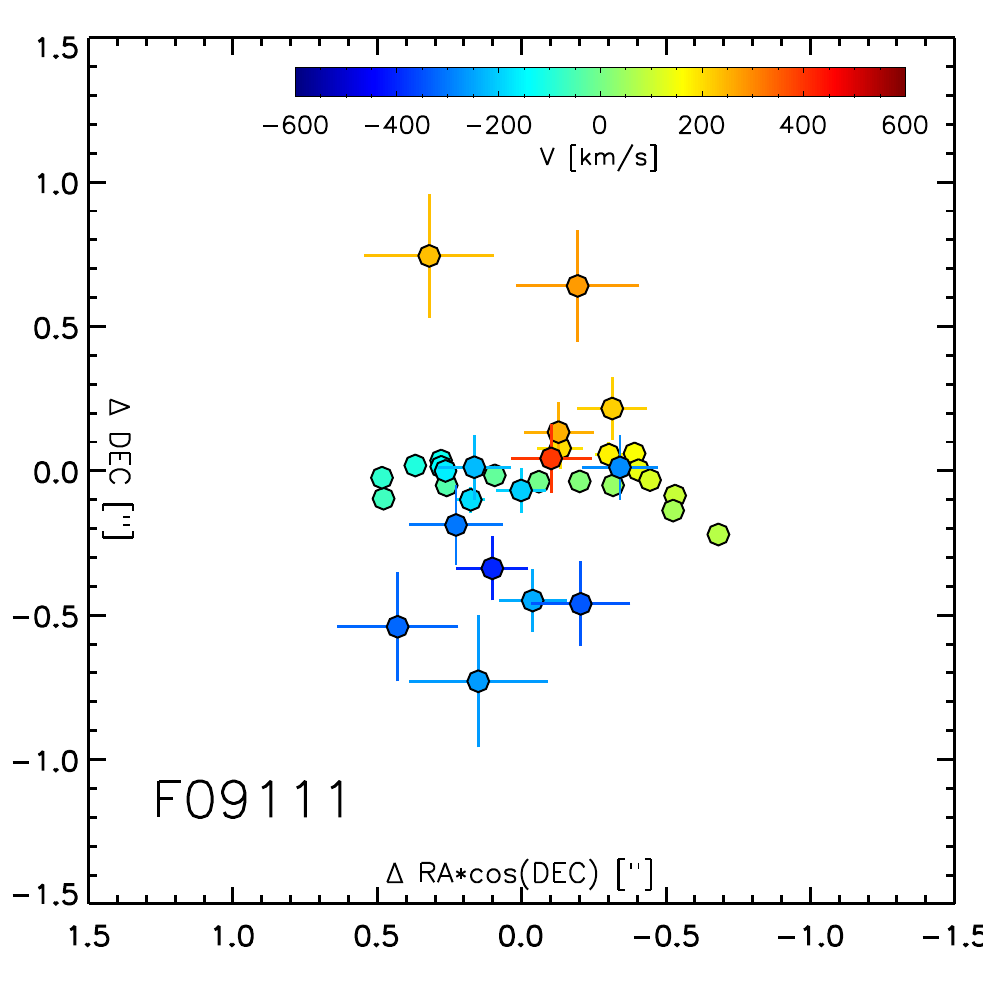}
\includegraphics[width=0.66\hsize]{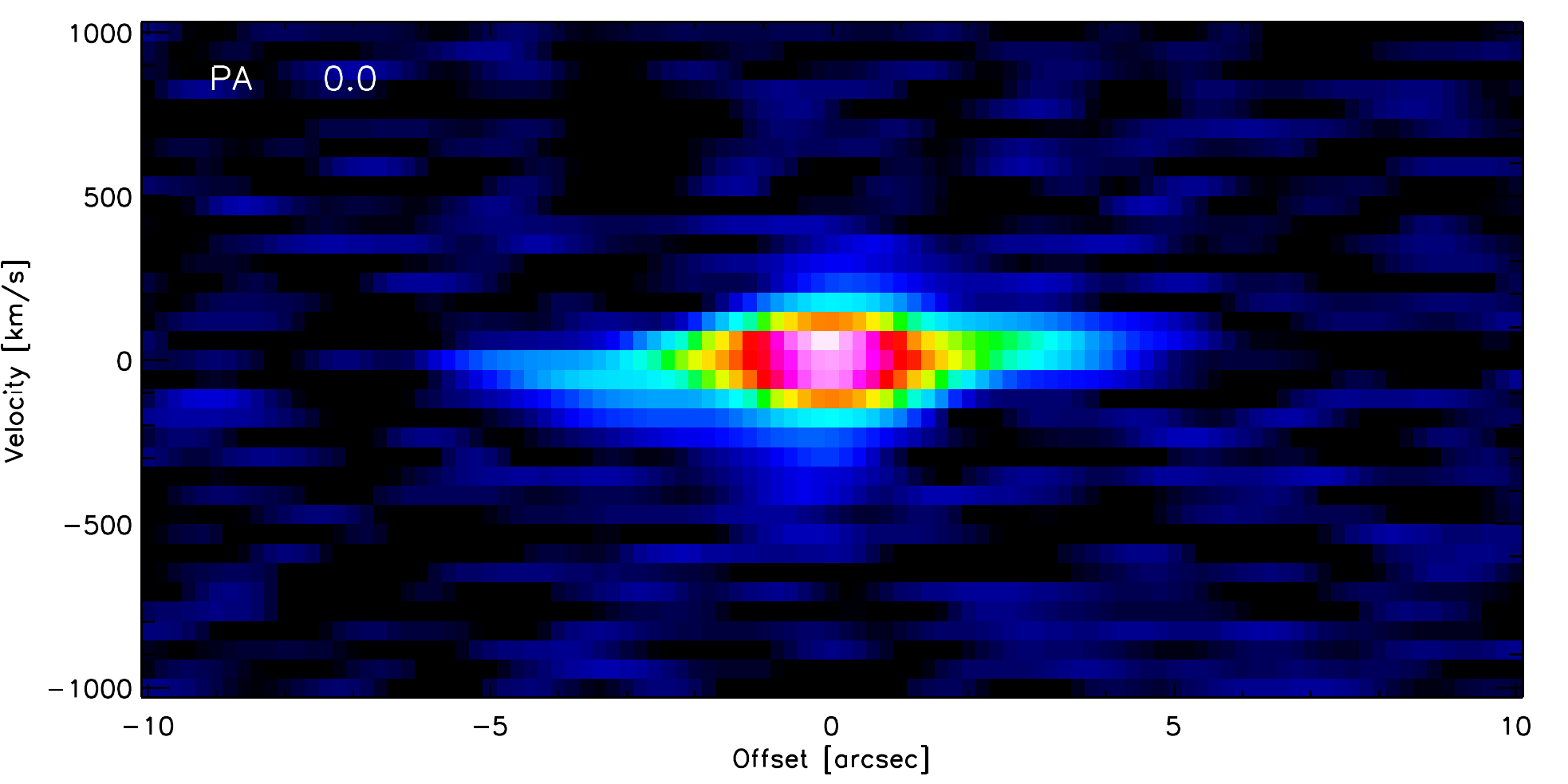}
\includegraphics[width=0.66\hsize,right]{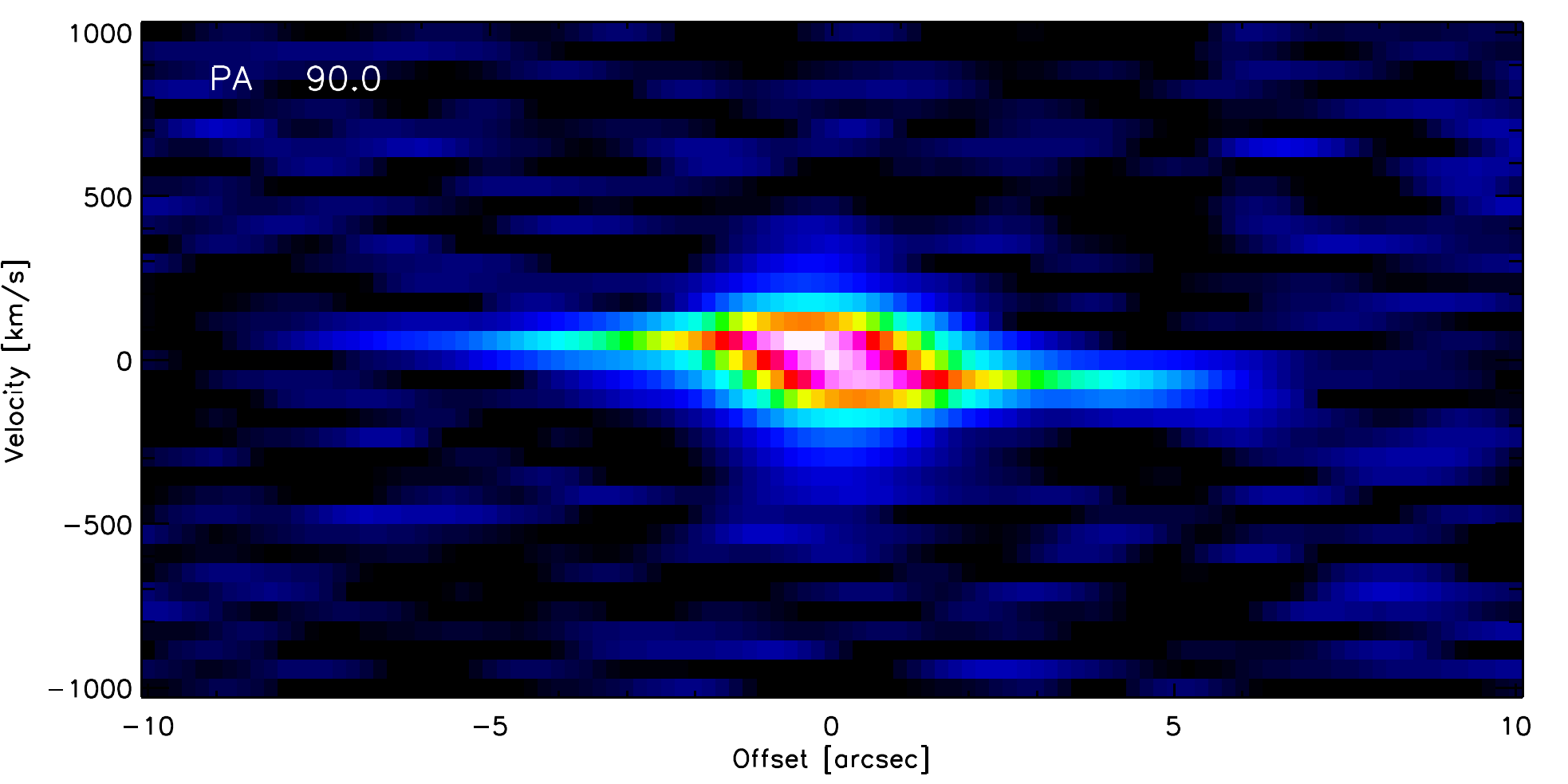}
\caption{ALMA CO(1-0) data for IRAS~F09111-1007W.
Top left: Moment 0 map. Overplotted are contours for outflow in the range
[-510,-210]~\kms\ (blue) and [210,510]~\kms\ (red), 
with contours at [0.35, 0.6, 1.1]~mJy/beam. Top right: Spectrum
centered on the continuum nucleus. The 
spectrum is decomposed into three Gaussians, two for the
host line profile and one for outflow (green). The black line includes 
all three Gaussian components. 
Outflow wings are detected on both sides of the line core.
Center left: Center positions and their 
errors, from UV fitting a Gaussian model to individual velocity channels. 
Center and bottom right: Position-velocity diagrams. PA~0\degr\ is close to 
the bipolar outflow orientation, and orthogonal to the velocity gradient of 
the line core at PA$\sim$90\degr\/.
}
\label{fig:f09111}
\end{figure*}

\begin{figure*}
\centering
\includegraphics[width=0.33\hsize]{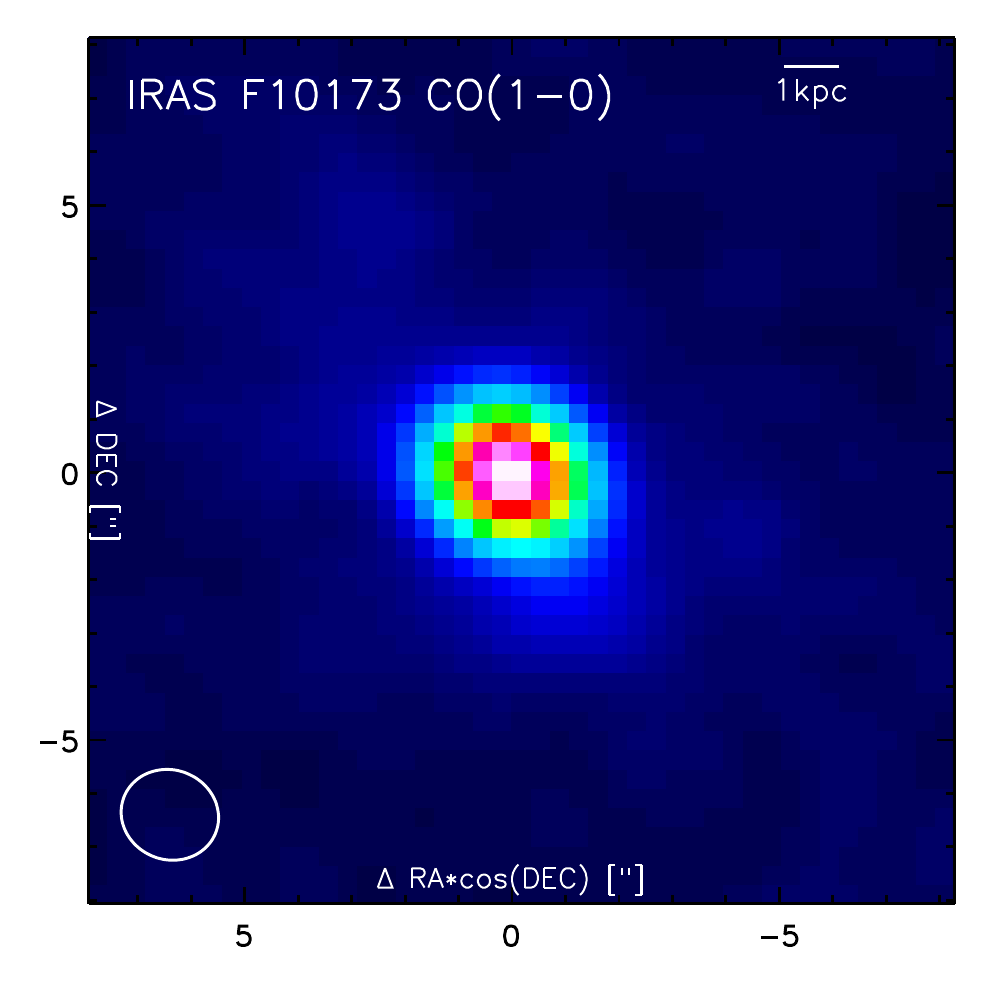}
\includegraphics[width=0.66\hsize]{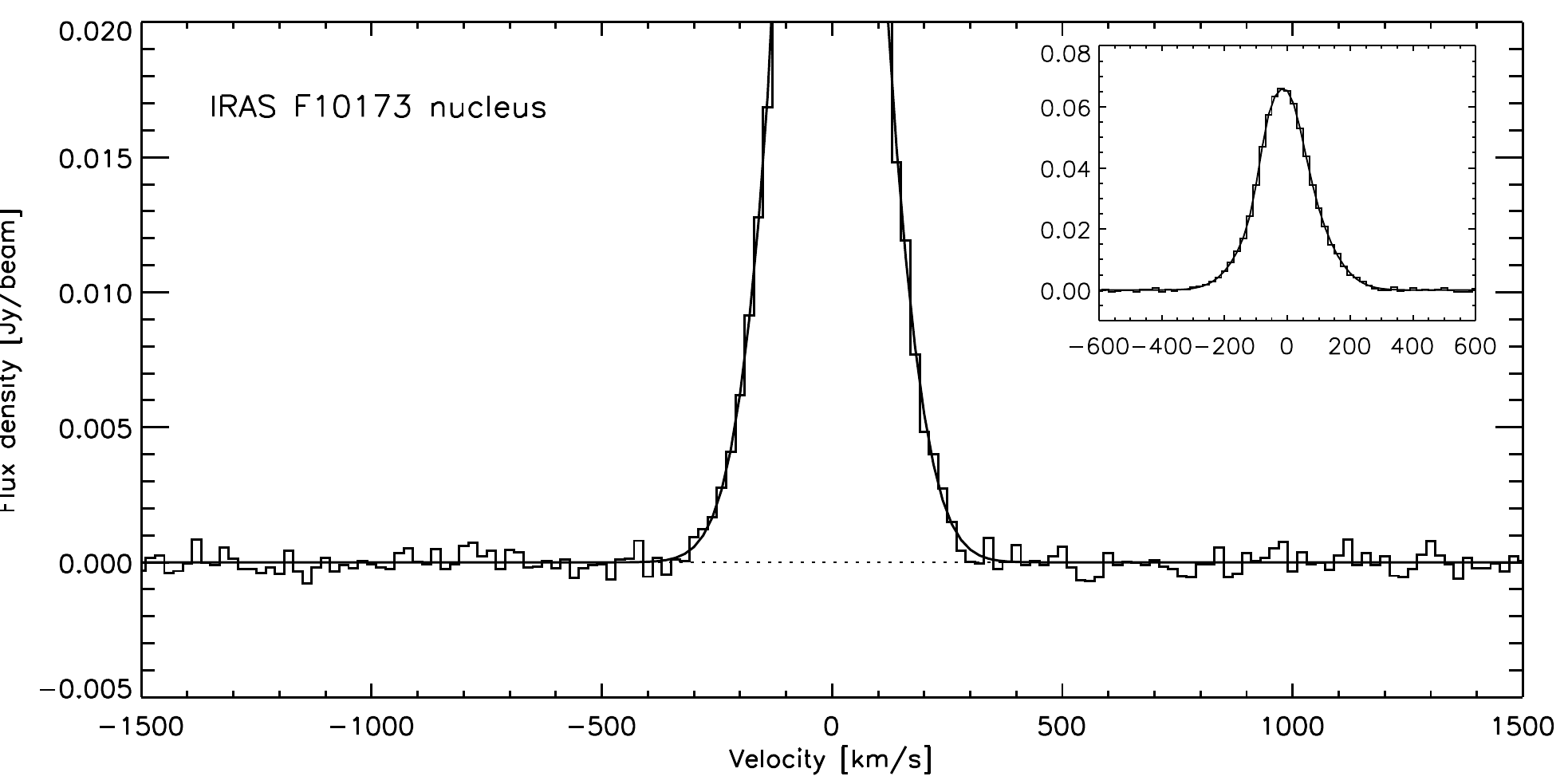}
\includegraphics[width=0.33\hsize]{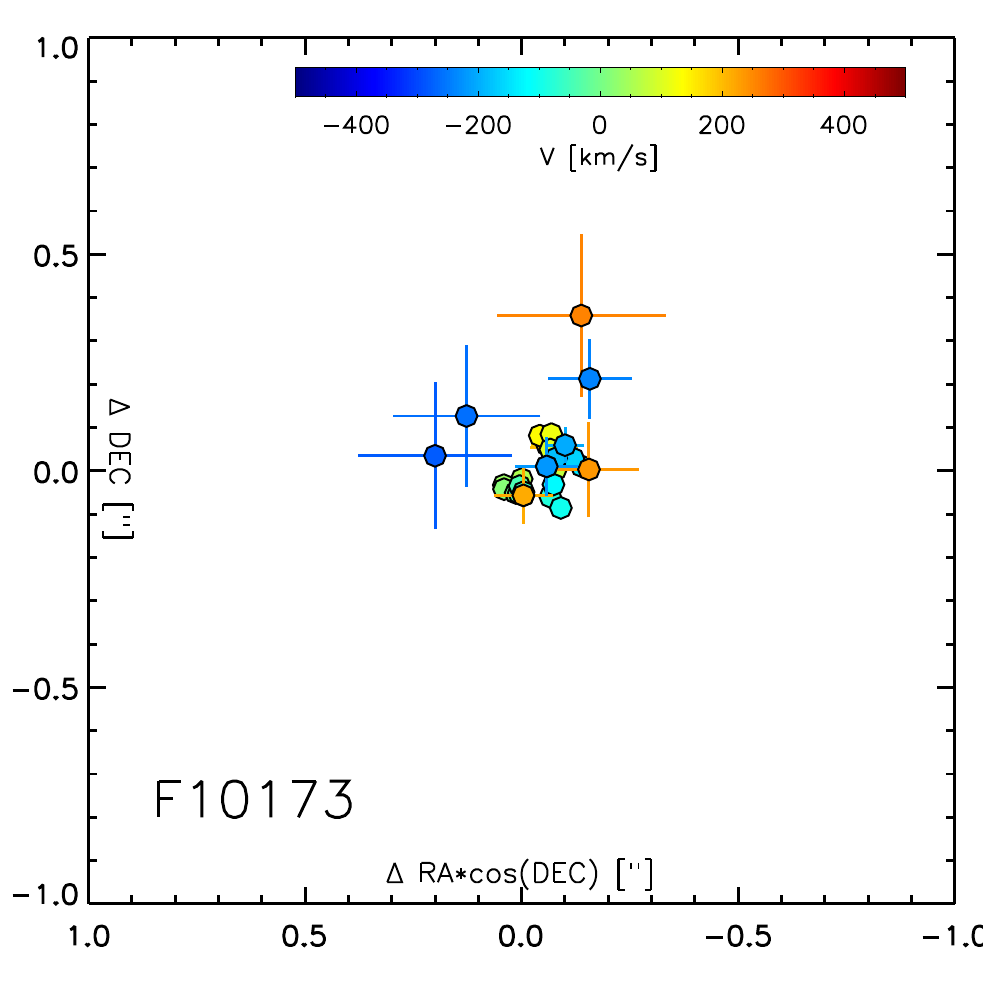}
\includegraphics[width=0.66\hsize]{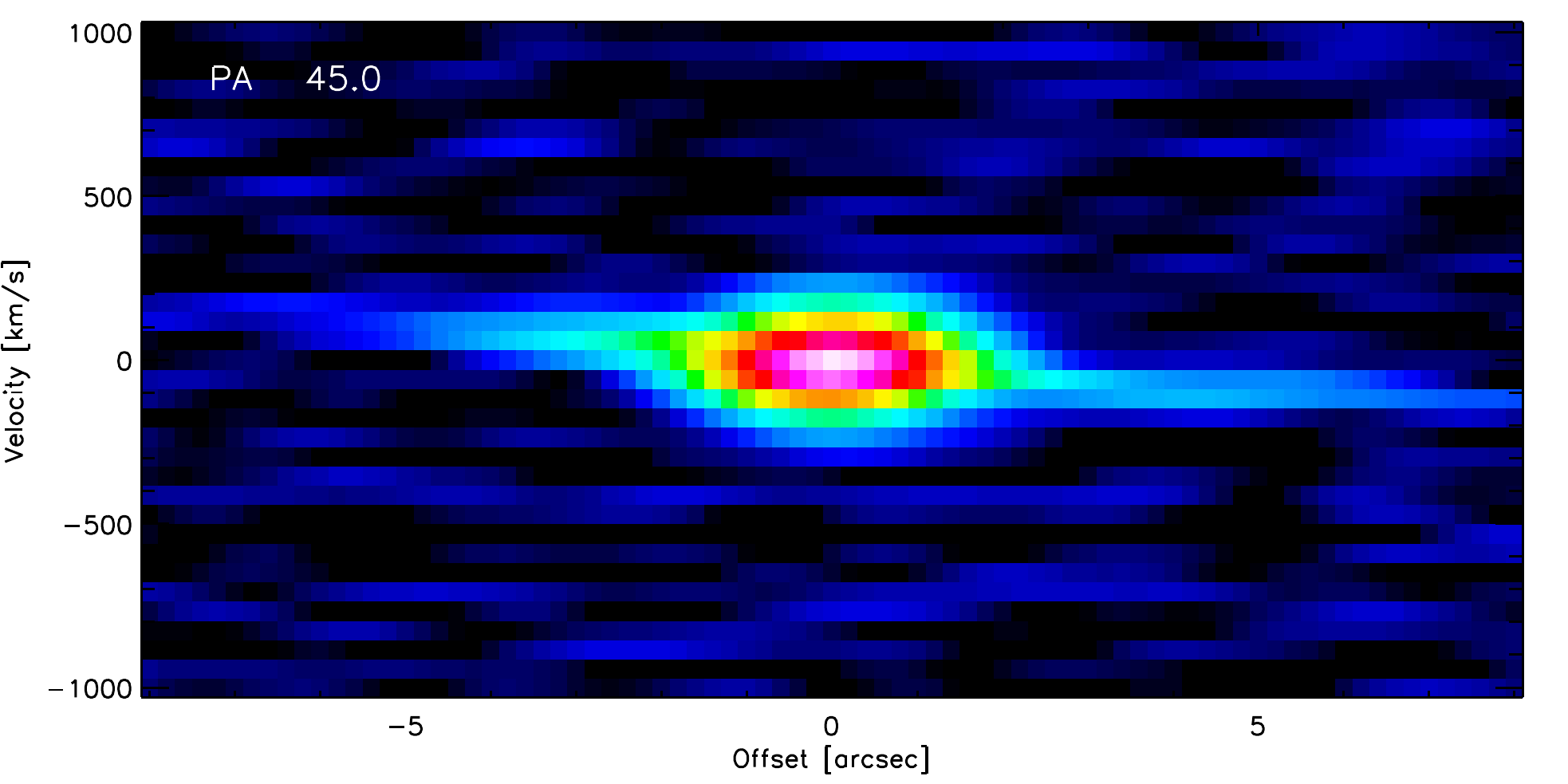}
\includegraphics[width=0.66\hsize,right]{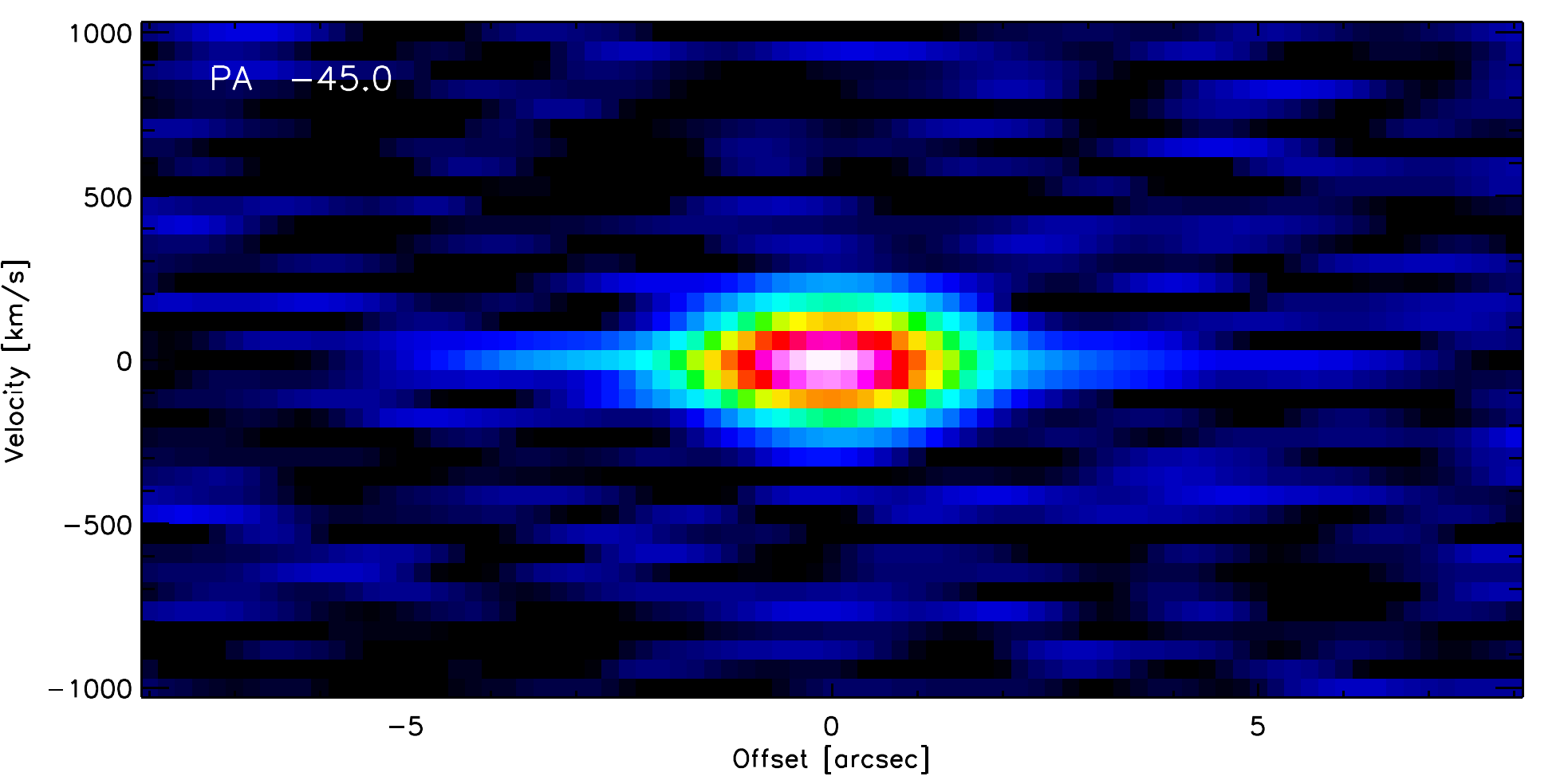}
\caption{ALMA CO(1-0) data for IRAS~F10173+0828.
Top left: Moment 0 map. Top right: Spectrum
centered on the continuum nucleus. The 
spectrum is fitted with three Gaussians which we all assign to the host. The 
black line includes all three Gaussian components. Center left: 
Center positions and their errors, from UV 
fitting a Gaussian model to individual velocity channels. 
Center and bottom right: Position-velocity diagrams. In the absence of a clear 
outflow signature or of a clear velocity gradient in the line core, 
we simply plot two orthogonal directions of which PA~+45\degr\ follows a 
velocity gradient on scales of several arc seconds.
We conservatively adopt an upper limit for 
outflow in this source but cannot exclude that part of the line wings to 
$\pm$300~\kms\ are caused by outflow.}
\label{fig:f10173}
\end{figure*}

\begin{figure*}
\centering
\includegraphics[width=0.33\hsize]{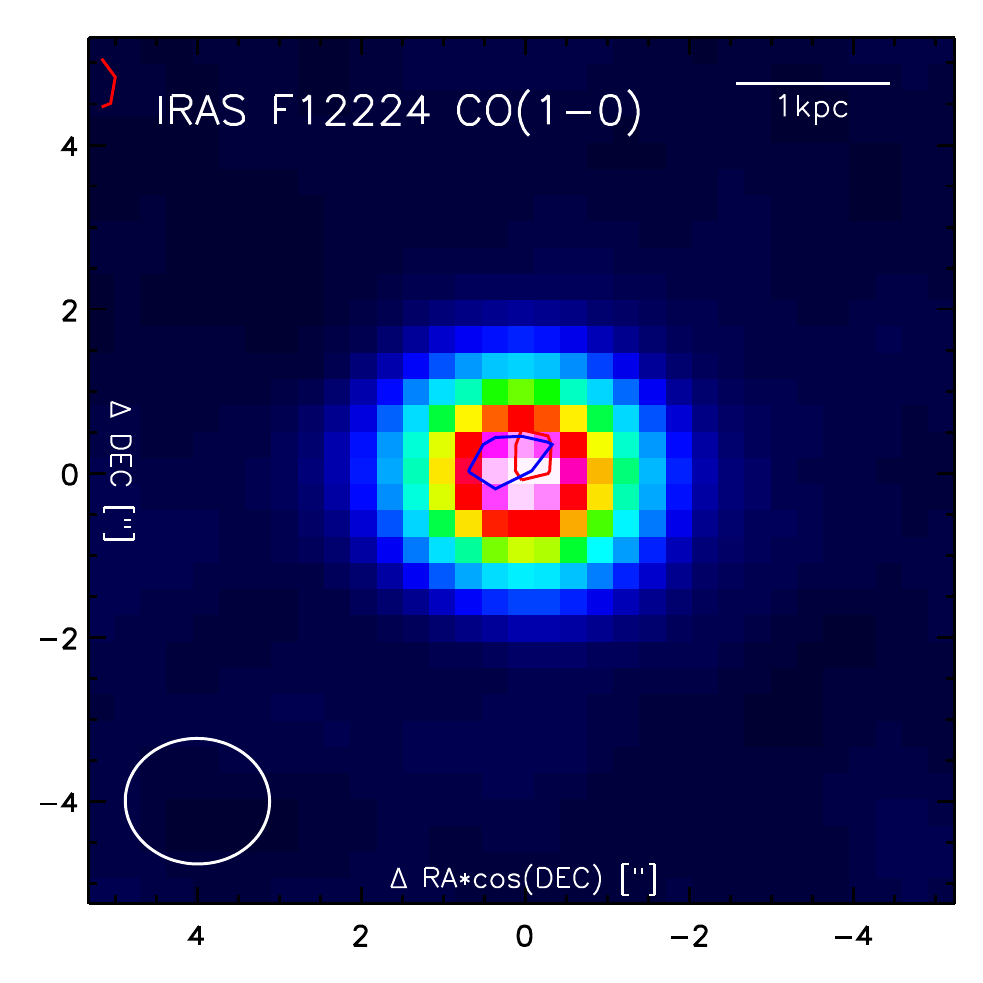}
\includegraphics[width=0.66\hsize]{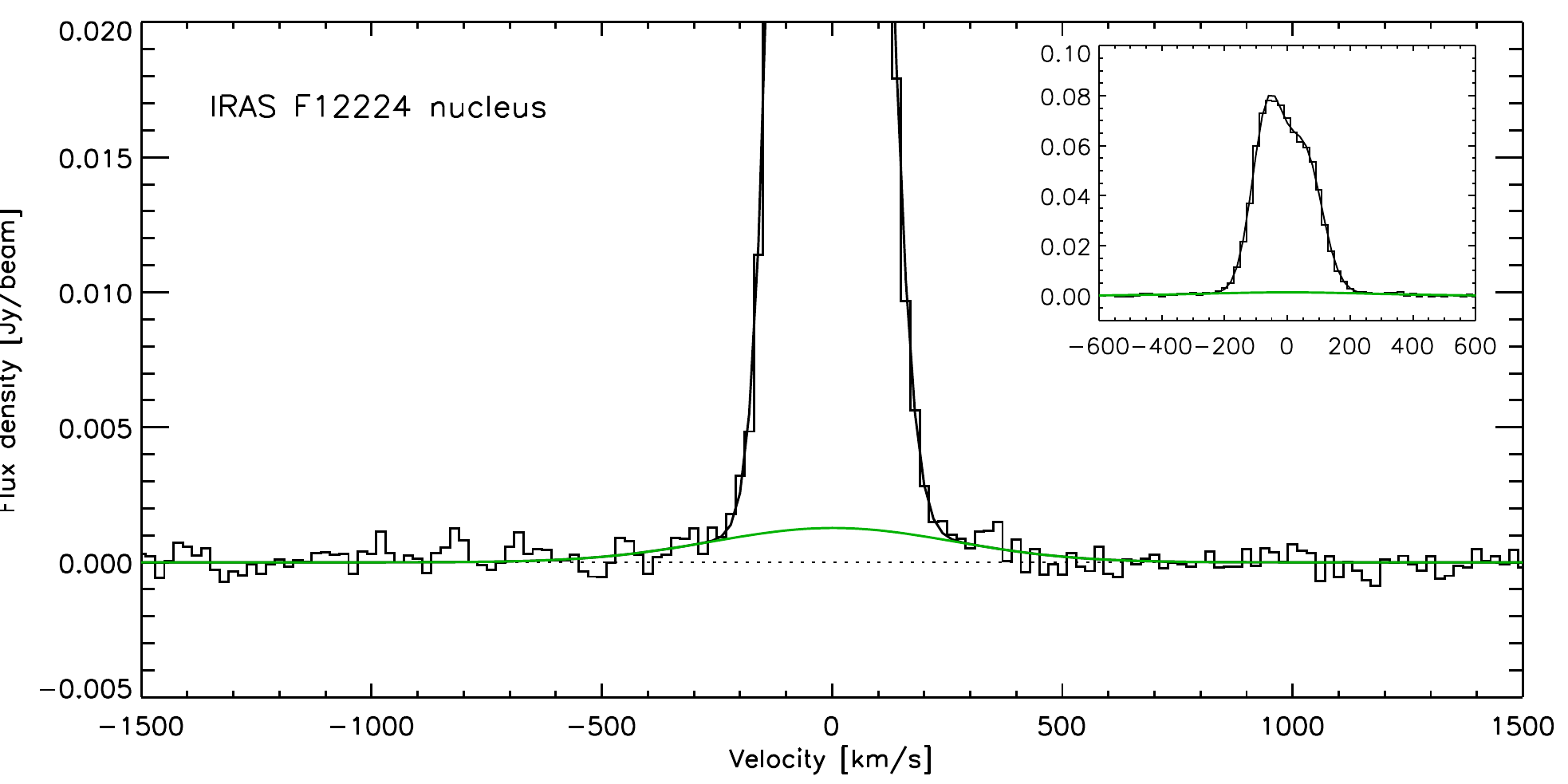}
\includegraphics[width=0.33\hsize]{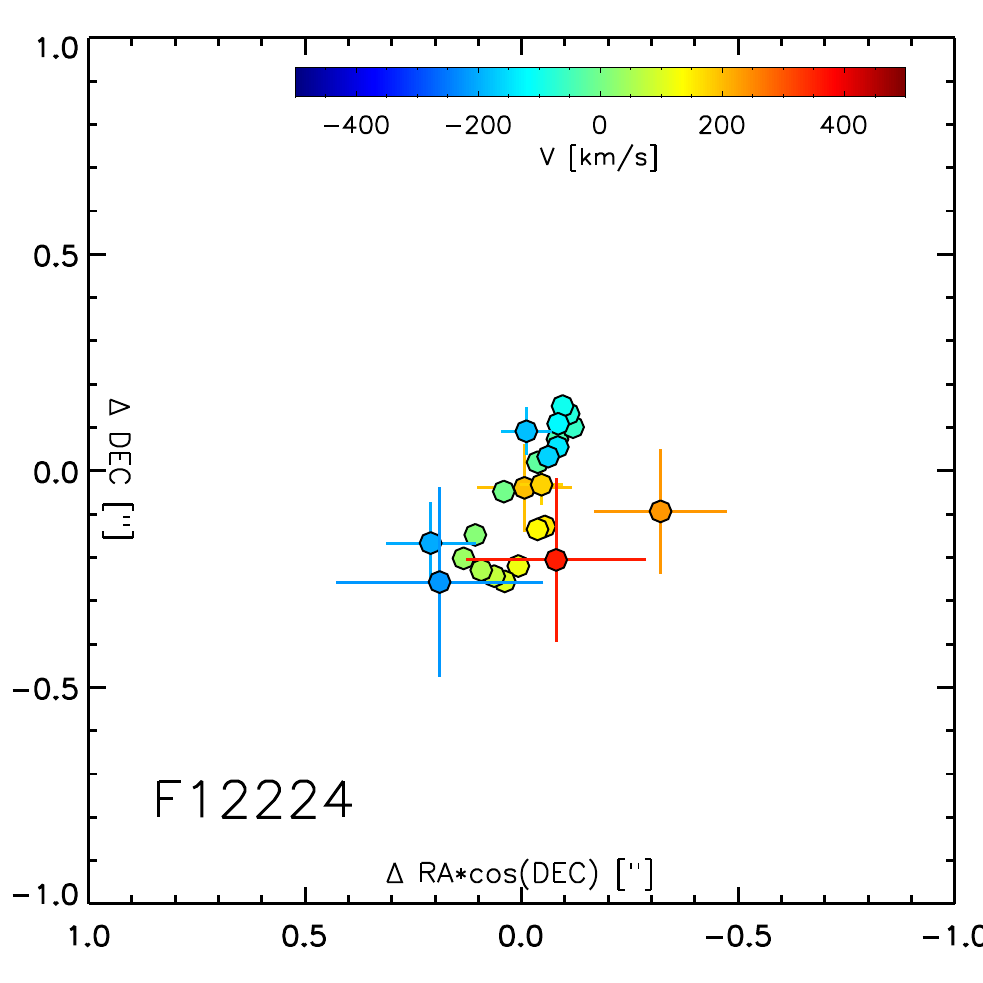}
\includegraphics[width=0.66\hsize]{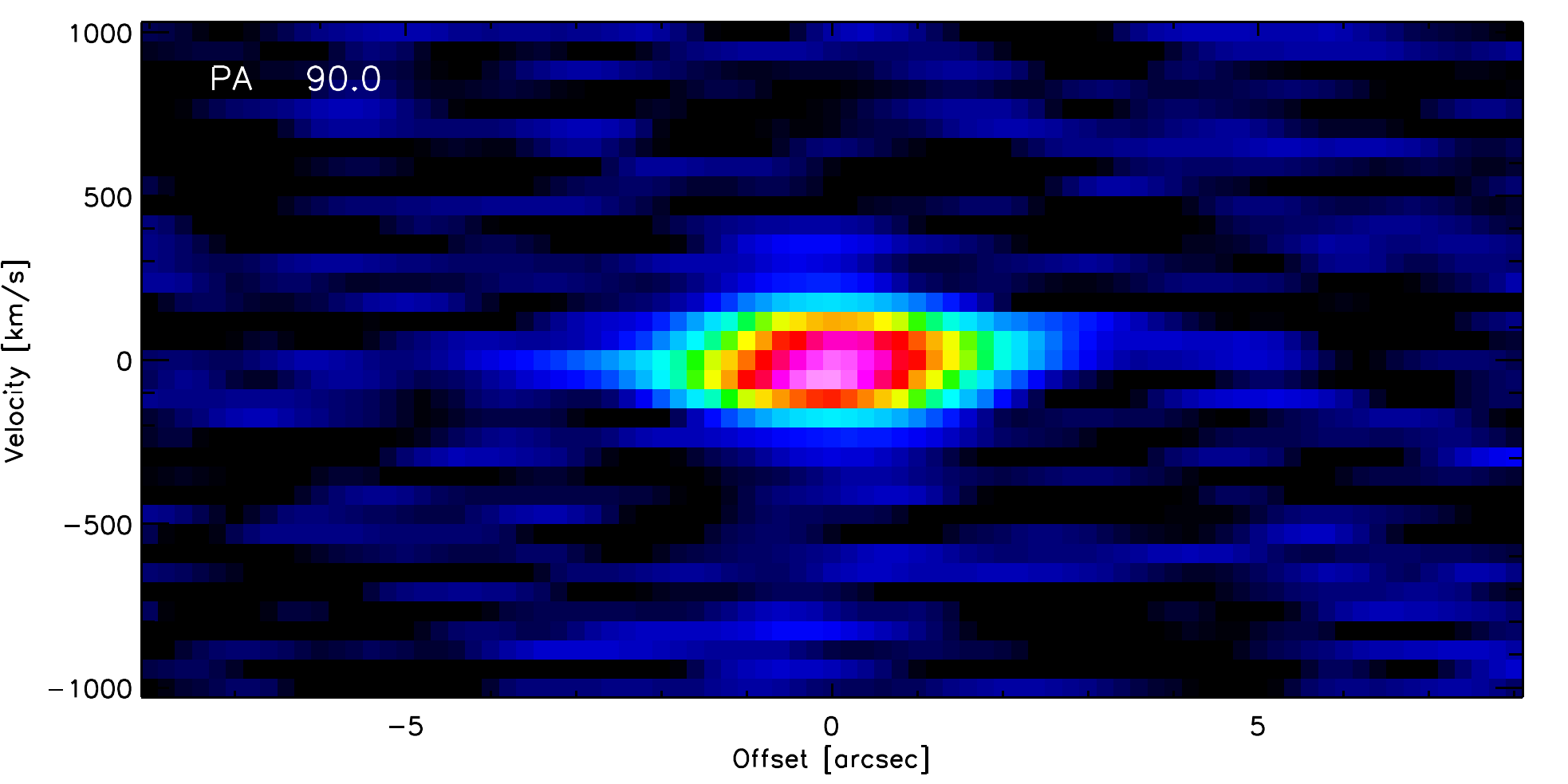}
\includegraphics[width=0.66\hsize,right]{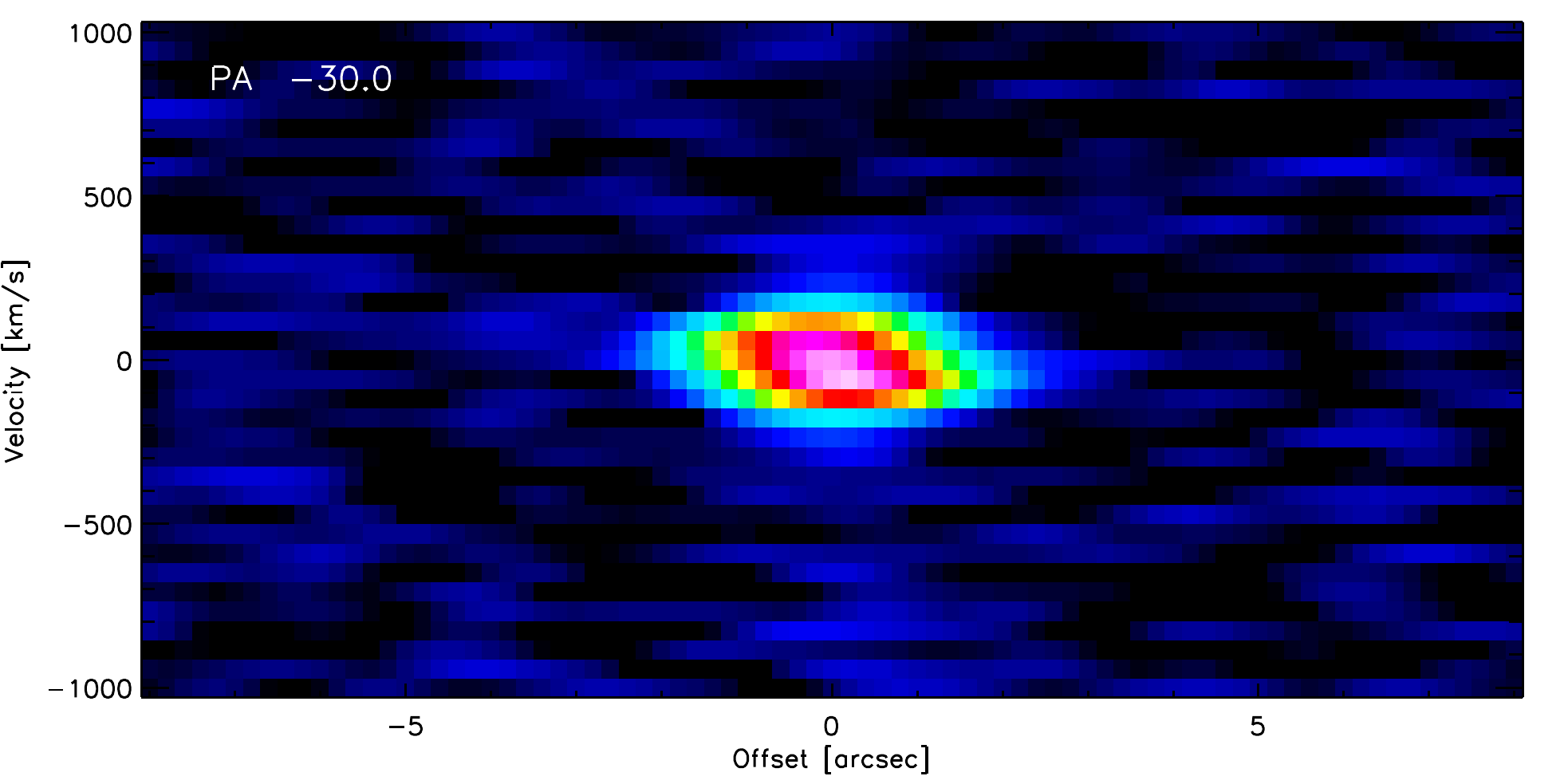}
\caption{ALMA CO(1-0) data for IRAS~F12224-0624.
Top left: Moment 0 map. Overplotted are contours (0.4~mJy = 3.8$\sigma$) 
for outflow in the range [-510,-270]~\kms\ (blue) and [270,510]~\kms (red), 
with contours at 0.4~mJy/beam. 
Top right: Spectrum centered on the continuum nucleus. The 
spectrum is decomposed into three Gaussians, two for the
host line profile and one for outflow (green). The black line includes 
all three Gaussian components. Weak outflow wings are detected on the blue and 
red side of the line core.
Center left: Center positions and their errors, from UV 
fitting a Gaussian model to individual velocity channels. 
Center and bottom right: Position-velocity diagrams. PA~+90\degr\ follows the
tentative E-W offset of blue and red outflow wings, while PA~-30\degr\ 
represents the velocity gradient in the line core. 
}
\label{fig:f12224}
\end{figure*}

\begin{figure*}
\centering
\includegraphics[width=0.33\hsize]{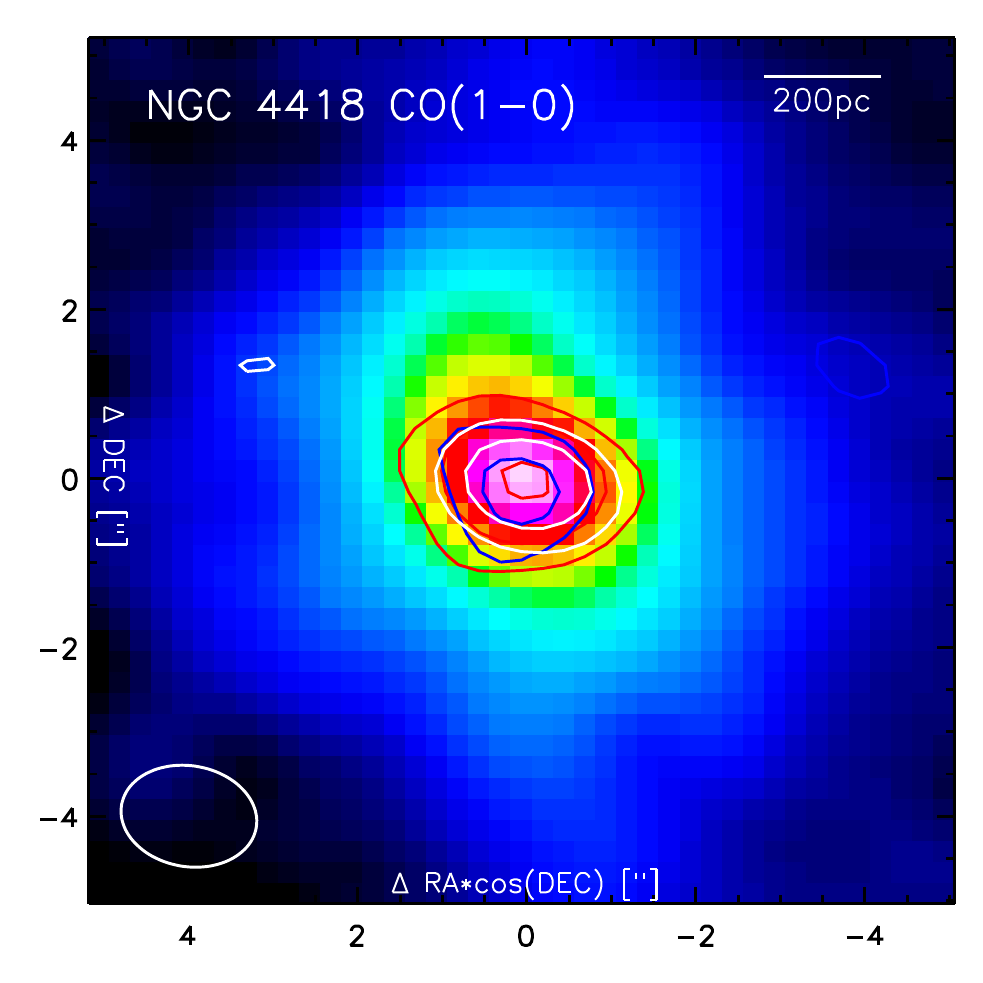}
\includegraphics[width=0.66\hsize]{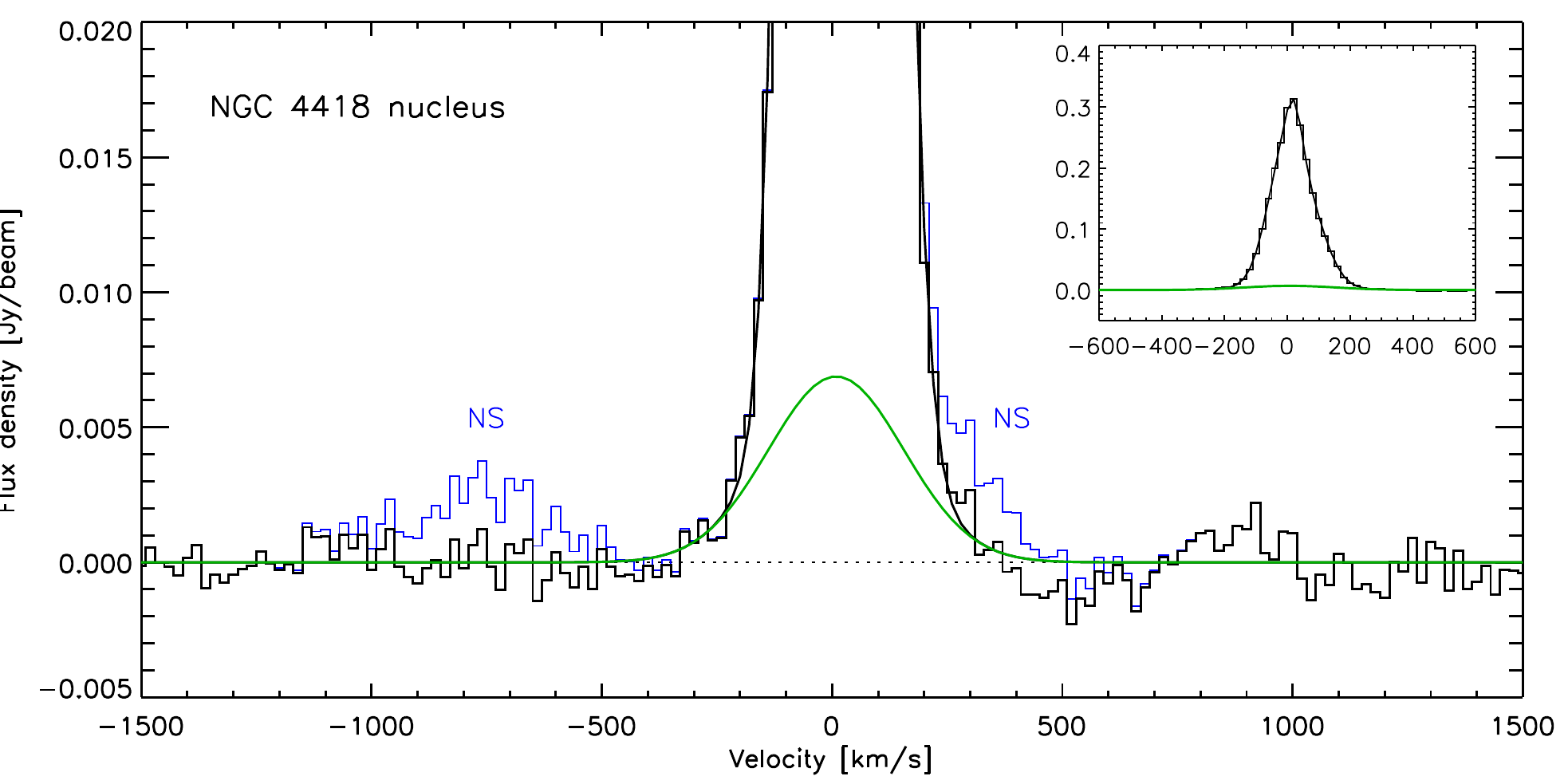}
\includegraphics[width=0.33\hsize]{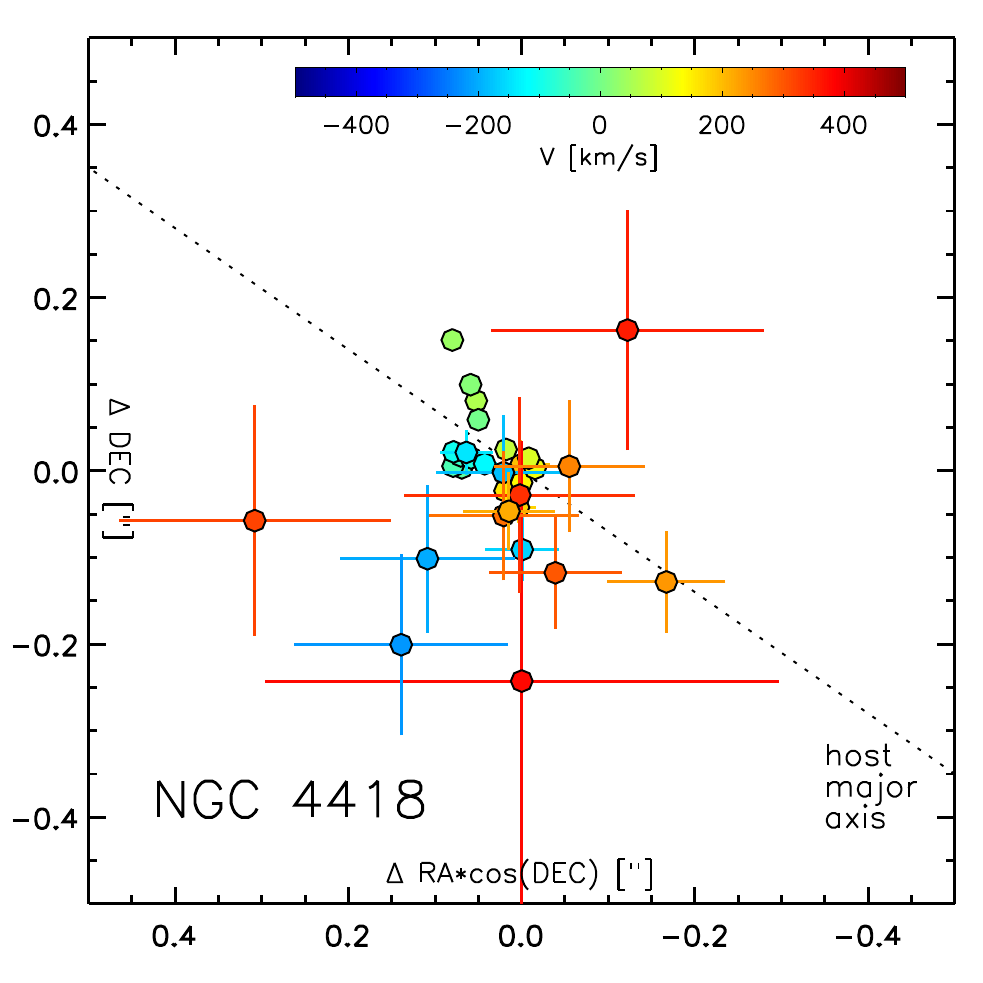}
\includegraphics[width=0.66\hsize]{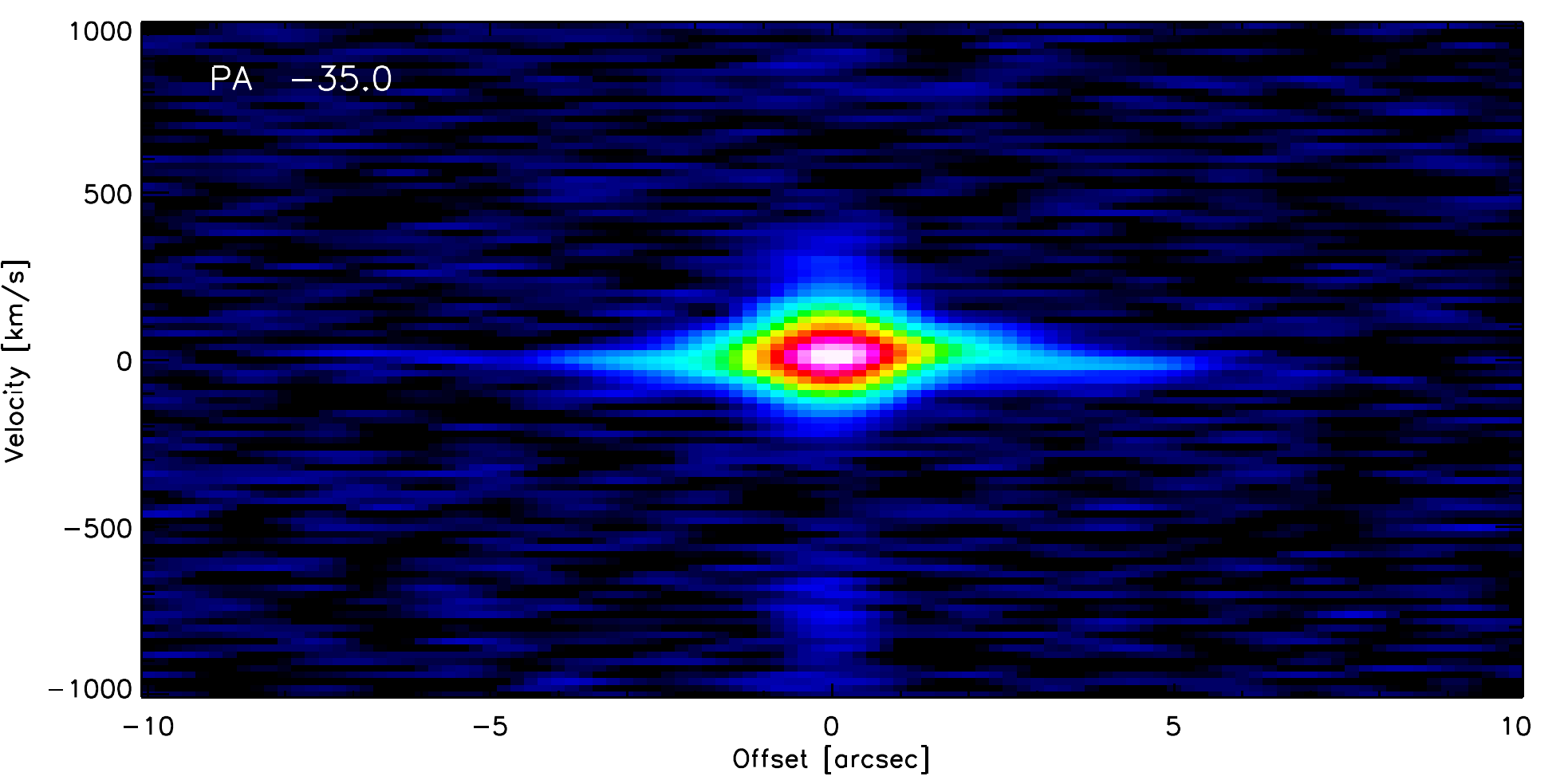}
\includegraphics[width=0.66\hsize,right]{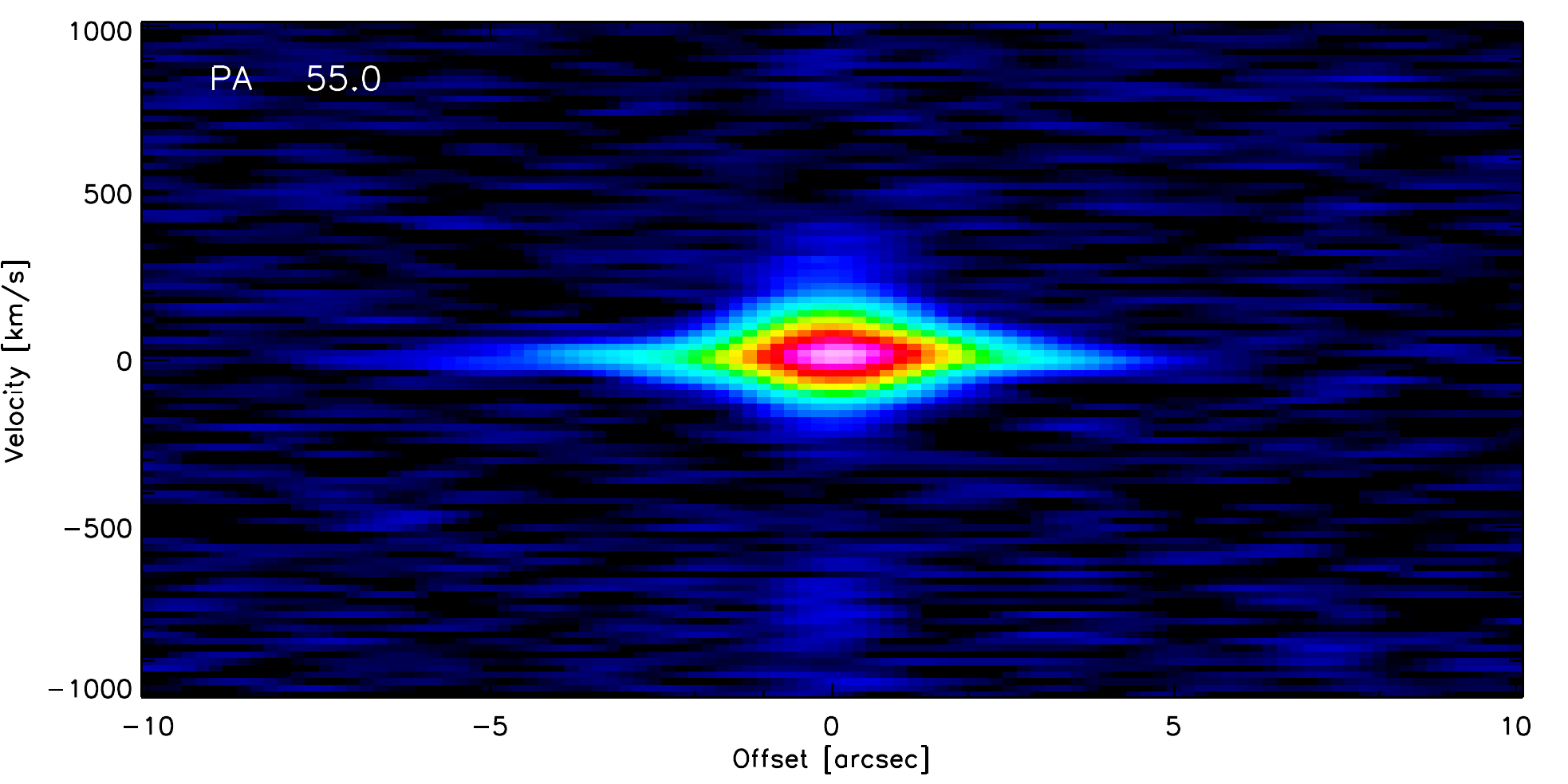}
\caption{ALMA CO(1-0) data for NGC~4418.
Top left: Moment 0 map. Overplotted are contours 
for the ranges [-930,-570]~\kms\ (white, this range is dominated
by emission of nitrogen sulfide),
[-330,-190]~\kms\ (blue, blue wing of CO, [0.8, 1.6]~mJy/beam) and 
[230,510]~\kms\ (red, nitrogen sulfide NS superposed on red wing of CO, 
[0.6, 1.2, 2.4]~mJy/beam).
Top right: Spectrum centered on the continuum nucleus. Before fitting,
we have subtracted emission from nitrogen sulfide (blue)
as described in the text, to obtain the black histogram CO(1-0) spectrum.
CO outflow wings remain present in the line profile 
after subtraction of nitrogen sulfide.
Nitrogen sulfide emission is still present in the other panels of this
figure. The CO spectrum is decomposed into four Gaussians, three for the
host line profile and one for outflow (green). The black line includes 
all four Gaussian components. Center left: Center positions and their errors, 
from UV fitting a Gaussian model to individual velocity channels. The
dotted line indicates the orientation of the major axis of the large scale 
disk of NGC~4418, according to 2MASS. Center and bottom right: 
Position-velocity diagrams. PA~-35\degr\ is orthogonal to the host galaxy's
major axis PA~+55\degr\/. No outflow orientation is firmly established given
signal-to-noise and contamination by NS, but the lesser contaminated blue wing 
may prefer outflow orthogonal to the disk.
}
\label{fig:ngc4418}
\end{figure*}

\begin{figure*}
\includegraphics[width=0.33\hsize]{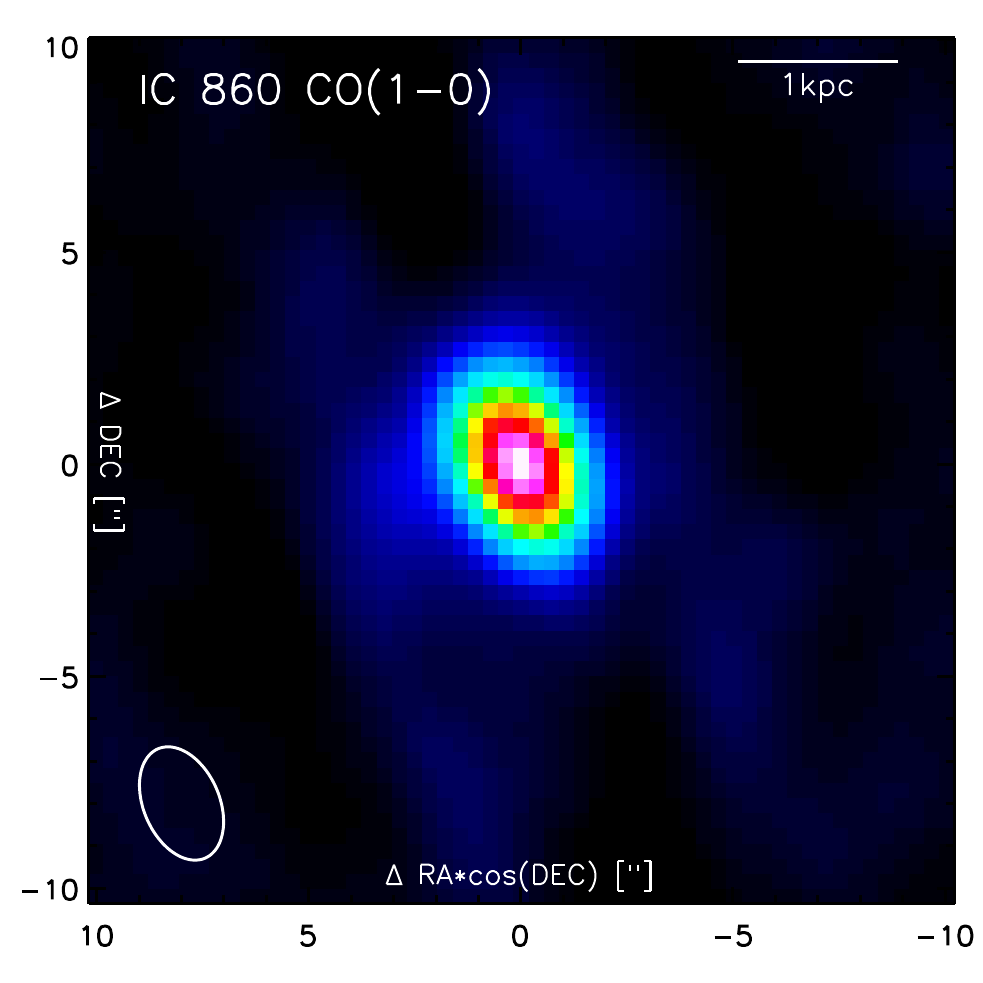}
\includegraphics[width=0.66\hsize]{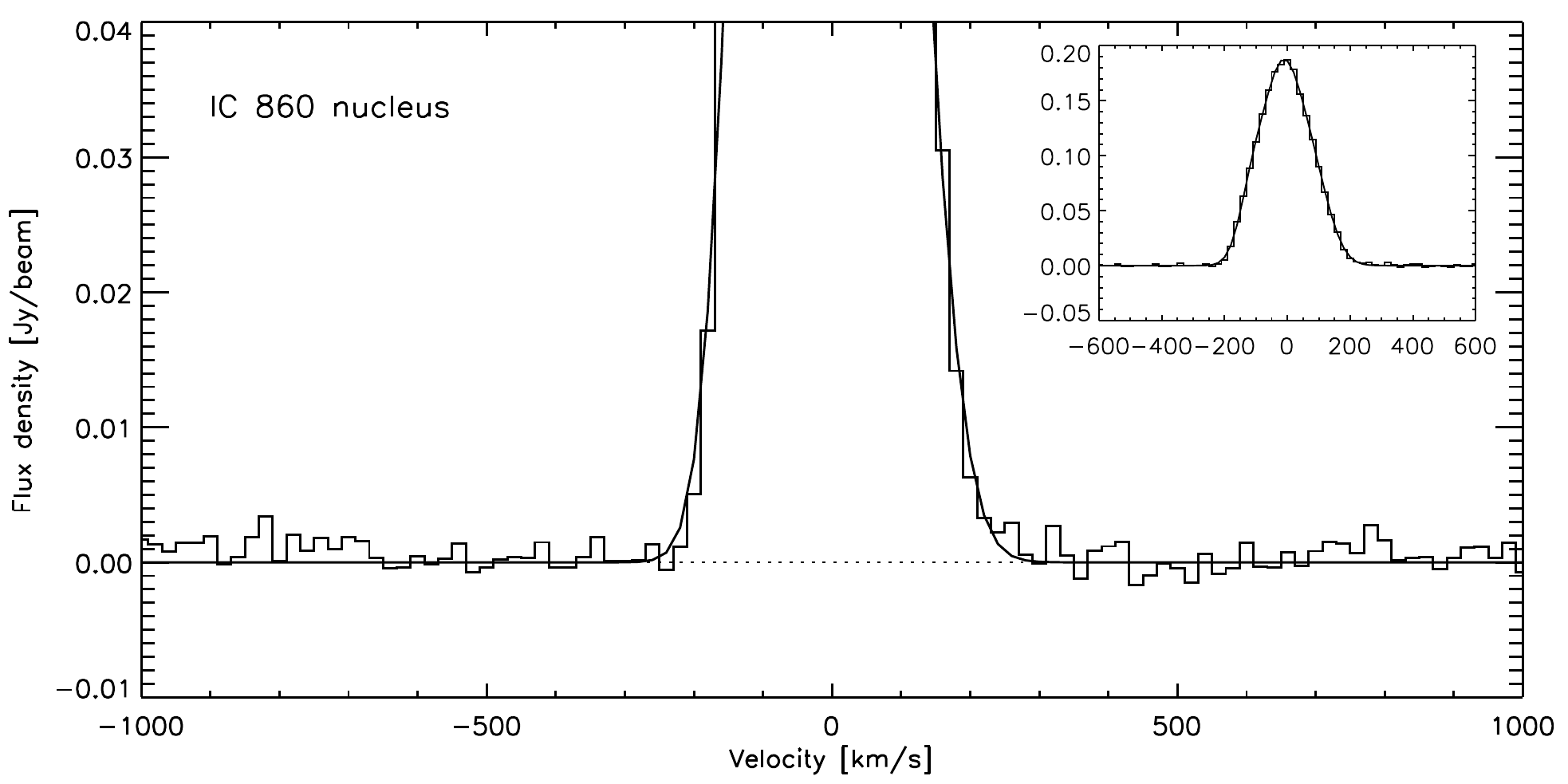}
\includegraphics[width=0.33\hsize]{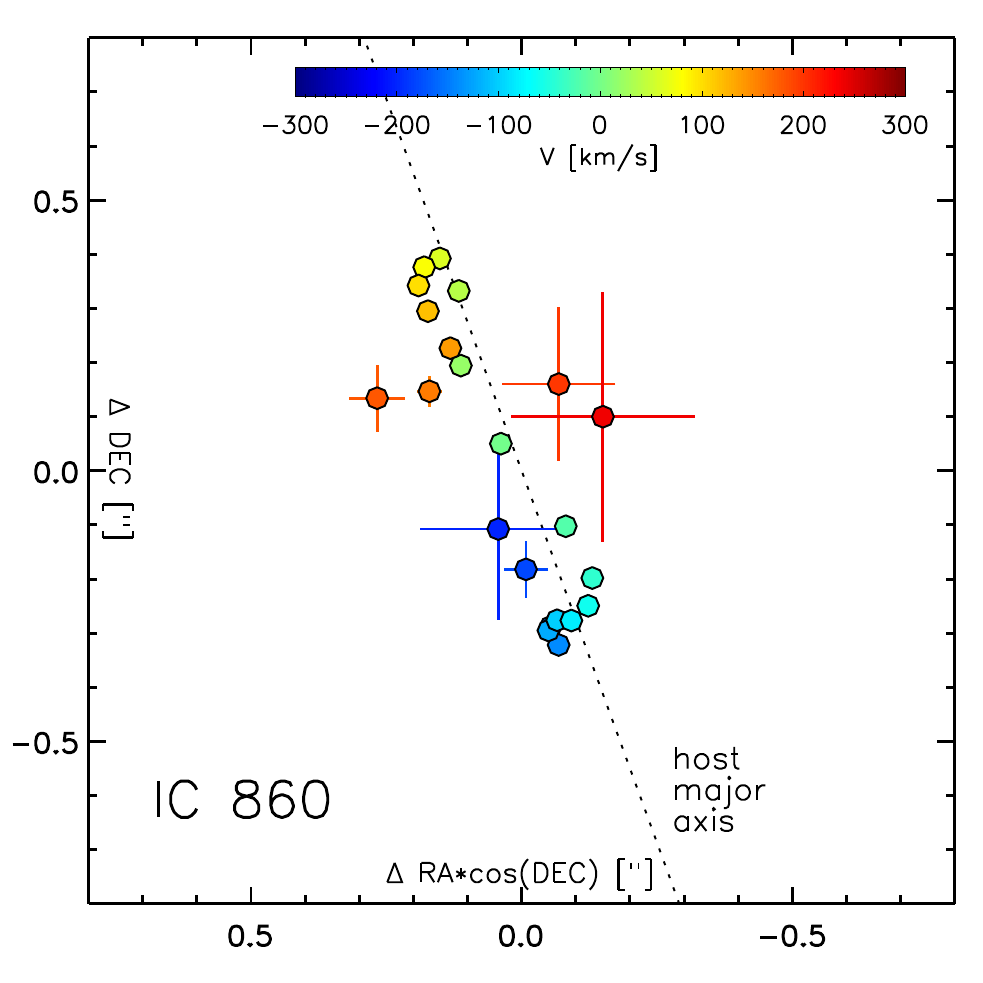}
\includegraphics[width=0.66\hsize]{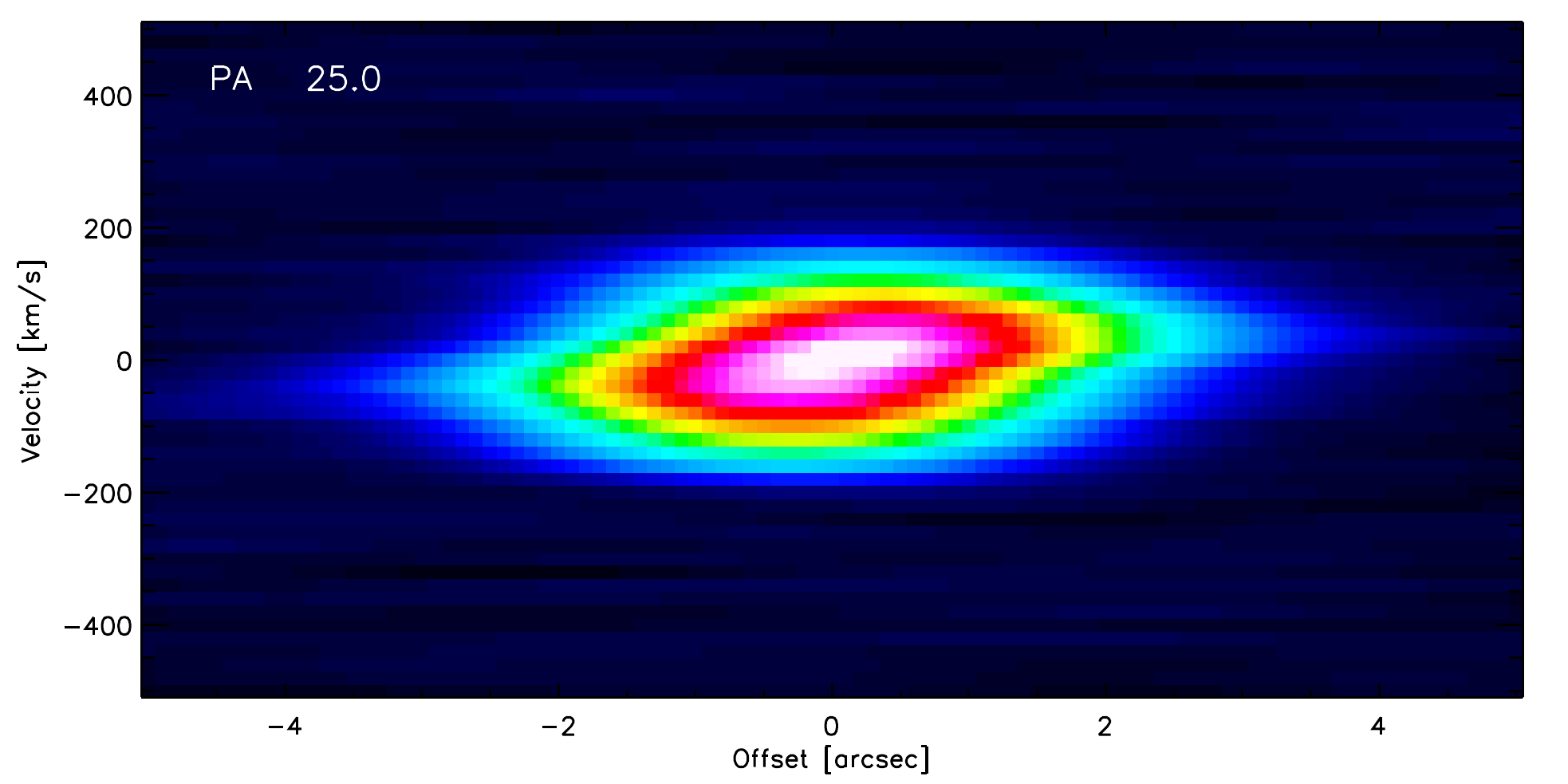}
\caption{NOEMA CO(1-0) data for IC 860.
Top left: Moment 0 map.
Top right: Spectrum centered on the continuum nucleus. The 
host spectrum is fit with three Gaussians, their sum is 
overplotted in black. We conservatively assign an upper limit to a CO 
outflow. Bottom left: Center positions and their errors, from UV 
fitting a Gaussian model to individual velocity channels. The dotted line indicates
the 2MASS extended source catalog K-band major axis PA 20\degr\ of the 
stellar host. Bottom right: Position-velocity diagram at PA~+25deg, 
close to the velocity gradient of the line core and to the host major axis. 
}
\label{fig:ic860}
\end{figure*}

\begin{figure*}
\centering
\includegraphics[width=0.33\hsize]{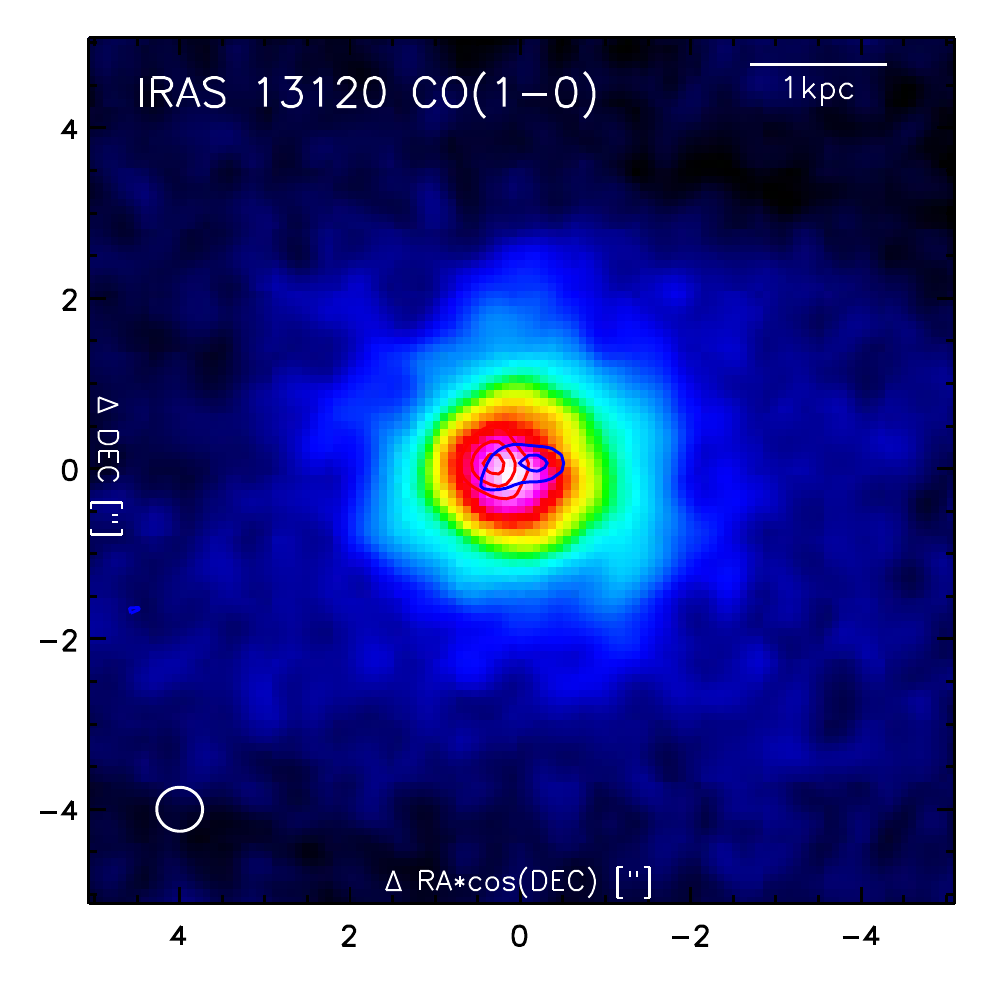}
\includegraphics[width=0.66\hsize]{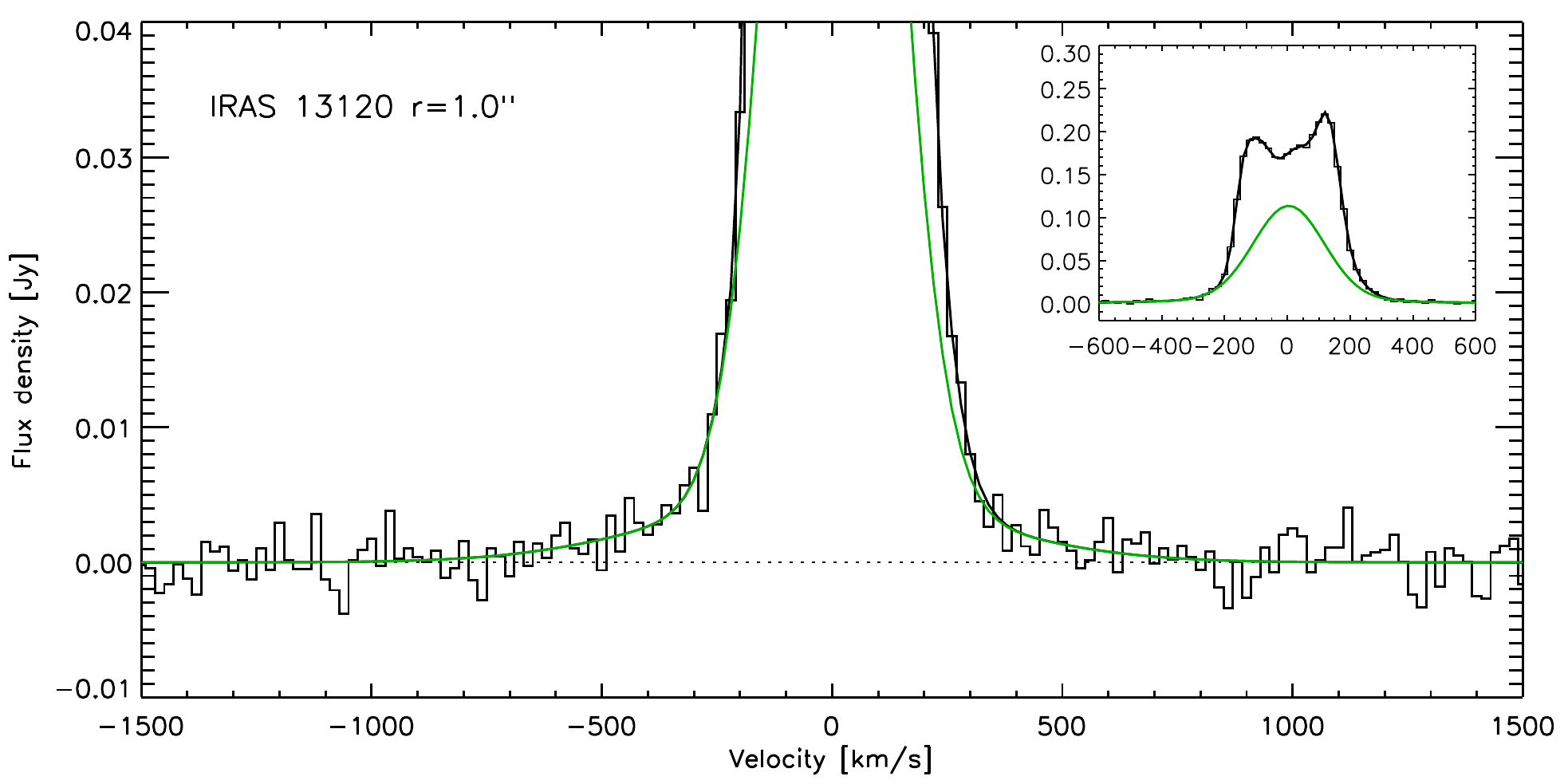}
\includegraphics[width=0.33\hsize]{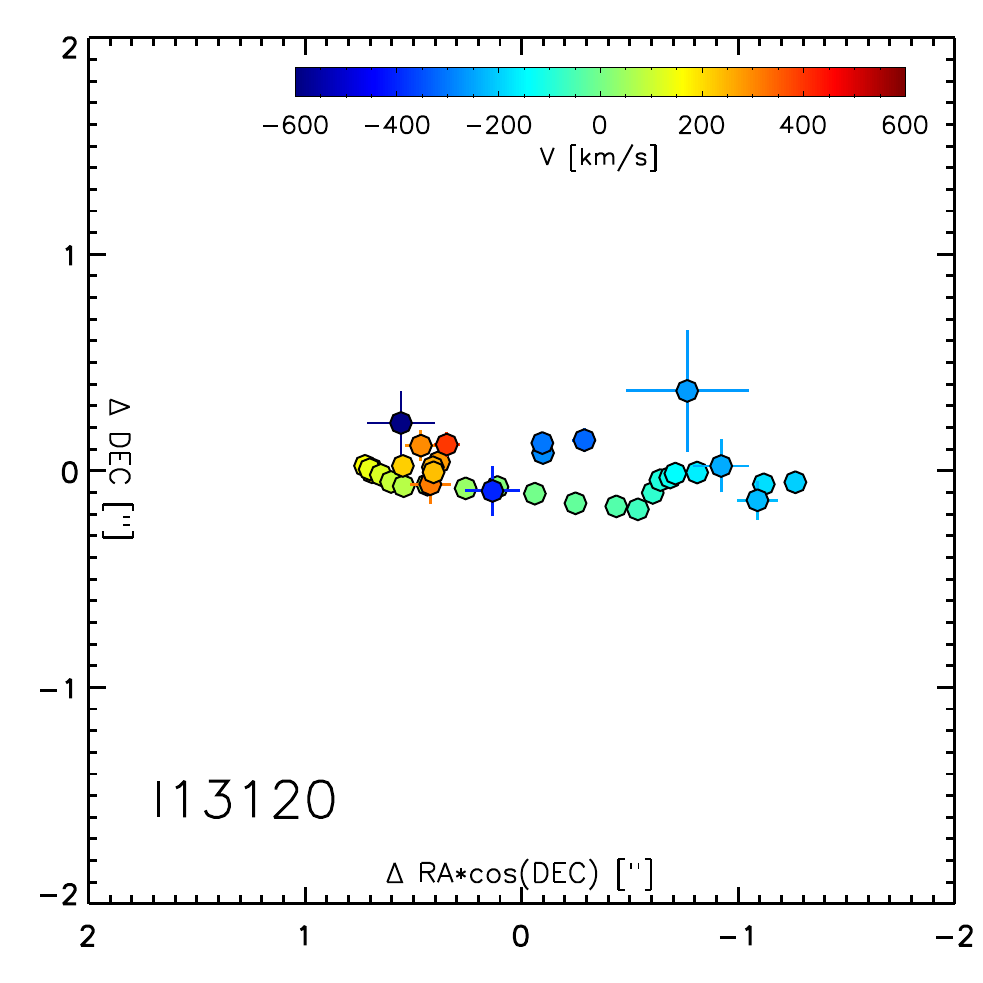}
\includegraphics[width=0.66\hsize]{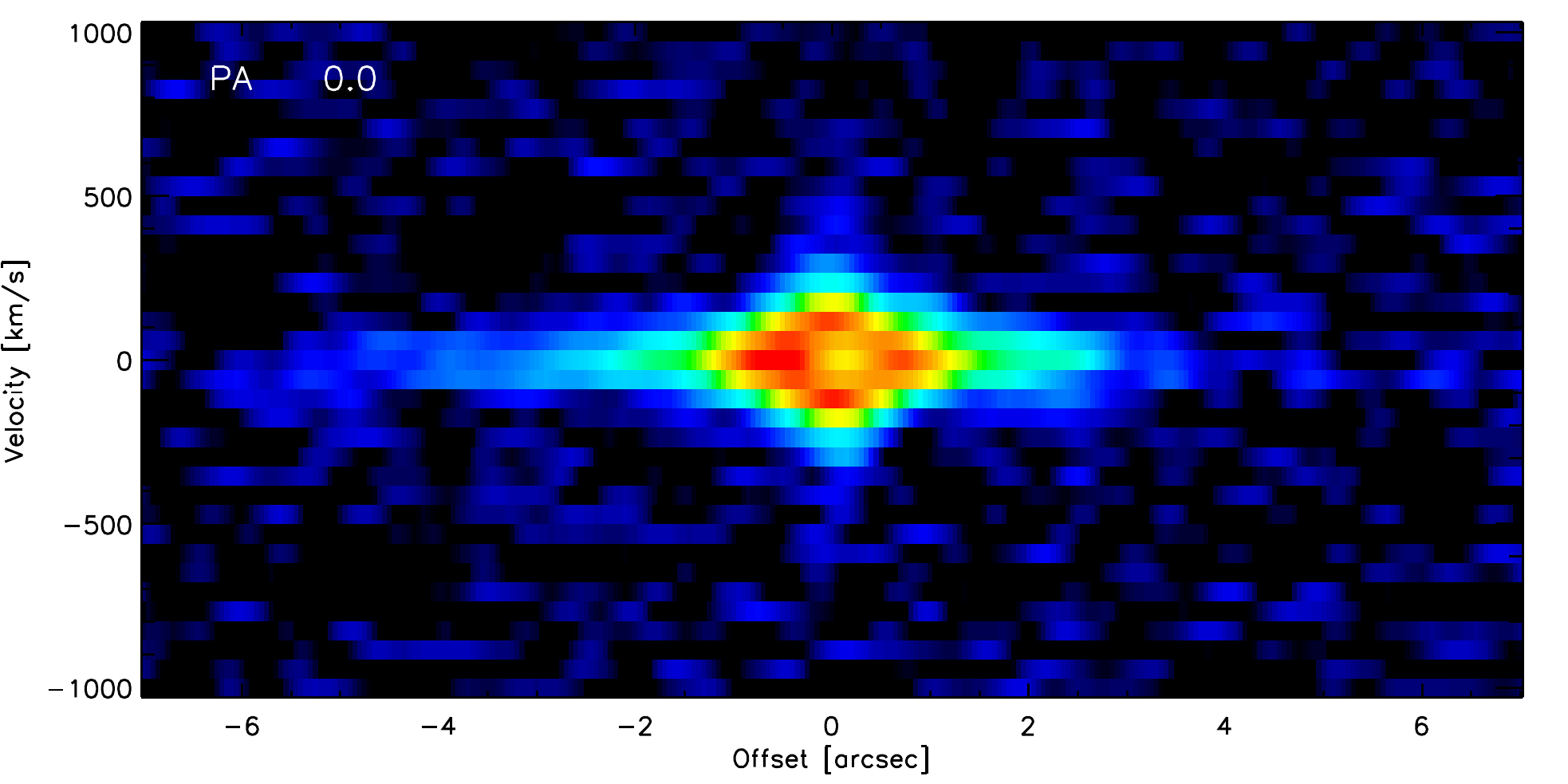}
\includegraphics[width=0.66\hsize,right]{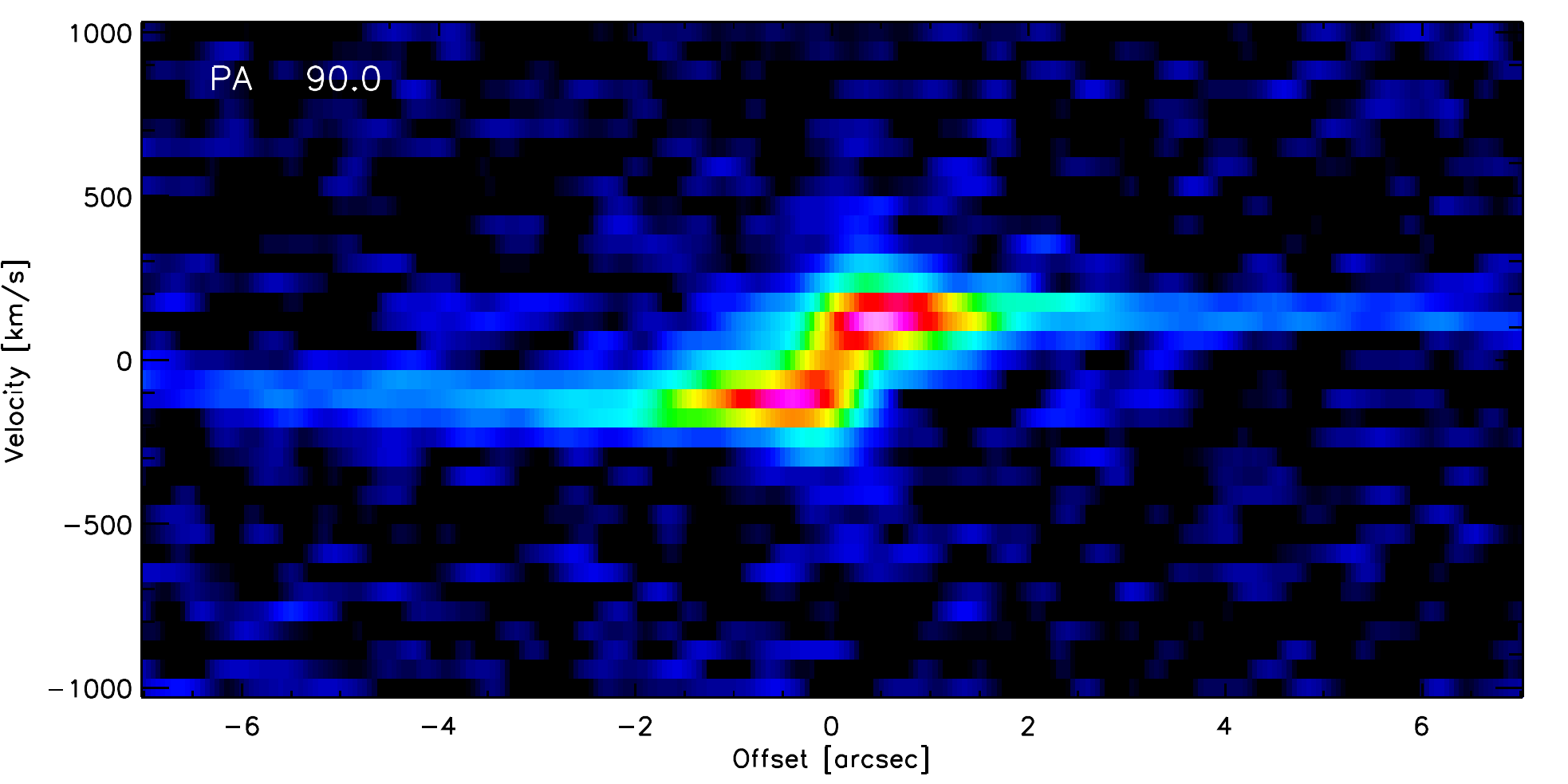}
\caption{ALMA CO(1-0) data for IRAS~13120-5453.
Top left: Moment 0 map. Overplotted are contours 
for outflow in the range [-690,-270]~\kms\ (blue) and [270,510]~\kms\ (red), 
with contours at [0.55, 0.9, 1.2]~mJy/beam. 
Top right: Spectrum in a r=1\arcsec\ aperture centered on the continuum 
nucleus. The 
spectrum is decomposed into six Gaussians, four for the double-horned
host line profile and two for outflow (green). The black line includes 
all six Gaussian components. The source hosts a strong outflow component. 
Center left: Center positions and their errors, from UV 
fitting a Gaussian model to individual velocity channels. Center and bottom right: Position-velocity diagrams, along the line core velocity gradient at 
PA$\sim$90\degr\ and orthogonal to it. 
}
\label{fig:i13120}
\end{figure*}

\begin{figure*}
\includegraphics[width=0.33\hsize]{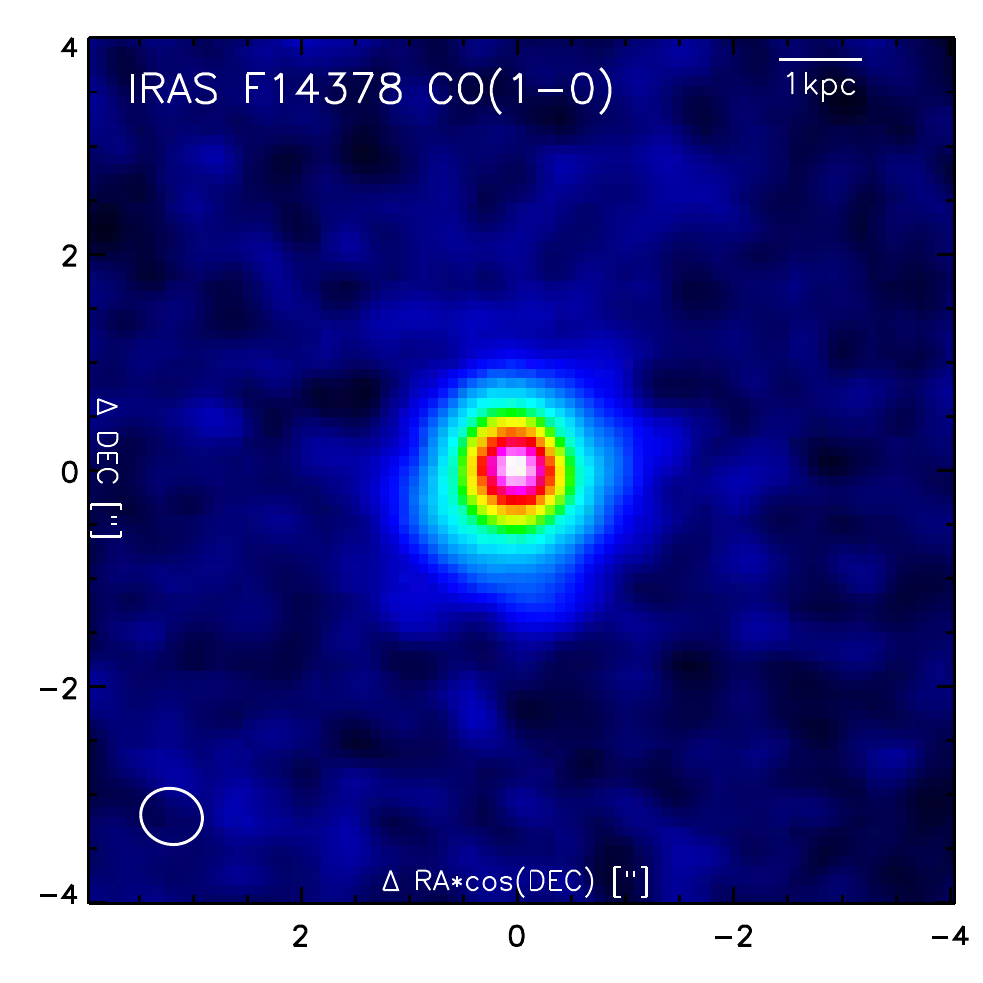}
\includegraphics[width=0.66\hsize]{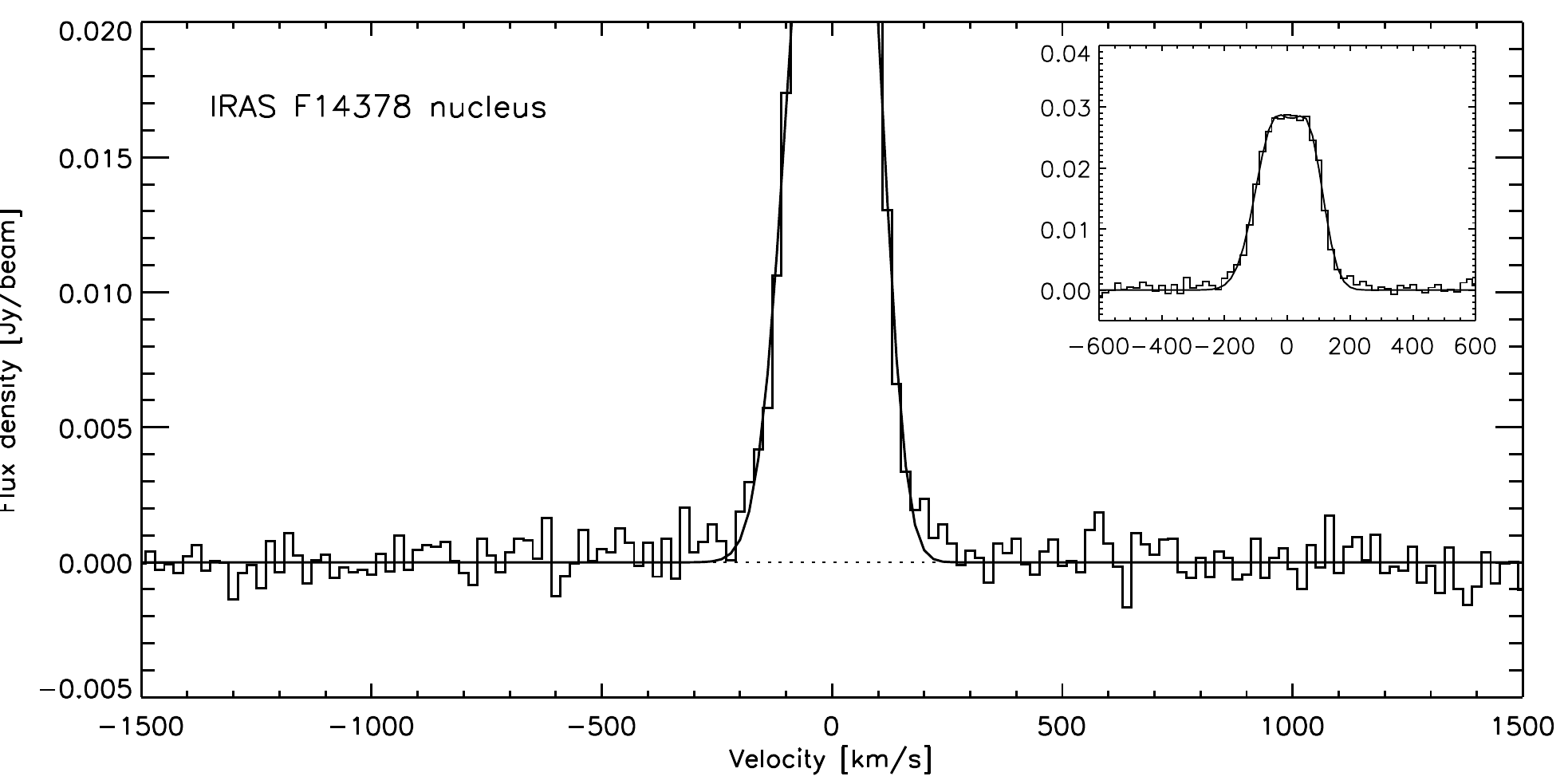}
\includegraphics[width=0.33\hsize]{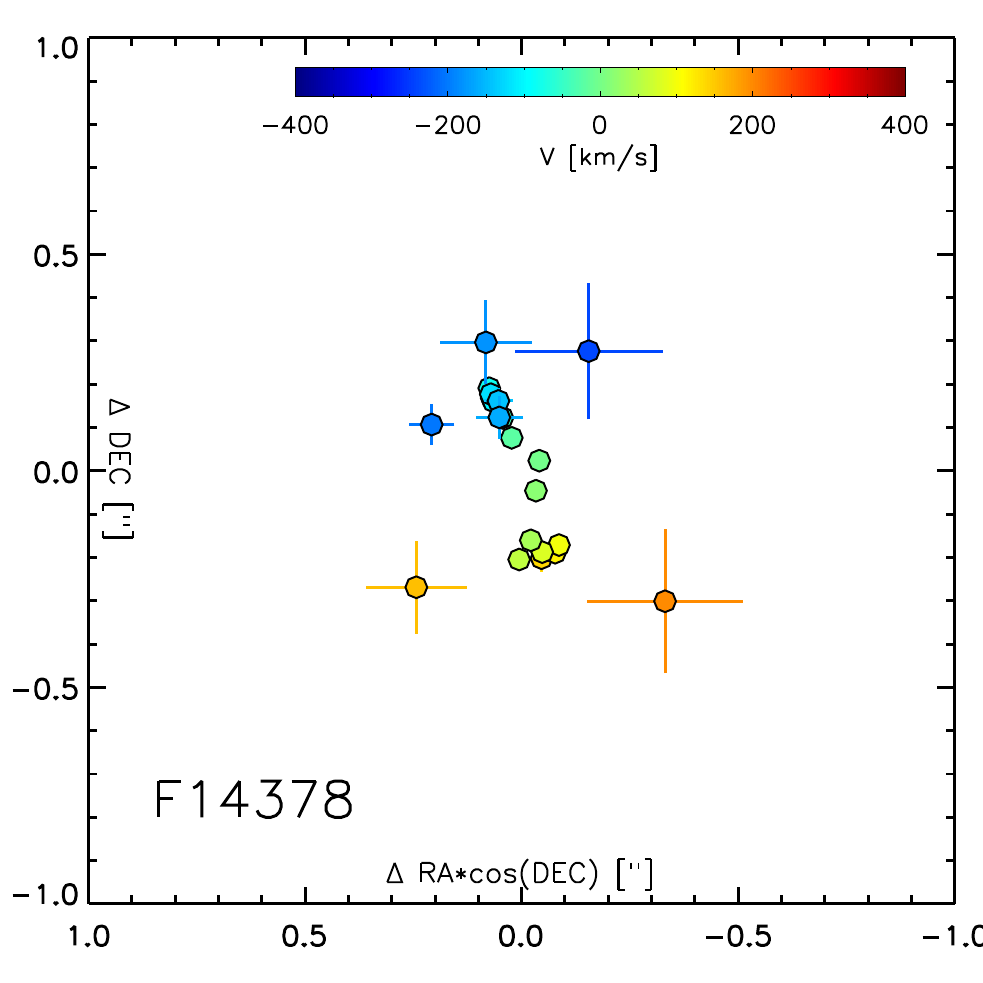}
\includegraphics[width=0.66\hsize]{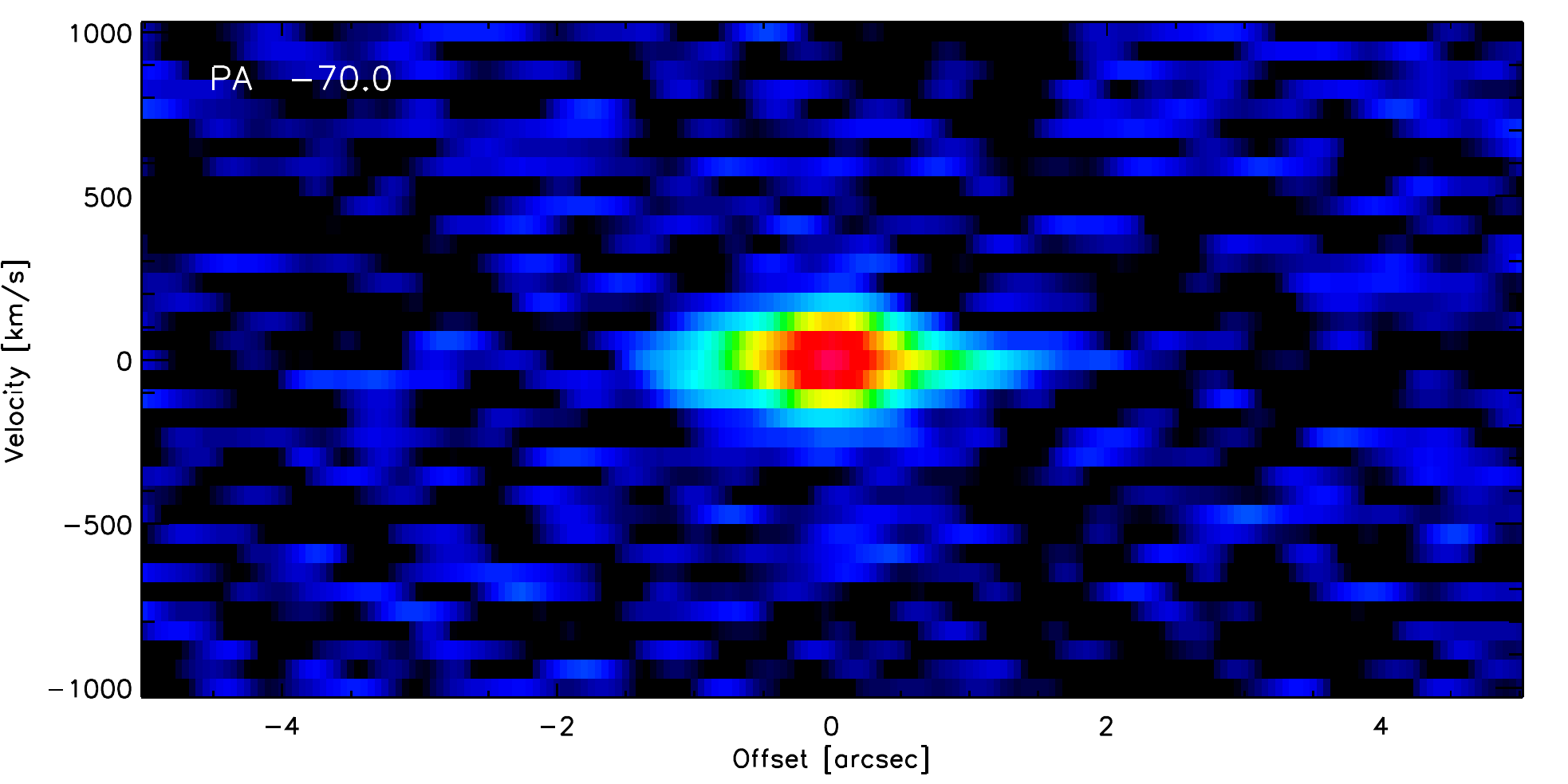}
\includegraphics[width=0.66\hsize,right]{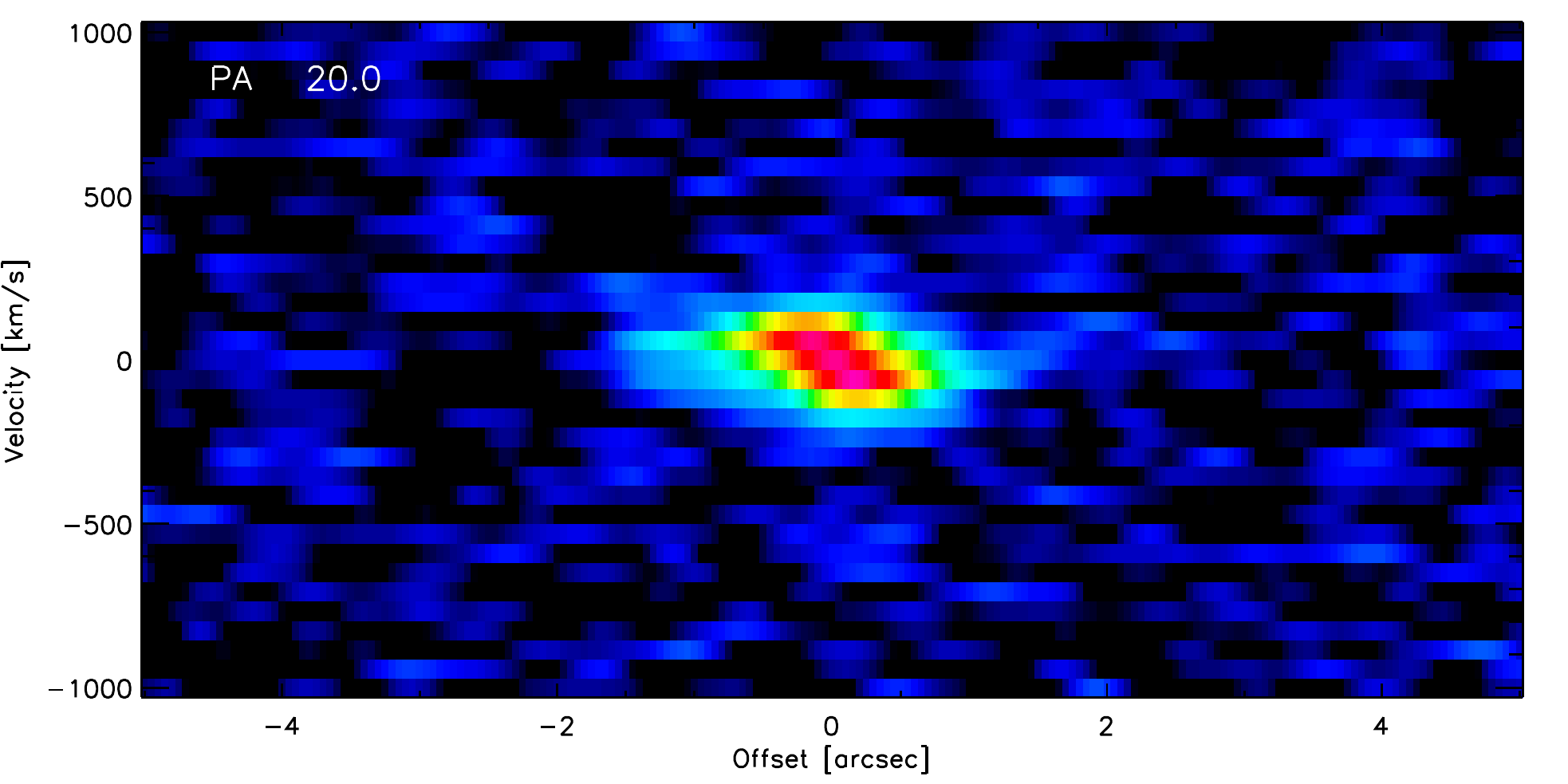}
\caption{ALMA CO(1-0) data for IRAS F14378-3651.
Top left: Moment 0 map.
Top right: Spectrum centered on the continuum nucleus. The 
host spectrum is fit with two Gaussians, their sum is 
overplotted in black. We conservatively assign an 
upper limit to molecular outflow, since the nature of the weak line wings
at $\pm$250~\kms\ is unclear.
Center left: Center positions and their errors, from UV 
fitting a Gaussian model to individual velocity channels. Center and
bottom right: Position-velocity diagrams, along the line core velocity 
gradient at PA$\sim$20\degr\ and orthogonal to it. 
}
\label{fig:f14378}
\end{figure*}

\begin{figure*}
\includegraphics[width=0.33\hsize]{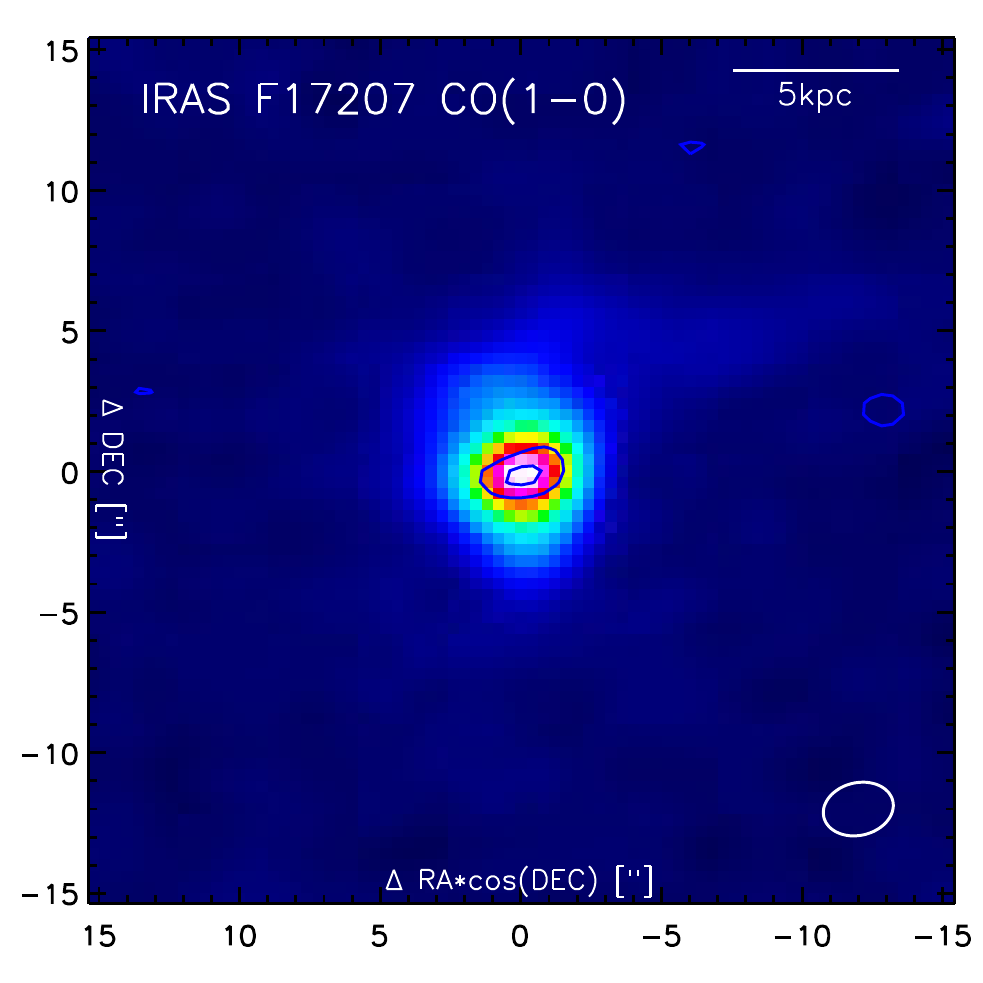}
\includegraphics[width=0.66\hsize]{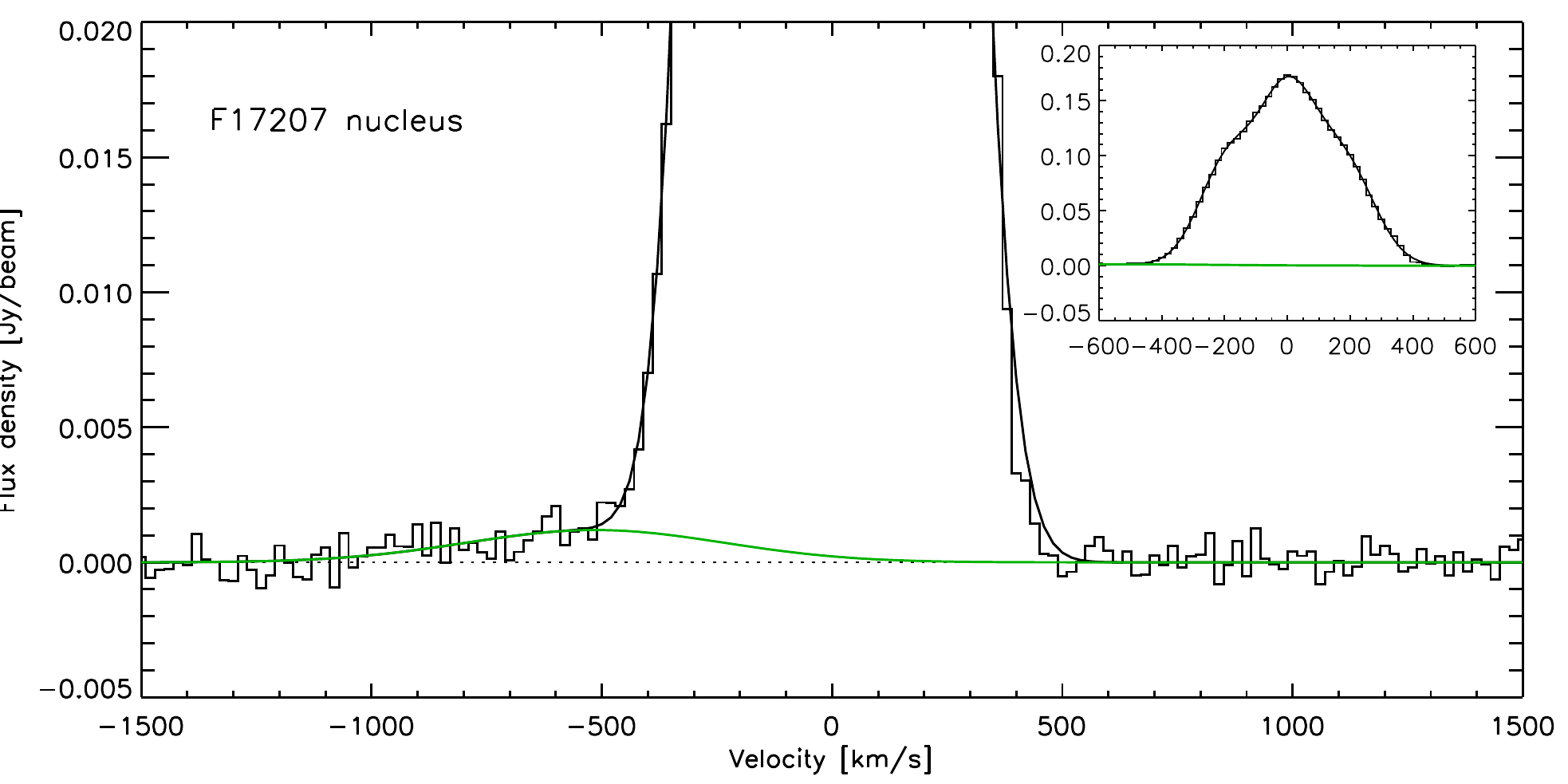}
\includegraphics[width=0.33\hsize]{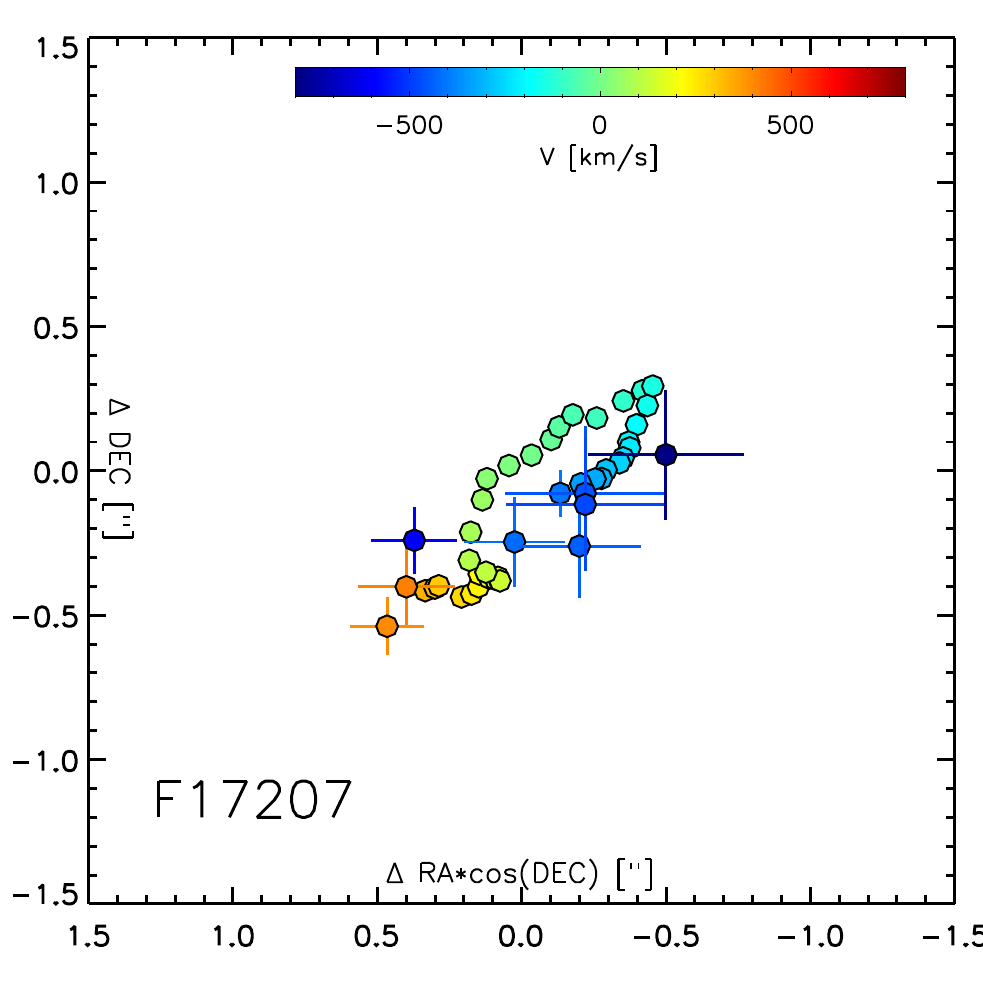}
\includegraphics[width=0.66\hsize]{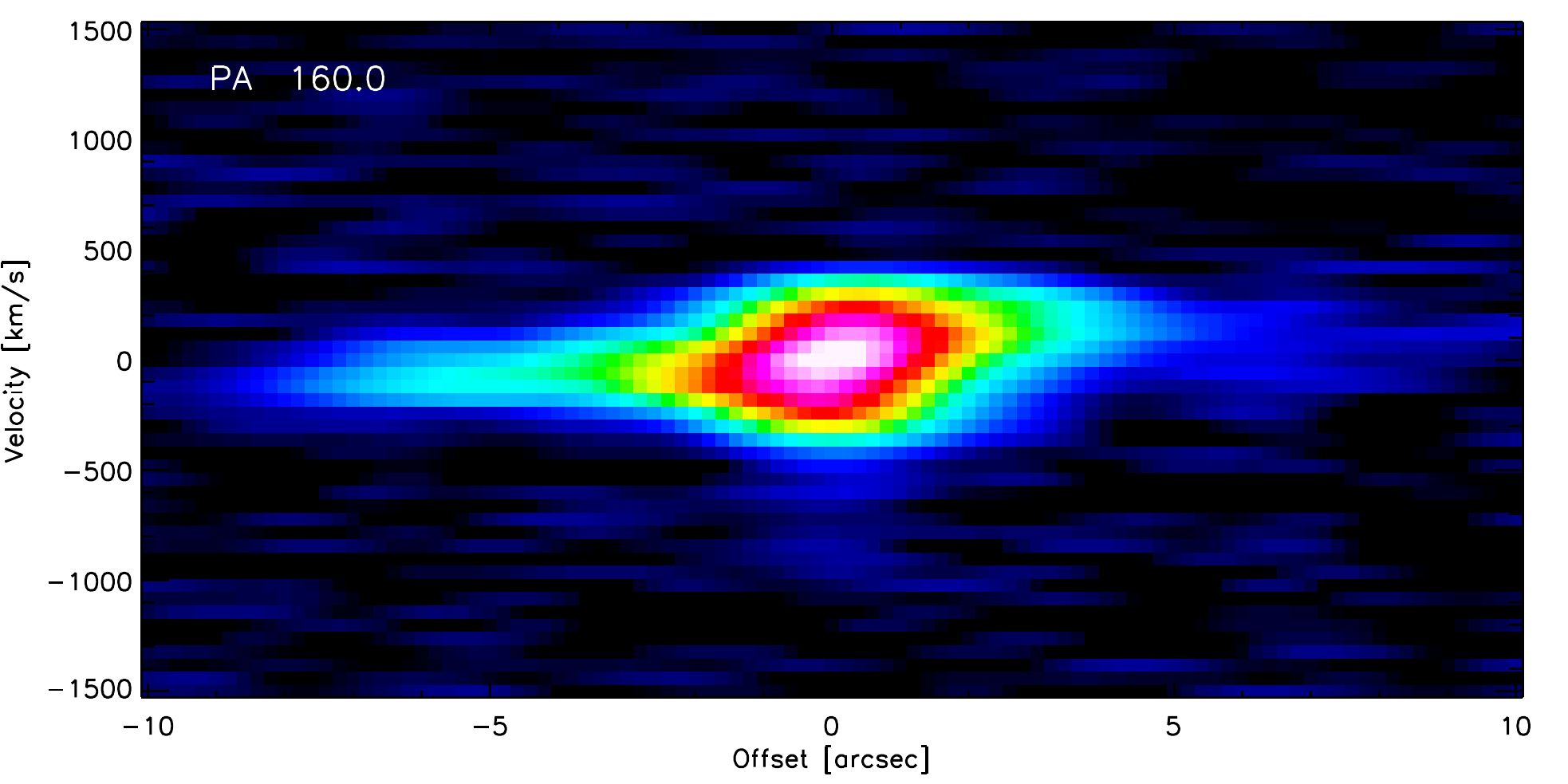}
\includegraphics[width=0.66\hsize,right]{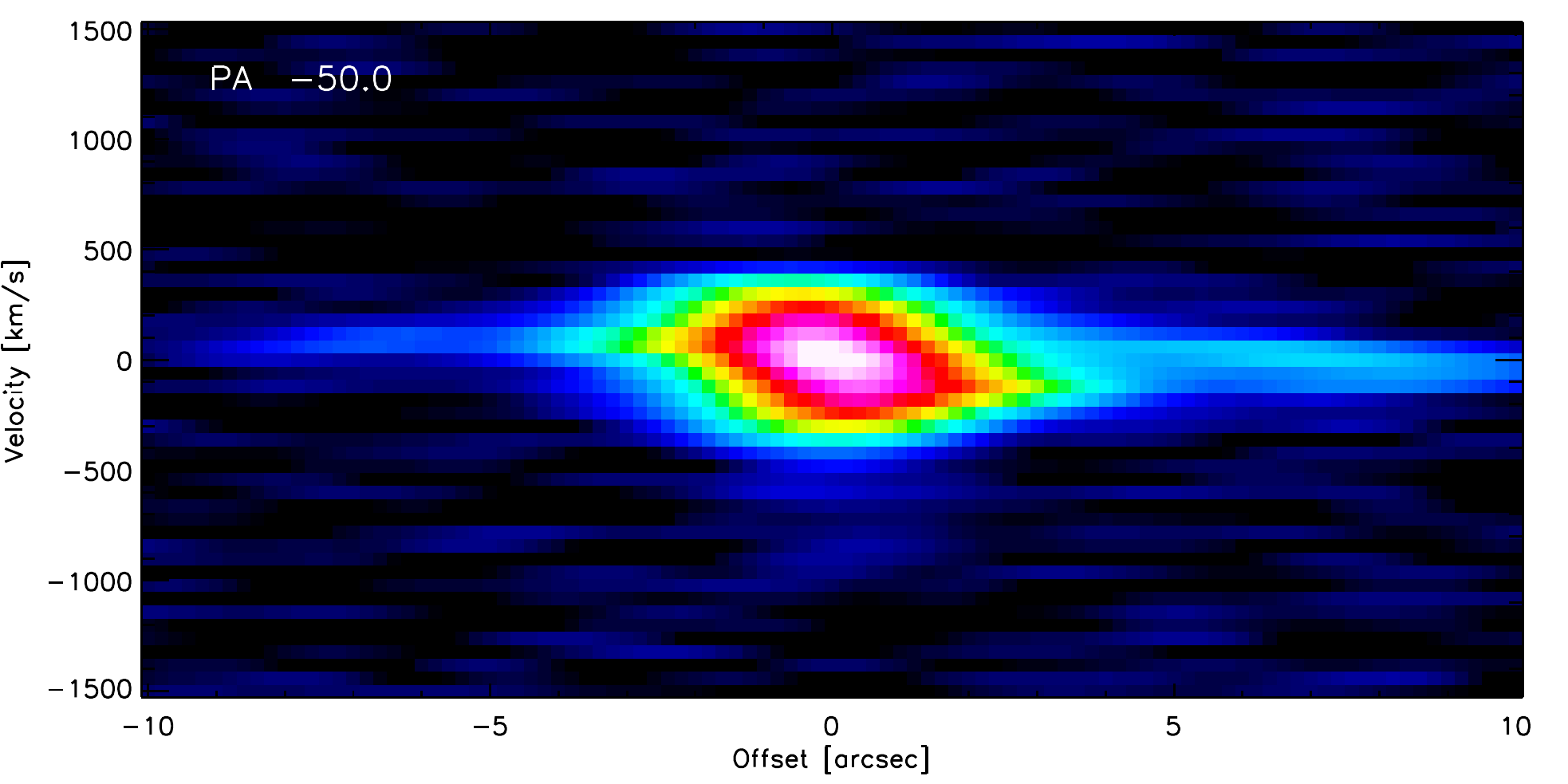}
\caption{ALMA CO(1-0) data for IRAS F17207-0014.
Top left: Moment 0 map. Overplotted are contours 
for outflow in the range [-900,-500]~\kms\ (blue, 
contours at [0.6, 0.9]~mJy/beam). Top right: Nuclear spectrum. The 
spectrum is decomposed into four Gaussians, three for the
host line profile and one for outflow (green). The black line includes 
all four Gaussian components. Center left: Center positions and their 
errors, from UV fitting a Gaussian model to individual velocity channels. 
Center and bottom right: Position-velocity diagrams. Fast outflowing 
molecular gas is seen on the blue side only.  PA +160\degr\ is the 
orientation of the blue outflow component measured in higher resolution
CO(2-1) data by \citet{garcia-burillo15}, our lower resolution data are 
consistent with this by placing the blue outflow slightly south of the 
nucleus (see center left panel). PA -50\degr\ is the approximate 
orientation of the central arc second scale gas disk.
}
\label{fig:f17207}
\end{figure*}

\begin{figure*}
\includegraphics[width=0.33\hsize]{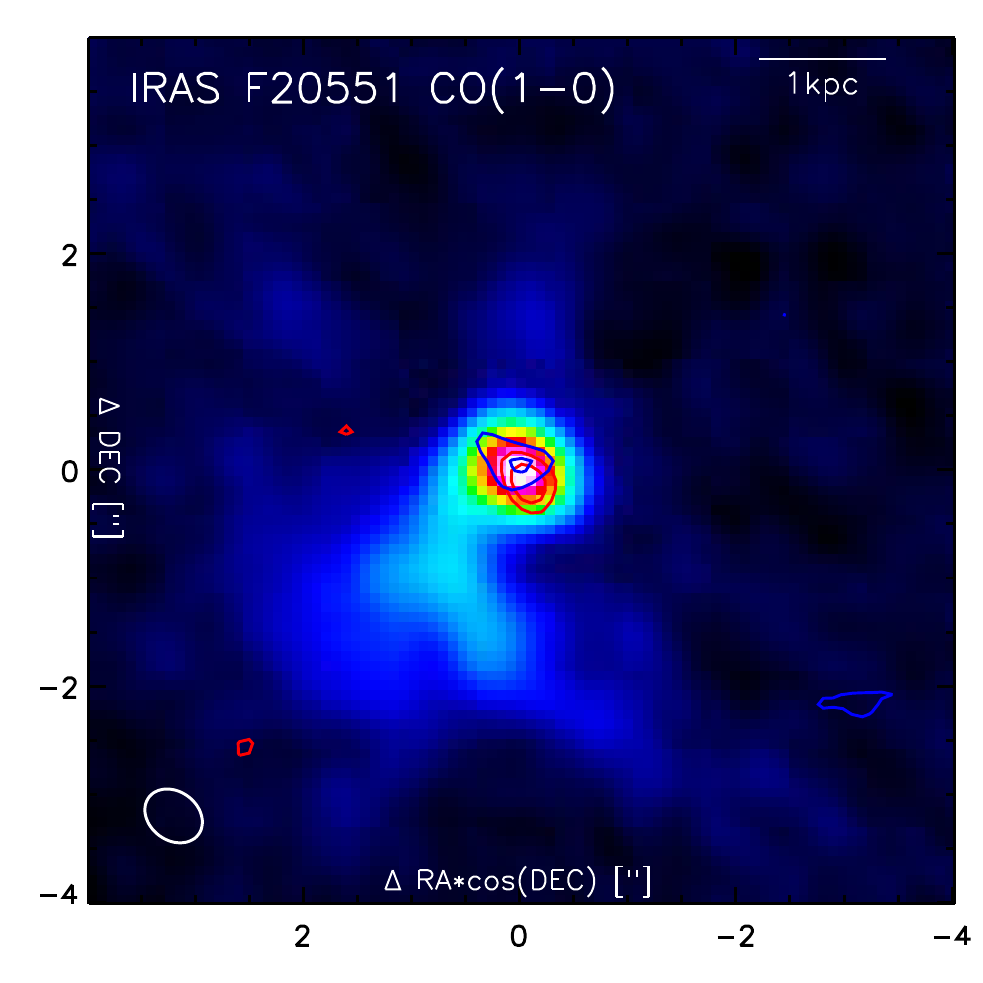}
\includegraphics[width=0.66\hsize]{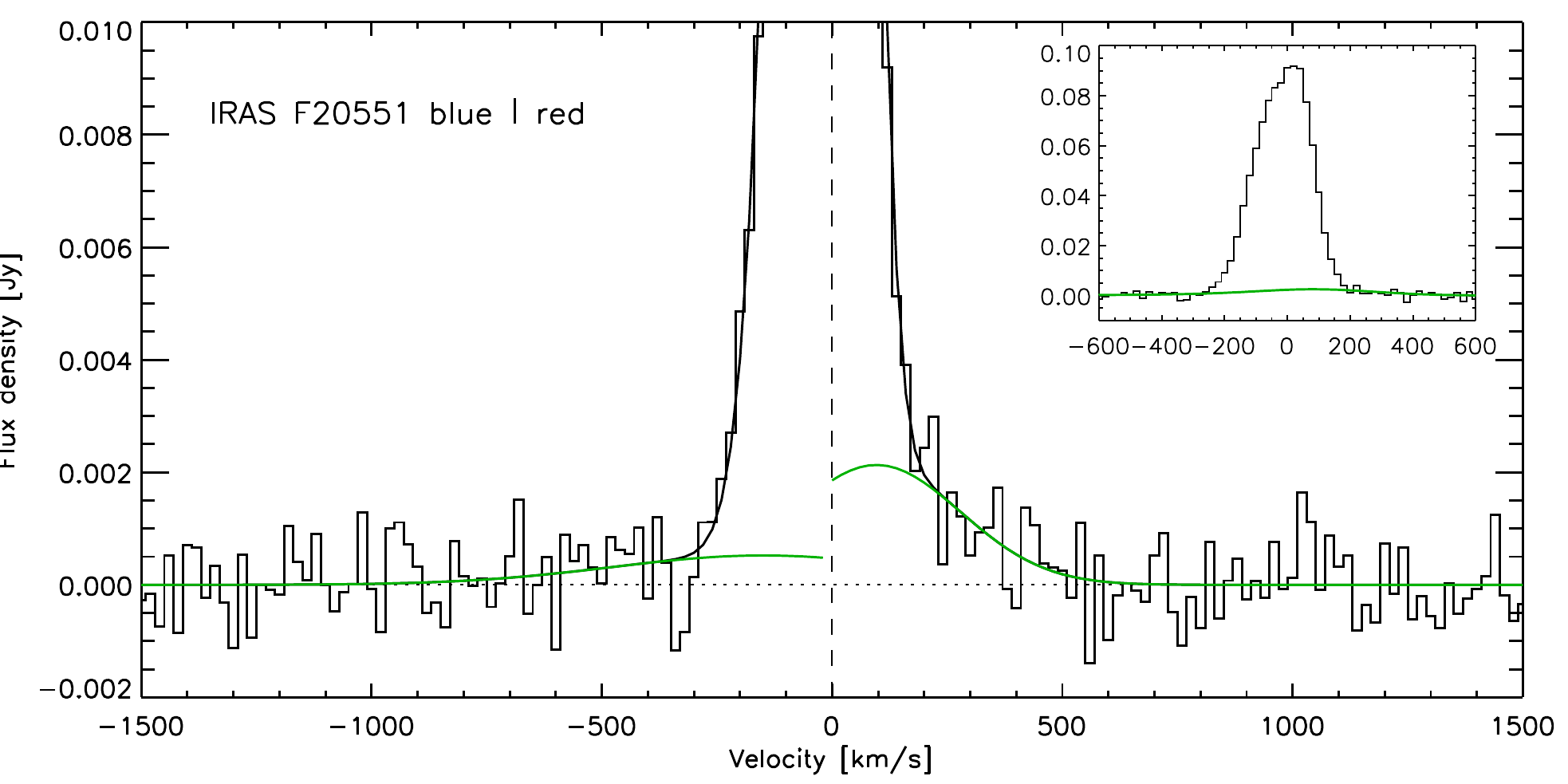}
\includegraphics[width=0.33\hsize]{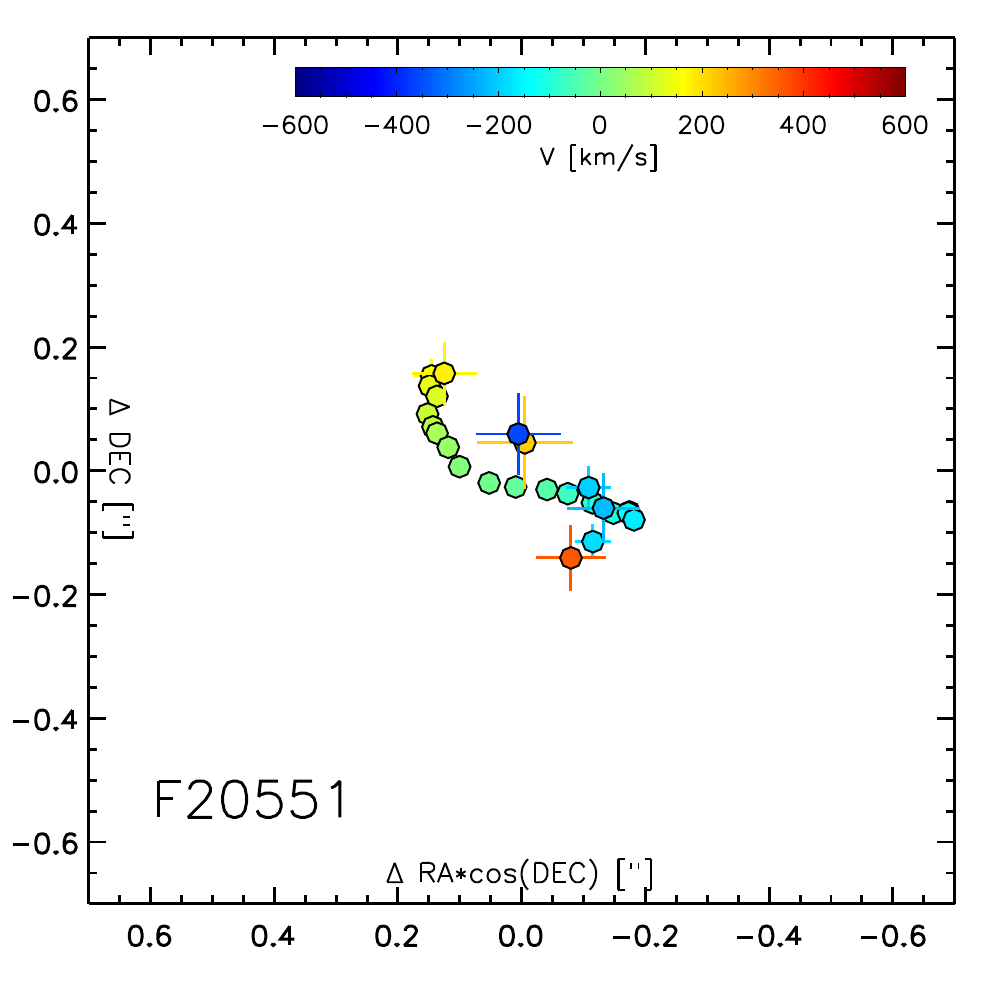}
\includegraphics[width=0.66\hsize]{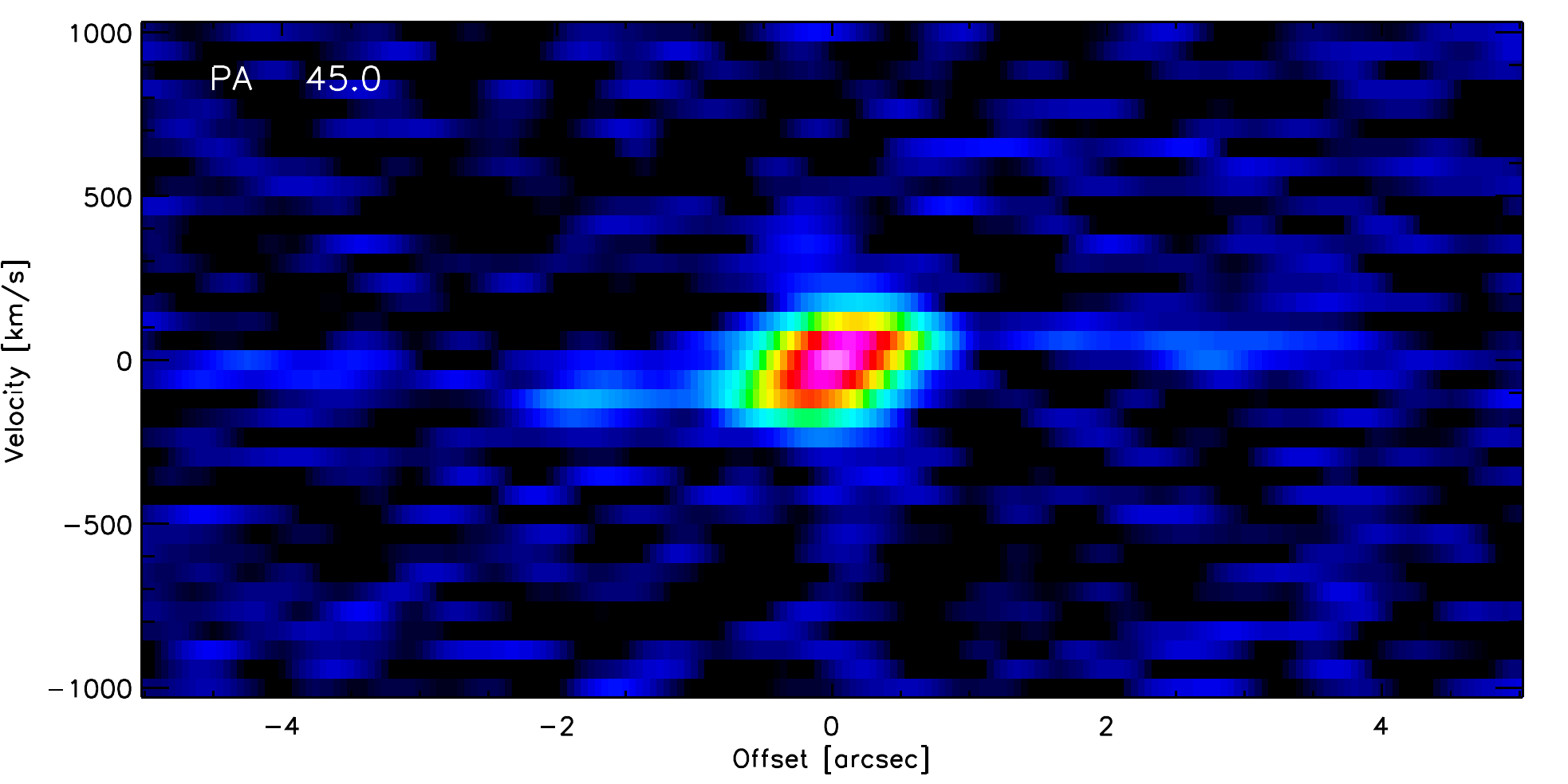}
\includegraphics[width=0.66\hsize,right]{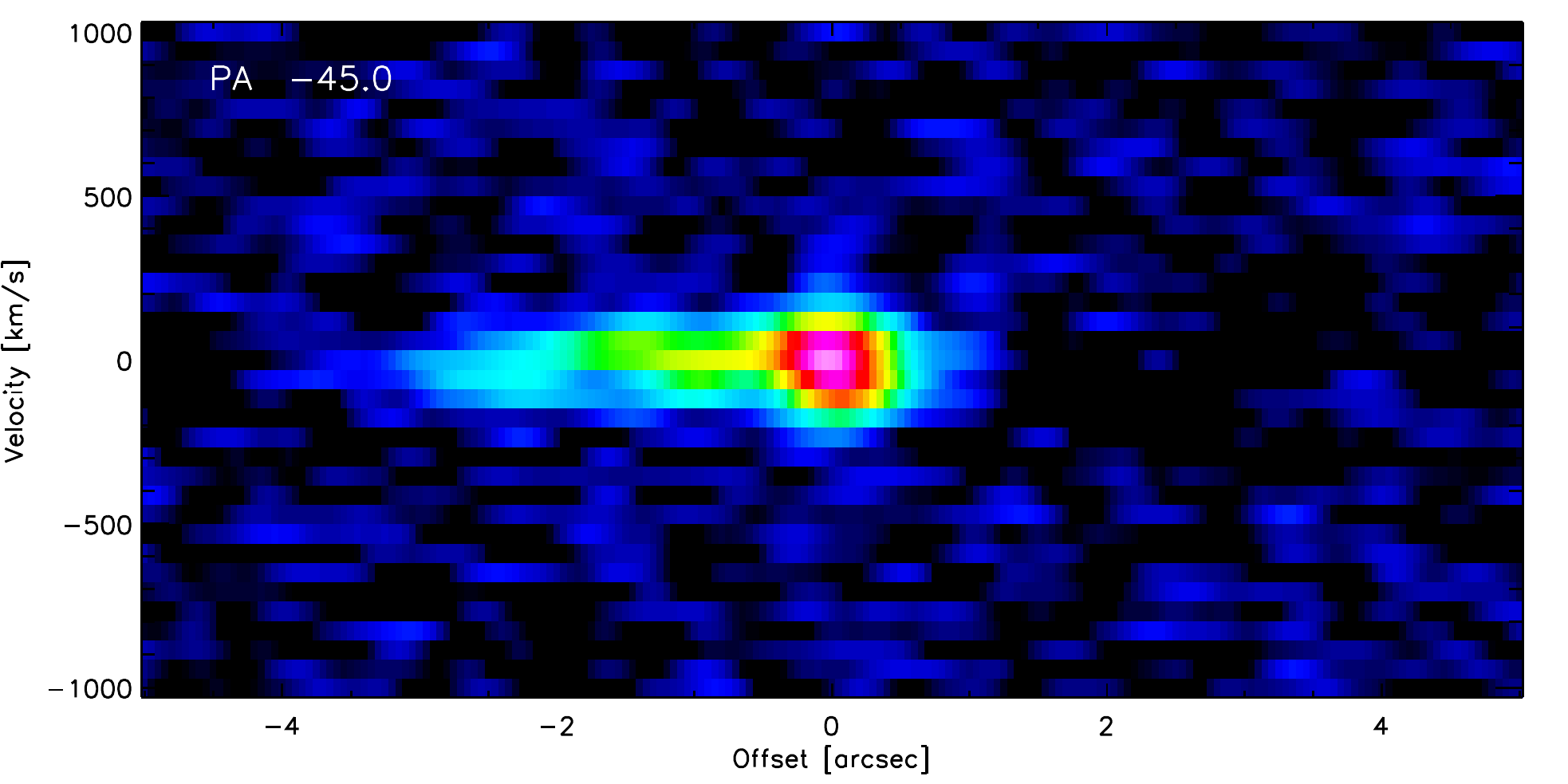}
\caption{ALMA CO(1-0) data for IRAS F20551-4250.
Top left: Moment 0 map. Overplotted are contours 
for outflow in the range [210,510]~\kms\ (red) and the tentative 
outflow in [-510,-210]~\kms\ (blue), 
with contours at [0.5, 0.7, 1.]~mJy/beam. 
Top right: Spectrum at the position of red outflow (right half) and
blue outflow (left half). Each 
spectrum, including the part not shown here, is decomposed into four 
Gaussians, three for the host line profile and one for outflow (green). 
The black line includes 
all four Gaussian components. The insert compares the sum of the outflow fits 
to an r=0.6\arcsec\ aperture spectrum. Center left: Center positions and their errors, from UV 
fitting a Gaussian model to individual velocity channels. 
Center and bottom right: Position-velocity diagrams for two orthogonal 
position angles. Molecular outflow is 
seen strongly on the red and weakly on the blue side of the line core, with 
spatial shifts better matching the PA~+45\degr\ PV diagram.
}
\label{fig:f20551}
\end{figure*}

\begin{figure*}
\includegraphics[width=0.33\hsize]{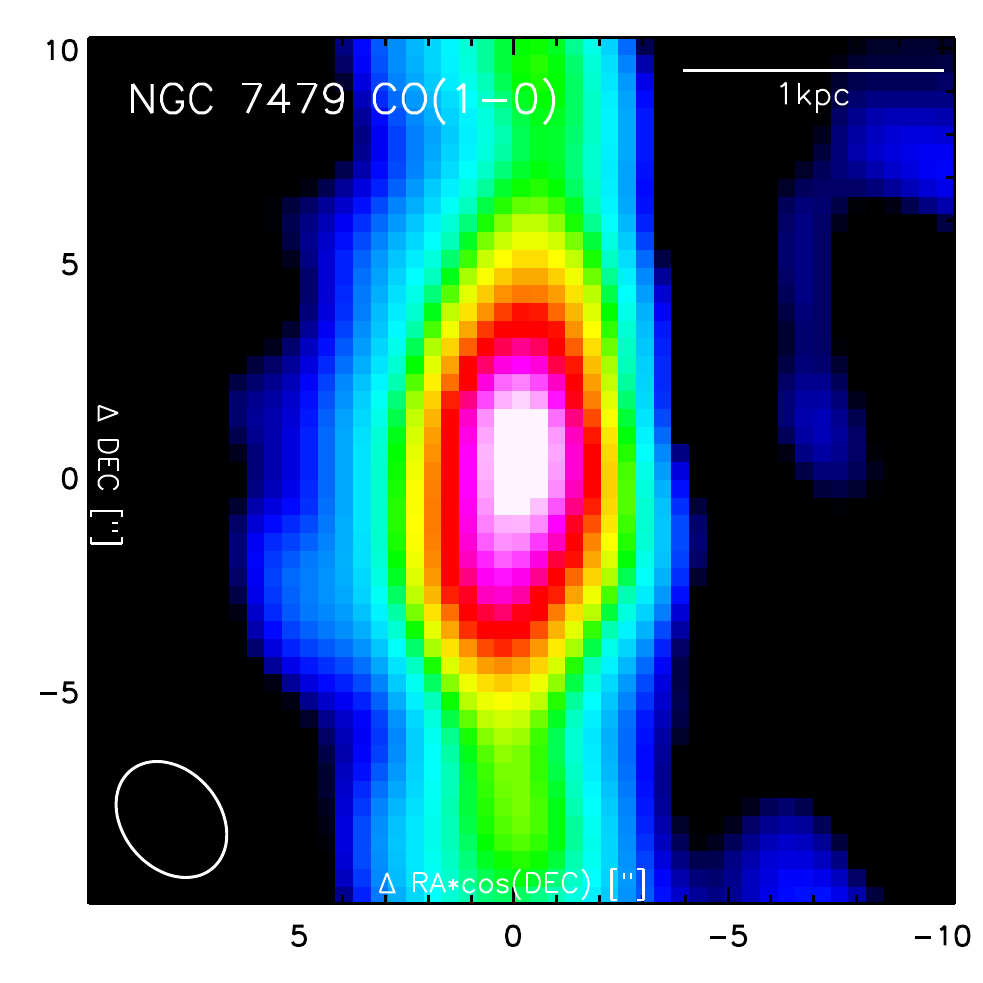}
\includegraphics[width=0.66\hsize]{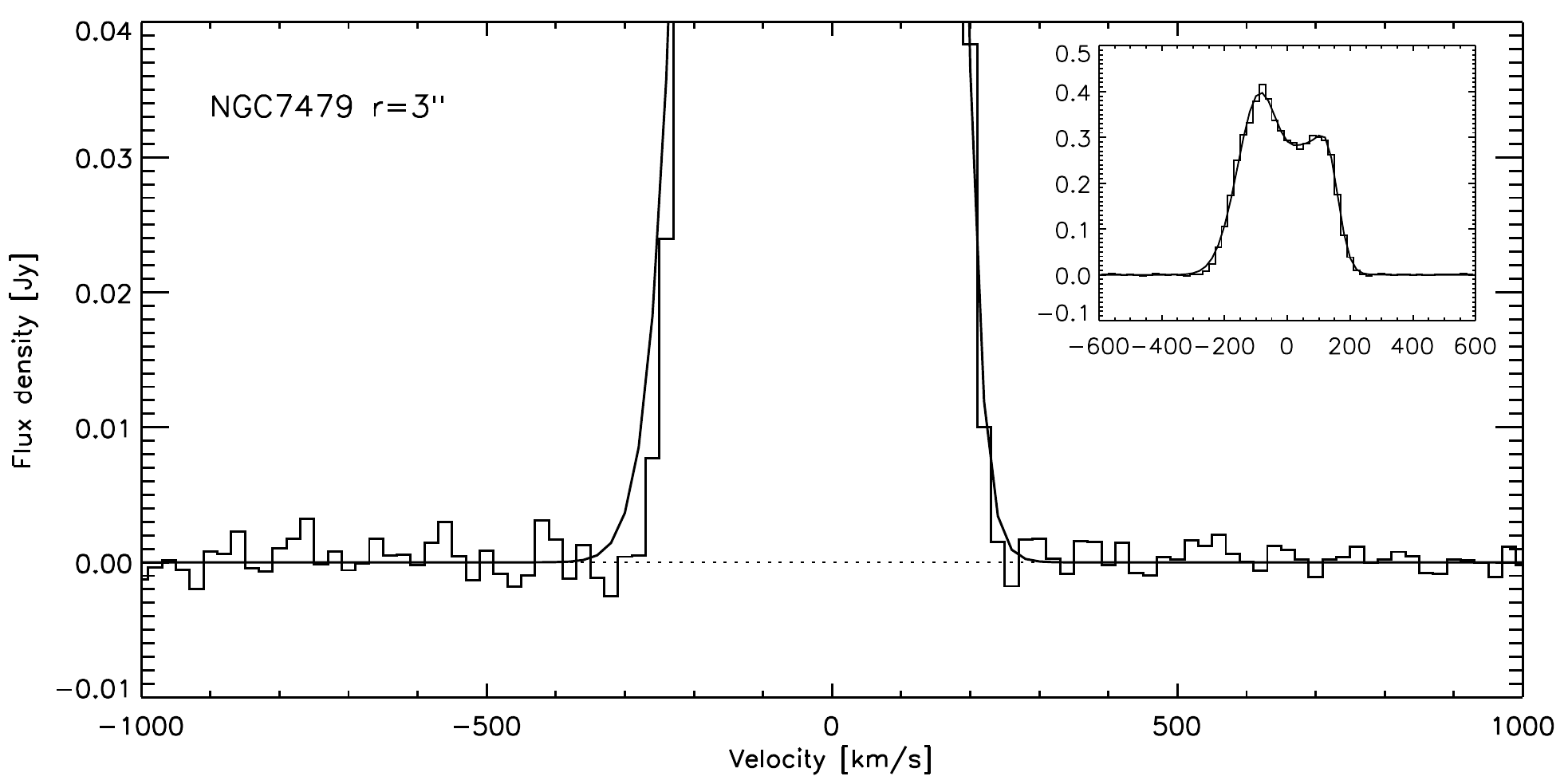}
\includegraphics[width=0.33\hsize]{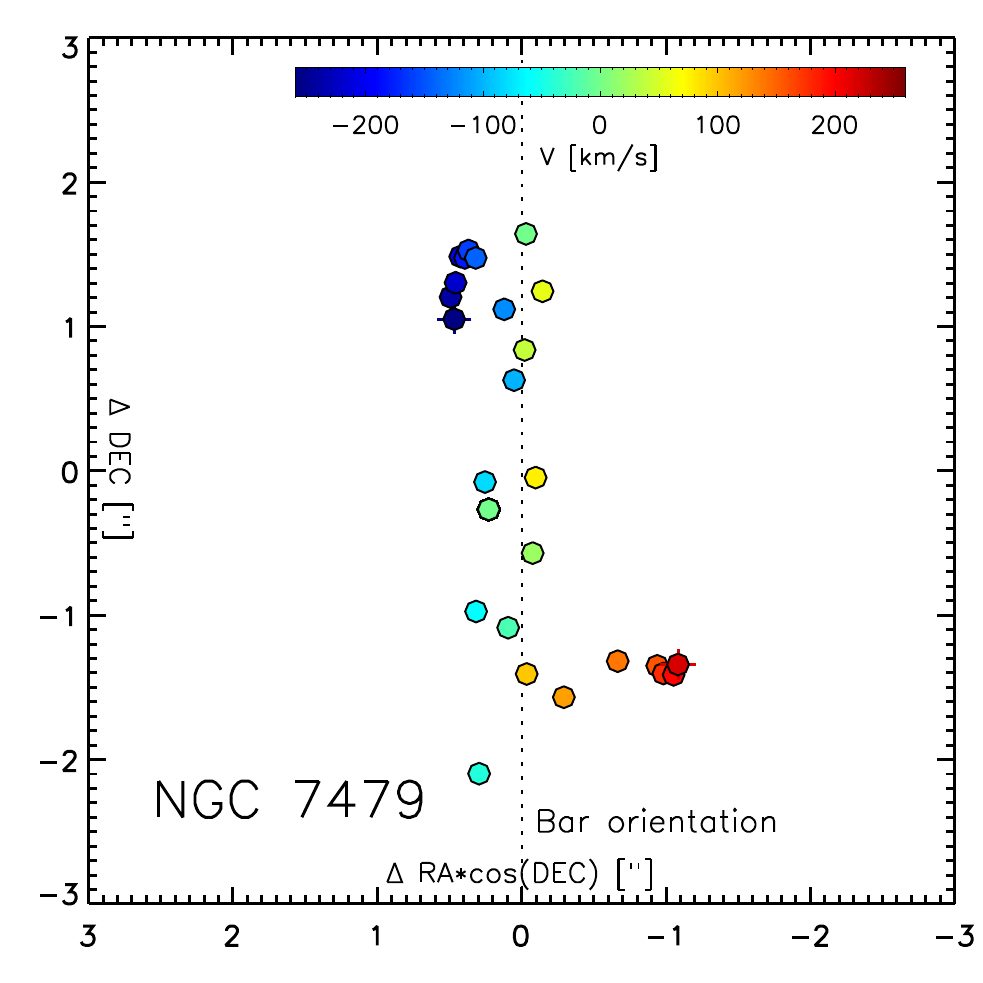}
\includegraphics[width=0.66\hsize]{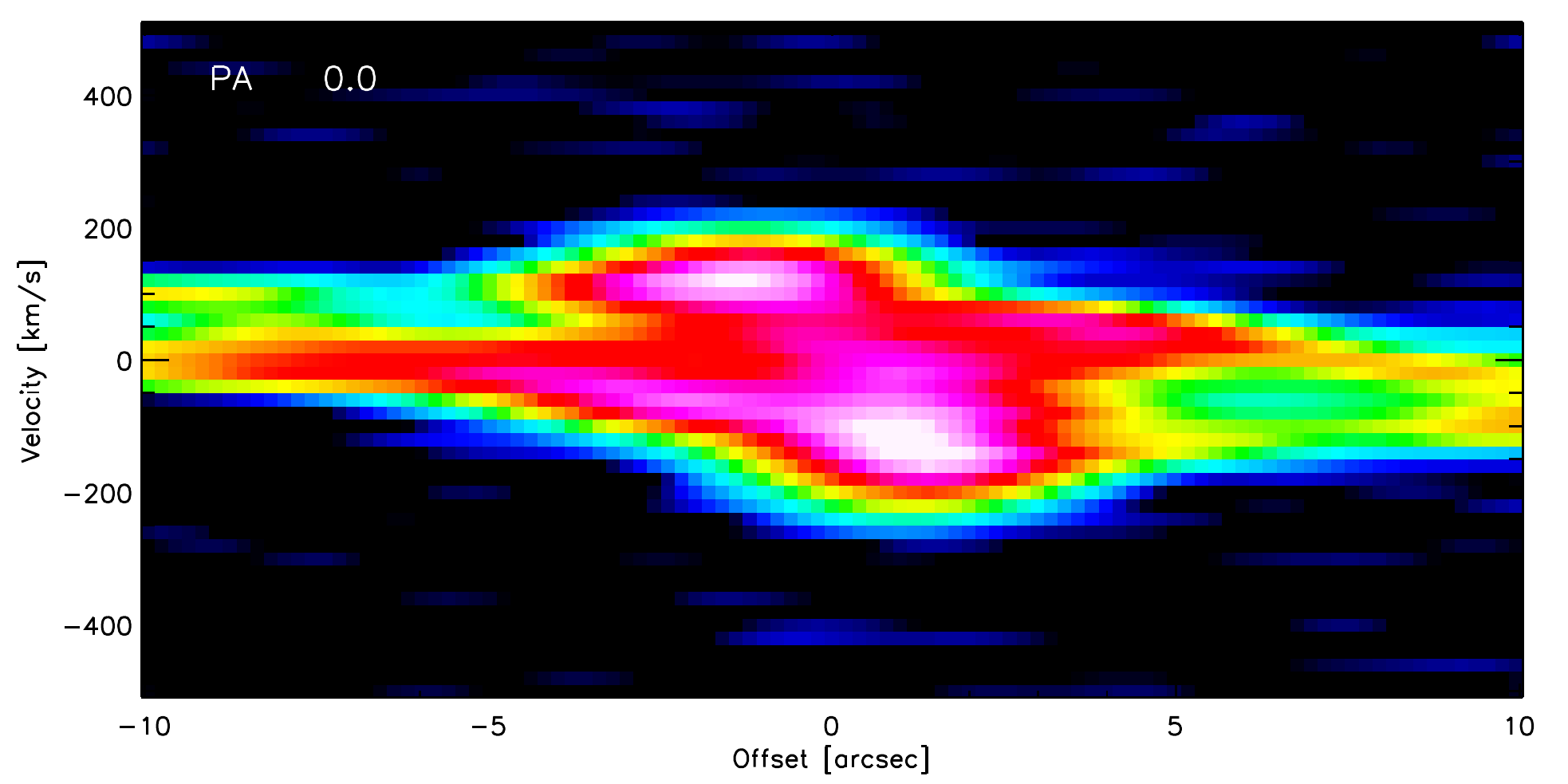}
\caption{NOEMA CO(1-0) data for NGC 7479. Top left: Moment 0 map. 
Top right: Spectrum in a r=3\arcsec\ aperture centered on the continuum 
nucleus. The spectrum is fitted with three Gaussians for the host line 
profile. There is no outflow detection. 
Bottom left: Center positions and their errors, from UV fitting a Gaussian model to individual velocity
channels. The dotted line indicates the N-S direction of NGC~7479's bar. 
Bottom right: Position-velocity diagram. The pronounced
bipolar structure in the PV-diagram is ascribed to flow in the galaxy's bar,
and we adopt an upper limit for molecular outflow. 
}
\label{fig:ngc7479}
\end{figure*}
\newpage

\section{Detected lines}
The observations used in this paper were targeted at CO(1-0) but some 
additional lines are detected in the wide NOEMA or ALMA bands. 
Table~\ref{tab:applines} lists these lines and their measured total fluxes.

\begin{table*}
\caption{Detected lines}
\begin{tabular}{lrllr}\hline
Source           &$\nu_{\rm Obs}$&Line     &$\nu_{\rm Rest}$&Flux\\ 
                 &GHz            &         &GHz             &\jykms\\ \hline
III Zw 035       &112.21         &CO(1-0)  &115.27          & 70.6\\
III Zw 035       &110.45         &CN N=1-0 &113.49 m        &  5.8\\
III Zw 035       &110.15         &CN N=1-0 &113.19 m        &  2.7\\
IRAS F05189-2524 &110.54         &CO(1-0)  &115.27          & 25.7\\
IRAS F05189-2524 &108.83         &CN N=1-0 &113.49 m        &  4.4\\
IRAS F05189-2524 &108.52         &CN N=1-0 &113.19 m        &  2.3\\
IRAS F05189-2524 & 95.96         &HC$_3$N 11-10&100.08      &  0.5\\
NGC 2623         &113.18         &CO(1-0)  &115.27          &161.0\\
NGC 2623         &111.44         &CN N=1-0 &113.49 m        &  8.5\\
NGC 2623         &111.13         &CN N=1-0 &113.19 m        &  4.9\\
IRAS F09111-1007W&109.33         &CO(1-0)  &115.27          & 51.0\\
IRAS F09111-1007W&107.64         &CN N=1-0 &113.49 m        &  4.0\\
IRAS F09111-1007W&107.33         &CN N=1-0 &113.19 m        &  2.6\\
IRAS F09111-1107W& 94.92         &HC$_3$N 11-10&100.08      &  0.3\\
IRAS F09111-1007E&109.28         &CO(1-0)  &115.27          & 19.7\\
IRAS F10173+0828 &109.89         &CO(1-0)  &115.27          & 24.2\\
IRAS F10173+0828 &108.19         &CN N=1-0 &113.49 m        &  1.4\\
IRAS F10173+0828 &107.89         &CN N=1-0 &113.19 m        &  0.7\\
IRAS F10173+0828 & 95.41         &HC$_3$N 11-10&100.08      &  0.7\\
IRAS F12224-0624 &112.30         &CO(1-0)  &115.27          & 22.6\\
IRAS F12224-0624 &110.56         &CN N=1-0 &113.49 m        &  2.0\\
IRAS F12224-0624 &110.25         &CN N=1-0 &113.19 m        &  1.1\\
IRAS F12224-0624 & 98.02         &?        &100.62 m        &  0.4\\
IRAS F12224-0624 & 97.50         &HC$_3$N 11-10&100.08      &  0.8\\
NGC 4418         &114.47         &CO(1-0)  &115.27          &117.0\\
NGC 4418         &112.70         &CN N=1-0 &113.49          & 10.5\\
NGC 4418         &112.38         &CN N=1-0 &113.19          &  4.6\\
NGC 4418         &101.81         &CH$_3$CCH&102.54          &  0.7\\
IC 860           &113.80         &CO(1-0)  &115.27          & 73.1\\
IC 860           &112.05         &CN N=1-0 &113.49 m        &  4.6\\
IRAS 13120-5453  &111.80         &CO(1-0)  &115.27          &234.0\\
IRAS 13120-5453  &110.07         &CN N=1-0 &113.49 m        & 24.6\\
IRAS 13120-5453  &109.75         &CN N=1-0 &113.19 m        & 14.6\\
IRAS 13120-5453  & 97.58         &?        &100.61 m        &  0.8\\
IRAS 13120-5453  & 97.08         &HC$_3$N 11-10&100.08      &  1.0\\
IRAS F14378-3651 &107.92         &CO(1-0)  &115.27          & 24.7\\
IRAS F14378-3651 &106.26         &CN N=1-0 &113.49 m        &  2.4\\
IRAS F14378-3651 &105.95         &CN N=1-0 &113.19 m        &  1.3\\
IRAS F17207-0014 &110.54         &CO(1-0)  &115.27          &171.0\\
IRAS F17207-0014 &108.83         &CN N=1-0 &113.49 m        & 10.0\\
IRAS F17207-0014 &108.53         &CN N=1-0 &113.19 m        &  4.5\\
IRAS F17207-0014 & 98.33         &CH$_3$CCH&102.54          &  0.2\\
IRAS F17207-0014 & 96.50         &?        &100.63 m        &  0.5\\
IRAS F17207-0014 & 95.97         &HC$_3$N 11-10&100.08      &  2.9\\
IRAS F20551-4250 &110.52         &CO(1-0)  &115.27          & 54.9\\
IRAS F20551-4250 &108.82         &CN N=1-0 &113.49          &  2.3\\
IRAS F20551-4250 &108.50         &CN N=1-0 &113.19          &  1.0\\
NGC 7479 (30\arcsec\/)&114.39    &CO(1-0)  &115.27          &270.0\\
NGC 7479 (30\arcsec\/)&112.60    &CN N=1-0 &113.49 m        &  6.4\\ \hline
\end{tabular}
\tablefoot{Total fluxes, or fluxes in the indicated aperture diameter, for 
lines detected in the spectral ranges covered
by NOEMA and ALMA. For the very rich spectrum of NGC~4418 we list only a 
few bright lines that are also detected in other targets. 
The listed frequencies are indicative for identification purposes 
only, in particular for lines with multiple components marked 'm'.} 
\label{tab:applines}
\end{table*}

\end{appendix}

\end{document}